\documentclass{iopart}
\usepackage{bm}

\newcommand{\fopa}{\hat{\Psi}}              
\newcommand{\fopc}{\hat{\Psi}^{\dagger}}    
\newcommand{\fopart}{\fopa ({\bf r}, t)}    
\newcommand{\fopcrt}{\fopc ({\bf r}, t)}    
\newcommand{\spa}{\hat{a}}                  
\newcommand{\spc}{\hat{a}^{\dagger}}        
\newcommand{\spwa}{\varphi}                 
\newcommand{\spop}{\hat{h}_0}               
\newcommand{\copa}{\hat{\phi}}              
\newcommand{\nopa}{\hat{\delta}}            
\newcommand{\nopc}{\hat{\delta}^{\dagger}}  
\newcommand{\wfn}{\phi}                     
\newcommand{\bldr}{{\bf r}}			    
\def\un{\leavevmode\hbox{\normalsize1\kern-4.6pt\large1}}
\newcommand{\be}{\begin{equation}}
\newcommand{\ee}{\end{equation}}
\newcommand{\bea}{\begin{eqnarray}}
\newcommand{\eea}{\end{eqnarray}}

\usepackage{rotating} 
\usepackage{amssymb} 

\begin{document}
\title[Finite Temperature Models of BEC]{Finite Temperature Models of Bose-Einstein condensation}

\author{Nick P.\ Proukakis and Brian Jackson\footnote{Brian Jackson died unexpectedly on 30 August 2007, in the early stages of the preparation of this manuscript.}}
\address{School of Mathematics and Statistics, Newcastle University, Newcastle-upon-Tyne, NE1 7RU, United Kingdom}
\ead{Nikolaos.Proukakis@ncl.ac.uk}

\begin{abstract}
 
The theoretical description of trapped weakly-interacting Bose-Einstein condensates is characterized by a large number of seemingly very different approaches which have been developed over the course of time by researchers with very distinct backgrounds. Newcomers to this field, experimentalists and young researchers all face a considerable challenge in navigating through the `maze' of abundant theoretical models, and simple correspondences between existing approaches are not always very transparent.
This Tutorial provides a generic introduction to such theories, 
in an attempt to single out common features and deficiencies of certain `classes of approaches' identified by their physical content, rather than their particular mathematical implementation.

This Tutorial is structured in a manner accessible to a non-specialist with a good working knowledge of quantum mechanics. Although some familiarity with concepts of quantum field theory would be an advantage, key notions such as the occupation number representation of second quantization are nonetheless briefly reviewed. Following a general introduction, the complexity of models is gradually built up, 
starting from the basic zero-temperature formalism of the Gross-Pitaevskii equation.
%
This structure enables readers to probe different levels of theoretical developments (mean-field, number-conserving and stochastic) according to their particular needs. In addition to its `training element', we hope that this Tutorial will prove useful to active researchers in this field, both in terms of the correspondences made between different theoretical models, and as a source of reference for existing and developing finite-temperature theoretical models.

\end{abstract}
\submitto{\JPB}

\newpage
\tableofcontents
\newpage

\section{Introduction}

\subsection{Bose-Einstein Condensation}

One of the most exciting developments in modern physics has been 
the increasing sophistication of experimental techniques to cool, confine and 
manipulate atoms with optical and magnetic fields. In recent years this has
culminated in the achievement of quantum degeneracy in dilute ultracold gases 
of bosons \cite{BEC_Exp_1,BEC_Exp_2,BEC_Exp_3,BEC_Exp_4,BEC_Exp_5} and fermions \cite{Fermion_Exp_1,Fermion_Exp_2}, allowing the creation of novel 
many-body systems with unprecedented control, tunability, and versatility \cite{General_Review_1,General_Review_2}. 
 
Indistinguishable particles with integer (bosons) and half-integer (fermions)
spin differ in how they occupy quantum states. While no more than 
one fermion can occupy each state (known as the Pauli exclusion principle), 
the number of bosons in a state is unrestricted. The difference becomes most 
apparent when one 
cools down a gas to low temperatures $T$, where the de Broglie wavelength 
$\lambda_{dB} \propto 1/\sqrt{T}$, becomes the same order or larger than the 
distance between particles. The system then enters a {\it quantum degenerate} 
regime. The most spectacular manifestation of this occurs for bosons, which 
below a critical temperature $T_c$ undergo a phase transition, or {\it 
Bose-Einstein Condensation} (BEC), where particles tend to macroscopically occupy a 
single quantum state--- the condensate.

For the purposes of the discussion, it is interesting to consider the number 
of particles in the condensate, $N_0$, as function of temperature $T$.
We consider an ideal gas of total atom number $N$ confined in a harmonic 
potential well of the 
form 
$V_{\rm ext} ({\bm r}) = m(\omega_x^2 x^2+\omega_y^2 y^2+\omega_z^2 z^2)/2$, 
which closely approximates the traps typically used in the experiments. A 
straightforward calculation \cite{Pethick_Book,Stringari_Book} gives 
$
N_0/N = 1-(T/T_c)^3,
$
where $T_c \simeq 0.94 \hbar \omega_{\rm ho} N^{1/3}$ with 
$\omega_{\rm ho}=(\omega_x \omega_y \omega_z)^{1/3}$.
One sees that at $T=T_c$, $N_0=0$, but as $T$ decreases the condensate
fraction $N_0/N$ increases, until at $T=0$, $N=N_0$ and all of the atoms
are in the condensate.
The inclusion of interactions and the finite size 
of the system changes matters quantitatively, but this simple picture remains qualitatively useful \cite{Stringari_Review,Leggett_Review}. 
  
Thus, at all temperatures below $T_c$ (apart from the physically unattainable
$T=0$) a Bose-Einstein condensate co-exists with non-condensed particles, 
which collectively make up a {\it thermal cloud}. For temperatures very close
to zero the thermal cloud can be, to some extent, neglected, leading to a 
relatively simple description in terms of a nonlinear Schr\"{o}dinger 
equation, also known as the Gross-Pitaevksii Equation \cite{Gross,Pitaevskii}.
Despite its evident simplicity, 
this equation nonetheless contains much interesting
physics and provides a good description of many experiments, with much of the 
early theoretical work in Bose-Einstein condensation focusing on 
solving the Gross-Pitaevskii Equation, an effort that continues to the present day \cite{Frantzeskakis_Book}.

Nevertheless, it is important to remember that experiments actually take place 
at finite temperatures. A thermal cloud is always present, and as one
increases the temperature towards $T_c$ the influence of this on the
system behaviour will become progressively more 
important. In some situations the thermal cloud is absolutely central, for
instance in the problem of condensate growth, or the heating of the gas under
strong external perturbations. 
Future applications of Bose-Einstein 
condensation, such as precision measurements based on matter wave interferometry, would also benefit from a good understanding of the behaviour of the system at finite 
temperatures.
The effects of finite temperature also become
particularly important for low dimensional systems, where 
the condensate exhibits fluctuations in its phase, and the usual picture of 
Bose-Einstein condensation needs to be revisited.

Theoretical models of BEC at finite temperature are therefore of tremendous value.
In order to be useful, such models should ideally 
accurately represent the important physics in a problem, while remaining
feasible to solve either analytically, or numerically, using current computing 
resources. The
development of dynamical finite temperature models that satisfy these conditions
has proven to be a great 
challenge.
Despite significant progress, research remains active, and much work remains
to be done.

A major problem that confronts newcomers to the field of Bose-Einstein condensation at finite temperatures, 
is the large number of seemingly very different approaches which
have been applied to the problem \cite{MyReview}. This is exacerbated by the fact that 
the major researchers have arrived from different backgrounds, and therefore
speak ``different languages''. 
The aim of this Tutorial
is to introduce the basic theoretical tools and approximations employed, with the emphasis on the physics on which these approximations are based.
This is done systematically, starting from a brief discussion of the zero temperature theory. In our presentation, we use the simplest possible notation which requires minimal prior knowledge, and we also give numerous references to equivalent approaches based on mathematically distinct formulations. Although space constraints do not allow us to
derive all different approaches
in detail,
we nonetheless provide an overview
which should allow the reader to arrive at a more general understanding of the 
field as a whole. 



\subsection{Basic Formalism}

We begin by reviewing the formalism which forms the basis of our 
discussion in this Tutorial (a more detailed introduction
can be found in \cite{Pethick_Book,Stringari_Book,Fetter_Walecka}). 
Throughout this Tutorial, we shall assume that we are dealing with relatively dilute weakly-interacting Bose gases, in the sense that the relevant interactions are binary collisions between two atoms. 
Although initial experiments were limited to this regime, more recent experiments with attractive condensates \cite{BEC_Attractive_Exp_1,BEC_Attractive_Exp_2}, Feshbach resonances \cite{BEC_Feshbach} and molecular BECs \cite{BEC_Molecular_1,BEC_Molecular_2,BEC_Molecular_3} require the inclusion of three-body collisions, which generally lead to modified scattering and loss of atoms from the system. This Tutorial does not include such effects, and for further information the reader is referred to Ref. \cite{Thorsten_Review} and references therein. 

Within this approximation, the hamiltonian of a closed system of $N$ atoms can be written as a sum of two contributions, one ($\hat{h}_0$) arising from single-particle effects, and the other ($\hat{V}$) arising from binary collisions. In the `coordinate representation', this takes the form
\be
\hat{H} = \sum_{k=1}^{N} \hat{h}_0(\bldr_k)+ \frac{1}{2} \sum_{k,l=1}^{N} \hat{V}(\bldr_k,\bldr_l),
\label{Original_H_Pos}
\ee
where the factor of $(1/2)$ ensures that the interactions between every pair of particles is only counted once.
Actually, in order to formulate a finite temperature theory for ultracold gases, it turns out to be much simpler to formulate the problem in terms of a different representation, known as the `occupation number representation' of second quantization. A brief review of this, following closely the discussion of Ref.\ \cite{Fetter_Walecka}, is included below for completeness - readers who are familiar with this should proceed directly to Sec.\ \ref{Basic_SH}.

\subsubsection{Occupation Number Representation of Second Quantization:}

A system consisting of $N$ particles can be fully described by an $N$-body wavefunction $\tilde{\Psi}(\bldr_1 \cdots \bldr_N,t)$ which obeys the well-known Schr\"{o}dinger equation
\begin{equation}
i \hbar \frac{ \partial}{ \partial t} \tilde{\Psi}(\bldr_{1} \cdots \bldr_{N}, t) = \hat{H} \tilde{\Psi}(\bldr_{1} \cdots \bldr_{N}, t)
\label{Schr_Eqn}
\;.
\end{equation}
Directly solving the Schr\"{o}dinger equation in the usual way would be very complicated for a system consisting of many particles, as it would require one to keep track of particle statistics at each level by using appropriately symmetrized products of single-particle wavefunctions, a procedure which is neither practical nor necessary.
Instead of dealing directly with the many-body wavefunction, one typically reformulates the problem in a different representation which greatly simplifies the mathematical handling.

The main idea in this `new representation' is to exploit the indistinguishability of particles in order to keep track only of the {\em number} of particles in each energetically accessible state of the system. 
Instead of the usual expansion of the many-body wavefunction in terms of a complete orthonormal time-independent basis of single-particle wavefunctions, one now defines a new complete orthonormal basis set $| n_1 , n_2 , \cdots n_{\infty} \rangle$, where $n_i$ denotes the number of particles in state $i$ (corresponding for example to a state with energy $\varepsilon_i$); while this basis set is infinite, the fact that the system has a (fixed) finite number of $N$ atoms implies that (for bosons) there are at most $N$ states populated, with all remaining states $j$ unoccupied (i.e.\ $n_j = 0$).

The $N$-body wavefunction is thus expanded into this `occupation number' basis set, via the mapping
\begin{equation}
\tilde{\Psi}(\bldr_{1} \cdots \bldr_{N}, t) \rightarrow | \tilde{\Psi}(t) \rangle = \sum_{n_{1} \cdots n_{\infty}} c(n_{1} \cdots n_{\infty}) |n_{1} \cdots n_{\infty} \rangle 
\label{Schr_Eqn_OccNum}
\;,
\end{equation}
where $c(n_{1} \cdots n_{\infty})$ denote appropriate expansion coefficients in the new basis.
Such coefficients must satisfy the following two properties: (i) they must be appropriately normalized, such that the probability of finding the system somewhere in configuration space is equal to one; (ii) they must incorporate the particle statistics, i.e.\ be symmetric (for bosons) under the interchange of two quantum numbers, to reflect the underlying symmetry of the many-body wavefunction.
Note that the problem of summing over all sets of quantum numbers consistent with a given set of occupation numbers is in fact
equivalent to the problem of putting $N$ objects into boxes, with $n_{1}$ objects in Box 1, $n_{2}$ objects in Box 2, $\cdots$ $n_{i}$ objects in Box $i$ and so on.
The main advantage of this formulation is that one does not need to explicitly consider symmetrized products of single-particle wavefunctions, as particle statistics are automatically incorporated within the new formalism.
The basis set $|n_{1} \cdots n_{\infty} \rangle$ is time-independent, and thus all system dynamics is incorporated in the evolution of the (normalized) expansion coefficients $c(n_{1} \cdots n_{\infty})$; the latter can be shown to reduce to equations for each set of values of the occupation numbers  $\{ n_1 \cdots n_{\infty} \}$ - i.e.\ the problem still remains very complicated.

The essence of this approach actually lies in the following realization: 
The equations for the expansion coefficients $c(n_{1} \cdots n_{\infty})$ generally describe the likelihood of particles moving around within the accessible levels, e.g. one particle changing its energy by moving from level $l$ to level $i$. Mathematically, this can be visualized as the result of the destruction of a particle in level $l$ and the simultaneous creation of a particle in state $i$; this analogy is very convenient, as it enables us to use the well-known single-particle annihilation ($\hat{a}$) and creation ($\hat{a}^{\dag}$) operators of quantum mechanics \cite{Schiff,Isham} in the suitably generalized form
\bea
\hat{a}_{l} |n_{1} \cdots n_{i} \cdots n_{l} \cdots n_{\infty} \rangle = 
\sqrt{n_{l}} |n_{1} \cdots n_{i} \cdots (n_{l}-1) \cdots n_{\infty} \rangle \nonumber \\
\hat{a}_{i}^{\dag} |n_{1} \cdots n_{i} \cdots n_{l} \cdots n_{\infty} \rangle = 
\sqrt{n_{i}+1} |n_{1} \cdots (n_{i}+1) \cdots n_{l} \cdots n_{\infty} \rangle\;.
\eea
These operators can be shown to obey the `standard' bosonic commutation relations
\be
\left [\hat{a}_{l}, \hat{a}_{i}^{\dag} \right] = \delta_{li}
\hspace{2.0cm} 
\left [\hat{a}_{l}, \hat{a}_{i} \right] = \left [\hat{a}_{l}^{\dag}, \hat{a}_{i}^{\dag} \right] = 0\;.
\ee
The number of atoms occupying a particular level $i$ can be obtained within the occupation representation formalism via 
$n_i=\langle \hat{N}_i \rangle$, where $\hat{N}_i=\hat{a}_{i}^\dag \hat{a}_i$ is the number operator, and $\langle \cdots \rangle$
denotes averaging over all single-particle states. This is evident from
\bea
\hat{N}_i | n_1 \cdots n_{i-1} (n_i) n_{i+1} \cdots n_{\infty} \rangle 
= \hat{a}_{i}^\dag \hat{a}_i | n_1 \cdots n_{i-1} (n_i) n_{i+1} \cdots n_{\infty} \rangle \nonumber \\
= \hat{a}_{i}^\dag \left[  \sqrt{n_i} | n_1 \cdots n_{i-1} (n_i-1) n_{i+1} \cdots n_{\infty} \rangle \right]
\nonumber \\
= \sqrt{(n_i-1)+1} \sqrt{n_i} | n_1 \cdots n_{i-1} (n_i) n_{i+1} \cdots n_{\infty} \rangle \nonumber \\
= n_i | n_1 \cdots n_{i-1} (n_i) n_{i+1} \cdots n_{\infty} \rangle \;,
\eea
leading in the orthonormal occupation number basis to
\bea
\langle \hat{N}_i \rangle &=& \langle  n_1 \cdots n_{i-1} (n_i) n_{i+1} \cdots n_{\infty} | n_i | n_1 \cdots n_{i-1} (n_i) n_{i+1} \cdots n_{\infty} \rangle \nonumber \\
&=& n_{i}\;.
\eea

Single-particle effects, such as the action of the potential or kinetic energy operator on a many-particle system, can only change the state of individual atoms, e.g.\ shift an atom from one accessible state (say $l$) to another one of different energy (say $i$). Such a change of state is clearly described by the combined action of one annihilation ($\hat{a}_l$) and one creation ($\hat{a}_i^\dag$) operator, i.e.\ by the product 
$\hat{a}_i^\dag \hat{a}_l$. Collisions included within our theoretical discussion involve only two atoms, with both atoms typically emerging from the collision in different states to the ones they were in prior to the collision. Such a collision is thus associated with the annihilation of an atom in state $l$ ($\hat{a}_l$) and creation in state $i$ ($\hat{a}_i^\dag$) and the simultaneous annihilation of an atom in state $k$ ($\hat{a}_k$)and creation in state $j$ ($\hat{a}_j^\dag$), i.e. the product of two creation and two annihilation operators
($\hat{a}_{i}^{\dag} \hat{a}_{j}^{\dag} \hat{a}_{k} \hat{a}_{l}$).

As a result, the original problem concerning the evolution of the many-body wavefunction (Eq.\ (\ref{Schr_Eqn})), has now been mapped onto the mathematically equivalent form
\begin{equation}
i \hbar \frac{ \partial}{ \partial t} | \tilde{\Psi} \rangle  = \hat{H} | \tilde{\Psi} \rangle
\label{Schr_Schr}
\end{equation}
where $| \tilde{\Psi} \rangle$ is defined by Eq.\ (\ref{Schr_Eqn_OccNum}), and the hamiltonian is expressed as
\begin{equation}
\hat{H} = \sum_{il} \langle i | \hat{h}_0 | l \rangle \hat{a}_{i}^{\dag} \hat{a}_{l} +
\frac{1}{2} \sum_{ijkl} \langle i j | \hat{V} | k l \rangle 
\hat{a}_{i}^{\dag} \hat{a}_{j}^{\dag} \hat{a}_{k} \hat{a}_{l}
\label{H_OccNum_SP}
\;.
\end{equation}
Here we have introduced the notation
\bea
\langle i | \hat{h}_0 | l \rangle = \int d\bldr \spwa_i^*(\bldr) \hat{h}_0(\bldr) \spwa_l(\bldr) 
\eea
with the symmetrized form of the non-local interaction potential taking the form
\be
\langle ij | V | kl \rangle = \frac{1}{2} \left[ (ij|\hat{V}|kl) + (ij|\hat{V}|lk) \right] \;,
\label{matrixelement}
\ee
where
\be
(ij |\hat{V} | kl) = \int d\bldr \int d \bldr' \spwa_i^*(\bldr) \spwa_j^*(\bldr') \hat{V}(\bldr-\bldr') \spwa_l(\bldr') \spwa_k(\bldr)\;.
\ee
This symmetrization accounts implicitly for all particle statistics, and demonstrates the power of this alternative mathematical technique.
Comparing the hamiltonian in the occupation number representation, Eq.\ (\ref{H_OccNum_SP}), 
with the original `position-space' Hamiltonian, Eq.\ (\ref{Original_H_Pos}), 
we note the following: Operators appearing in Eq.\ (\ref{Original_H_Pos}), e.g.\ $\hat{h}_0(\bldr_k)$,
are now replaced by complex numbers, which are obtained by taking matrix elements 
$\langle i | \hat{h}_0 | l \rangle $ of the original operators over the single-particle eigenstates $\spwa_i(\bldr)$ - and similarly for $\hat{V}(\bldr_k,\bldr_l)$.
The operator nature of the system hamiltonian must be however maintained, and this is achieved by the
single-particle operators ($\hat{a}_i^{(\dag)}$) appearing in the definition of Eq.\ (\ref{H_OccNum_SP}), whose role is to remove or add particles in particular states of the occupation number representation.

One may wonder where the averaging $\langle \cdots \rangle$ over the initial single-particle eigenstates comes from: 
We have argued that the original Schr\"{o}dinger equation (Eq.\ (\ref{Schr_Eqn})) is mapped onto an equation for the normalized and appropriately symmetrized expansion coefficients $c(n_1 \cdots n_{\infty})$ in the occupation number basis, with all time-dependence included in these coefficients. 
These expansion coefficients are precisely the link between the original `first quantized' position representation formulation of the problem and the equivalent `second-quantized' representation.
In order to obtain an equation for the evolution of these coefficients, starting from the original formulation of the problem in terms of a complete basis of single-particle wavefunctions $\{ \spwa_i(\bldr) \}$, we must first `project out' the original basis by averaging over single-particle eigenstates. Upon mapping the $N$-body wavefunction onto the occupation number representation basis, the expansion coefficients thus depend on matrix elements over the initial eigenfunctions - more details on this procedure can be found in \cite{Fetter_Walecka}. 

The hamiltonian of Eq.\ (\ref{H_OccNum_SP}) thus represents, in the occupation number representation of second quantization the basic system hamiltonian corresponding to the original `position representation' hamiltonian of Eq.\ (\ref{Original_H_Pos}).
Rather than working explicitly with the individual levels in the occupation number representation, it is often more convenient for brevity to construct linear combinations of such operators
\begin{equation}
 \fopa (\bldr,t) = \sum_i \spa_i (t) \spwa_i (\bldr,t), \ \ \ \ \ \ \ 
 \fopc (\bldr,t) = \sum_i \spc_i (t) \spwa_i^* (\bldr,t),
\label{single-wave}
\end{equation}
which are summed over the complete set of single-particle quantum numbers. These quantities, which are operators in the abstract occupation-number Hilbert space, are called `field operators'.
Here $\fopcrt$ represents the addition of a particle at 
point $\bldr$ and time $t$, while $\fopart$ removes a particle. Hence they
are known as creation and annihilation operators respectively.
For bosons, these operators satisfy the commutation relations
$[\fopa (\bldr,t),\fopc(\bldr',t)]=\delta(\bldr-\bldr')$ and
$[\fopa (\bldr,t),\fopa(\bldr',t)]=[\fopc (\bldr,t),\fopc(\bldr',t)]=0$.
All properties of the full quantum problem are actually contained within these Bose field operators,
and the problem reduces to identifying suitable techniques for extracting the desired information.

Using Eq.\ (\ref{single-wave}) we can thus show that the transition from the coordinate representation to the occupation number representation of the system hamiltonian is achieved via the transformations
\bea
\sum_{k=1}^{N} \hat{h}_0(\bldr_k) \rightarrow \int  d\bldr \fopcrt \spop(\bldr) \fopart 
\\
\sum_{k,l=1}^{N} \hat{V}(\bldr_k,\bldr_l) \rightarrow 
\int \int d\bldr d\bldr' \fopcrt \fopc (\bldr',t)  V(\bldr-\bldr') \fopa (\bldr',t) \fopart 
\nonumber
\;.
\eea

Our subsequent discussion will be mainly given in terms of these Bose field operators, as this allows for more compact expressions. However, in order to deal with some of the more subtle issues, we will occasionally expand the field operator in a suitable basis via Eq.\ (\ref{single-wave}).

\subsubsection{Basic System Hamiltonian}
\label{Basic_SH}

In the occupation number representation of second quantization, the system Hamiltonian can thus be written in terms of Bose field operators as:
\begin{eqnarray}
{\hat H} =  \int  d\bldr \fopcrt \spop(\bldr) \fopart 
\nonumber \\
+ \frac{1}{2} \int \int d\bldr d\bldr' \fopcrt \fopc (\bldr',t) 
 V(\bldr-\bldr') \fopa (\bldr',t) \fopart \;,
\label{H}
\end{eqnarray}
where $\spop = - \hbar^2 \nabla^2/(2m)  + 
V_{\rm ext}(\bldr,t)$ is an operator for a single particle in the 
external potential $V_{\rm ext}(\bldr,t)$ (typically a harmonic
or periodic potential), and
$V(\bldr-\bldr')$ is the exact two-body interatomic potential. 

This Hamiltonian is the starting point for {\em all} theoretical treatments of dilute Bose gases, and includes both thermal and quantum fluctuations. All theories appearing in the literature arise from distinct approximations to and mathematical approaches for dealing with this Hamiltonian.
In this Tutorial, we start from the simplest such approach, which corresponds to the lowest order mean field theory, effectively representing zero temperature. We then gradually build up the complexity of treatments by incorporating additional effects step-by-step.

For dilute
gases at very low temperature, the usual procedure is to make a contact 
interaction approximation
\begin{equation}
V (\bldr-\bldr')=g\delta (\bldr-\bldr')\;,
\label{pseudopotential}
\end{equation}
i.e. to assume that the two atoms undergo perfectly elastic local collisions, like two billiard balls.
The strength of the interaction is given by $g=4\pi \hbar^2 a/m$,
where $a$ is the $s$-wave scattering length for a particular atomic species, which can be determined from experiments. 
This is a somewhat idealized scenario, which can nonetheless be put on firm ground by a more careful treatment - the origin and validity of this replacement will be further discussed in Sec.\ \ref{effect-int}.

Substitution into the Hamiltonian
(\ref{H}) then gives:
\begin{equation}
 \hat{H} = \int d\bldr \ \fopcrt \spop \fopart + \frac{g}{2} 
 \int d\bldr \ \fopcrt \fopcrt \fopart \fopart. 
\label{H-g}
\end{equation}
The starting point for the dynamics of ultracold gases is to consider the equation governing the evolution of the bosonic field operator $\hat{\Psi}(\bldr,t)$, i.e.\ the so-called equation of motion.
The second-quantized form of the appropriate equation can be analyzed in one of three distinct pictures, known as the `Schr\"{o}dinger', `Interaction' and `Heisenberg' pictures, depending on whether the state-vectors, the operators corresponding to system observables, or both are time-dependent \cite{Fetter_Walecka,Schiff}.
Our discussion so far (see Eq.\ (\ref{Schr_Eqn})) has been given in the so-called `Schr\"{o}dinger' picture, in which the state vectors are time-dependent and the operators are time-independent; in this picture, the solution to the Schr\"{o}dinger equation at time $t$, given the initial solution at $t_0$, is given by the unitary transformation
$| \tilde{\Psi}_{\rm S}(t) \rangle = e^{-i\hat{H}(t-t_0)/\hbar} | \tilde{\Psi}_{\rm S}(t_0) \rangle$ 
where $\hat{H}$ does not contain any {\em explicit} time-dependence, and the subscript `S' has been introduced to denote the Schr\"{o}dinger picture.
In the interaction picture, both operators and state vectors depend on time. 
Finally, in the Heisenberg picture the state vectors, which can be constructed from the corresponding Schr\"{o}dinger picture state vectors via
$| \tilde{\Psi}_{\rm H}(t) \rangle = e^{i\hat{H}t/\hbar} | \tilde{\Psi}_{\rm S}(t) \rangle$, 
 are time-independent and all time-dependence is contained in the operators.

While all three pictures can be used interchangeably, most subsequent discussion will be given in the Heisenberg picture in which the equation of motion of a general operator $\hat{O}_{\rm H}$ can be shown to obey
\be
i \hbar \frac{\partial \hat{O}_{\rm H}}{\partial t} = \left[ \hat{O}_{\rm H}(t), \hat{H} \right]\;.
\ee
In studying the system dynamics, we will actually be concerned with the equations of motion of the
Bose field operator $\hat{\Psi}$.
For the hamiltonian of Eq.\ (\ref{H-g}), this evolves according to the Heisenberg equation of motion (henceforth suppressing the subscript `H' for compactness) 
%
\begin{equation}
 i \hbar \frac{\partial \fopart}{\partial t} = \left[ \fopart , \hat{H}  \right] 
= \hat{h}_{0}  \fopart + g \fopcrt \fopart \fopart\;.
\label{Heisenberg} 
\end{equation} 
This equation contains all the information we can hope to extract from the system.


\subsubsection{Separating Condensate and Non-condensate Contributions:}

In order to extract information from this equation, it is convenient to separate the condensate contribution, which corresponds to the macroscopic occupation of a single quantum state, from the remaining part of the Bose field operator.
 
For example, assuming initially for simplicity that 
a single-particle basis corresponds to a good basis set, one can approximate the expectation value $\langle \spc_i \spa_l \rangle = \delta_{il} n_i$, 
where
$n_i$ is the occupancy of the single particle state. Bose-Einstein condensation would then
correspond to the macroscopic occupation of one of these states, $\spwa_0$,
such that $n_0 \equiv N_0 \gg 1$. 
In general, interactions between atoms induce mixing between different single-particle states. As a result, it is often beneficial to describe the system in terms of dressed basis states, known as quasiparticle states, which will be discussed later.

Assuming just one (suitably identified) state of the system to be occupied macroscopically,
it is natural to re-arrange the Bose-field operator into two parts \cite{Fetter_1972},
%
\begin{equation}
 \fopart = \copa (\bldr,t) +\nopa (\bldr,t),
\label{sep-fieldops}
\end{equation}
corresponding respectively 
to a field operator for the condensate, $\copa$, and one for the non-condensed atoms, $\nopa$. These
could either correspond to thermally-excited atoms, quantum-mechanical fluctuations, or atoms promoted into higher energy states due to interactions.
%
Although such a split is mathematically exact, approximations are inevitably required in the subsequent analysis and when assigning direct physical interpretations to these operators.
This split is essentially equivalent to the separation of the zero-momentum mode in the usual textbook discussion of Bose-Einstein condensation in a homogeneous system.

The above operators can be re-expressed in a general basis set $\{ \spwa_0, \spwa_i\}$ as
%
\begin{equation}
 \copa(\bldr,t) = \spa_0 (t) \spwa_0 (\bldr,t), \ \ \ \ \ \ 
 \nopa(\bldr,t) = \sum_{i \neq 0} \spa_i (t) \spwa_i (\bldr,t).
\label{cn-fieldops}
\end{equation} 
%

At this point it is worth noting the plethora of different symbols used in the literature to denote condensate and non-condensate operators, the most common alternatives to those used here being $\hat{\psi}$ for the condensate, and $\hat{\psi}'$, $\tilde{\psi}$, or $\delta \hat{\psi}$ for the non-condensate.

\subsubsection{Concept of Symmetry breaking:}

The condensate operator, $\hat{\wfn}(\bldr,t)$,
corresponds to the annihilation operator for atoms in
a state macroscopically occupied by $N_0 \gg 1$ particles.
Since the operator for the number of atoms in the condensate is given by $\hat{N}_0 = \hat{a}_0^\dag \hat{a}_0$, we find $\hat{N}_0 |N_0 \rangle = N_0 | N_0 \rangle$, i.e this operator acting upon a state with a definite number of $N_0$ condensate atoms returns an eigenvalue $N_0$.
Now, since 
$( \hat{a}_0 \hat{a}_0^\dag - \hat{a}_0^\dag \hat{a}_0 )/N_0 |N_0 \rangle = (1/N_0) |N_0 \rangle$, in the limit of a large number of condensate atoms,
$1/N_0 \rightarrow 0$, and the condensate single-particle operators can be thought of as approximately commuting.
One could now make the Bogoliubov replacement
(often referred to as the Bogoliubov approximation) \cite{Bogoliubov}, whereby $\spa_0 \simeq \sqrt{N_0}$.
The operator $\copa(\bldr,t)$ appearing in Eq.\ (\ref{sep-fieldops}) is thus replaced by a complex
number $\wfn(\bldr,t) = \sqrt{N_0} \spwa_0 (\bldr,t)$, often
named the ``condensate wavefunction''. In this approximation, 
the field operator is simply decomposed as
\begin{equation}
 \fopart = \wfn (\bldr,t) +\nopa (\bldr,t)\;,
\label{sep-broken}
\end{equation}
i.e. all operator dependence is contained in the fluctuation operator $\hat{\delta}(\bldr,t)$.

This is a somewhat drastic approximation which has the direct consequence that the physical state described by such a decomposition does not satisfy the same symmetries as the original hamiltonian. 
In particular, although the system hamiltonian is invariant under a gauge transformation in the phase of the Bose field operator (as it is expressed in terms of an equal number of annihilation and creation operators), the wavefunction $\wfn(\bldr,t)$ no longer shares this symmetry; technically one says that this symmetry has been `broken'. Breaking of the $U(1)$ global phase symmetry (i.e.\ fixing the condensate phase) leads to a non-conservation of the total number of particles (since these two quantities are canonically conjugate). The consequence of violation of particle number conservation is evident, since in this approximation one assumes that the addition or removal of a particle in the condensate does not affect the state of the system,  i.e. $N_0 \pm 1 \approx N_0$ for large atom numbers $N_0$.
This approximation is equivalent to the statement that
the ensemble average of the Bose field operator is well-defined and non-zero, i.e. $\langle \fopart \rangle = \wfn (\bldr,t) \neq 0$. This leads directly to $\langle \nopa (\bldr,t) \rangle = 0$, a desirable property for a fluctuation operator.
Given the typical large system size, such a subtle issue can, in first instance, be overlooked, thus leading to the simplified mathematical formalism reviewed in Secs. \ref{zeroT}-\ref{finitet:dynamic}. This provides significant insight into the underlying physics, while still yielding excellent agreement with experiments. However, from a fundamental point of view, this is an important inconsistency that is revisited in Secs. \ref{Phase_Number}-\ref{scho-class}.

We should further remark here that, defined this way, $\wfn (\bldr,t)$ is a 
classical field; such an approximation bears close resemblance to the conventional treatment of the
electric field in the theory of electrodynamics, where
the quantum description in terms of photons is replaced by a classical field. 
Although we shall here identify $\wfn (\bldr,t)$ with the condensate, 
in principle this can also include excitations of the system, as long as
their occupation $n_i \gg 1$ and quantum fluctuations are negligible -
we shall return to this point in Sec.\ \ref{Classical_Fields}. 

Using Eq.\ (\ref{sep-broken}), we can approximate the
atom density $n(\bldr,t) = \langle \fopcrt \fopart \rangle = n_c (\bldr,t) + \tilde{n} (\bldr,t)$ into two contributions, namely a condensate density where 
$n_c(\bldr,t)=|\wfn(\bldr,t)|^2$ and a non-condensate density
$\tilde{n}(\bldr,t) = \langle \nopc(\bldr,t) \nopa (\bldr,t) \rangle$. 
However, in most current experiments, quantum fluctuations are largely overwhelmed by thermal fluctuations.
%
For example, even at $T=0$, the typical quantum depletion in 3D is given by $n-n_{0} = (8/3 \sqrt{\pi}) \sqrt{na^{3}}$ \cite{Stringari_Review}, 
which is typically only a small fraction of the total number of atoms, as the `diluteness parameter' $na^{3}$ takes typical experimental values around $na^{3} \sim 10^{-3}$
(where $n$ denotes the density and $a$ the s-wave scattering length discussed later). 
One thus tends to identify $\hat{\delta}(\bldr,t)$ as the operator for the {\em thermal cloud}, in which case $ \tilde{n}(\bldr,t)$ describes the density of the thermal atoms.
Note however that quantum fluctuations may become important in low-dimensional geometries giving rise to novel phenomena that lie beyond the scope of this Tutorial \cite{Gangardt_LesHouches}.

\subsubsection{Basic Hamiltonian Contributions:}

The `standard' procedure in modelling Bose gases theoretically is to separate the Bose field operator into condensate and non-condensate parts and then break down the full system hamiltonian into various contributions based on the number of condensate and non-condensate factors contained in each of them. In particular, substitution of
Eq.\ (\ref{sep-broken}) into the system Hamiltonian
(\ref{H-g}) leads to
\begin{equation}
 \hat{H} = H_0 + \hat{H}_1 + \hat{H}_2 + \hat{H}_3 + \hat{H}_4.
\label{H-sep}
\end{equation}
%
where
\begin{eqnarray}
 H_0 = \int d\bldr \left[ \wfn^* \spop \wfn + \frac{g}{2} 
 |\wfn|^4 \right], 
\label{H_0} \\
 \hat{H}_1 = \int d\bldr \left[ \nopc (\spop + g |\wfn|^2) \wfn
 + \wfn^* (\spop + g|\wfn|^2) \nopa \right],
\label{H_1} \\
 \hat{H}_2 = \int d\bldr \left[ \nopc \left( \spop + 2g |\wfn|^2 \right) \nopa 
 + \frac{g}{2} ( (\wfn^*)^2 \nopa \nopa + \wfn^2 \nopc \nopc) \right],
\label{H_2} \\
 \hat{H}_3 = g \int d\bldr \left[ \wfn \nopc \nopc \nopa + \wfn^*
 \nopc \nopa \nopa \right],
\label{H_3} \\
 \hat{H}_4 = \frac{g}{2} \int d\bldr \ \nopc \nopc \nopa \nopa \;,
\label{H_4} 
\end{eqnarray}  
%
%
where $H_0$ has no `hat' as there are no operators within it (it is a purely classical quantity represented by a complex function).
Consideration of these separate contributions forms the basis of our subsequent discussion. In particular, we shall show how inclusion of different contributions to the Hamiltonian 
(\ref{H-sep}-\ref{H_4}) can be used to derive progressively more sophisticated
treatments.

We shall present various approaches, which can be roughly grouped into three different `classes': (i) perturbative approaches based on symmetry-breaking; (ii) perturbative number-conserving approaches; and (iii) unified number-conserving approaches which include fluctuations around the mean field.

Within the context of mean-field theory treatments, there are basically two equivalent ways of formulating the problem: When one is interested in static properties, one can either write down the corresponding equation of motion and solve it in the time-independent limit, or, equivalently, diagonalize the appropriate hamiltonian in the grand canonical ensemble $\hat{H} - \mu \hat{N}$, where $\mu$ is the chemical potential of the system and $\hat{N}$ the total number operator. Note that as symmetry-breaking approaches violate particle number conservation, any subsequent calculations should be performed within the grand canonical ensemble \cite{Huang_Book}. For dynamical properties, one either works with the full equations of motion, or after diagonalizing the hamiltonian, one can consider linearized equations arising from the addition of time-dependent fluctuations around the equilibrium values.

The description of finite temperature Bose gases has historically relied on diverse approaches, with some of the key early papers in this field given in Refs. 
\cite{Landau,Bogoliubov,Key_Paper_1,Penrose_Onsager,Key_Paper_2,Beliaev1,Beliaev2,Key_Paper_3,Key_Paper_4,Baym_Kadanoff_Paper,Takano,Key_Paper_5,Kadanoff_Baym,Martin,Gavoret,Keldysh,Kane_Kadanoff,Key_Paper_6,Popov,Mermin,Hohenberg,Fetter_1972} (although this list is by no means exhaustive).
Existing theoretical models are based on similar underlying principles, and the aim of this Tutorial is to 
focus on the underlying physics while developing the theoretical understanding of such systems from first principles.
While this Tutorial is aimed at the non-expert, active researchers in this field may also find it useful, as we have placed particular emphasis on demonstrating the relation between the different theoretical formulations currently employed.


\subsection{Overview}

We present here a somewhat detailed overview of this Tutorial:

Sec.\ \ref{zeroT} focuses on the
simplest possible mean field theory. This is based on the approximation $\hat{H} \approx H_0$, in which the non-condensate field operator is 
neglected completely. 
This gives rise to the well-known Gross-Pitaevskii equation, which is valid at $T=0$ and in the limit of negligible quantum fluctuations.
By linearization we then derive equations
for the collective modes around the ground state. These are shown to be identical to the modes of
Bogoliubov quasiparticles found by diagonalization of the first three 
terms of the Hamiltonian, i.e.  $\hat{H} \approx (H_0 + \hat{H}_{1} + \hat{H}_{2})$.

Consideration of the finite 
temperature case necessitates the 
identification of suitable `generalized' mean fields presented in Sec.\ \ref{finitet:static}. This requires the
inclusion of contributions from $( \hat{H}_3 + \hat{H}_4 )$ to the system Hamiltonian.  `Suitable approximations' can be used to reduce these additional hamiltonian contributions to quadratic form, thus introducing finite temperature mean field corrections which lead to diverse theories in suitably identified limits. 
In the so-called Hartree-Fock-Bogoliubov approximation,
the applications and limitations of which will be discussed and assessed,
one typically assumes that the Bogoliubov quasiparticle states are thermally 
populated, but their collisions are neglected and the thermal cloud remains 
in static equilibrium. Technical subtleties associated with the inclusion of effective contact interactions are also briefly addressed in our formalism.
 
Sec.\ \ref{finitet:dynamic} generalizes this treatment to a dynamic thermal cloud. This requires a more careful consideration of the contributions $( \hat{H}_3 + \hat{H}_4 )$, beyond the usual mean field. Such a description is important, as in most experiments, the condensate and the thermal cloud are not in equilibrium, but continuously exchange particles, within the constraint of a constant total atom number. This leads to a more realistic description
in terms of a dissipative Gross-Pitaevskii equation for the condensate coupled to a Quantum Boltzmann Equation for the non-condensate; such an approach
includes
collisions between the 
quasiparticles and thus accurately describes the known damping mechanisms.

Sec.\ \ref{Phase_Number} addresses some subtle  issues associated with fluctuations in the condensate phase and number-conservation. The absence of Bose-Einstein condensation in low-dimensional geometries is briefly discussed and an appropriately modified mean field theory is highlighted which can describe the equilibrium properties of such systems at finite temperatures.
The discussion of Secs. \ref{zeroT}-\ref{finitet:dynamic} is then revisited under the physically-relevant constraint of a constant total particle number.

This Tutorial would not be complete without a brief presentation of alternative more advanced techniques for dealing with both thermal and quantum fluctuations given in
Sec.\ \ref{scho-class} (see also \cite{Davis_Review}). These include classical field methods (already reviewed in \cite{Polish_Review}) and stochastic methods.
Sec.\ \ref{applications} discusses the existing implementation of the presented approaches in two key areas of ultracold Bose gases,  thus clearly highlighting the benefits and limitations of the diverse approaches presented throughout this Tutorial.
Sec.\ \ref{conclusions} summarizes and links the different methodologies, with emphasis on the current status of the field, and our progress towards a `complete' description of ultracold Bose gases. 

This Tutorial also contains three fully-linked appendices which discuss certain lengthy theoretical models, whose physical content and/or simplified limits are dealt with in the main body of the Tutorial.
For maximum clarity, the end of each section features a brief summary of the main results presented in that section, with the essential content of the different theoretical models additionally summarized in appropriate tables.

To aid the reader in selecting the most appropriate sections for their study, we note that
researchers who merely wish to obtain an overview of the underlying physical issues without the need for precision and in-depth understanding, should focus their attention on Secs. \ref{zeroT}-\ref{finitet:dynamic} which constitute the main part of this Tutorial.

\section{Zero Temperature Mean Field Theory}
\label{zeroT}

We start by briefly reviewing the zero temperature formalism in order to facilitate easier comparison to our subsequent finite temperature discussion.

\subsection{Consideration of $H_0$: The Gross-Pitaevskii Equation}
\label{Zero_Temp}

In the $T=0$ limit (essentially) all of the particles are in the 
condensate, so that $N=N_0$ and the noncondensate operator can be neglected ($\nopa=\nopc=0$); in other words, we set $\fopa(\bldr,t) \rightarrow \wfn (\bldr, t)$. Hence, $\fopcrt = \wfn^* (\bldr,t)$, and the exact Heisenberg equation of motion (\ref{Heisenberg}) reduces
to the so-called Gross-Pitaevskii Equation (GPE) \cite{Gross,Pitaevskii},
\begin{eqnarray}
 i\hbar \frac{\partial \wfn(\bldr,t)}{\partial t} &=& 
\left [\hat{h}_0(\bldr,t) + g |\wfn (\bldr,t)|^2 \right] \wfn (\bldr,t) \nonumber \\
&=& \left [-\frac{\hbar^2}{2m} \nabla^2  + V_{\rm ext}(\bldr,t) + g |\wfn (\bldr,t)|^2 \right] 
 \wfn (\bldr,t)\;.
\label{GP-T0}
\end{eqnarray}

This equation is a nonlinear Schr\"{o}dinger equation, corresponding to the zero temperature hydrodynamic description of Bose gases, first introduced to study vortex lines in an imperfect Bose gas \cite{Gross,Pitaevskii}. This equation is analogous to the equation describing the electric field in Kerr nonlinear media, only in our present context
the nonlinearity arises from atomic interactions. 
Moreover, this equation is mathematically analogous to a Ginzburg-Landau-type approach \cite{Fetter_Walecka}, valid near the critical regime - although the origin of the various contributions and the physical interpretation here is actually quite distinct. 
Remarkably, the GPE, whose first numerical implementation to dilute weakly-interacting trapped Bose gases was undertaken by Mark Edwards, Keith Burnett and collaborators \cite{Burnett_GPE_1,Burnett_GPE_2,Burnett_GPE_3}, provides a good
description of the dynamics of a Bose-Einstein condensate for a large range of problems at temperatures as high as $T \approx T_{c}/2$. 

One can find static 
solutions by  eliminating the `trivial' time-dependence via the replacement:
\begin{equation}
 \wfn(\bldr,t) = \wfn_0 (\bldr) e^{-i\mu t/\hbar},
\label{time-dep}
\end{equation} 
where $\mu$ is the chemical potential (which is equal to the energy per particle only for non-interacting particles). Substituting (\ref{time-dep}) into
(\ref{GP-T0}) yields the time-independent GPE:
\begin{equation}
 \mu \wfn_0 (\bldr)= \left [-\frac{\hbar^2}
 {2m} \nabla^2  + V_{\rm ext}(\bldr) + g |\wfn_0 (\bldr)|^2 \right] 
 \wfn_0 (\bldr).
\label{GP-indie}
\end{equation}
%
%
In the form given above,  the wavefunction is normalized to the total number of particles, i.e.
$ \int d \bldr | \wfn(\bldr,t) |^2 = N\;.$
However, various authors work instead with a re-defined wavefunction,
$\wfn(\bldr,t)/\sqrt{N}$, which is normalised to 1, for which the prefactor of the nonlinear term in the GPE becomes $gN$.

Note that the GPE can also be derived from
variational considerations as follows:
In general, the condensate energy can be written as a function of the condensate wavefunction $\wfn(\bldr)$ 
, which is termed a `functional', denoted by $E[\wfn]$. In the limit of a large number of atoms, this takes the approximate form
\be
E[\wfn]= \int d\bldr \left[ \frac{\hbar^{2}}{2m} \left| \nabla \wfn(\bldr) \right|^{2} 
+ V_{\rm ext}(\bldr) | \wfn(\bldr) |^{2} + \frac{1}{2}g |\wfn(\bldr)|^{4} \right]\;.
\label{GL}
\ee
The desired optimal form for the condensate wavefunction is actually obtained by minimizing this energy functional, $E[\wfn]$,
with respect to the wavefunction $\wfn(\bldr)$; this is achieved by the process of functional differentiation, denoted by $\delta E/ \delta \wfn$, whereby one examines how the entire functional $E[\wfn]$ changes as a result of small changes in the function $\wfn(\bldr)$. As the wavefunction is a complex quantity, this minimization should consider both real and imaginary contributions as independent, i.e.\ consider changes in $E[\wfn]$ arising from independent variations of $\wfn(\bldr)$ and $\wfn^*(\bldr)$ \cite{Negele_Book}.
In performing this minimization, one should impose the constraint of a fixed total number of atoms $N$; this can be included by the technique of Lagrange multipliers: in other words, instead of minimizing $E$, one should minimize (or find saddle points of) the quantity $(E-\mu N)$, where the Lagrange multiplier $\mu$ is identified as the chemical potential of the system \cite{Pethick_Book}.

\subsubsection{Hydrodynamic Description}

\label{Hydro}

Since the wavefunction is a complex quantity, one can use the Madelung transformation $\wfn(\bldr,t) = \sqrt{n_0(\bldr,t)}e^{i \theta(\bldr,t)}$, along with the identification of ${\bf v}(\bldr,t) = (\hbar/m) [ \nabla \theta(\bldr,t)]$ as the superfluid velocity to re-express
Eq.\ (\ref{GP-T0}) in the form of a `conservation of mass' (or continuity) equation 
\be
\frac{ \partial n_0}{\partial t} + {\bf \nabla} \cdot (n_0 {\bf v}) = 0 
\label{mass} 
\ee
coupled to a `generalized Euler equation' given by
\be
m \left( \frac{\partial}{\partial t} + {\bf v} \cdot {\bf \nabla} \right) {\bf v} 
= - {\bf \nabla} \mu_0  = - {\bf \nabla} \left( -\frac{ \hbar^2 \nabla^2 \sqrt{n_0}}{2m \sqrt{n_0}} + V_{\rm ext} +g n_0 \right).
\label{Euler}
\ee
In the above equations, the subscript `0' has been used to highlight the fact that all the atoms are in the condensate.
Eq.\ (\ref{Euler}) depends on the gradient of the $T=0$ chemical potential, $\mu_0$, and can be re- expressed in terms of a force ${\bf F} = -(1/m) {\bf \nabla} V_{\rm ext}$, and the gradient of a `quantum pressure' contribution $P_Q$; the latter, defined by $P_Q=(1/2)gn_0^{2} - (1/4)n_0 \nabla^{2} ({\rm ln}n_0)$, contains two contributions: the
first corresponds to the usual pressure term for a gas, while the second term describes the kinetic energy contribution arising from spatial variations of the wavefunction occurring within a small scale \cite{Pethick_Book}.

\subsection{The Bogoliubov equations}
\label{Bog-dG}

The GPE, Eq.\ (\ref{GP-T0}) also admits time-dependent solutions, that 
correspond to collective modes or elementary excitations of the system. These
can be found using:
\begin{equation}
 \wfn(\bldr,t) = e^{-i \mu t/\hbar} [\wfn_0 (\bldr) + \delta \wfn 
 (\bldr,t) ],
\label{excite}
\end{equation}
where $\delta \wfn$ represents fluctuations of the condensate 
wavefunction around the ground state $\wfn_0$. If these fluctuations are 
small, $\delta \wfn \ll \wfn_0$, then one can neglect powers of 
$\delta \wfn$ greater than one, a procedure known as linearization. 
Substituting (\ref{excite}) into (\ref{GP-T0}) and subtracting 
(\ref{GP-indie}) then yields:
\begin{equation}
 i\hbar \frac{\partial}{\partial t}  \delta\wfn (\bldr,t) = \left[ \spop 
 + 2g |\wfn_0|^2 - \mu \right] \delta \wfn (\bldr,t)+ g \wfn_0^2 
 \, \delta \wfn^* (\bldr,t)\;.  
\label{linear}
\end{equation}

The linearized equation (\ref{linear}) can then 
be solved to find collective modes of the system. This is usually done by 
looking for solutions of the form:
\begin{equation}
 \delta \phi (\bldr,t)= \sum_i \left[ u_i (\bldr) e^{-i\omega_i t} + 
 v_i^* (\bldr) e^{i\omega_i t} \right],
\label{u-v}
\end{equation}
where $i$ labels the different modes, each with frequency $\omega_i$. 
Substituting (\ref{u-v}), together with its complex conjugate, into 
(\ref{linear}), and collecting together prefactors of $e^{-i\omega_i t}$ 
and $e^{i\omega_i t}$ yields the two coupled equations:
\begin{eqnarray}
\left[\spop + 2g |\wfn_0(\bldr)|^2- \mu \right] 
 u_i (\bldr)+ g [\phi_0 (\bldr)]^2 v_i (\bldr) &=&  \epsilon_i u_i (\bldr)
\label{bog-u} \\
 \left[ \spop + 2g|\wfn_0(\bldr)|^2 - \mu 
 \right]  v_i (\bldr) + g [\phi_0^* (\bldr)]^2 u_i (\bldr)
&=& -\epsilon_i v_i (\bldr)
\label{bog-v}
\end{eqnarray}
%
where $\epsilon_i = \hbar \omega_i$ denotes the dressed (quasiparticle) energy of level $i$;
as discussed explicitly below, one actually views the excitations above the ground state as dressed particles, or {\it quasiparticles}, where the dressing arises from interactions via mean field coupling. 
These are known as the (zero-temperature) {\it 
Bogoliubov} equations, and were first used by Pitaevskii in 1958 \cite{Pitaevskii} to discuss vortex excitations in a superfluid.
Similar coupled equations arise in the Bardeen-Cooper-Schrieffer (BCS) theory of superconductivity for the description of Fermi quasiparticles \cite{deGennes_Book};  in this context, these equations are often referred to as Bogoliubov-de Gennes equations, a terminology which is sometimes also used for their bosonic counterparts.
 
Eqs.\ (\ref{bog-u})-(\ref{bog-v}) can also be derived by an alternative, yet equivalent, procedure of diagonalizing the quadratic hamiltonian $(H_0 + \hat{H}_1 +\hat{H}_2)$, as discussed in Sec.\ \ref{diagonalize}.

The Bogoliubov equations are often cast in matrix notation as
\cite{Burnett_GPE_3}
\begin{eqnarray}
\left( 
\begin{array}{cc} 
\hat{L}(\bldr)  & \hat{M}(\bldr) \\ -\hat{M}^{*}(\bldr) & -\hat{L}^*(\bldr)
\end{array}
\right)
\left(
\begin{array}{l}
u_i(\bldr) \\ v_i(\bldr)
\end{array}
\right)
= \epsilon_i
\left(
\begin{array}{l}
u_i(\bldr) \\ v_i(\bldr)
\end{array}
\right)\;,
\label{BdG_Matrix_T0}
\end{eqnarray}
with 
\bea
\hat{L}= \hat{h}_0 + 2g|\wfn_0(\bldr)|^2-\mu \\
\hat{M}(\bldr) = g [ \wfn_0 (\bldr) ]^2
\eea
such that $\hat{L}^*(\bldr) = \hat{L}(\bldr)$.
Self-consistent numerical solution of the GPE (Eq.\ (\ref{GP-indie})) and the Bogoliubov equations (Eq.\ (\ref{BdG_Matrix_T0})) fully describes the static properties of the system at this level of approximation
\cite{Burnett_GPE_3}. 


\subsubsection{Uniform condensates:} 

In order to explain the notion of a bosonic quasiparticle,
it is instructive to examine the solution of
Eqs.\ (\ref{GP-indie}) and (\ref{BdG_Matrix_T0}) for a uniform condensate.
In this case,
$V_{\rm ext}=0$, and the density $n_{c0}=|\wfn_0|^2$ thus becomes independent of position, so that Eq.\ 
(\ref{GP-indie}) gives $\mu = gn_{c0}$. Substituting the
plane wave solutions $u_i (\bldr)= u_{\bf p} 
e^{i {\bf p}\cdot\bldr/\hbar}$ 
and $v_i (\bldr) = v_{\bf p} e^{i {\bf p}\cdot\bldr/\hbar}$ into 
(\ref{BdG_Matrix_T0}) yields:
\begin{equation}
 \epsilon ({\bf p}) = \hbar \omega_{\bf p} = \sqrt{ \frac{|{\bf p}|^2}{2m} \left [ 
 \frac{|{\bf p}|^2}{2m} + 2 gn_{c0} \right]}
= \sqrt{ \varepsilon_{\bf p}^2 + 2 g n_{c0} \varepsilon_{\bf p} }
\label{Bog_Spectrum}
\end{equation} 
Here $ \varepsilon_{\bf p} = |{\bf p}|^2/2m$ denotes the energy of a free particle and $\epsilon({\bf p}) = \hbar \omega_{\bf p}$ introduces a new dressed energy corresponding to the energy of a quasiparticle excitation.
In particular,  Eq.\ (\ref{Bog_Spectrum}) yields 
the Bogoliubov dispersion relation \cite{Bogoliubov}. 
At large momenta $p > mc$
(or equivalently, $\epsilon({\bf p}) > \mu$),  this excitation energy approaches that of a free particle, 
$\epsilon ({\bf p})=|{\bf p}|^2/2m + gn_{c0} = \varepsilon_{\bf p} + gn_{c0}$.
However, in the opposite regime of small
$p$, the spectrum approximates a phonon dispersion $\epsilon ({\bf p}) \rightarrow c p$, where 
$c = \sqrt{gn_{c0}/m}$ is the speed of sound.

\subsubsection{Non-uniform condensates:} 

Solving
Eqs.\ (\ref{GP-indie}) and (\ref{BdG_Matrix_T0}) numerically for the case of trapped condensates \cite{Burnett_GPE_3},
leads to a discrete spectrum of excitations, which at low energies 
correspond to bulk oscillations of the condensate. For example, in harmonic 
traps there exists 
a set of dipole modes, at $\omega = \omega_x = \omega_y = \omega_z$, which
correspond to a centre-of-mass oscillation of the condensate within the 
trap. Other common modes involve changes in the condensate widths, either 
in-phase in the different directions (``breathing modes'') or out-of-phase (``quadrupole modes'').
However, for excitations whose wavelength is smaller than the length-scale of variations of the condensate, this leads to a locally homogeneous condensate with a quasi-continuous spectrum, resulting in sound wave propagation, as observed in \cite{Ketterle_Sound}.
A more complete discussion of the different collective modes can be found in 
\cite{Stringari_Review,Stringari_Modes,Ketterle_Varenna}. 
As discussed in Sec.\ \ref{Excitations}, the study of such low-energy modes has provided historically  a systematic test for the development of a consistent theoretical formalism for finite temperature effects, with Eqs.\ (\ref{GP-indie}) and (\ref{BdG_Matrix_T0}) providing an excellent description at sufficiently low temperatures.


\subsubsection{Diagonalization of $\left(H_0 + \hat{H}_1 + \hat{H}_2 \right)$:}
\label{diagonalize}

The Bogoliubov equations (\ref{BdG_Matrix_T0}) 
can also be derived from a diagonalization of the 
first three terms of the Hamiltonian of Eq.\ (\ref{H-sep}), i.e.\ upon approximating 
$\hat{H} \approx (H_0 + \hat{H}_1 + \hat{H}_2)$.
Due to the non-conservation of total atom number, this diagonalization should be performed
in the grand-canonical ensemble, i.e.\ when the system hamiltonian $\hat{H}$ is replaced by
$\hat{K}=\hat{H}-\mu \hat{N}$, where 
$\hat{N}=\int d\bldr \fopc \fopa$ is the number operator. We thus repeat the procedure of Eqs.\ (\ref{H_0})-(\ref{H_2}), and separate $\hat{K}$ into its constitutent parts $K_0+\hat{K}_1+\hat{K}_2$. Upon noting that the wavefunction $\wfn_0(\bldr)$ is now time independent, i.e.  $\fopart = \wfn_0 (\bldr) + \nopa (\bldr,t)$, we obtain:
\begin{eqnarray}
 K_0 = \int d\bldr \left[ \wfn_0^* \left( \spop-\mu \right) \wfn_0 + 
 \frac{g}{2}  |\wfn_0|^4 \right], 
\label{K_0} \\
 \hat{K}_1 = \int d\bldr \left[ \nopc \left( \spop+g|\wfn_0|^2-\mu \right) 
 \wfn_0 + \wfn_0^* \left( \spop + g |\wfn_0|^2-\mu \right) \nopa \right],
\label{K_1} \\
 \hat{K}_2 = \int d\bldr \left[ \nopc \left( \spop + 2g |\wfn_0|^2  -
 \mu \right) \nopa + \frac{g}{2} \left( (\wfn_0^*)^2 \nopa \nopa + 
 \wfn_0^2 \nopc \nopc \right) \right],
\label{K_2}
\end{eqnarray}
Minimization of $K_0$ at constant chemical potential is equivalent to the minimization of the energy functional performed in Sec.\ \ref{Zero_Temp} under the constraint of a fixed total atom number, and thus leads to the time-independent GPE of Eq.\ (\ref{GP-indie}). Note that, in performing such minimization, we have already assumed that a unique minimum exists in which the condensate chooses a definite (yet random) phase, consistent with our assumption of symmetry-breaking in Eq.\ (\ref{sep-broken}).


How does the inclusion of further contributions modify this picture? Firstly, regarding $\hat{K}_1$,  substitution of the GPE into Eq.\ (\ref{K_1}), leads to $\hat{K}_1=0$, upon noting that $\spop$ is hermitian.
Moreover, the $\hat{K}_2$ term (\ref{K_2}) can be diagonalized with the 
Bogoliubov transformation \cite{Fetter_Walecka,Blaizot_Ripka,Bogoliubov,Fetter_1972}:
\begin{equation}
 \nopa (\bldr,t) = \sum_i \left [u_i (\bldr) \hat{\beta_i} (t) 
 + v_i^* (\bldr) \hat{\beta}_i^{\dagger} (t) \right ],
\label{bog-transform}
\end{equation}
where $\hat{\beta}$ and $\hat{\beta}^{\dagger}$ are annihilation and 
creation operators 
for quasiparticles\footnote{The quasiparticle operators are often denoted in the literature as $\hat{\alpha}_i$ and $\hat{\alpha}_i^\dag$.} which obey the Bose commutation relations 
$[\beta_i,\beta_j^\dagger]=\delta_{ij}$ and $[\beta_i,\beta_j]=[\beta_i^\dagger,\beta_j^\dagger]=0$. 
Expressing the non-condensate operators in terms of quasiparticles via Eq.\ (\ref{bog-transform}) is analogous to the expansion around the condensate mean field given by Eq.\ (\ref{u-v}).
From the commutation relations for the noncondensate field operator 
$[\nopa (\bldr), \nopc (\bldr')]=\delta(\bldr-\bldr')$ and
$[\nopa (\bldr), \nopa (\bldr')] =[\nopc (\bldr), \nopc (\bldr')]=0$,  one can then 
derive the orthonormality relations
(see also Sec.\ \ref{MorganGardiner}):
\begin{eqnarray}
 \int d\bldr \left[ u_i^* (\bldr) u_j (\bldr) - 
 v_i^* (\bldr) v_j (\bldr) \right] = \delta_{ij}, \\
 \int d\bldr \left[ u_i (\bldr) v_j (\bldr) - 
 v_i (\bldr) u_j (\bldr)
 \right] = 0
\label{ortho}
\end{eqnarray} 
Substituting (\ref{bog-transform}) into (\ref{K_2}), one can show that if 
$u_i(\bldr)$ and $v_i (\bldr)$ satisfy the Bogoliubov
equations then:
\begin{equation}
 \hat{K}_2 = \sum_i \epsilon_i \hat{\beta}_i^\dagger \hat{\beta}_i - 
 \sum_i \epsilon_i \int d\bldr |v_i (\bldr)|^2.
\label{diagonalized}
\end{equation}
Since the second contribution in Eq.\ (\ref{diagonalized})  is typically small (assuming negligible quantum depletion), the problem essentially reduces to a system of non-interacting quasiparticles with an energy spectrum obtained from the Bogoliubov Eq.\ (\ref{BdG_Matrix_T0}).

\subsection{Brief Summary}

{\em
The $T=0$ behaviour of dilute weakly-interacting Bose-Einstein condensates is fully described by the hamiltonian $H_0$, whose purely mean field nature gives rise to a nonlinear Schr\"{o}dinger equation known as the Gross-Pitaevskii equation.
This is given in full by Eq.\ (\ref{GP-T0}) with its corresponding time-independent expression given by Eq.\ (\ref{GP-indie}). Excitations on top of the condensate ground state can either be described by linearizing around the condensate wavefunction via Eq.\ (\ref{u-v}), or equivalently by diagonalizing the quadratic hamiltonian 
$(H_0+\hat{H}_1+\hat{H}_2)$ in the grand canonical ensemble by means of the Bogoliubov transformation of Eq.\ (\ref{bog-transform}).
}\\

Having established our essential notation, we now proceed to discuss how the effect of temperature can be introduced into our treatment.
A significant part of our subsequent discussion will be to examine how the GPE and Bogoliubov  equations are modified at finite temperatures.
Importantly, while the GPE is modified by additional mean field potentials and kinetic contributions, the structure of the Bogoliubov equations is essentially maintained, with modifications taking the form of additional contributions to the operators $\hat{L}$ and $\hat{M}$.

\section{Finite Temperature Mean Field Theory: Static Case}
\label{finitet:static}

\subsection{Approximate Inclusion of 
$\left( \hat{H}_3 + \hat{H}_4 \right)$}


We now extend the formalism to finite temperatures, by 
explicitly retaining the non-condensate operator $\nopa(\bldr,t)$
in Eq.\ (\ref{sep-broken}).
We proceed via the technique of hamiltonian
diagonalization already discussed in Sec.\ \ref{diagonalize}. 
In order to go beyond the zero-temperature discussion, we must additionally include the remaining contributions $(\hat{H}_3 + \hat{H}_4)$ of Eqs.\ (\ref{H_3})-(\ref{H_4}) to the system hamiltonian. This Section discusses their approximate mean field inclusion, which amounts to a static thermal cloud approximation. In Sec.\ \ref{finitet:dynamic} we shall see that a more careful perturbative analysis of such terms leads to the additional incorporation of the  dynamics of the thermal cloud.


\subsubsection{Conventional Mean Field Approximations:}

In order to reduce the full hamiltonian of Eq.\ (\ref{H-sep}) to a desired quadratic form, one may consider imposing the following mean-field approximation for the non-condensate operators in $\hat{H}_4$,
\begin{equation}
 \nopc \nopc \nopa \nopa \simeq 4 \langle \nopc \nopa \rangle \nopc \nopa 
 + \langle \nopc \nopc \rangle \nopa \nopa + \langle \nopa \nopa \rangle
 \nopc \nopc - [ 2 \langle \nopc \nopa \rangle \langle \nopc \nopa \rangle
  + \langle \nopa \nopa \rangle \langle \nopc \nopc \rangle].
\label{mf-quad}
\end{equation}
This approximation is
motivated from Wick's theorem \cite{Fetter_Walecka,Blaizot_Ripka}, which states that at equilibrium, an average over multiple operators can be approximated by sums of averages of pairwise contracted operators, i.e.
\begin{equation}
\langle \nopc \nopc \nopa \nopa \rangle = 2 \langle \nopc \nopa \rangle \langle \nopc \nopa \rangle + \langle \nopa \nopa \rangle \langle \nopc \nopc \rangle 
\label{mf-wick} \;.
\end{equation}
The above approximation maintains correlations of non-condensate operators only to quadratic order.
One may thus also wish to approximate products of three non-condensate operators appearing in $\hat{H}_3$ by their corresponding `quadratic forms'
\begin{equation}
 \nopc \nopa \nopa \simeq 2 \langle \nopc \nopa \rangle \nopa + 
 \nopc \langle \nopa \nopa \rangle, \ \ \ \ 
 \nopc \nopc \nopa \simeq 2 \nopc \langle \nopc \nopa \rangle + 
 \langle \nopc \nopc \rangle \nopa\;.
\label{mf-trip}
\end{equation}
Since by construction $\langle \hat{\delta}^{(\dag)} \rangle = 0$, the approximation of Eq.\ (\ref{mf-trip}) implies that $\langle \nopc \nopa \nopa \rangle = 0$. 

In Sec.\ \ref{Beyond_HFB} we shall show that imposing the approximation of Eq.\ (\ref{mf-quad}) is equivalent to ignoring collisions between thermal atoms, while Eq.\ (\ref{mf-trip}) amounts to ignoring particle-exchanging collisions between condensed and thermal atoms.
%
%
%
We have already seen that the system may be split into two sub-components, namely the `condensate' contribution, $n_{c}(\bldr,t)=|\wfn(\bldr,t)|^2$ and the `thermal cloud' $\tilde{n}(\bldr,t)=\langle \nopc (\bldr,t) \nopa (\bldr,t) \rangle $, satisfying 
$n_{\rm TOTAL}(\bldr,t) = \langle \fopcrt \fopart \rangle 
= n_{c} (\bldr,t) + \tilde{n}(\bldr,t)\;.
$
Similarly, the approximations of Eqs.\ (\ref{mf-quad}), (\ref{mf-trip}) motivate the definition of 
an additional mean field contribution
$\tilde{m} (\bldr,t) = \langle \nopa (\bldr,t) \nopa (\bldr,t) \rangle$.
This is often referred to as the pair
anomalous average, and bears its name from the fact that there is an unequal number of creation and annihilation operators being averaged over. 
An analogous correlation plays a dominant role in the BCS theory of superconductivity \cite{deGennes_Book}, where fermionic atoms pair up to form so-called Cooper pairs. In the case of Bose-Einstein condensation of neutral bosonic atoms, where the condensate mean field $\wfn(\bldr)$ is the dominant parameter, 
the anomalous average plays a more minor role for effectively repulsive interactions between condensate atoms;
such a contribution, which does actually become crucial in the presence of attractive effective interactions and molecular BECs, should however not be dismissed a priori even for repulsive interactions, where it will be shown to lead to modifications of the atomic scattering.

%


With 
the mean field approximations (\ref{mf-quad}) and (\ref{mf-trip}), the 
$\hat{H}_3$ and $\hat{H}_4$ parts of the Hamiltonian can be 
rewritten in terms linear or quadratic in $\nopa$, as
\begin{eqnarray}
 \hat{H}_3 &=& g \int d\bldr \left[ \wfn \nopc \nopc \nopa + \wfn^*
 \nopc \nopa \nopa \right] \nonumber \\
&\rightarrow&
\delta \hat{H}_{1} = \delta \hat{H}_{1}^{HF} + \delta \hat{H}_{1}^{BOG} \nonumber \\
&=&
g \int d\bldr \left( 2 \wfn \langle \nopc \nopa \rangle \nopc + {\rm h.c.} \right)
+ g \int d\bldr \left( \wfn \langle \nopc \nopc \rangle \nopa + {\rm h.c.} \right)  
\label{dH_1}
\\
 \hat{H}_4 &=& \frac{g}{2} \int d\bldr \ \nopc \nopc \nopa \nopa 
\rightarrow \delta H_0 + \delta \hat{H}_{2}
\end{eqnarray}
\begin{eqnarray}
\delta H_0 &=& \delta H_0^{HF} + \delta H_0^{BOG}
= - g \int d\bldr  \langle \nopc \nopa \rangle \langle \nopc \nopa \rangle
- \frac{g}{2} \int d\bldr \langle \nopa \nopa \rangle \langle \nopc \nopc \rangle \\
%
\delta \hat{H}_{2} &=& \delta \hat{H}_{2}^{HF} + \delta \hat{H}_{2}^{BOG}
\nonumber \\
&=& 2g \int d\bldr \langle \nopc \nopa \rangle \nopc \nopa 
+ \frac{g}{2} \int d\bldr  \left[ \langle \nopc \nopc \rangle \nopa \nopa + \langle \nopa \nopa \rangle
 \nopc \nopc \right]\;.
\label{dH_2}
\end{eqnarray}  
Note that as the following discussion relies a lot on the inclusion of these and other subsequently identified related contributions to the quadratic system hamiltonian, the definition, origin and interpretation of all such contributions is summarized in Table \ref{Table_Hamiltonians}.
The inclusion of the above contributions into the system hamiltonian 
lead to modifications to the original hamiltonians $H_0$, $\hat{H}_{1}$ and $\hat{H}_{2}$ of Eqs.\ (\ref{H_0})-(\ref{H_2}):
$\delta H_0$ merely introduces a shift in the overall system energy, and is therefore often neglected, 
whereas $\delta \hat{H}_{1}  $ and $ \delta \hat{H}_{2} $ introduce crucial modifications to the governing system equations.

\begin{figure}[t]
\centering \scalebox{0.5}
 {\includegraphics{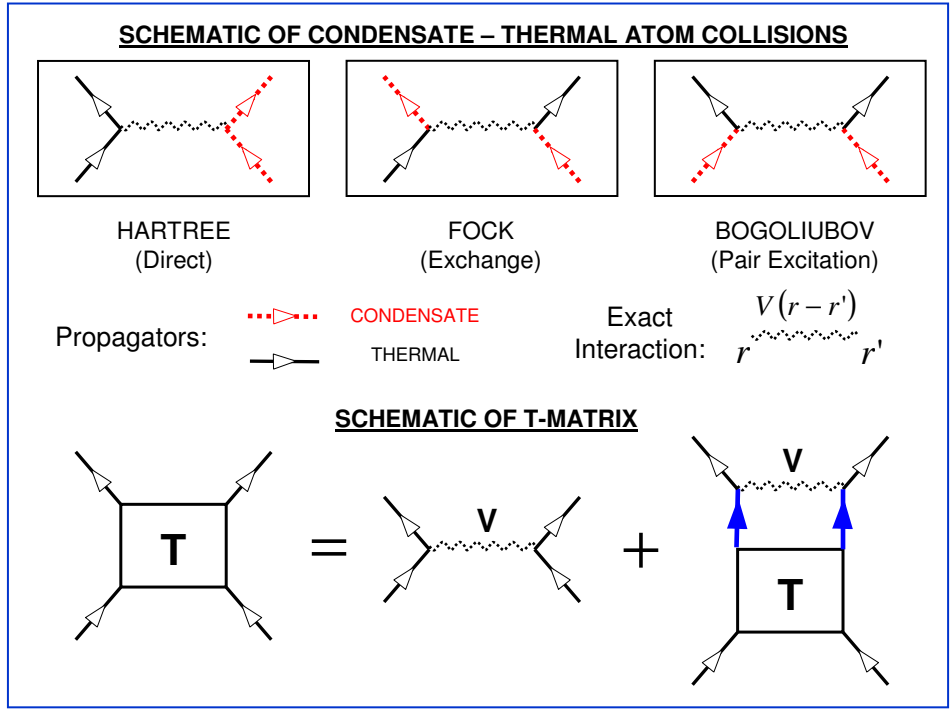}}
 \caption{
(colour online) Top: Schematic of `Hartree', `Fock' and `Bogoliubov' collisional processes involving both condensate and thermal atoms. 
Bottom: Schematic of the definition of the Transition (T) matrix in terms of the exact interatomic potential $V(\bldr-\bldr')$. In the simplest case, the upgrade is to the two-body T-matrix, which can then be justifiably approximated by the pseudopotential (along with an upper momentum cut-off). However, both the intermediate propagators (denoted by filled blue arrows in the latter term) and their corresponding energies may be dressed, either due to thermal occupation affecting the scattering into these states, or due to the states themselves being dressed to quasiparticle ones.
}
\label{Schematic_Processes}
\end{figure}

In the above expressions h.c.\ stands for hermitian conjugate, and we have separated off the so-called Hartree-Fock (HF) terms involving the collision of one condensate and one thermal atom from the Bogoliubov (BOG) terms which correspond to the collisional promotion of two condensate atoms to thermal modes (and its inverse process).
%
The physical significance of such contributions is displayed diagrammatically in Fig.\ \ref{Schematic_Processes} (top).
Such a separation enables us to identify various distinct approximations.

\subsubsection{The Hartree-Fock (HF) Limit:}

\label{HF_Regime}

In the Hartree-Fock limit, one first discards from the quadratic hamiltonian 
$(H_0+\hat{H}_1+\hat{H}_2)$ the latter contribution of Eq.\ (\ref{H_2}) which contains two annihilation or two creation operators. One is thus left with the reduced zero-temperature hamiltonian
$(H_0+\hat{H}_1+\hat{H}_2^{H})$, where we have introduced the notation
\be
\hat{H}_2^{H} = \int d\bldr \hat{\delta}^\dag \left( \hat{h}_0 + 2 g |\wfn(\bldr)|^2 \right) \hat{\delta} \;.
\ee
This hamiltonian is then generalized by the additional inclusion of the contributions ($\delta H_0^{HF} + \delta \hat{H}_1^{HF} + \delta \hat{H}_2^{HF}$) arising from the `higher-order' hamiltonians $\hat{H}_3$ and $\hat{H}_4$. 
Diagonalization of the resulting hamiltonian in the grand canonical ensemble leads essentially to 
\be
\hat{K}_{HF} 
= \int d \bldr \nopc \left[ \hat{h}_0  + 2g \left( |\wfn_0|^2 + \tilde{n}_0 \right) -\mu \right] \nopa \
\;,
\label{HF_Ham}
\ee
with the condensate wavefunction satisfying the generalized time-independent GPE
\begin{equation}
 \left[ \spop  + g|\wfn_0|^2 + 2g\tilde{n}_0 \right] \wfn_0  = \mu \wfn_0\;.
\label{GPE_HF_Static}
\end{equation}
The absence of any contributions involving two like (creation or annihilation) operators in the system hamiltonian in this limit thus implies that the system is still described by {\em single-particle} energies; however, these energies are modified both by the condensate mean field $|\wfn_0|^2$ and by the thermal atoms $\tilde{n}_0$, leading to dressed `Hartree-Fock' energies 
of the form
\be
\tilde{\varepsilon}_i(\bldr) = \varepsilon_{i} + 2g \left[ |\wfn_0(\bldr)|^2 + \tilde{n}_0(\bldr) \right] -\mu
\label{HF_Energy_mu}
\;,
\ee
where $\varepsilon_i$ is the corresponding single-particle energy in a trap.
The equilibrium thermal cloud density is given by $\tilde{n}_0(\bldr)=\sum_{i}|\spwa_i(\bldr)|^2 \langle \hat{a}_i^\dag \hat{a}_i \rangle$ where $\spwa_i(\bldr)$ are the corresponding eigenfunctions,
$\langle \hat{a}_i^\dag \hat{a}_i \rangle=[e^{\beta \tilde{\varepsilon}_i(\bldr)}-1]^{-1}$, and $\beta=1/(k_BT)$ is the inverse temperature ($k_B$ is Boltzmann's constant). 

Let us now briefly remark on the origin of the terminology `Hartree-Fock': The symmetrization of the matrix elements performed within the second quantization formalism via Eq.\ (\ref{matrixelement}) implies that in considering collisions involving thermal atoms, one is automatically including both `direct', or Hartree, contributions of the form $(0j|\hat{V}|0j)$ and `exchange', or Fock, terms   $(0j|\hat{V}|j0)$ (see Fig.\ \ref{Schematic_Processes} (top)). Careful consideration (for a detailed discussion see \cite{Pethick_Book}) shows that the appearance of the exchange term, which is not present for a pure condensate, implies that for a system with a fixed number of atoms, the interaction energy above the critical temperature is double that at $T=0$. Hence, the relative factor of two in the interaction contributions appearing in Eq.\ (\ref{GPE_HF_Static}), whereas such a distinction does not arise in the respective equation (Eq.\ (\ref{HF_Energy_mu})) for the excitation energies above the ground state, where the mean fields can be thought of as providing an additional potential for the atoms.

\begin{figure}
  \begin{center}
    \begin{tabular}{cc}
      \resizebox{65mm}{!}{\includegraphics{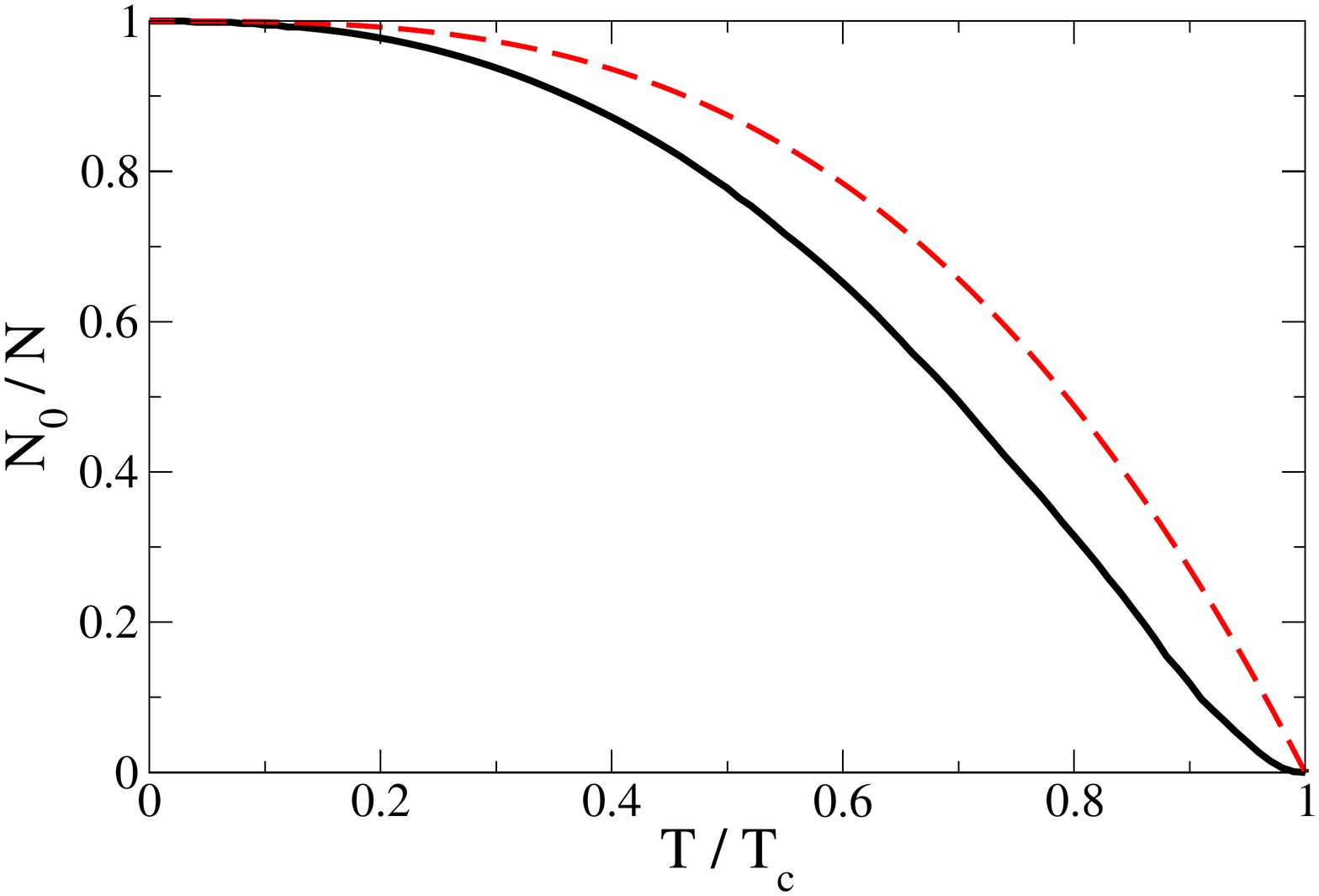}} &
      \resizebox{65mm}{!}{\includegraphics{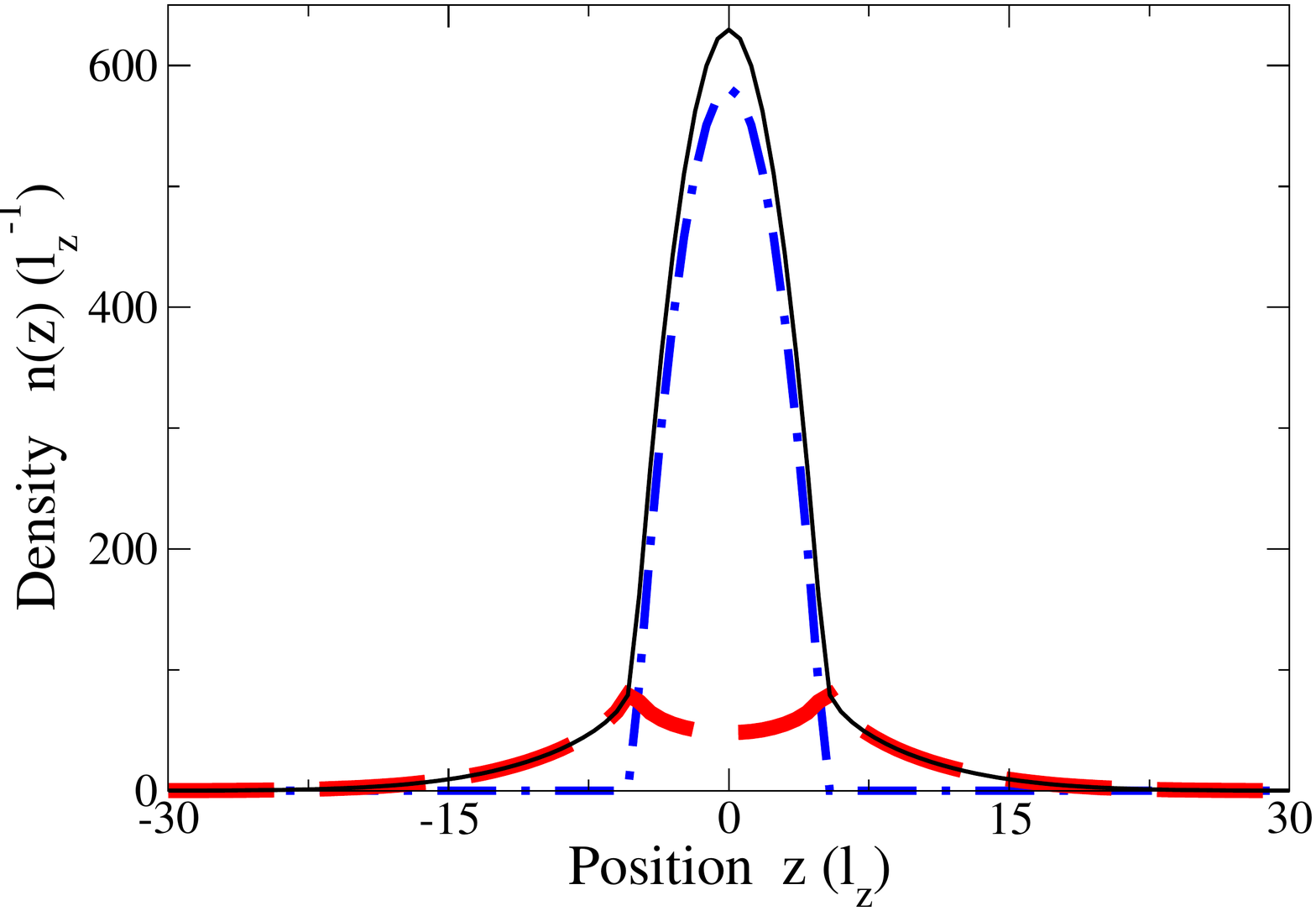}} 
    \end{tabular}
    \caption{
(colour online) Left: Condensate fraction as a function of reduced temperature as predicted by Hartree-Fock theory (solid, black) and for the ideal trapped Bose gas via $N_0/N=1-(T/T_c)^3$ (dashed, red), where $T_c$ is the transition temperature for an ideal gas.
Right: Total atomic density profile (solid, black) consisting of a condensate (dot-dashed, blue - calculated here within the Thomas-Fermi approximation \cite{Pethick_Book,Stringari_Review}) 
and a thermal (dashed, red) contribution. Shown is the column density of a dilute isotropically confined trapped 3D $^{23}$Na Bose gas consisting of $640,000$ atoms, with a condensate fraction of $20\%$. (Figure created by Stuart Cockburn).
}
    \label{HF_Figure}
  \end{center}
\end{figure}


The Hartree-Fock theory was used in numerous early studies of homogeneous condensates \cite{HF_1,HF_2}. To proceed further with using this theory in trapped gases, one often imposes an additional semi-classical approximation \cite{Pethick_Book,Giorgini_Semiclassical} which assumes
that all relevant quantities of the trapped system (such as densities) vary slowly on the typical length scale of the confining potential (i.e.\ the system locally resembles a homogeneous gas). The excitation spectrum can then be described in terms of momentum ${\bf p}$ via
\be
\tilde{\varepsilon}(\bldr,{\bf p}) = \varepsilon({\bf p}) + V_{\rm ext}(\bldr) + 2g \left[ |\wfn_0(\bldr)|^2 + \tilde{n}_0(\bldr) \right] -\mu
\label{HF_Energy}
\;,
\ee
where $\varepsilon({\bf p})=|{\bf p}|^2/2m$ is the energy of a free particle of momentum ${\bf p}$.
This semiclassical approximation enables a direct calculation of the non-condensate density via 
\be
\tilde{n}(\bldr) = \int \frac{d {\bf p}}{(2 \pi \hbar)^{3}} \frac{1}{e^{\beta \tilde{\varepsilon}(\bldr,{\bf p})}-1}
\label{HF_Density} 
\;,
\ee
as discussed further in \cite{Pethick_Book,Giorgini_JLTP}. 
Predictions of the Hartree-Fock theory for condensate fraction and atomic density profiles are shown in Fig.\ \ref{HF_Figure}, showing clearly the separation into condensate (dot-dashed, blue) and thermal cloud (dashed, red). In the absence of interactions, the thermal cloud in a harmonic trap acquires a gaussian density profile; however, the presence of effectively repulsive atomic interactions between the atoms, and thus also between the condensate and the thermal cloud, leads to the appearance of a local dip in the thermal density at the centre of the trap, where the condensate has its maximum density.

The Hartree-Fock basis will be used in our subsequent analysis (Sec.\ \ref{ZNG}) as the suitable basis for developing more advanced dynamical treatments. Before doing so, however, let us first present various alternative (static) formulations, which are more advanced in the sense that they include the dressing of particles into quasiparticles, and thus appropriately generalize the Bogoliubov equations of Sec.\ \ref{Bog-dG}.

\subsubsection{The Hartree-Fock-Bogoliubov (HFB) Limit:}

\label{HFB_Static}

This is a generalization of the Hartree-Fock regime, in which one includes {\em all quadratic} non-condensate operators to the hamiltonian, and also explicitly maintains correlations of pairs of like non-condensate operators, i.e. the anomalous average $\tilde{m}(\bldr,t) = \langle \nopa (\bldr,t) \nopa (\bldr,t) \rangle$ and its conjugate $\tilde{m}^*(\bldr,t) = \langle \nopc (\bldr,t) \nopc (\bldr,t) \rangle$. 
Inclusion of these additional contributions ($\delta H_0^{BOG} + \delta \hat{H}_1^{BOG} + \delta \hat{H}_2^{BOG}$) reduces the system hamiltonian to the following
quadratic form, known as the Hartree-Fock-Bogoliubov hamiltonian:
\begin{equation}
\hat{H} \approx \hat{H}_{HFB} = \left( H_0 + \delta H_0 \right) + \left( \hat{H}_1 + \delta \hat{H}_1 \right) + \left( \hat{H}_2 + \delta \hat{H}_2 \right)\;.
\label{H_HFB}
\end{equation}
%
%
This effective hamiltonian can be diagonalized via a Bogoliubov transformation of Eq.\ (\ref{bog-transform}).
Using the splitting 
$\copa = \wfn_0 + \nopa$ introduced earlier, 
the grand canonical Hamiltonians (\ref{K_1}) and 
(\ref{K_2}) take the following generalized `mean field' forms
\begin{eqnarray}
 \hat{K}_1' &=& \int d\bldr \, \nopc \left[ \left( \spop+
 g|\wfn_0|^2+2g\tilde{n}_0-\mu \right) \wfn_0 + g \tilde{m}_0 \wfn_0^* \right] + 
 {\rm h.c.},
\label{K_1mf} \\
\hat{K}_2' &=& \int d\bldr \left\{ \nopc \left( \spop + 2g 
 \left( |\wfn_0|^2  + \tilde{n}_0 \right) - \mu \right) \nopa \right. \nonumber \\
 &+& \left. \frac{g}{2} \left[ \left( (\wfn_0^*)^2 + \tilde{m}_0^* \right)  \nopa \nopa 
 + \left( \wfn_0^2 + \tilde{m}_0 \right) \nopc \nopc \right] \right\}\;.
\label{K_2mf}
\end{eqnarray}
As before, the equation describing the condensate is obtained by the requirement that $\hat{K}_1' =0$, which leads to the following time-independent form of the generalized GPE
\begin{equation}
 \left[ \spop  + g|\wfn_0|^2 + 2g\tilde{n}_0 \right] \wfn_0 + g\tilde{m}_0 
 \wfn_0^*= \mu \wfn_0\;.
\label{genGP-indie}
\end{equation}
Diagonalizing the 
$\hat{K}_2'$ term (\ref{K_2mf}) with the Bogoliubov transformation
(\ref{bog-transform}) leads to the generalized Bogoliubov equations
(c.f.\ Eqs.\ (\ref{BdG_Matrix_T0})):
\begin{eqnarray}
\left( 
\begin{array}{cc} 
\hat{\cal L}(\bldr)  & \hat{\cal M}(\bldr) \\ -\hat{\cal M}^{*}(\bldr) & -\hat{\cal L}^*(\bldr)
\end{array}
\right)
\left(
\begin{array}{l}
u_i(\bldr) \\ v_i(\bldr)
\end{array}
\right)
= \epsilon_i
\left(
\begin{array}{l}
u_i(\bldr) \\ v_i(\bldr)
\end{array}
\right)\;.
\label{HFB-uv}
\end{eqnarray}
where 
\bea
\hat{\cal{L}} (\bldr) = \hat{L}(\bldr) +2g \tilde{n}_0(\bldr) = \spop (\bldr)+ 2g |\wfn_0(\bldr)|^2 + 
2g \tilde{n}_0 (\bldr) -\mu 
\label{L_HFB}
\\
\hat{\cal M}(\bldr) = \hat{M}(\bldr) + g \tilde{m}_0(\bldr) = g [ \wfn_0 (\bldr) ]^2 + g \tilde{m}_{0}(\bldr)
\label{M_HFB} \;.
\eea
%
%
Thus, the excitations of the system are now Bogoliubov quasiparticles, whose energies are obtained from Eq.\ (\ref{HFB-uv}).
At finite temperature the Bogoliubov excitations will be thermally populated.
Since we are considering a static thermal cloud, the occupation of the 
quasiparticle states is diagonal, i.e.\ $\langle \hat{\beta}_i^\dagger 
\hat{\beta}_j \rangle_0= \delta_{ij} f_i^0$ and $\langle \hat{\beta}_i 
\hat{\beta}_j \rangle_0 =0$.
Here we have defined the so-called quasiparticle distribution function $f_i$, defined by
\begin{equation}
 f_i^0 = \frac{1}{e^{\beta \epsilon_i}-1}
\label{BE-dist}
\end{equation}
where 
the subscript $0$ denotes a static value. 
Thus, the equilibrium values of normal and anomalous averages can be expressed in terms of these distribution functions and Bogoliubov amplitudes $u_i(\bldr)$ and $v_i(\bldr)$ via \cite{MyJPhysB_Review}
\begin{eqnarray}
 \tilde{n}_0 (\bldr)= \langle \hat{\delta}^{\dag}(\bldr) \hat{\delta}(\bldr) \rangle =
\sum_i (|u_i (\bldr)|^2 + |v_i (\bldr)|^2) f_i^0  + |v_i (\bldr)|^2,
\label{tilden_0} \\
 \tilde{m}_0 (\bldr) = \langle \hat{\delta}(\bldr) \hat{\delta}(\bldr) \rangle =
\sum_i u_i (\bldr) v_i^* (\bldr) (1+2f_i^0 ).
\label{tildem_0}
\end{eqnarray}

The set of equations (\ref{genGP-indie})-(\ref{tildem_0}) are collectively termed the time-independent Hartree-Fock-Bogoliubov (HFB) equations. They can, in principle, be solved self-consistently \cite{Blaizot_Ripka}. Hence the
energies and eigenfunctions of the quasiparticles can be found
, as well
as the equilibrium condensate, `normal' and `anomalous' averages ($n_{c0}=|\wfn_0|^2$,
$\tilde{n}_0$ and $\tilde{m}_0$ respectively) - see e.g \cite{MyJPhysB_Review} and references therein. 
%


The HFB equations can be easily seen to reduce to the HF theory in the appropriate limit. To visualize this, consider setting the quasiparticle operators $\hat{b}_i$ to be equal to the single-particle operators $\hat{a}_i$, which can be achieved by setting $v_i(\bldr)=0$ in Eq.\ (\ref{bog-transform}). From Eq.\ (\ref{tildem_0}) we then see that $\tilde{m}_0(\bldr)=0$, which implies that $u_i(\bldr)$ reduces to a (dressed) single-particle eigenstate. The Bogoliubov equations thus reduce to
\be
\hat{\cal{L}}(\bldr) u_i(\bldr) = \epsilon_i u_i(\bldr) \rightarrow 
\left( \hat{h}_0 +2g |\wfn_0|^2+2g \tilde{n}_0 \right) u_i(\bldr) = \epsilon_i u_i(\bldr) \;,
\ee
thus mapping the excitation energies $\epsilon_i$ onto the HF energies $\tilde{\varepsilon}_i$ of Eq.\ (\ref{HF_Energy_mu}).

From one perspective, the inclusion of the anomalous average within HFB is desired, as it provides a better approximation to the (many-body) wavefunction, and hence it lowers the free energy of the system\footnote{Another appealing feature of the HFB 
approach is that it is `conserving', in the sense that the response functions generated by this approach can be shown to satisfy various conservation laws. A discussion of conserving vs. gapless approximations can be found in the seminal paper by Hohenberg and Martin \cite{Key_Paper_6}, with more recent summative discussions made by Griffin \cite{Griffin_HFB}, and additional related insight presented in the reviews of Andersen \cite{Andersen_Review} and Yukalov \cite{Yukalov_SB}.}. 
However,
as formulated above, these equations have two fundamental limitations which restrict their applicability. Firstly, the anomalous average defined by Eq.\ (\ref{tildem_0}) diverges as $i \rightarrow \infty$, which is a direct consequence of our somewhat careless handling of the effective interaction, thus requiring us to revisit the imposed pseudopotential approximation of Eq.\ (\ref{pseudopotential}).
Secondly, the homogeneous spectrum of elementary excitations predicted by the HFB equations does not vanish, as it should, in the zero momentum limit, i.e. it exhibits a `gap'. Both these subtle issues can be circumvented by more careful considerations, as discussed in the subsequent section.

    \begin{sidewaystable}
    \begin{center}
      \begin{tabular}{|c|c|c|c|c|}
        \hline
	         &
        EXPRESSION &
        ORIGIN &
        INTERPRETATION &
	DUE TO
        \\
	\hline   
           
	$H_0$ &
        $\int d \bldr \wfn^* \left( \hat{h}_0+(g/2)|\wfn|^2 \right) \wfn$ &
        $T=0$ Quadratic System Hamiltonian &
	$T=0$ Energy &
	$\wfn$
	\\
	\hline

	$\delta H_0^{HF}$ &
        $-g\int d \bldr \tilde{n}^2$ &
        From $\hat{H}_4$ via Eq.\ (\ref{mf-quad}) &
	$T>0$ Energy Shift &
	$\tilde{n}$
	\\
	\hline

	$\delta H_0^{BOG}$ &
        $-(g/2)\int d \bldr |\tilde{m}|^2 $ &
        From $\hat{H}_4$ via Eq.\ (\ref{mf-quad}) &
	$T>0$ Energy Shift &
	$\tilde{m}$
	\\
	\hline

	$\hat{H}_1$ &
        $\int d \bldr \nopc \left( \hat{h}_0+g|\wfn|^2 \right)\wfn +{\rm h.c.}$ &
        $T=0$ Quadratic System Hamiltonian &
	Defines $T=0$ GPE &
	$\wfn$
	\\

	&
	&
	&
	i.e.\ identifies chemical potential &
	\\
	\hline

	$\delta \hat{H}_1^{HF}$ &
        $2g\int d \bldr \nopc \tilde{n} \wfn +{\rm h.c.}$ &
        From $\hat{H}_3$ via Eq.\ (\ref{mf-trip}) &
	Additional $T>0$ GPE Contribution &
	$\tilde{n}$
	\\
	\hline
   
	$\delta \hat{H}_1^{BOG}$ &
        $g\int d \bldr \nopc \tilde{m} \wfn^* +{\rm h.c.}$ &
        From $\hat{H}_3$ via Eq.\ (\ref{mf-trip}) &
	Additional $T>0$ GPE Contribution &
	$\tilde{m}$
	\\
	\hline

	&
        &
        &
	Defines Spectrum of &
	$\wfn$
	\\

	$\hat{H}_2^{H}$ &
	$\int d \bldr \nopc \left( \hat{h}_0+2g|\wfn|^2 \right) \nopa$ &
	$T=0$ Quadratic System Hamiltonian &
	Elementary Excitations &
	\\

	&
	&
	&
	(Trap Eigenstates Modified &
	\\

	&
	&
	&
	by BEC Mean Field) &
	\\
	\hline

	$\hat{H}_2^{BOG}$ &
        $(g/2) \int d \bldr \left[ \nopc \nopc \wfn^2 + {\rm h.c.} \right]$  &
        $T=0$ Quadratic System Hamiltonian &
	Dresses Particles to Quasiparticles &
	$\wfn$
	\\
	\hline

	&
        &
        &
	$T>0$ Correction to &
	$\tilde{n}$
	\\

	$\delta \hat{H}_2^{HF}$ &
	$2g \int d \bldr \nopc \tilde{n} \nopa$ &
	From $\hat{H}_4$ via Eq.\ (\ref{mf-quad}) &
	Single-Particle Excitation Spectrum &
	\\

	&
	&
	&
	(Hartree-Fock Energies) &
	\\
	\hline

	$\delta \hat{H}_2^{BOG}$ &
        $(g/2) \int d \bldr \left[ \nopc \nopc \tilde{m} + {\rm h.c.} \right]$  &
        From $\hat{H}_4$ via Eq.\ (\ref{mf-quad}) &
	$T>0$ Correction to &
	$\tilde{m}$
	\\

	&
	&
	&
	Quasiparticle Excitation Spectrum &
	\\
	\hline

	&
	&
	(For later use)&
	&
	\\
	\hline

	$\delta H_0^{TRIP}$ &
        $g \int d \bldr \left[ \langle \nopc \nopa \nopa \rangle \wfn^* + {\rm c.c.} \right]$  &
        From $\hat{H}_3$ via Appropriate &
	See Sec.\ \ref{ZNG} &
	$\langle \nopc \nopa \nopa \rangle$
	\\

	&
	&
	Generalized Decoupling Approximations &
	&
	\\
	\hline

	$\delta \hat{H}_1^{TRIP}$ &
        $g \int d \bldr \left[ \nopc \langle \nopc \nopa \nopa \rangle + {\rm h.c.} \right]$  &
        From $\hat{H}_4$ via Eq.\ (\ref{dH_1_Trip}) &
	See Sec.\ \ref{ZNG} &
	$\langle \nopc \nopa \nopa \rangle$
	\\
	\hline


      \end{tabular}
      \caption{\label{Table_Hamiltonians}
Potential characteristic contributions to the effective generalized quadratic system hamiltonian at finite temperatures; the origin and physical interpretation of each term is clearly indicated:
in brief, contributions $H_0$, $\hat{H}_1$ and $\hat{H}_2=(\hat{H}_2^{H}+\hat{H}_2^{BOG})$ arise in the full original system hamiltonian (Eqs.\ (\ref{H-sep})-(\ref{H_2}), whereas contributions $\delta \hat{H}_i^{\cdots}$ arise from terms $\hat{H}_3$ and $\hat{H}_4$ involving three or four operators, which are only reduced to quadratic form by the generalized mean field approximations of Eqs.\ (\ref{mf-quad}), (\ref{mf-trip}). For completeness, we have also included here contributions arising from more general approximations which explicitly include the quantity $\langle \nopc \nopa \nopa \rangle$
(see Sec. \ref{HFBT_Dynamics}).
      }
    \end{center}
  \end{sidewaystable}


\subsection{Introduction of an Effective Interaction:}
\label{effect-int}

In our discussion so far, we have simply replaced the interatomic potential $V(\bldr - \bldr')$ by an effective contact potential $g \delta( \bldr - \bldr')$, via Eq.\ (\ref{pseudopotential}), where $g$ is an effective constant strength of binary interactions; as this approximation is actually related to the above-mentioned problems arising with the HFB formulation, we should now justify its use and discuss how to overcome any associated limitations.
 
It is natural to wonder in what sense the exact non-local interatomic potential $V(\bldr-\bldr')$ between two atoms can be meaningfully
replaced by a contact potential. The answer actually lies in the details of the particular system under consideration: For the low temperature dilute systems we are considering here\footnote{Typical conditions for this are $n a^3 \ll 1$ and $a \ll \lambda_{dB}$, where $a$ is the s-wave scattering length and $\lambda_{dB} = \sqrt{2 \pi \hbar^2 / m k_B T}$ is the thermal de Broglie wavelength of the atoms.},
atoms spend most of their time far apart from each other, i.e.\ at distances much larger than the typical range of the interatomic potential, such that short-range correlations are unimportant. Typical atomic interactions can thus for most practical purposes be described as scattering processes, and one is only interested in the effect that the full potential has at large distances $|\bldr-\bldr'|$, i.e. in the asymptotic scattering states. The only effect of atomic interactions on these states is a change of phase of the quantum-mechanical wavefunction
\cite{Pethick_Book,Huang_Book,Fetter_Varenna,Dalibard_Varenna}.


It turns out that such a phase shift can be well-reproduced by means of a so-called `pseudopotential'. 
To explain this,
we consider the scattering of two atoms located at a relative distance $r$ from each other, where $r$ is much larger than the effective range of their interatomic potential. At sufficiently low energies, only one scattering channel is energetically accessible, the so-called s-wave scattering one. In this case, the net effect of the potential can be well-approximated by a pseudopotential of the form \cite{Huang_Book,Key_Paper_2}
\be
V_{\rm pseudo}(r) = g \delta(r) \frac{\partial}{\partial r} r \rightarrow g \delta(r)
\label{pseudo_delta}\;.
\ee
%
%
It is important to note that the above pseudopotential is actually an operator which acts on the particle wavefunction, thus leading to terms of the form
$V_{\rm pseudo}(r)\wfn(r)=g \delta(r) \partial (r \wfn)/\partial r$. 
However, the problem simplifies considerably under the assumption that the
pseudopotential acts on unperturbed free-particle wavefunctions; in this limit, the operator $(\partial/\partial r)$ can simply be replaced by $1$, reducing the problem as indicated by the arrow in Eq.\ (\ref{pseudo_delta}).
However, extreme care is needed when using a hamiltonian based on Eq.\ (\ref{pseudo_delta}), and there are in particular two related issues which should be considered:
firstly, under what conditions can the presence of the operator $\partial/\partial r$ be overlooked, and secondly what is a suitable value (or function) for $g$ which would then correctly describes the scattering process?

To address both of these, we note that a scattering process can be viewed as the end result of a collisional process which in general includes the effect of repeated virtual collisions between the two atoms; mathematically, this corresponds to the repeated action of the exact interatomic potential $V(r)$. While the effect of each repeated collision may be very large, one can actually construct an infinite series over such virtual collisions, i.e.\ a series in powers of the exact interatomic potential and their corresponding propagators. This series actually converges, and such a mathematical construction amounts physically to upgrading the exact interatomic potential to an effective one, known as the Transition (or simply T) matrix; the defining relation of the T-matrix in terms of the actual interatomic potential, the so-called Lippmann-Schwinger equation \cite{Pethick_Book,Stoof_NIST}, is depicted diagrammatically in Fig.\  \ref{Schematic_Processes} (bottom). 
It is important to realize that the pseudopotential approximation $g \delta(r)$ is only strictly meaningful when imposed onto this upgraded effective interaction. 

So, what would happen if one were to overlook this, and replace the exact interatomic potential in the full system hamiltonian of Eq.\ (\ref{H}) by $g \delta(\bldr -\bldr')$ as hinted by Eq.\ (\ref{pseudopotential})? As long as the subsequent treatment is restricted to first order perturbation theory, this would be fine. However, imagine one chose instead to use this hamiltonian to compute properties of the system within second order perturbation theory. Considering for simplicity the case of a homogeneous Bose gas, second order perturbation theory would generate terms containing the interaction strength contribution $ g^{(2)}=g[1+g \sum_{p}' (1/\varepsilon_p)]$ where the prime implies that the summation is restricted to nonzero momenta - see \cite{Pitaevskii_StatisticalPhysics_2} for more details.
However, this contribution is easily seen to diverge for large momenta $p$ (so-called ultraviolet divergence). Such a divergence has appeared here because, without realizing it, the above procedure has resulted in double-counting (due to the use of a constant interaction strength combined with an unrestricted summation over momenta). 
The reason is that, in this second order perturbation theory, the effective interaction strength $g$
should appear for the total interaction strength to second order, and not
for each of its individual contributions as implicitly assumed in the above $ g^{(2)}$ expression. 
In this context (i.e.\ when the pseudopotential approximation has already been made in the original hamiltonian), the problem can nonetheless be solved by the technique of `renormalization' whereby one simply replaces the second-order expression for the effective interaction strength $g^{(2)}$ by $g$ \cite{Lee_Huang_Yang,Pitaevskii_StatisticalPhysics_2}.
However, one can avoid the need for such renormalization by ensuring that the pseudopotential approximation is correctly implemented in the calculations; in other words, the pseudopotential approximation should only be made after the exact interatomic potential has been upgraded to the effective interaction given by the T-matrix. 
In fact the problem above arose because the pseudopotential approximation was imposed on the first order contribution (Born approximation) in the series expansion of the T-matrix in terms of the exact interatomic potential.

We should still address the related issue of the strength of the effective interaction given by the T-matrix; in general, this should depend on the momentum of the incoming particles \cite{Stoof_NIST,Shi_Thesis}.
For very dilute gases, a natural first assumption would be that the scattering process is taking place in vacuum; in that case, the resulting effective interaction between two atoms is known as the two-body T-matrix. 
As long as one is only considering atoms whose relative momentum is small, one can ignore the momentum dependence of the T-matrix, thus approximating it by a constant, $g$, equal to the zero-energy, zero-momentum limit of the full T-matrix. The value of this latter quantity
is fixed by a single parameter, namely the shift of the projection of the asymptotic wavefunction onto the relative coordinate axis from the origin - a quantity known as the s-wave scattering length, $a$. The scattering length can be either positive, or negative, depending on the intricate details of the exact interatomic potential, with the respective sign indicating effective repulsion, or attraction, between the atoms.
%
%
It is important to note that this constant value is only a good approximation when describing collisions between low-momentum states, and it can be thought to arise by the elimination of high-lying modes from the problem \cite{Proukakis_Burnett_Stoof,Morgan_JPhysB}. In fact, for momenta larger than $\hbar/a$, the full two-body T-matrix would rapidly decrease to zero, an effect not reflected in our choice of a constant strength $g$, which would thus need to be implemented in conjunction with an upper momentum cut-off \cite{Beliaev1,Beliaev2,Popov}.

We have thus discussed how a contact interaction potential arises in the context of a microscopic theory of ultracold Bose gases, which is actually a crucial point in assessing the validity of any ab initio theory. 
As long as this replacement can be routinely performed, one does not need to worry about the above subtle issues, i.e.\ one can start off from the hamiltonian of Eq.\ (\ref{H-g}); this is, for example, true in arriving at the HF theory of Sec.\ \ref{HF_Regime}.
However, in trying to construct more advanced theories - such as the HFB - from a microscopic perspective, one should be more cautious, and investigate the following issues, as expounded clearly by Burnett \cite{Burnett_LesHouches}:
%
\begin{enumerate}
\item Is the introduction of the relevant effective interaction (T-matrix), which is then replaced by a contact pseudopotential, performed consistently at the level of the considered approximation?
\item How does the fact that collisions in a BEC take place within a medium, rather than in vacuum, affect the above considerations?
\end{enumerate}


\subsection{Problems of the HFB Formulation:}
\label{HFB_Popov}

HFB differs from the simpler HF by the explicit consideration of
the static anomalous average, $\tilde{m}_0(\bldr)$ of Eq.\ (\ref{tildem_0}). Since the HFB equations, Eqs.\ (\ref{genGP-indie})-(\ref{tildem_0}), are formulated in terms of the constant interaction strength $g$, the summation in Eq.\ (\ref{tildem_0}) should exclude high-lying modes, as these have already been implicitly considered in replacing $V(\bldr-\bldr')$ by the effective interaction in vacuum (two-body T-matrix). If this were not done, then the resulting expression for $\tilde{m}_0(\bldr)$ would blow-up as $i \rightarrow \infty$, i.e. it would exhibit an ultraviolet divergence. Having identified the origin of this divergence, we can routinely eliminate it from the problem, by regularizing the anomalous average. This is achieved via the subtraction of the high energy part, whose contribution corresponds precisely to the difference between the actual and the effective interatomic potential (which would otherwise be double-counted). We thus write \cite{Burnett_LesHouches,Morgan_JPhysB}
\be
\tilde{m}_0(\bldr) \rightarrow \tilde{m}_0^R(\bldr) 
= \tilde{m}_0(\bldr) - lim_{i \rightarrow \infty} u_i(\bldr) v_i^*(\bldr)\;.
\label{tilde_m0R}
\ee
Any subsequent treatment including the anomalous average has to be performed with this regularized form, so we will henceforth always quote $\tilde{m}_0^R(\bldr)$ as the anomalous average, implicitly assuming that such an essential `renormalization' has been performed.

The second related problem arising from the inclusion of the anomalous average is the appearance of a gap in the excitation spectrum of the homogeneous gas.
As already done for $T=0$ in Sec.\ \ref{Bog-dG}, we consider here the predictions of the HFB equations in the case of a uniform gas, for which
$V_{\rm ext} (\bldr) =0$ and 
$\wfn_0 (\bldr)$, $\tilde{n}_0 (\bldr)$ and $\tilde{m}_0^R (\bldr)$ are 
constant. From Eq.\ (\ref{genGP-indie}) the chemical potential becomes 
%
$ \mu = g(|\wfn_0|^2+2 \tilde{n}_0 + \tilde{m}_0^R)$.
%
On the other hand, the energy spectrum is given by
$ \epsilon(p) = [ p^2/2m + 2g(|\wfn_0|^2+\tilde{n}_0) - \mu ]^2
- g^2  (|\wfn_0|^2+\tilde{m}_0^R  )^2$.
%
%
The dispersion relation for the quasiparticles is obtained by substituting the value of $\mu$ into this expression. However, for $p\rightarrow 0$ 
the energy spectrum has a finite value (proportional to $\tilde{m}_0^R$), or in other words, it has a gap.    
This is however in direct contradiction with the Goldstone theorem; the latter requires that the energy spectrum arising from a theory based on symmetry-breaking should be gapless, which is equivalent to the statement that it should cost zero energy to excite the lowest mode (known as the Goldstone mode) \cite{Critical_Book}.
%
%
The HFB theory must thus be modified to be in accordance with this theorem.
The simplest, yet somewhat heuristic, way to solve this problem is to completely ignore the 
`anomalous average' $\tilde{m}_0$. Such an approximation was routinely made in the early literature, and is often called the 
{\it HFB-Popov approximation}, as discussed by Griffin in \cite{Griffin_HFB,Shi_Griffin}, although objections have been noted as to the use of this terminology \cite{Yukalov_Popov}.

\subsubsection{The HFB-Popov (HFBP) Limit:}

\label{HFB_Popov}

The Hartree-Fock-Bogoliubov-Popov (HFBP) limit is an intermediate regime between Hartree-Fock and Hartree-Fock-Bogoliubov.
In comparison to HF, it additionally includes the contribution $\hat{H}_{2}^{BOG}$ which dresses the finite-temperature single-particle energies to quasiparticle energies, whereas it can also be obtained from the full HFB theory by discarding {\it all} 
$\delta \hat{H}_i^{BOG}$ contributions ($i=0,1,2$).
Mathematically, this can be expressed as 
\bea
\hat{H}_{HFBP}&=&\hat{H}_{HF}+\hat{H}_2^{BOG} \nonumber \\
&=& \hat{H}_{HFB} - \left( \delta \hat{H}_0^{BOG}+ \delta \hat{H}_1^{BOG}+\delta \hat{H}_2^{BOG}\right)\;,
\eea
with the system hamiltonian in this regime given by
\be
\hat{H}_{HFBP}= \left( \hat{H}_0+\delta \hat{H}_0^{HF} \right) + \left( \hat{H}_1+\delta \hat{H}_1^{HF} \right)
+ \left( \hat{H}_2+\delta \hat{H}_2^{HF} \right)\;.
\ee
This leads to the same GPE as in the HF limit, i.e.\ the condensate is described by Eq.\ (\ref{GPE_HF_Static}). However, contrary to the HF limit, quasiparticle dressing is actually included here via the Bogoliubov equations of Eq.\ (\ref{HFB-uv}), in which the respective operators are now modified according to:
\bea
\hat{\cal{L}} (\bldr) \rightarrow \hat{\cal{L}}_{P} (\bldr) = \hat{\cal{L}} (\bldr) = \hat{L}(\bldr) +2g \tilde{n}_0(\bldr) \nonumber \\
\hspace{3.4cm} = \spop (\bldr)+ 2g |\wfn_0(\bldr)|^2 + 2g \tilde{n}_0 (\bldr) -\mu 
\label{L_HFBP}
\\
\hat{\cal M}(\bldr) \rightarrow \hat{\cal M}_{P}(\bldr)= \hat{\cal M}(\bldr) -g \tilde{m}_0(\bldr) = \hat{M}(\bldr) = g [ \wfn_0 (\bldr) ]^2
\label{M_HFBP}\;.
\eea
It is easy to verify that 
this set of equations leads to a gapless energy spectrum (and hence corresponds to a better theory for calculating frequencies of elementary excitations). However, this approximation may also be problematic at very low temperatures $T \ll T_c$, since then $\tilde{m}_0$ is of
the same
order as $\tilde{n}_0$ \cite{Yukalov_Popov}.
 
Takano \cite{Takano} was, to the best of our knowledge, the first person to point out (using slightly different arguments to those presented above) that the gap problem can actually be removed by going beyond the mean-field approximations of Eqs.\ (\ref{mf-quad}), (\ref{mf-trip}) via the consistent consideration of correlations of three fluctuation operators. The next section presents two possible (yet somewhat heuristic) generalized approaches along those lines.

\subsection{Static Generalized Many-Body Theories:}
\label{many-body}

Having identified the difficulties with the HFB model, one can now attempt to construct an improved theory which includes the effect of the pair anomalous average, but simultaneously also yields a gapless homogeneous excitation spectrum.

The strength of the effective interaction employed thus far corresponds to collisions taking place in vacuum, whereas in our system there is actually an active atomic medium present. This will in general affect the scattering in two ways \cite{Burnett_LesHouches}: (i) firstly, the intermediate states - denoted by thick (blue) lines and filled arrows in Fig.\ \ref{Schematic_Processes} -  may actually be thermally occupied, thus providing a bosonic enhancement for the transfer rates into those states; (ii) moreover, the intermediate states may actually be dressed (quasiparticle) states, instead of single-particle ones. 
The anomalous average $\tilde{m}_0(\bldr)$ contains information about correlations between two nearby atoms. We have already argued that its contribution over high-lying modes is related to the difference between the actual and the effective interatomic potential. It should therefore come as no surprise that its static value, $\tilde{m}_0^R$, over low-lying modes contains information about modifications to the effective interaction due to the presence of a medium; in technical jargon, inclusion of $\tilde{m}_0(\bldr)$
%
has the effect of upgrading the previously-considered two-body T-matrix describing scattering in vacuum to the so-called many-body T-matrix \cite{Stoof_NIST,Shi_Griffin}, which describes the additional modifications due to the presence of the medium (see also \cite{Morgan_T_Matrix_1,Morgan_T_Matrix_2} for a more detailed handling of many-body theories which also includes the regimes of lower dimensionality); these theories thus clearly extend beyond the HFB limit.


Following Proukakis, Morgan, Choi and Burnett, 
one can thus define a generalized effective interaction as \cite{Proukakis_Morgan}:
\begin{equation}
g(\bldr) = g \left( 1 + \frac{\tilde{m}_{0}^R}{\wfn_{0}^{2}} \right) \label{U_eff}\;.
\end{equation}
Such a definition enables us to express the
interaction term between two condensate atoms in the form 
$g(\bldr) |\wfn_0|^2 \wfn_0 = g \left[ |\wfn_0|^2 + \tilde{m}_{0}^R \right] \wfn_0$. 
A careful consideration of the HFB equations reveals that such a generalized interaction does not enter in all terms of the HFB equations; in particular, even though $\tilde{g}(\bldr)$ appears explicitly in the finite temperature GPE of Eq.\ (\ref{genGP-indie}), there is nonetheless
no $\tilde{m}_{0}^R$ contribution within the $\hat{\mathcal{L}}$ operator of Eq.\ (\ref{L_HFB}).
This leads to an inconsistency in the manner in which
interactions are handled within the HFB formalism, and can be identified as the cause for the appearance of a gap in the homogeneous excitation spectrum at low momenta. 
This problem is a direct consequence of the truncation of the coupled equations of motion to some particular order \cite{Andersen_Review}, and has already been identified in \cite{Stoof_Variational}. Nonetheless, it has been suggested \cite{Proukakis_Morgan} that this problem can be `fixed by hand', by forcing the interaction strength in the HFB equations to take the modified form given in Eq.\ (\ref{U_eff}).

Although the above arguments hold for the collisions between two condensate atoms,
an immediate issue arises regarding the form of the interaction strength when one of the atoms belongs to the thermal cloud. Consistency requires this interaction to be also computed in the many-body limit. However, due to the large range of values of the relative momenta between the two colliding atoms, a full treatment, expounded analytically in \cite{Stoof_NIST,Shi_Thesis}, is numerically quite involved, and one typically resorts to approximations. In fact, an interpretation of the generalized effective interaction $g(\bldr)$ can be made by resorting to the homogeneous limit where the role of many-body effects is well-known. Such a comparison reveals that the generalized effective interaction of Eq.\ (\ref{U_eff}) indeed includes one aspect of the many-body effects, namely the population of intermediate states, but it does not include the dressing of these states to quasiparticles. Formally, this is equivalent to the zero-energy zero-momentum limit of the full (many-body) T-matrix.
In this case, the many-body T-matrix describing the collision between a condensate and a thermal atom reduces to the corresponding two-body expression in vacuum, i.e. the interaction strength is simply equal to $g$. In the other extreme, one can approximate the effective interaction strength between a condensate and a thermal atom by the same expression as for two condensate atoms, i.e. $g(\bldr)$ - the latter expression was also subsequently derived by rigorous consideration of pseudopotentials by Olshanii and Pricoupenko \cite{Olshanii_Pricoupenko}. 

\begin{figure}
  \begin{center}
    \begin{tabular}{cc}
      \resizebox{60mm}{!}{\includegraphics{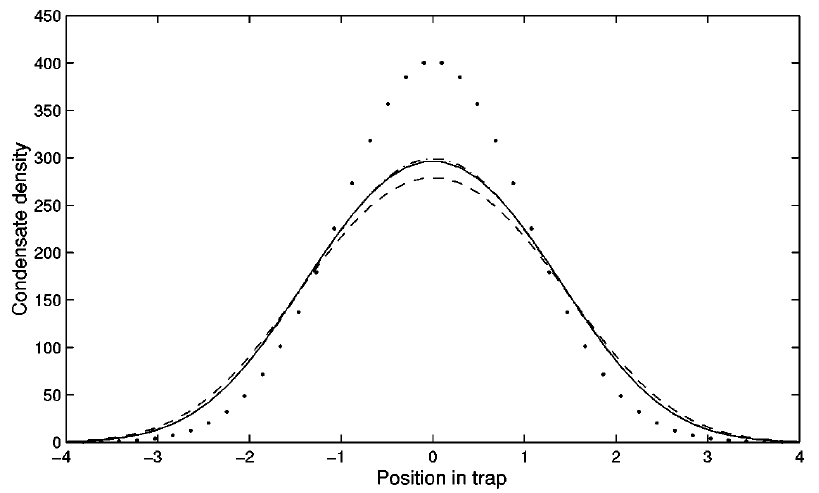}} &
      \resizebox{60mm}{!}{\includegraphics{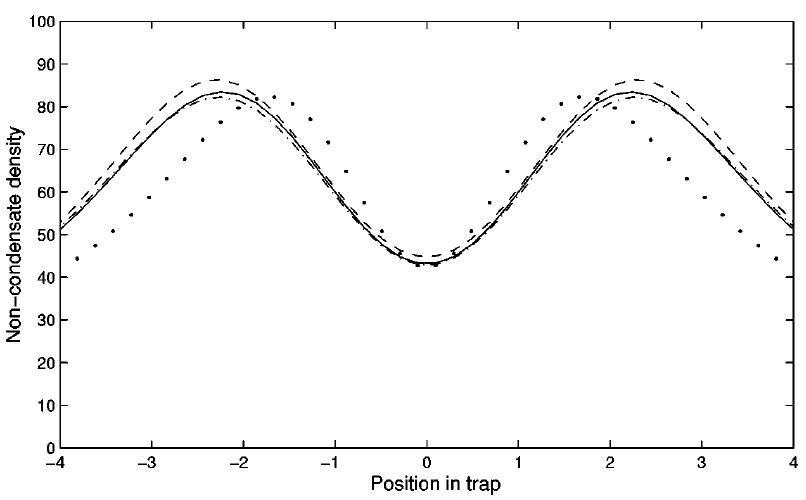}} \\
      \resizebox{60mm}{!}{\includegraphics{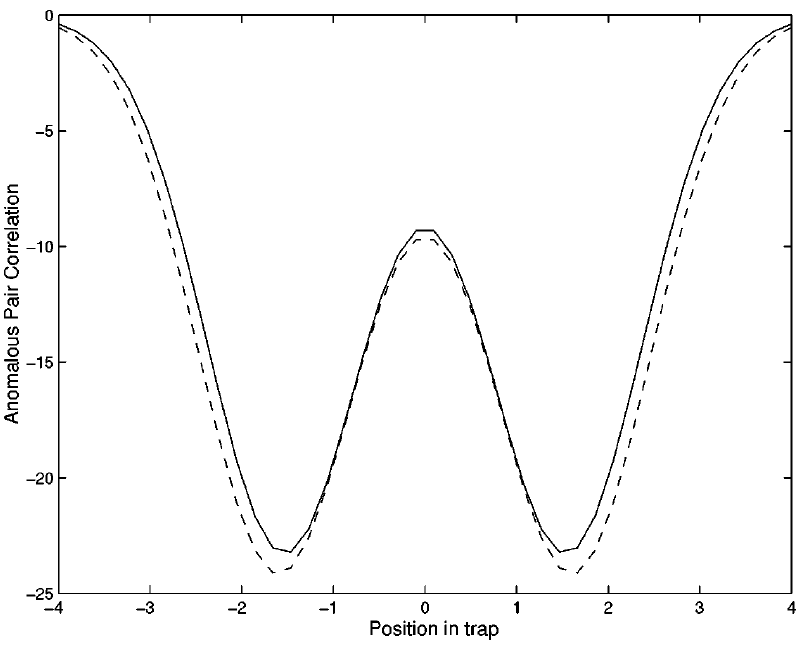}} &
      \resizebox{60mm}{!}{\includegraphics{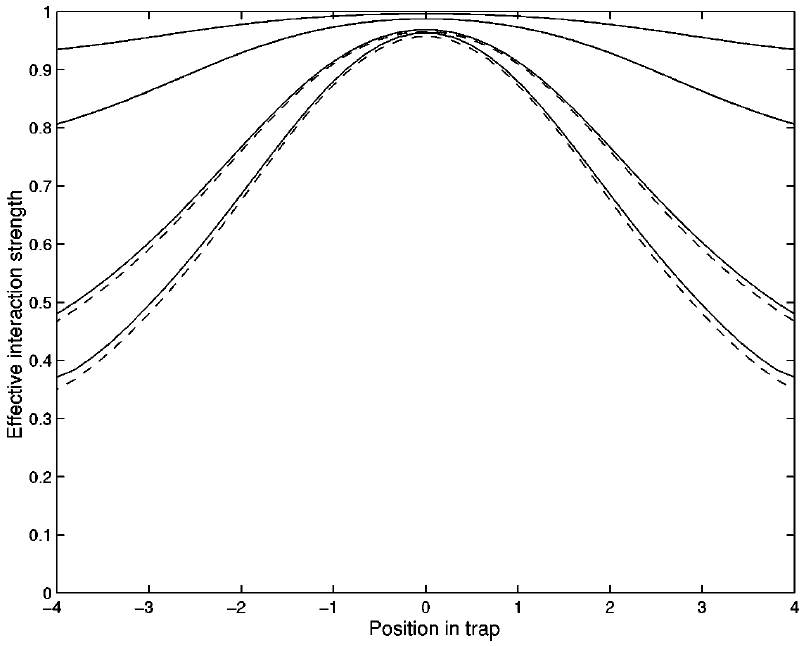}} \\
    \end{tabular}
    \caption{
Equilibrium spatial dependence (near the trap centre) of the generalized mean fields for a one-dimensional trapped gas consisting of 2000 atoms at a temperature $T/T_c \approx 0.6$, where there is an approximate condensate fraction of $50\%$ (here $T_c$ denotes an approximate ideal gas degeneracy temperature \cite{Ketterle_vanDruten_PRA}); densities are plotted in units of $l_z^{-1}$ and position in units of $l_z$. 
Top Row: Condensate $|\wfn_0(z)|^2$ (left) and thermal cloud $\tilde{n}_0(z)$ (right) computed self-consistently for the generalized HFB approaches ($g_c=g(z)$) with $g_t=g(z)$ (dashed) or $g_t=g$ (solid) and for HFB-Popov (dot-dashed); dotted lines indicate corresponding results in the limit when interactions between condensate atoms are ignored, and only interactions between condensate and thermal atoms are maintained.
Bottom Row: Corresponding profile for the pair anomalous average $\tilde{m}_0^R(z)$ within the two generalized HFB approaches (left); Typical spatial dependence of the generalized effective interaction $g(z)$ within the two generalized HFB theories, plotted in units of the constant strength $g$, at different temperatures - from top to bottom: $T/T_c \approx 0.1$, $0.2$, $0.6$ and $0.9$. 
(Reprinted figures with permission from N.P.\ Proukakis, S.A.\ Morgan, S.\ Choi and K.\ Burnett, {\it Phys. Rev. A} {\bf 58}, 2435 (1998). Copyright (1998) by the American Physical Society.)
}
    \label{Fig:HFB}
  \end{center}
\end{figure}


The above physically-motivated, yet mathematically ad hoc procedure leads in general to the following somewhat generalized set of Bogoliubov equations
%
\begin{eqnarray}
\left( 
\begin{array}{cc} 
\hat{\cal L'}(\bldr)  & \hat{\cal M'}(\bldr) \\ -\hat{\cal M'}^{*}(\bldr) & -\hat{\cal L'}^*(\bldr)
\end{array}
\right)
\left(
\begin{array}{l}
u_i(\bldr) \\ v_i(\bldr)
\end{array}
\right)
= \epsilon_i
\left(
\begin{array}{l}
u_i(\bldr) \\ v_i(\bldr)
\end{array}
\right)\;.
\label{BdG_GHFB}
\end{eqnarray}
where 
\bea
\hat{\cal L'} (\bldr) = \spop (\bldr)+ 2 g_{c}(\bldr) |\wfn_0(\bldr)|^2 + 
2 g_{t}(\bldr) \tilde{n}_0 (\bldr) -\mu \\
\hat{\cal M'} (\bldr) = g_c [ \wfn_0 (\bldr) ]^2\;.
\eea
Here $g_{c}(\bldr)$ denotes the interaction strength between two condensate atoms given by $g_{c}(\bldr) = g(\bldr)$, whereas $g_{t}(\bldr)$ expresses the effective interaction strength between a condensate and a thermal atom.
%
Clearly these equations are also coupled to a suitably-generalized GPE, given by
\be
\left[ \hat{h}_{0}  + g_{c}(\bldr) | \wfn_0 |^{2}  + 2 g_{t}(\bldr) \tilde{n}_0(\bldr) \right] \wfn_0(\bldr) = \mu \wfn_0(\bldr)\;.
\label{GPE_GHFB}
\ee
The above discussion leads naturally to two different theories, which we shall henceforth collectively refer to as generalized HFB, depending on whether 
$g_{t}(\bldr)=g_{c}(\bldr)=g(\bldr)$, or $g_{t}(\bldr) = g$ .
Clearly the limiting case $g_c(\bldr)=g_t(\bldr)=g$ corresponds preciselt to the HFB-Popov limit discussed in Sec.\ \ref{HFB_Popov}. 
Such generalized theories were applied with some success to the study of finite temperature excitation 
frequencies \cite{Hutchinson_GHFB} (see Sec.\ \ref{Excitations}), 
to vortices \cite{GHFB_Vortex}, and to the coherence of 2D condensates \cite{HFB_2D_1,HFB_2D_2}.
Note that a related approach for the self-consistent calculation of the coupling constant appearing in the GPE was suggested in \cite{Cherny_Brand}, whereas
alternative approaches to overcome the problem of the gap of the HFB theory are discussed in \cite{Yukalov_Popov,HFB_Tommasini,HFB_Kita}. 



It is appropriate to make a brief comparison of the approaches mentioned above, as discussed in \cite{Proukakis_Morgan} (see also \cite{Hutchinson_GHFB,MyJPhysB_Review,GHFB_Vortex}). Fig.\ \ref{Fig:HFB} displays the position dependence of the generalized mean fields for the previously discussed theories. Focusing initially on the normal averages $\wfn_0$ and $\tilde{n}_0$, we note that the predictions of all such theories differ considerably from the corresponding ones in the limit when condensate-condensate interactions are completely ignored, and the only interactions considered are those between condensate and thermal atoms (in the various approximations); in particular, inclusion of repulsive condensate-condensate interactions leads to the condensate wavefunction being more spread out and having a lower central density; in comparison, the differences in the respective profiles predicted by the various interacting theories are generally small, with the only noticeable deviation arising for the case $g_t=g(\bldr)$.
Investigation of the change in the temperature dependence of the condensate fraction again reveals marginal differences, with the corresponding predictions for excitation frequencies only being significantly different for the case $g_t=g(\bldr)$ (see also Sec.\ \ref{Excitations}).
The anomalous average $\tilde{m}_0^R$ is found to be negative and to exhibit minima at opposite off-centred points, with its value exhibiting a local maximum at the trap centre. As a result, the generalized effective interaction $g(\bldr)$ is consistently smaller than, or equal to, $g$. In particular, $g(\bldr)$ exhibits a local maximum at the centre of the trap (where the condensate density is largest) and approaches the value of $g$ asymptotically far from the trap centre; moreover, its central value $g(0)$ is temperature dependent and reaches its minimum value at the transition temperature.

    \begin{sidewaystable}
    \begin{center}
      \begin{tabular}{|c|c|c|c|c|c|c|}
        \hline
	         &
        Single- &
        \multicolumn{2}{c|}{Quasiparticle}&
        Respective &
\multicolumn{2}{c|}{Bogoliubov Equation Operators}
        \\
                 
       & Particle & \multicolumn{2}{c|}{Energies} & Hamiltonian & \multicolumn{2}{c|}{}  \\ [0.3ex]
        & Energies & $T=0$ & $T>0$ & Contributions & $\hat{\cal L}(\bldr)$  & $\hat{\cal M}(\bldr)$ \\
	& (Dressed) & & & & & \\ [0.3ex]
        \hline

        Bogoliubov  &  &     &          &       &       &      \\
        ($T=0$) & & \checkmark & & $H_0+\hat{H}_1 +\hat{H}_2$ & $\hat{h}_0(\bldr) + 2 g | \wfn_0(\bldr) |^2  - \mu$ & $g [ \wfn_0(\bldr) ]^2$ \\
        & & & & & & \\
        \hline 
        & & & & & & \\
        Hartree-Fock & \checkmark & &           &  $H_0+\hat{H}_1 +\hat{H}_2^{H}$       &   &      \\
        & & & & $+\left( \delta H_0^{HF} +\delta \hat{H}_1^{HF} +\delta \hat{H}_2^{HF}\right)$ & $\hat{h}_0(\bldr) + 2 g | \wfn_0(\bldr) |^2 + 2 g \tilde{n}_0(\bldr) - \mu$  & 0 \\
        & & & & & & \\
        \hline
        & &  &  &  &        &     \\
           Hartree-Fock   &            &          & \checkmark     
         &  $H_0+\hat{H}_1 +\hat{H}_2$  
         & & $g [ \wfn_0(\bldr) ]^2$\\
        Bogoliubov  & & & & $+\left( \delta H_0 +\delta \hat{H}_1 +\delta \hat{H}_2\right)$
        &   $\hat{h}_0(\bldr) + 2 g | \wfn_0(\bldr) |^2 + 2 g \tilde{n}_0(\bldr) - \mu$ 
         &  + $g \tilde{m}_0^R(\bldr)$  \\
        & & & & & & \\
        \hline
        & & & & & & \\
        Hartree-Fock & &               &  \checkmark      &      $H_0+\hat{H}_1 +\hat{H}_2$            &   &  \\
        Bogoliubov & & & & $+\left( \delta H_0^{HF} +\delta \hat{H}_1^{HF} +\delta \hat{H}_2^{HF}\right)$ & $\hat{h}_0(\bldr) + 2 g | \wfn_0(\bldr) |^2 + 2 g \tilde{n}_0(\bldr) - \mu$ & $g [ \wfn_0(\bldr) ]^2 $  \\
        Popov & & & & & & \\
        \hline
	Generalized  & &               &      &      Includes all HFB     &  &  \\
        Hartree-Fock & & & \checkmark & contributions plus  & $\hat{h}_0(\bldr) + 2 g(\bldr) | \wfn_0(\bldr) |^2 + 2 g_t \tilde{n}_0(\bldr) - \mu$ & $g(\bldr) [ \wfn_0(\bldr) ]^2 $  \\
        Bogoliubov & & & & heuristic additional ones  & & \\
        \hline
      \end{tabular}
      \caption{\label{Table_StaticTheories}
Direct comparison of common static mean field theories. With the exception of the single-particle excitations of the Hartree-Fock theory (for which $\hat{\cal M}(\bldr)=0$), all other approaches include quasiparticle dressing by means of the Bogoliubov equations
$\left( 
\begin{array}{cc} 
\hat{\cal L}(\bldr)  & \hat{\cal M}(\bldr) \\ -\hat{\cal M}^{*}(\bldr) & -\hat{\cal L}^*(\bldr)
\end{array}
\right)
\left(
\begin{array}{l}
u_i(\bldr) \\ v_i(\bldr)
\end{array}
\right)
= \epsilon_i
\left(
\begin{array}{l}
u_i(\bldr) \\ v_i(\bldr)
\end{array}
\right)\;,$
where $u_i(\bldr)$ and $v_i(\bldr)$ the Bogoliubov functions of Eq.\ (\ref{u-v}). Here we have also introduced a generalized `many-body' effective potential via $g(\bldr) = g \left( 1 + \frac{\tilde{m}_0^R(\bldr)}{(\wfn_0(\bldr))^2} \right)$. Many-body effects are only consistently included in the latter generalized theory, where the resulting equations have not been defined from suitable hamiltonians, but have been heuristically modified on physical grounds; in particular, the term $g_t$ describes collisions between a condensate and a thermal atom, and is either given by $g$, or by $g(\bldr)$ in different levels of approximation (to the momentum dependence of the many-body T-matrix).
      }
    \end{center}
  \end{sidewaystable}


\subsection{Brief Summary}

{\it 
The mean field effect of a static thermal cloud on the condensate can be calculated by an approximate inclusion of the $(\hat{H}_3+\hat{H}_4)$ contributions to the hamiltonian by means of the mean field approximations of Eqs.\ (\ref{mf-quad}) and (\ref{mf-trip}). These approximations give rise to a set of generalized mean fields consisting of the condensate mean field, $\wfn(\bldr)$, the normal average denoting the non-condensate or thermal density, 
$\tilde{n}(\bldr)=\langle \hat{\delta}^{\dag}(\bldr) \hat{\delta}(\bldr) \rangle$,
and the anomalous average,
$\tilde{m}(\bldr)=\langle \hat{\delta}(\bldr) \hat{\delta}(\bldr) \rangle$.
Implementation of Eqs.\ (\ref{mf-quad}) and (\ref{mf-trip}) thus reduces the system hamiltonian to an approximate quadratic form (more general than that considered in Sec.\ \ref{zeroT}) which enables the self-consistent evaluation of (the static values of) $\tilde{n}(\bldr)$ and $\tilde{m}(\bldr)$ in terms of the Bogoliubov functions $u_i(\bldr)$ and $v_i(\bldr)$ (via Eqs.\ (\ref{tilden_0})-(\ref{tilde_m0R})). These approximations cannot describe dynamical effects as they discard contributions describing both particle-exchange collisions between thermal and condensate atoms and collisions between pairs of thermal atoms which lead to thermal population redistribution.
There are four variants of such generalized mean field theories, as summarized below and in Table 1. Note that an alternative review of (most of) these theories using the Green's function formalism can be found in \cite{Shi_Griffin}.
\begin{itemize}
\item Hartree-Fock (Eqs.\ (\ref{HF_Energy})-(\ref{HF_Density})): Only maintains quadratic hamiltonian contributions of the form $\hat{H}_{HF}=\int d\bldr \hat{\delta}^{\dag} ( \cdots ) \hat{\delta}$, i.e.\ contributions containing one creation and one annihilation non-condensate operator. Usually implemented in the semiclassical approximation.
\item Hartree-Fock-Bogoliubov (Eqs.\ (\ref{genGP-indie})-(\ref{M_HFB})): Additionally includes terms of the form 
$\int d\bldr ( \cdots ) \hat{\delta} \hat{\delta} +{\rm h.c.}$, i.e. single-particle states are dressed to Bogoliubov quasiparticles. Produces lowest ground state energy, but 
suffers from an inconsistent treatment of atomic collisions, associated with the incorrect introduction of an effective interaction which does not treat all collisions consistently.
\end{itemize}
Two modified mean-field approaches have been proposed to address this issue and generate consistent theories in different levels of approximation:
\begin{itemize}
\item Hartree-Fock-Bogoliubov-Popov (Eqs.\ (\ref{GPE_HF_Static}), (\ref{HFB-uv}), (\ref{L_HFBP})-(\ref{M_HFBP})): Contributions of the form 
$\int d\bldr ( \cdots ) \hat{\delta} \hat{\delta} +{\rm h.c.}$ are only maintained in the limit $\tilde{m}_0(\bldr)=0$ - often referred to as the `Popov' approximation.
\item Generalized Hartree-Fock-Bogoliubov (Eqs.\ (\ref{BdG_GHFB})-(\ref{GPE_GHFB})): The $T>0$ Bogoliubov equations are amended `by hand' to include physics beyond the usual HFB quadratic hamiltonian, such that the effective collisional strength between two atoms becomes upgraded throughout the coupled equations from $g$ to 
$g(\bldr)=g(1+\tilde{m}_0^R/\wfn_0^2)$ for condensate-condensate collisions; a variant of this theory also includes a similar replacement for collisions involving one condensate and one thermal atom.
\end{itemize}
}

Having concluded our discussion of static theories, we next extend our formalism to incorporate effects introduced by the dynamics of the thermal cloud.

\section{Finite Temperature Mean Field Theory: Dynamic Case}
\label{finitet:dynamic}

Our discussion so far has been restricted to the study of static variables, aimed mainly at interpreting finite temperature properties of the ultracold gas, such as density profiles and coherence properties at steady state. 
This treatment can be systematically generalized to time-dependent variables.
We start by obtaining exact
coupled equations of motion for the condensate and the non-condensate components, and then discuss various approximate treatments which enable direct numerical simulation of the full coupled evolution.

\subsection{Exact Dynamical Evolution}
\label{Dynamics_Exact}

Using the Heisenberg equation of motion, Eq.\ (\ref{Heisenberg}) we readily obtain the following exact evolution for the condensate mean field \cite{Proukakis_NIST,Proukakis_Thesis}:
\begin{eqnarray}
i\hbar \frac{\partial \wfn(\bldr,t)}{\partial t} &=& 
\langle \left[ \fopart, \hat{H}  \right] \rangle 
\nonumber \\
& = &
\langle \hat{h}_0
\fopart \rangle + g \langle \fopcrt \fopart \fopart \rangle\;. 
\label{cond_HFB}
\end{eqnarray}
Using the symmetry-breaking decomposition into condensate and non-condensate contributions via 
Eq.\ (\ref{sep-broken}), and suppressing explicit spatial dependence for brevity, the latter term can be re-expressed as
\begin{eqnarray}
\fopc \fopa \fopa = \left| \wfn \right|^{2} \wfn + 2 \left| \wfn \right|^{2} \nopa + \wfn^2 \nopc 
+ \wfn^{*} \nopa \nopa + 2 \wfn \nopc \nopa + \nopc \nopa \nopa\;.
\end{eqnarray}
This then yields
\begin{eqnarray}
i\hbar \frac{\partial \wfn}{\partial t} 
= \left[
\hat{h}_0
+ g |\wfn|^{2} \right] \wfn  
+ 2g \langle \nopc \nopa \rangle \wfn %
+ g \langle \nopa \nopa \rangle  \wfn^* 
+g \langle \nopc \nopa \nopa \rangle\;.
\label{genGP}
\end{eqnarray}
Clearly the condensate evolution is dynamically coupled to the behaviour of the non-condensate via the evolution of normal and anomalous averages. Using similar arguments, we obtain the exact expression
\begin{eqnarray}
i\hbar \frac{\partial \nopa(\bldr,t)}{\partial t} 
 = \left[ \nopa(\bldr,t) , \hat{H} \right] 
= i\hbar \frac{\partial}{\partial t} \left( \fopart - \langle \fopa(\bldr,t) \rangle \right)
\label{delta_exact}
\end{eqnarray}
which can be written as \cite{ZNG}
\begin{eqnarray}
i\hbar \frac{\partial \nopa}{\partial t} 
& = & \hat{h}_{0} \nopa + g \left[ 2 |\wfn|^2 \nopa + \wfn^2 \nopc \right]
\nonumber \\
& + & 2g  \wfn \left( \nopc \nopa - \langle \nopc \nopa \rangle \right)
+ g\wfn^* \left( \nopa \nopa - \langle \nopa \nopa \rangle \right) \nonumber \\
& +&  g\left( \nopc \nopa \nopa - \langle \nopc \nopa \nopa \rangle \right).
\label{delta_evol}
\end{eqnarray}
One can thus obtain the exact equations for the evolution of normal and anomalous averages either directly from Eq.\ (\ref{delta_evol}), or equivalently via
\be
i \hbar \frac{\partial}{\partial t} \tilde{n}(\bldr,t)  =
\langle \left[ \nopc \nopa, \hat{H} \right] \rangle\;,
\hspace{1.0cm}
i \hbar \frac{\partial}{\partial t} \tilde{m}(\bldr,t) =
\langle \left[ \nopa \nopa, \hat{H} \right] \rangle
\label{nm_HFB}
\ee

While formally exact, without any truncation this procedure leads to an infinite hierarchy of coupled equations of motion for the higher order correlations.
This is easy to see, since for the given hamiltonian, the right-hand side of the equation of motion for a product of $n$ operators actually contains $n+2$ operators.
Thus, in order to solve such equations one must choose an appropriate set of generalized mean fields which are assumed to accurately describe the system to the desired level of approximation.
This Section discusses the various approximations that are typically made, and the physics that they contain. 
An alternative, yet closely related approach based on the method of non-commutative cumulants and pursued by K\"{o}hler, Burnett and collaborators is discussed in \ref{Cumulants}.

In order to formulate such theories, we occasionally need to resort to a basis set notation
in terms of single-particle eigenstates $\spwa_i(\bldr)$. Following the notation introduced in the book of Blaizot and Ripka \cite{Blaizot_Ripka} and implemented extensively in the group of Keith Burnett (Oxford) the condensate and non-condensate contributions are denoted by
\bea
\wfn(\bldr,t) = \sum_i \spwa_i(\bldr) z_i(t)\;, \hspace{2.0cm} z_i = \langle \hat{a}_i \rangle\;, \nonumber \\
\nopa(\bldr,t) = \sum_i \spwa_i(\bldr) \hat{c}_i(t)\;, \hspace{2.0cm}
 \hat{c}_{i} = \hat{a}_{i} - \langle \hat{a}_{i} \rangle\;. 
\label{zc}
\eea
Correspondingly, the
normal and anomalous averages are defined by
\be
\rho_{ij} = \langle \hat{c}_{j}^{\dag} \hat{c}_{i} \rangle\;, \hspace{2.0cm}
\kappa_{ij} =  \langle \hat{c}_{j} \hat{c}_{i} \rangle \;.
\label{rhokappa}
\ee
For convenience in our subsequent discussion we also define the `Hartree-Fock hamiltonian' ($h$) and the `Pairing Field' ($\Delta$), along with their `reduced forms' $h^c$ and $\Delta^c$, via their respective matrix elements
\be
h_{ij} = h_{ij}^{c} + \sum_{kl} V_{iklj} z_{k}^{*} z_{l}
= \langle i | \hat{h}_0 | j \rangle + 2 \sum_{kl} V_{ikjl} \left( z_{k}^{*} z_{l} + \rho_{lk} \right)\;,
\label{h}
\ee
\be
\Delta_{ij} = \Delta_{ij}^{c} + \sum_{kl} V_{ijkl} z_{k} z_{l}
= \sum_{kl} V_{ijkl} \left( z_{k} z_{l} + \kappa_{lk} \right)\;.
\label{delta}
\ee
Here, $V_{ikjl} = \langle ik | V | jl \rangle = (1/2) [ (ik|\hat{V}|jl) + 
(ik|\hat{V}|lj) ]$
is the symmetrized form of the interaction matrix element defined by Eq.\ (\ref{matrixelement}),
corresponding to the collision of two atoms initially in states $j$ and $l$, which emerge from the collision in states $i$ and $k$; 
we also note here that $h_{ij}^* = h_{ji}$.

\subsection{The Hartree-Fock Approximation}

\label{Hartree_Fock}

As in Sec.\ \ref{HF_Regime}, the simplest finite temperature approximation that one can consider is formulated in terms of the condensate $\wfn(\bldr,t)$ and the thermal $\tilde{n}(\bldr,t)$ components. 
In the position representation, Eq.\ (\ref{genGP}) leads to a generalized finite temperature Gross-Pitaevskii equation
\be
i \hbar \frac{\partial \wfn (\bldr,t)}{\partial t} 
= \left[ \hat{h}_0 + g \left( |\wfn(\bldr,t)|^2 + 2 \tilde{n}(\bldr,t) \right) \right] 
\wfn(\bldr,t)\;,
\label{GPE_HF}
\ee
which is the time-dependent generalization of Eq.\ (\ref{GPE_HF_Static}).

In order to express the coupled equations for condensate and non-condensate, we shift to our basis set notation, where  Eqs.\ (\ref{genGP}) and (\ref{delta_evol}) generate the following set of self-consistent equations \cite{Blaizot_Ripka}:
\be
i \hbar \frac{d z}{dt} = h^{c} z + \Delta^c  z^*\;,
\hspace{2.0cm}
i \hbar \frac{d \rho}{dt} = \left[ h \, , \, \rho \right]\;,
\ee
which constitute the time-dependent generalization of the `Hartree-Fock' equations of Sec.\ \ref{HF_Regime}. As before,
this level of approximation only includes mean field coupling between the two subsystems and will not be considered further here. We will however return to an appropriate generalization of these equations which additionally includes particle exchange in Sec.\ \ref{ZNG}.

\subsection{The Hartree-Fock-Bogoliubov Approximation}

In the Hartree-Fock-Bogoliubov (HFB) approximation, one works with slightly generalized mean fields: in addition to the condensate mean field and the normal average, one further includes the pair anomalous average $\tilde{m}(\bldr,t)$, i.e. one assumes the system is well-described by the HFB hamiltonian of Eq.\ (\ref{H_HFB}).

In first instance, we consider weak perturbations around the equilibrium solutions, which is sufficient to yield expressions for the dominant damping mechanisms both at zero and finite temperatures.
We then proceed to give the full dynamical equations.

\subsubsection{A Perturbative Linear Response Treatment:}
\label{Linear_Response}

One assumes that the system is close to steady-state, with both the condensate and the thermal cloud described by small deviations around their equilibrium values, via
\be
\tilde{\wfn}(\bldr,t) \simeq e^{-i \mu t/\hbar}\left[ \wfn_{0}(\bldr)+\delta \wfn(\bldr,t) \right] \;,
\ee
\be
\tilde{n}(\bldr,t) \simeq \tilde{n}_0(\bldr)+\delta \tilde{n}(\bldr,t) \;.
\ee
This is a generalization of the linear approximation used to derive the (zero-temperature) Bogoliubov equations in Sec.\ \ref{Bog-dG}. Consistency requires one to also look at small variations about the anomalous average $\tilde{m}(\bldr,t)$, via
\be
\tilde{m}(\bldr,t) \simeq e^{-2 i \mu t/\hbar}\left[ \tilde{m}_{0}^R(\bldr)+\delta \tilde{m}(\bldr,t) \right]
\;.
\ee
We will see that this approach leads to the inclusion 
of the dominant damping mechanisms for elementary excitations.
The wavefunction $\wfn_0 (\bldr)$ satisfies the static generalized GPE of Eq.\ (\ref{genGP-indie}). 
Linearization yields the following equation for the condensate perturbations \cite{Proukakis_Thesis,Rusch_Burnett,Giorgini_1,Giorgini_2}
\bea
i \hbar \frac{\partial}{\partial t} \delta \wfn (\bldr,t) &=& 
\left[ \hat{h}_0 - \mu +2g \left( |\wfn_0(\bldr)|^2 + \tilde{n}_0(\bldr) \right) \right] \delta \wfn (\bldr,t) 
\nonumber \\
& + & \left[ g |\wfn_0(\bldr)|^2 (\bldr) + g \tilde{m}_0^R (\bldr) \right] \delta \wfn^{*}(\bldr,t) 
\nonumber \\
& + & 2g \wfn_0 (\bldr) \delta \tilde{n} (\bldr,t) + g \wfn_0 (\bldr) \delta \tilde{m}(\bldr,t)\;.
\label{GPE-linearresponse}
\eea
As discussed earlier, the static values of $\tilde{n}_0(\bldr)$ and $\tilde{m}_0^R(\bldr)$ are obtained via a transformation to a quasi-particle basis - see Eqs.\ (\ref{tilden_0})-(\ref{tilde_m0R}).
Clearly the condensate perturbations are coupled to the dynamic change in the thermal cloud $\delta \tilde{n}(\bldr)$, and the anomalous average $\delta \tilde{m}(\bldr)$ which
respectively acquire the forms \cite{Rusch_Burnett,Giorgini_1,Giorgini_2}
\bea
\delta \tilde{n}(\bldr,t) = \sum_{ij} && \Big\{
\left[ 
u_{i}^{*}(\bldr) u_{j}(\bldr) + v_{i}^{*}(\bldr) v_j(\bldr) \right] f_{ij}(t) \Big. \nonumber \\
&& \Big. + u_{i}(\bldr) v_j(\bldr) g_{ij}(t) + u_{i}^{*}(\bldr) v_j^*(\bldr)  g_{ij}^*(t) \Big\} \;, 
\eea
\bea
\delta \tilde{m}(\bldr,t) = \sum_{ij} && \Big\{ 
 2  v_{i}^{*}(\bldr) u_{j}(\bldr) f_{ij}(t) + u_{i}(\bldr) u_j(\bldr) g_{ij}(t) \Big. \nonumber \\
&& \Big. + v_{i}^{*}(\bldr) v_j^*(\bldr)  g_{ij}^*(t) \Big\} \;.
\eea
Here both
$f_{ij}=\langle \hat{\beta}_{i}^{\dag} \hat{\beta}_{j} \rangle(t) - \delta_{ij} f_{i}^{0}$ and 
$g_{ij}$ 
correspond to non-equilibrium distribution functions ($f_{i}^{0}$ denotes the Bose-Einstein distribution of Eq.\ (\ref{BE-dist})).
The evolution of such functions can be easily obtained within our generalized quadratic (quasiparticle) hamiltonian. Here we
apply perturbation theory to lowest order, i.e. integrate the equation of motion for $f_{ij}$ and $g_{ij}$ which yields corresponding expressions to lowest order in the interaction strength $g$. Any contributions of $\delta \tilde{n}(t)$, and $\delta \tilde{m}(t)$ appearing on their {\em respective right-hand sides}, will thus, to this order, yield no contribution to the expressions of interest.
With that in mind, the evolution of the above non-equilibrium distribution functions is governed by the simplified equations:
\bea
i \hbar \frac{\partial}{\partial t} f_{ij}(t) =
(\epsilon_{j}-\epsilon_{i}) f_{ij}(t)  +2g A_{ij} (f_{i}^{0}-f_{j}^{0})\;, \\
\label{f_ij}
i \hbar \frac{\partial}{\partial t} g_{ij}(t) =
(\epsilon_{j}+\epsilon_{i}) f_{ij}(t)  +2g B_{ij} (1+ f_{i}^{0}+f_{j}^{0})\;, 
\label{g_ij}
\eea
where we have dropped from their respective right hand sides all terms containing $\delta \tilde{n}$ and $\delta \tilde{m}$. The coefficients $A_{ij}$ and $B_{ij}$ appearing here are sums of overlap integrals of the condensate with the respective quasi-particle amplitudes, given by \cite{Giorgini_1,Giorgini_2}
\bea
A_{ij} = \int d \bldr \wfn_0 && \Big\{
 \delta \wfn \left( u_{i} u_j^* +v_i v_j^* +v_i u_j^* \right) \Big. \nonumber \\
&& \Big. + \delta \wfn^* \left( u_{i} u_j^* +v_i v_j^* +u_i v_j^* \right) \Big\}
\eea
\bea
B_{ij} = \int d \bldr \wfn_0 
&& \Big\{  \delta \wfn \left( u_{i}^* v_j^* +v_i^* u_j^* +u_i^* u_j^* \right) \Big. \nonumber \\
&& \Big. + \delta \wfn^* \left( u_{i}^* v_j^* +v_i^* u_j^* +v_i^* v_j^* \right) \Big\}\;.
\eea

Importantly we find that the above procedure already incorporates the two main damping mechanisms encountered in experiments with ultracold gases. To visualize this, we consider oscillations of the condensate at some given frequency $\omega$, i.e. $\delta \wfn(\bldr,t) = \delta \wfn_0 (\bldr) e^{-i \omega t}$. These condensate fluctuations induce changes in $f_{ij}(t)$ and $g_{ij}(t)$, which modify 
$\delta \tilde{n}(\bldr,t)$ and $\delta \tilde{m}(\bldr,t)$ and in
turn act back onto the condensate via Eq.\ (\ref{GPE-linearresponse}). This leads both to a shift of the frequency of oscillation and to damping; the damping is determined by the coefficient 
$\gamma = \gamma_{L} + \gamma_{B}$, which consists of two terms:

(i) A thermal excitation (quasiparticle) of energy $\epsilon_i$ can interact with the collective mode of the condensate 
of typical energy $\hbar \omega_0$, and thereby become converted into a higher energy excitation, $\epsilon_j = \epsilon_i + \hbar \omega_0$ (i.e. energy is absorbed from the condensate). Such a process (Landau damping) damps the condensate oscillations at a rate
%
\be
\gamma_{L} = 4 \pi g^{2} \sum_{ij} \left| A_{ij} \right|^{2} \left( f_i^0-f_j^0 \right) \delta \left( \hbar \omega_0 + \epsilon_i - \epsilon_j \right) \;,
\ee
where the factor $(f_i^0-f_j^0)=\left[ f_i^0(f_j^0+1)-(f_i^0+1) f_j^0 \right]$ denotes the difference in the amplitudes for the destruction of a quasi-particle of energy $\epsilon_i$, and the simultaneous creation of a quasi-particle of energy $\epsilon_j$ and its inverse process.

(ii) The absorption of a quantum of oscillation can also lead to the creation of two excitations, of respective energies $ \epsilon_i$ and $ \epsilon_j$ with $\epsilon_i+\epsilon_j=\hbar \omega_0$. This process (Beliaev damping) leads to damping at a rate
\be
\gamma_{B} = 2 \pi g^{2} \sum_{ij} \left| B_{ij} \right|^{2} \left(1+ f_i^0+f_j^0 \right) \delta \left( \hbar \omega_0 - \epsilon_i - \epsilon_j \right) \;,
\ee
where the factor $(1+ f_i^0+f_j^0)=\left[ (f_i^0+1)(f_j^0+1)-f_i^0 f_j^0 \right]$ denotes the difference in the amplitudes for the simultaneous creation of two quasi-particles of energies $\epsilon_i$ and $\epsilon_j$ and its inverse process.

In the homogeneous limit, the `low-temperature' (quantum) regime $k_{B} T \ll \hbar \omega_0$ is dominated by Beliaev damping, while the `high-temperature' regime $k_{B} T \gg \hbar \omega_0$ is dominated by Landau damping
\cite{Morgan_JPhysB,Giorgini_1,Giorgini_2,Excitations_Fedichev_1,Excitations_Fedichev_2}.

\subsubsection{Self-Consistent Time-dependent Hartree-Fock-Bogoliubov Theory:}
\label{tdep_HFB}

Rather than making a linear response analysis for situations close to equilibrium, one can formulate the Hartree-Fock-Bogoliubov problem in a fully coupled dynamical manner.

We are thus interested in writing down full dynamical equations for the selected mean field variables $\wfn(\bldr,t) = \langle \hat{\wfn}(\bldr,t) \rangle$,
$\tilde{n}(\bldr) = \langle \nopc(\bldr,t) \nopa(\bldr,t) \rangle$, and 
$\tilde{m}^R(\bldr) = \langle \nopa(\bldr,t) \nopa(\bldr,t) \rangle$.
These can be obtained from 
Eqs.\ (\ref{cond_HFB}), (\ref{delta_exact}) and (\ref{nm_HFB}) by replacing $\hat{H}$ by $\hat{H}_{HFB}$. Alternatively, one arrives at the same equations using the full expressions of Eqs.\ (\ref{genGP}) and (\ref{delta_evol}) and {\em subsequently} imposing 
the decoupling approximations of Eqs.\ (\ref{mf-quad})-(\ref{mf-trip}), in order to reduce the expressions to a closed system of equations.
Note that in the HFB approximation $\langle \nopc \nopa \nopa \rangle = 0$, so that the equation for the condensate evolution differs from the exact equation, Eq.\ (\ref{genGP}), by the absence of the last term.

There are various ways to write down the precise form of the HFB equations. 
It is convenient to generalize the notation of Eqs.\ (\ref{zc})-(\ref{delta}) by defining
generalized density matrices for the condensed ($ R_C$) and uncondensed ($R_{NC}$) parts of the system by
\begin{eqnarray}
R_C = \left( 
\begin{array}{c} 
z \\ z^*
\end{array}
\right)\;,
%
\hspace{3.0cm}
%
R_{NC} = \left( 
\begin{array}{cc} 
\rho & \kappa \\ \kappa^{*} & \left( \rho^* + \un \right)
\end{array}
\right)\;,
\label{RC_RNC}
\end{eqnarray}
where $\un$ is the unit matrix.
The corresponding generalized hamiltonians are given by
\begin{eqnarray}
H_C = \left( 
\begin{array}{cc} 
h^{c} & \Delta^{c} \\ -\left( \Delta^{c} \right)^{*} & -\left(h^c \right) ^{*}
\end{array}
\right),
%
\hspace{0.3cm}
%
H_{NC} = \left( 
\begin{array}{cc} 
h & \Delta \\ -\left( \Delta \right)^{*} & -\left(h \right) ^{*}
\end{array}
\right).
\label{HC_HNC}
\end{eqnarray}
%
thus casting the equations in a notation which is closely related to the conventional Green's functions one, used for example in \cite{Shi_Thesis,Shi_Griffin}.

The time-dependent HFB equations thus take the compact form \cite{Proukakis_JPhysB}
\be
i \hbar \frac{d R_C}{dt} = H_C R_C \hspace{0.4cm}
{\rm and} \hspace{0.4cm}
i \hbar \frac{d R_{NC}}{dt} = H_{NC} R_{NC} - R_{NC} H_{NC}^{\dag}\;. 
\label{HFB_Temporal}
\ee
%
%
The above expressions are a generalization of the Hartree-Fock equations of Sec.\ \ref{Hartree_Fock}, and additionally include the damping processes discussed in Sec.\ \ref{Linear_Response}.
Although they contain the essential quasiparticle physics, these equations have nonetheless been formulated in terms of single-particle operators. As a result, their interpretation is perhaps not very transparent and one may wish to explicitly perform a Bogoliubov transformation to re-express these in terms of quasiparticles.
Such an approach was undertaken by Imamovic-Tomasovic and Griffin \cite{Milena_Griffin} using an equivalent but notationally distinct approach based on the Kadanoff-Baym formalism \cite{Kadanoff_Baym} in terms of non-equilibrium Green's functions.
This approach is well-documented in the literature \cite{Kadanoff_Baym} and we encourage readers who are familiar with the Green's function approach to consult Refs. \cite{Shi_Griffin} and \cite{Milena_Thesis}, as well as their  extended analysis  \cite{Milena_Griffin}. 
%
%
We do not discuss the HFB equations here any further, as they 
correspond to a purely mean field theory, 
which does not include the crucial process of particle exchange between the condensate and the thermal cloud. This latter issue is addressed in the following sections.

\subsection{Dynamical Perturbative Treatment of $\left( \hat{H}_3 + \hat{H}_4 \right)$} 
\label{Beyond_HFB}

The theories developed thus far include mean-field coupling between the selected generalized mean fields. As such they can describe 
collisions between two condensate atoms, or collisions between one condensate and one thermal atom for which the number of condensed and non-condensed atoms in the final state remains identical to that in the initial state. However, collisional processes which transfer atoms between the condensate and the thermal cloud have so far been ignored.
In order to include these into our treatment, we must thus
relax some of the approximations used in our previous discussion, essentially following along the lines of important early work by Kirkpatrick and Dorfman \cite{Kirkpatrick_1,Kirkpatrick_2,Kirkpatrick_3} and Eckern \cite{Eckern}.

To identify how to proceed, we note that
comparison of the exact evolution of the condensate mean field of Eq.\ (\ref{genGP}), to its corresponding evolution in the HFB basis, reveals the critical absence of the triplet contribution $\langle \nopc \nopa \nopa \rangle$. This term, whose importance has been stressed by one of us (NPP) \cite{Proukakis_NIST,Proukakis_Thesis}, 
can in fact be identified as describing particle transfer between the condensate and the thermal cloud, a process which is however prevented by the mean field approximation of Eq.\ (\ref{mf-trip}) which dictates that $\langle \nopc \nopa \nopa \rangle = 0$.
Similarly one finds that
collisions between two thermal atoms, corresponding to the usual scattering processes in a classical gas, can only be described if one considers corrections beyond the other mean-field approximation of Eq.\ (\ref{mf-quad}).

Collisional dynamics of this form can actually be introduced into the theory by perturbatively including the difference between the full system hamiltonian, $\hat{H}$, of Eq.\ (\ref{H-sep}), and an appropriately chosen generalized mean-field hamiltonian $\hat{H}_{\rm MF}$
\cite{ZNG,Morgan_JPhysB,JILA_Kinetic_1,Proukakis_JPhysB}, i.e. by 
separating the hamiltonian as
\be
\hat{H} = \hat{H}_{\rm MF} + \left( \hat{H} -\hat{H}_{\rm MF} \right)
\label{MF_Pert} \;.
\ee
The first contribution $ \hat{H}_{\rm MF}$ should be included self-consistently and defines the unperturbed basis of the system.
This is fixed by the choice of mean field approximations (like Eqs.\ (\ref{mf-quad}), (\ref{mf-trip})) imposed in reducing
$(\hat{H}_3+\hat{H}_4)$ to approximate quadratic form. 
For example, setting $\hat{H}_{\rm MF} = \hat{H}_{HF}$ defines the unperturbed system basis as the Hartree-Fock basis, whereas $\hat{H}_{\rm MF} = \hat{H}_{HFB}$ additionally includes the pair anomalous average in the basis. 
Diagonalization of the corresponding hamiltonian would define the excitation energies, respectively corresponding in the above examples to the dressed single-particle Hartree-Fock, or the Hartree-Fock-Bogoliubov quasiparticle energies.

Such perturbative treatments are presented below for various appropriately chosen basis; as the discussion  is restricted to second order in the effective interaction strength $g$, we should comment on the validity of such a methodology:
To appreciate this, it is important to note that the equations presented here already feature a consistent introduction of an effective (T-matrix) interaction; as a result, this perturbation theory can effectively be seen as an expansion in terms of the difference between scattering in the presence of a medium (namely a partially condensed gas) and in vacuum \cite{Morgan_JPhysB}. Thus, although the usual perturbative expansion in terms of the actual interatomic interaction would fail, perturbation theory to second order in terms of the effective interaction should be valid for a dilute gas sufficiently far from the critical region, i.e.\ when the condition $(k_B T / gn) \sqrt{na^3} \ll 1$ is satisfied \cite{Popov};
this is generally true for experimentally relevant parameters, except in a very narrow window near $T_c$.

The corresponding number-conserving perturbative expansion is presented in Sec.\ \ref{Number}, while a
%
%
related treatment based on the method of non-commutative cumulants, which provides a well-defined decorrelation scheme that leads to self-consistent expressions in different limits and can also handle `memory effects' is discussed in \ref{Cumulants}.

\subsubsection{Perturbative Formulation Beyond the HFB Hamiltonian:}
\label{Prouk_QK}

We start by extending the HFB theory of Sec.\ \ref{tdep_HFB}. 
%
Collisional dynamics can be introduced into the theory by the previous perturbative prescription.
Using the mean field approximations of Eqs.\ (\ref{mf-quad}), (\ref{mf-trip}) to simplify $(\hat{H}_3+\hat{H}_4)$ according to Eqs.\ (\ref{dH_1})-(\ref{dH_2}) identifies the mean field `basis hamiltonian', $\hat{H}_{\rm MF}$ as 
the HFB hamiltonian of Eq.\ (\ref{H_HFB}).
In our treatment we thus still focus on the same generalized mean fields $\wfn$, $\tilde{n}$ and $\tilde{m}$ as in HFB, but perform second order perturbation theory beyond the HFB basis via
\be
\hat{H} = \hat{H}_{HFB} + \hat{H}'\;. 
\label{H_Pert}
\ee
where the perturbing hamiltonian, $\hat{H}'$, is given by
\be
\hat{H}' =  \hat{H}_3'+\hat{H}_4' 
= \left[ \hat{H}_{3} - \delta \hat{H}_1 \right] + \left[ \hat{H}_{4} - \left( \delta H_0 + \delta \hat{H}_{2} \right) \right]\;.
\ee
%
We thus obtain the exact relations \cite{Proukakis_JPhysB}
\bea
i \hbar \frac{dR_{C}}{dt} = H_{C}R_{C} + I_{C}\;, \nonumber \\
i \hbar \frac{ dR_{NC}}{dt} = \left(H_{NC}R_{NC} - R_{NC} H_{NC}^{\dag} \right) + I_{NC}\;.
\label{Kinetic_HFB}
\eea
These expressions differ from their HFB couterparts (Eq.\ (\ref{HFB_Temporal})) by the inclusion of the `kinetic' matrices $I_{C}$ and $I_{NC}$. These matrices are expressed in terms of averages of three and four single-particle fluctuations operators $\langle \hat{c}^{\dag} \hat{c}^{(\dag)} \hat{c} \rangle$ and $\langle \hat{c}^{\dag} \hat{c}^{(\dag)} \hat{c} \hat{c} \rangle$ beyond their approximate mean field values, and
are responsible for the inclusion of all relevant collisional terms (detailed expressions can be found in \cite{Proukakis_JPhysB}). 
%
%
In principle, one can actually obtain a closed system of coupled equations for the evolutions of the matrices $R_C$, $R_{NC}$, $I_{C}$ and $I_{NC}$ (or equivalently for their corresponding generalized mean fields)
subject only to appropriate generalized decoupling approximations \cite{Proukakis_NIST,Proukakis_JPhysB}. A potential advantage of such a system of equations would be that it can be solved self-consistently using the {\em exact} non-local interatomic potential, and does not actually require one to resort to the pseudopotential approximation. However, computation with actual eigenstates and non-local potentials, although feasible, is numerically demanding, and one therefore seeks simplified alternatives which nonetheless capture the essential physics; the remainder of this Section thus discusses such suitably-constructed alternative approaches.

We start by discussing how these `kinetic' contributions $I_C$ and $I_{NC}$
lead to the introduction of collisional terms into the formalism:
Firstly, one should obtain
the exact equations of motion for both $I_{C}$ and $I_{NC}$ by means of the  effective hamiltonian of Eq.\ (\ref{H_Pert}). Then, one should formally integrate these expressions, assuming their characteristic evolution in the basis specified by the unperturbed hamiltonian $\hat{H}_{HFB}$ can be described in terms of suitable dressed eigenenergies (here HFB quasiparticle ones); only at that stage should one impose any  decoupling approximations.
In other words, while the mean field approximations of Eqs.\ (\ref{mf-quad}), (\ref{mf-trip}) 
are used to define the unperturbed hamiltonian, in the final expressions one imposes generalized decoupling schemes (such as the one of Eq.\ (\ref{mf-wick})) to reduce correlations of the form $\langle \hat{c}^{\dag} \hat{c}^{\dag} \hat{c}^{\dag} \hat{c} \hat{c} \hat{c} \rangle$ 
to pair operator averages, and thus obtain
a closed system of equations 
in second order perturbation theory.
Following the usual procedure implemented in the kinetic theory of dilute gases, one simultaneously simplifies the resulting expressions by assuming that collisions are well-separated in time, so that the mean fields evolve slowly compared to the collisional duration; the pseudopotential approximation is also made on the appropriately introduced effective interaction (i.e.\ consistent with restricting the anomalous averages to low-lying modes to avoid double-counting certain interaction effects). 
for details regarding this procedure and for the precise expressions of the generalized decoupling approximations the reader is referred to \cite{Proukakis_JPhysB}.
This approach yields a set of self-consistent equations for the variables $\wfn$, $\tilde{m}$ and $\tilde{n}$ which include both damping mechanisms and collisional integrals.

We are now ready to give our final second order theory in the effective interaction strength $g$. However, as Eqs.\ (\ref{Kinetic_HFB}) have been explicitly formulated in terms of single-particle operators, the final expressions for the collisional terms given in \cite{Proukakis_JPhysB} are quite lengthy. We have therefore chosen not to give these expressions here in full, but to focus instead on the simplest illustrative application of this formalism which highlights the essential physics of such contributions.
%

Although one of the authors (NPP) was instrumental in the preceeding development (in collaboration with Burnett and Stoof) \cite{Proukakis_Burnett_Stoof,Proukakis_NIST,Proukakis_Thesis,Proukakis_JPhysB}, the final expressions for the full collisional integrals mentioned above were first given by Walser {\it et al.} in \cite{JILA_Kinetic_1,JILA_Kinetic_2}. This latter treatment is based on more `conventional' formulations of a quantum kinetic theory, whereby one defines a set of suitable slowly-varying gaussian `master variables' (corresponding precisely to the condensate and the normal and anomalous averages introduced earlier) whose evolution is subsequently studied. 
The full second order collisional integrals are given in \ref{JILA_Theory} using the notation introduced in these latter works.
%
Note that despite their differences in formal development and notation,
the theory presented in the Appendix is {\em identical} to the presentation given in this section, as explicitly demonstrated in \cite{Proukakis_JPhysB};
furthermore, both theories are formally equivalent \cite{JILA_Kinetic_3} to the non-equilibrium Green's function approach
originally put forward by Kadanoff and Baym \cite{Kadanoff_Baym} and subsequently applied to trapped gases by Imamovic-Tomasovic and Griffin \cite{Milena_HFB,Milena_Griffin,Milena_Thesis}.
The full version of this theory has been applied (in the ergodic approximation) to the study of spherically symmetric harmonic traps \cite{JILA_Kinetic_2}, with appropriate generalizations introduced to study quasiparticle damping and finite collisional duration \cite{JILA_NonMarkovian}, and recent extensions to the study of one-dimensional gases \cite{Walser_1D_1,Walser_1D_2}.

\begin{figure}[t]
\centering \scalebox{0.5}
 {\includegraphics{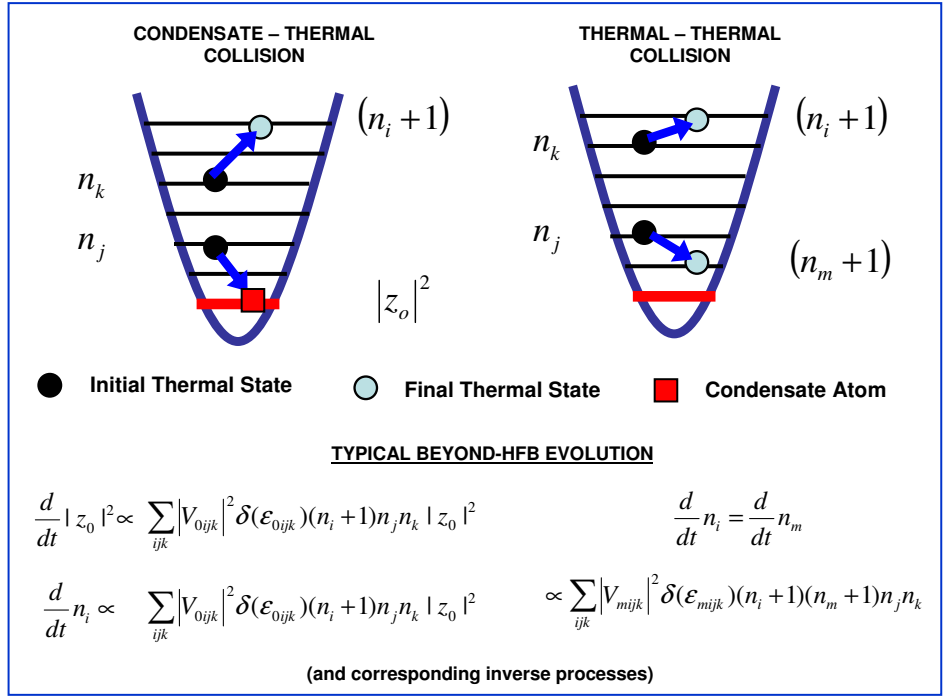}}
 \caption{
(colour online)
Schematic of characteristic collisional processes involving either the transfer of a thermal atom into the condensate (left), or collisions within the thermal cloud leading to redistribution among thermal energy levels (right). Characteristic rates for these processes are also given; note that the inverse processes are also allowed, with their combined effect corresponding to Eqs.\ (\ref{cond_growth})-(\ref{thermal_growth}).
}
\label{Schematic_Collisions}
\end{figure}

In order to highlight the main physical issues of such theories, we now focus on a simple model. Following the discussion of Proukakis and Burnett, we recall
that the exact equation of motion for the condensate mean field amplitude in level $i$, corresponding to Eq.\ (\ref{HFB_Temporal}) is given by \cite{Proukakis_NIST}
\bea
i \hbar \frac{dz_n}{dt} &=& \sum_k \langle n| \hat{h}_0 | k \rangle z_k
\nonumber \\
&+& \sum_{ijk} V_{nijk} \left( z_i^* z_j z_k + \kappa_{jk} z_i^* + 2 \rho_{ji} z_k 
+ \langle \hat{c}_i^\dag \hat{c}_j \hat{c}_k \rangle \right)\;.
\label{zn_triplet}
\eea
We also assume
an idealized weakly-interacting Bose gas, for which 
(i) the condensate occupies only the lowest mode of the trap, denoted by the label `0', and (ii) the thermal particles are diagonal in the bare particle basis, i.e $\langle n|\hat{h}_0| k \rangle = \varepsilon_n \delta_{nk}$, where $\varepsilon_n$ is the eigenenergy of trap level $n$. 
Although the theories of Walser {\em et al.} \cite{JILA_Kinetic_1,JILA_Kinetic_2} and Proukakis \cite{Proukakis_JPhysB} explicitly contain the pair anomalous average $\tilde{m}(\bldr)=\sum_{ij} \spwa_i(\bldr) \spwa_j(\bldr) \kappa_{ij}$ in a self-consistent manner (see \ref{JILA_Theory}) we simplify our current presentation even further by assuming that (the static values of) such averages can be ignored.
In this idealized limit we find 
the following contributions to the evolution of the condensate and the non-condensate (see, e.g.\ \cite{Proukakis_JPhysB,Proukakis_Lambropoulos,PGPE_T}), which are shown diagramatically in Fig.\ \ref{Schematic_Collisions}:\\

\noindent {\bf Condensate Evolution:}
Beyond the usual HFB contributions 
(condensate mean field, normal and pair anomalous averages) 
of Eq.\ (\ref{HFB_Temporal}),
we find the additional term
\be
\frac{dz_0}{dt} = \cdots - \frac{i}{\hbar} \sum_{ijk} V_{0ijk} \langle \hat{c}_i^{\dag} \hat{c}_{j} \hat{c}_k \rangle\;. 
\label{z0_triplet}
\ee
This latter quantity, which has no contribution to the condensate mean field to first order in the interatomic potential, i.e.\ no
static value within HFB, acquires a non-zero value in second order perturbation theory.  In the context of the
simplified model considered here, its respective equation of motion contains 
(after imposing suitable generalized decoupling approximations to reduce correlations of the form $\langle \hat{c}^{\dag} \hat{c}^{\dag} \hat{c} \hat{c} \rangle$ and $\langle \hat{c}^{\dag} \hat{c}^{\dag} \hat{c}^{\dag} \hat{c} \hat{c} \hat{c} \rangle$ to simpler pair correlations $\langle \hat{c}^{\dag} \hat{c}  \rangle$),
among other terms, the contribution
(see \cite{Proukakis_Burnett_Stoof,Proukakis_NIST,Proukakis_Thesis} for details)
\bea
i \hbar \frac{d}{dt} & \langle \hat{c}_i^\dag \hat{c}_j \hat{c}_k \rangle & =  
\left( \varepsilon_j + \varepsilon_k - \varepsilon_i \right) + \cdots \nonumber \\
& & + 2 V_{jki0} \left[ n_i (n_j +1)(n_k +1) - (n_i +1)n_j n_k \right] z_0.
\eea
We can now formally integrate this equation of motion. In doing so, we assume that the chosen mean fields do not vary appreciably on the timescale of a single collision;
here we implicitly make an assumption, common in kinetic theories of gases, that collisions are short in duration and well-separated in time, and that we are only interested in `long-time' evolution of the system, i.e. we can neglect the effect of any coherences present in the initial state of the system on the subsequent system dynamics.
This enables us to approximate the quantity $z_0(t')$ appearing in the integrand by
$z_0(t') = e^{i \varepsilon_0 (t-t')/\hbar} z_0(t)$, which is called the `Markov' approximation.
In lowest order perturbation theory, one thus obtains \cite{Proukakis_Thesis}
\bea
\langle \hat{c}_i^\dag \hat{c}_j \hat{c}_k \rangle & = & \cdots 
- \frac{i}{\hbar} \int dt' e^{-(i/\hbar) (\varepsilon_j + \varepsilon_k - \varepsilon_i - \varepsilon_0 )(t-t')} V_{jki0}
\nonumber \\
& \times & 2 \left[ n_i (n_j +1)(n_k +1) - (n_i +1)n_j n_k \right](t) z_0(t)
\label{triplet} \;.
\eea
Assuming no dependence on the initial state and introducing the shorthand notation $\varepsilon_{jki0} = \varepsilon_j + \varepsilon_k - \varepsilon_i - \varepsilon_0$ we can evaluate the above integral (by introducing a convergence factor) to obtain
\be
\int_{-\infty}^{t} dt' e^{-(i/\hbar) \varepsilon_{jkio} (t-t')} = \pi \delta(\varepsilon_{jki0}) 
+ i{\cal P}\left( \frac{1}{\varepsilon_{jki0}} \right)\;,
\label{Integral}
\ee
where ${\cal P}$ denotes the principal part, and the `loss' of memory effects has enabled us to extend the lower limit of integration to $-\infty$.
The presence of both a real and an imaginary contribution in the expression for 
$\langle \hat{c}_i^\dag \hat{c}_j \hat{c}_k \rangle $ implies that the appearance of this term in Eq.\ (\ref{z0_triplet}) leads both to a change in the amplitude of the condensate mean field $z_0$ (arising from the $\delta$-function term) and a change in the system's frequency (principal value term).
The former contribution reads
\bea
\frac{dz_0}{dt} = \cdots +&& 
 \left( \frac{2 \pi}{\hbar^2} \right) z_0 \sum_{ijk} \left| V_{0ijk} \right|^{2}
\delta \left( \varepsilon_{0ijk} \right) \nonumber \\
&& \times \left[ (n_i+1)n_j n_k - n_i (n_j+1)(n_k +1) \right]\;.
\label{cond_growth}
\eea
Here $V_{0ijk}$ describes a process leading to the creation of a condensate atom (level $0$) by the collision of two thermal particles (levels $j$ and $k$), with the excess energy carried by a thermal particle in level $i$ (see Fig.\ \ref{Schematic_Collisions}). Such a process leads to a `production factor' which is proportional to $(n_i +1) n_j n_k$ with its inverse process, also included in Eq.\ (\ref{cond_growth}), yielding a corresponding factor of $n_i(n_j +1)(n_k +1)$. 
%
Finally, consistent elimination of the high-lying states (see e.g.\ \cite{Morgan_JPhysB,PGPE_T}) enables us to replace the exact interatomic potential $V$ by an effective T-matrix; in that case,
%
the interaction strength appearing here can be treated as a contact potential of effective strength $g$ (combined with a self-consistent high-energy truncation in the definition of the anomalous average).\\


\noindent {\bf Non-condensate Evolution:}
Following similar arguments,
the evolution of the thermally excited population $n_{i}$ is found to contain (in addition to other terms) the following two contributions (shown schematically in Fig.\ \ref{Schematic_Collisions}):

(i) The factor
\bea
\frac{ dn_i}{dt} = \cdots &+& \left( \frac{4 \pi}{\hbar^{2}} \right) \sum_{jk} \left| V_{0ijk} \right|^{2} \delta \left( \varepsilon_{0ijk} \right) \nonumber \\
&\times& |z_0|^2 \left[ (n_i+1)n_j n_k - n_i (n_j+1)(n_k +1) \right] 
\label{thermal_exchange}
\eea
clearly corresponds to the same collisional process as described earlier for the condensate, Eq.\ (\ref{cond_growth}), and displays the effect of particle exchange collisions between the condensate and the thermal cloud on the population of a given single-particle thermal level.

Observation of these equations shows that while the scattering of a particle into state $i$ is bosonically enhanced by the factor $(n_i+1)$, the corresponding scattering into the condensate does not feature spontaneous growth; in other words, one obtains $|z_0|^2$ both in Eq.\ (\ref{thermal_exchange}) and in the evolution of the condensate population, $d|z_0|^2/dt$, arising from Eq.\ (\ref{cond_growth}). This highlights the fact that the condensate mean field cannot grow from zero initial value, a well-known limitation of any mean field theory. (This issue can be cured within the context of stochastic approaches, briefly reviewed in Secs.\ \ref{Stoof_Theory}-\ref{Gardiner_Zoller_Theory}.)

(ii) The factor
\bea
\frac{ dn_i}{dt} &=& \cdots + \left( \frac{4 \pi}{\hbar^{2}} \right) \sum_{mjk} \left| V_{mijk} \right|^{2} \delta \left(\varepsilon_{mijk} \right) \nonumber \\
&\times& \left[ (n_i+1)(n_m +1)n_j n_k - n_i n_m (n_j+1)(n_k +1) \right] 
\label{thermal_growth}
\eea
arises from scattering of particles from states $j$ and $k$ into states $i$ and $m$, and the inverse process, and has associated with it the usual bosonic enhancement ($(n_.+1)$).
This term is present irrespective of the existence of a condensate and corresponds to the usual scattering factors and amplitudes appearing in the collisional integrals of the (classical) Boltzmann equation for scattering of particles in a thermal gas \cite{QK_II,QBE_Luiten,QBE_Holland}.

Let us now revisit the above discussion in a more realistic context:
Our preceding toy model has been expressed in terms of bare single-particle energies. In general, populations $\langle \hat{c}_i^\dag \hat{c}_j \rangle$ are however not diagonal in this basis, resulting in a plethora of additional terms in the corresponding dynamical equations within such a basis (see \ref{JILA_Theory}). We stress that the appearance of such contributions is formally correct, and any self-consistent treatment which includes them is not restricted to the extremely weakly-interacting regime $gn \ll \hbar \omega$; nonetheless, when one chooses such a basis,  the interpretation of the resulting equations may not be as straightforward.
In order to meaningfully interpret all contributions arising in this approach, one should ideally transform all expressions in terms of quasiparticle populations, which can be physically identified as the corresponding appropriate dressed particles due to interactions. Such a reformulation
was performed (in the so-called Popov limit) by Imamovic-Tomasovic and Griffin in the context of non-equilibrium Green's functions \cite{Milena_Griffin}, while the theory of Walser {\em et al.} \cite{JILA_Kinetic_1} was also recast in more compact form using similar notation by Wachter in \cite{Wachter_Masters,JILA_Unpublished}.

\subsubsection{Perturbative formulation beyond an appropriately generalized basis:}

\label{HFBT_Dynamics}

In our mean field approximations (\ref{mf-quad}), (\ref{mf-trip}) which led to the HFB hamiltonian, we have only considered up to quadratic mean fields, i.e. averages of two non-condensate operators, namely $\tilde{n} = \langle \nopc \nopa \rangle$ and $\tilde{m} = \langle \nopa \nopa \rangle$; the unperturbed hamiltonian $\hat{H}_{MF}$ of Eq.\ (\ref{MF_Pert}) was thus limited to the HFB hamiltonian. 
However, the exact equations of motion for $\wfn$ (Eq.\ (\ref{genGP})) and $\nopa$ (Eq.\ (\ref{delta_evol})) identify the crucial role of the triplet  $\langle \nopc \nopa \nopa \rangle$ which was ignored in much of the early mean-field literature (but see also \cite{Key_Paper_6}).
In order to also include this, one could thus consider resorting to a more general decoupling approximation for $\nopc \nopc \nopa \nopa$ than the one given by Eq.\ (\ref{mf-quad}).
In particular, one could impose an approximation of the form \cite{Proukakis_Thesis,Proukakis_JPhysB}
\bea
 \nopc \nopc \nopa \nopa &\simeq& 4 \langle \nopc \nopa \rangle \nopc \nopa 
 + \langle \nopc \nopc \rangle \nopa \nopa + \langle \nopa \nopa \rangle
 \nopc \nopc 
- [ 2 \langle \nopc \nopa \rangle \langle \nopc \nopa \rangle
  + \langle \nopa \nopa \rangle \langle \nopc \nopc \rangle].
\nonumber \\
&& + 2 \left[ \langle \nopc \nopc \nopa \rangle \nopa + \langle \nopc \nopa \nopa \rangle \nopc \right]\;.
\label{mf-quad-new}
\eea

Following our previously introduced notation, substitution of this generalized mean-field approximation into the expression for $\hat{H}_4$ (Eq.\ (\ref{H_4})) would generate, in addition to the contributions $(\delta \hat{H}_1^{HF} + \delta \hat{H}_1^{BOG})$ of Eq.\ (\ref{dH_1}),
an extra contribution to the linear part of the hamiltonian $\int d\bldr \nopc ( \cdots ) + {\rm h.c.}$, of the form
\be
\delta \hat{H}_1^{TRIP} = g \int d \bldr \nopc \langle \nopc \nopa \nopa \rangle +{\rm h.c.}
\label{dH_1_Trip} \;.
\ee

We shall now follow our earlier perturbative prescription of Eq.\ (\ref{MF_Pert}) ($\hat{H} = \hat{H}_{\rm MF} + ( \hat{H} -\hat{H}_{\rm MF} )$);
however, this time we choose a slightly more general unperturbed hamiltonian which
also includes $ \delta \hat{H}_1^{TRIP}$, i.e.\ 
we choose $\hat{H}_{MF}=\hat{H}_{HFBT}$
where we have identified a new `Hartree-Fock-Bogoliubov-Triplet' basis via (making explicit the split into Hartree-Fock and Bogoliubov contributions)
\bea
\hat{H}_{HFBT}&=& \hat{H}_{HFB} + \delta \hat{H}_1^{TRIP} 
= \left( H_0 + \delta H_0^{HF} +\delta H_0^{BOG} +\delta H_0^{TRIP}\right) \nonumber \\
&+& \left[ \hat{H}_1 + \left( \delta \hat{H}_1^{HF} + \delta \hat{H}_1^{BOG} + \delta \hat{H}_1^{TRIP} \right) \right] \nonumber \\
&+& \left[ \hat{H}_2 + \left( \delta \hat{H}_2^{HF} + \delta \hat{H}_2^{BOG} \right) \right]
\label{H_HFBT} \;.
\eea
The above expression additionally includes a mean field correction 
$\delta H_0^{TRIP} = g \int d\bldr \langle \nopc \nopa \nopa \rangle \wfn^* +{\rm c.c.}$
to the zeroth-order hamiltonian arising from an appropriately generalized form of Eq.\ (\ref{mf-trip}) which additionally includes $\langle \nopc \nopa \nopa \rangle$ in its right hand side.
Since such mean field contributions merely shift the ground state energy of the system and have no additional effects on the dynamics of interest, we shall henceforth disregard in our discussion all non-operator mean field contributions $\delta H_0^{HF}$, $\delta H_0^{BOG}$ and $\delta H_0^{TRIP}$.

It is easy to see that the above expression still constitutes a generalized quadratic hamiltonian.
Hence, this hamiltonian can in principle be diagonalized by means of the Bogoliubov transformation discussed earlier; the first such discussion was presented by one of us (NPP) in early work \cite{Proukakis_Thesis} in an attempt to justify from a microscopic basis the generalized HFB theory of Sec.\ \ref{many-body}, in which pair and triplet anomalous averages play a key role in upgrading the effective interaction between two atoms into a many-body one. 
However, it was found in \cite{Proukakis_Thesis} that such a theory cannot be microscopically derived by variational methods, as had been previously argued (on slightly different and more general grounds) by Bijlsma and Stoof \cite{Stoof_Variational}; the reason is that one runs into the same difficulties regarding the inconsistent treatment of atomic interactions, as in the full HFB theory. (This is in fact what led to the `heuristic' addition of the generalized effective interaction $g(\bldr)$.)
One way to overcome this complication is based on a minor redefinition and handling of the hamiltonian; this is effectively what was done by Zaremba, Nikuni and Griffin, as discussed below - although their arguments were actually given in a slightly different manner.

For ease of subsequent comparison, we give here the full expression for the above perturbing hamiltonian (ignoring energy shifts of the form $\delta H_0^{\cdots}$), which takes the form
\bea
\hat{H}''&=& \hat{H}-\hat{H}_{HFBT} \nonumber \\
&=&
\left[ \hat{H}_3 - \left( \delta \hat{H}_1^{HF} + \delta \hat{H}_1^{BOG} \right) \right] \nonumber \\
&+& \left[ \hat{H}_4 
- \left(
\delta \hat{H}_2^{HF} + \delta \hat{H}_2^{BOG} + \delta \hat{H}_1^{TRIP}\right) \right]
\label{Beyond_HFBT} \;.
\eea

\subsubsection{Perturbative Distribution Function Formulation: Treatment of Zaremba-Nikuni-Griffin:}

\label{ZNG}

The preceeding discussion demonstrated how use of perturbation theory facilitates a generalization beyond the static thermal cloud approximation, even after imposing the mean field approximations.
One can actually use this approach to formulate equations which are similar in spirit to those encountered in the kinetic theory of classical gases, but are appropriately generalized below the transition temperature by the additional inclusion of the condensate mean field.

Such an approach was formulated in the early 1980's in seminal work by Kirkpatrick and Dorfman \cite{Kirkpatrick_1,Kirkpatrick_2,Kirkpatrick_3}, with related work undertaken by Eckern \cite{Eckern}.
In particular, Kirkpatrick and Dorfman derived a closed kinetic equation for the quasiparticle distribution function of an inhomogeneous Bose gas below the transition temperature. The evolution of such a distribution function contains both `streaming' and `collisional' terms, and momentum enters explicitly as a system variable.
This approach was recently employed and extended by Zaremba, Nikuni and Griffin \cite{ZNG}, as outlined below. In terms of the `classifications' introduced earlier, this theory can be thought of as 
 the consistent time-dependent extension of Hartree-Fock-Bogoliubov-Popov of Sec.\ \ref{HFB_Popov} which additionally includes collisions within the thermal cloud and particle-exchange collisions between condensate and thermal atoms.


We start our discussion by reformulating the preceeding problem in terms of a suitable density matrix as follows:
we assume that the system is initially described by a density matrix $\hat{\rho}(t_0)$, such that the expectation value of a general operator $\hat{O}$ is given by
\be
\langle \hat{O} \rangle 
= {\rm Tr} \hat{\rho}(t_0) \hat{O}(t) = {\rm Tr} \tilde{\rho}(t,t_0) \hat{O}(t_0)\;.
\label{exp_value}
\ee
Here ${\rm Tr}$ denotes the trace and we have introduced the density matrix
$\tilde{\rho}(t,t_0) = \hat{U}(t,t_0) \hat{\rho}(t_0) \hat{U}^{\dag}(t,t_0)$
where $\hat{U}$ is a unitary operator defined by
\be
i \hbar \frac{d}{dt} \hat{U}(t,t_0) = \hat{H}_{\rm eff}(t) \hat{U}(t,t_0)\;.
\label{U_evol}
\ee
The corresponding density matrix evolves according to the equation of motion
\be
i \hbar \frac{d}{dt} \tilde{\rho}(t,t_0)  = \left[ \hat{H}_{\rm eff}(t) , \tilde{\rho}(t,t_0) \right]\;.
\label{rho_evol}
\ee
and the problem thus reduces to the identification of the appropriate effective system hamiltonian $\hat{H}_{\rm eff}$.

In this section, we follow closely the discussion given by Zaremba, Nikuni and Griffin in \cite{ZNG}: In their work, they chose $\hat{H}_{\rm eff}$ such that
the selected hamiltonian reproduces accurately the {\em exact} equations of motion for the condensate and the non-condensate operator, given respectively by Eqs.\ (\ref{genGP}) and (\ref{delta_evol}).
As before, the effective hamiltonian was split into an unperturbed hamiltonian ($\hat{H}_{\rm unp}$) and a contribution to be treated in second order perturbation theory, via
$\hat{H}_{\rm eff} = \hat{H}_{\rm unp} + \hat{H}'''$, with
the perturbing hamiltonian, $\hat{H}'''$, defined as \cite{ZNG}
\bea
\hat{H}''' 
&=& \frac{g}{2} \int d \bldr \left[ \wfn^2 \nopc \nopc + (\wfn^*)^2 \nopa \nopa \right] 
+g \int d \bldr \left[ \wfn^* \nopc \nopa \nopa + \nopc \nopc \nopa \wfn \right] \nonumber \\
&& -g \int d\bldr \left[ \nopc \left( 2 \tilde{n} \wfn + \tilde{m} \wfn^* \right)
+ \left( 2 \tilde{n} \wfn^* + \tilde{m}^* \wfn \right) \nopa \right] \nonumber \\
&+& \frac{g}{2} \int d \bldr \nopc \nopc \nopa \nopa 
-2g \int d\bldr \tilde{n} \nopc \nopa  \nonumber \\
&& -g \int d\bldr \left[ \nopc \langle \nopc \nopa \nopa \rangle 
+ \langle \nopc \nopc \nopa \rangle \nopa \right] 
\;.
\eea

It is instructive to compare this perturbing hamiltonian $\hat{H}'''$ to the one introduced earlier in
 Eq.\ (\ref{Beyond_HFBT}). To enable a more direct comparison, we separate here the original $\hat{H}_2$ hamiltonian of Eq.\ (\ref{H_2}) into two contributions (see also Table \ref{Table_Hamiltonians}), which we shall henceforth refer to as `Hartree' and `Bogoliubov' terms, via
\bea
\hat{H}_2 &=& \hat{H}_2^{H}+\hat{H}_2^{BOG} \nonumber \\
&=& \int d\bldr  \nopc \left( \spop + 2g |\wfn|^2 \right) \nopa 
 + \frac{g}{2} \int d\bldr \left[ \nopc \nopc \wfn^2 + (\wfn^*)^2 \nopa \nopa \right]
\;.
\eea
We find that
$\hat{H}'''$, is related to the perturbing hamiltonian $\hat{H}''$ of Sec. \ref{HFBT_Dynamics} via
\bea
\hat{H}''' 
&=& \hat{H}_{2}^{BOG} + \left[ \hat{H}_3 - \left( \delta \hat{H}_1^{HF} + \delta \hat{H}_1^{BOG} \right) \right] \nonumber \\
&&+ \left[ \hat{H}_4 - \left( \delta \hat{H}_2^{HF} + \delta \hat{H}_1^{TRIP} \right) \right] \nonumber \\ 
&=& \hat{H}'' + \left( \hat{H}_2^{BOG} + \delta \hat{H}_2^{BOG} \right) 
\;.
\eea

The particular choice of the perturbing hamiltonian directly identifies the unperturbed hamiltonian which defines the spectrum of elementary excitations;
here we clearly see that the main difference to the discussion of Sec.\ \ref{HFBT_Dynamics} is the contribution
\be
\hat{H}_2^{BOG} + \delta \hat{H}_2^{BOG} = \frac{g}{2} \int d \bldr 
\left\{ \nopc \nopc \left( \wfn^2 + \tilde{m} \right) + \left( (\wfn^*)^2+\tilde{m}^* \right) \nopa \nopa \right\} \;,
\ee
whose role is to change the excitation energies from single-particle to (finite temperature) quasiparticle ones.


If one were to literally follow the above argument,
the unperturbed hamiltonian within the context of the Zaremba-Nikuni-Griffin approach should be given by
\bea
\hat{H}_{\rm unp} = \hat{H} - \hat{H}''' &=& H_0 + \left[ \hat{H}_1 
+ \left( \delta \hat{H}_1^{HF} + \delta \hat{H}_1^{BOG} + \delta \hat{H}_1^{TRIP} \right) \right]
\nonumber \\
&+& \left[ \hat{H}_2^{H} + \delta \hat{H}_2^{HF} \right]
\label{H_ZNG_Un}\;.
\eea
%
%
%
Writing the single-operator contributions within $\hat{H}_{\rm unp}$ explicitly as
\be
\int d \bldr \nopc \left( \hat{h}_0 \wfn + g |\wfn|^2 \wfn + 2 g \tilde{n} \wfn + g \tilde{m} \wfn^* + g \langle \nopc \nopa \nopa \rangle \right) + {\rm h.c.} \;,
\ee
and observing the exact time-dependent equation for the condensate mean field given by Eq.\ (\ref{genGP}), we immediately see that these terms 
provide us with as expression for the time-dependent chemical potential. To make direct contact with the original presentation Zaremba, Nikuni and Griffin \cite{ZNG}, we use the Madelung transformation $\wfn(\bldr,t)=|\wfn(\bldr,t)|e^{i \theta(\bldr,t)}$ to obtain the hydrodynamic equations (given in full in Sec.\ \ref{Hydro_T}) for the condensate density $n_0 = |\wfn(\bldr,t)|^2$ and ${\bf v_c}(\bldr,t) = (\hbar/m) \nabla \theta(\bldr,t)$. 
This procedure identifies a local chemical potential $\mu_c$, of the form \cite{ZNG}
\be
\mu_c =  \mu_0 + 2 g \tilde{n} 
+ \frac{1}{|\wfn|^2} Re \left\{ \tilde{m} (\wfn^*)^2 + \langle \nopc \nopa \nopa \rangle \wfn^* \right\}
\;;
\ee
here we see that that contributions arising from the anomalous averages provide a net real contribution to $\mu_c$, as physically relevant.
These `anomalous' contributions can change the chemical potential by introducing modifications to the collisional amplitude due to the medium in which the collisions are taking place (see discussion of $\tilde{m}_0(\bldr)$ in Sec.\ \ref{many-body}).
To avoid consistency issues with the unified treatment of interactions when simultaneously imposing the pseudopotential approximation (already implicit in this treatment), Zaremba, Nikuni and Griffin limited their analysis to the `Popov' limit \cite{Griffin_HFB}, in which all anomalous averages are assumed to be zero;
this should not be misinterpreted as indicating that such terms have no effect, as in fact an important part of their role is taken into account in introducing particle-exchange collisions into the theory.
This procedure is essentially equivalent to maintaining interaction effects in the chemical potential and excitation energies only to first order in the effective interaction strength $g$, whereas the effect of interactions in calculating collision integrals is treated, as required, to second order in $g$ \cite{ZNG}. In this limit, the condensate chemical potential acquires the form
\be
\mu_c =  \mu_0 + 2 g \tilde{n}
= -\frac{ \hbar^2 \nabla^2 \sqrt{n_0}}{2m \sqrt{n_0}} + V_{\rm ext} +g (n_0 + 2 \tilde{n}) \;.
\label{mu_T}
\ee

We now also consider the quadratic hamiltonian 
$(\hat{H}_2^{H}+\delta \hat{H}_2^{HF})$
appearing in Eq.\ (\ref{H_ZNG_Un}). It is easy to see that, as argued by Zaremba, Nikuni and Griffin \cite{ZNG}, this is directly equivalent to the Hartree-Fock hamiltonian,
namely
\bea
\hat{H}_{HF} 
= \int d \bldr \nopc \left[ - \frac{\hbar^2 \nabla^2}{2m}  + \left[ V_{\rm ext}(\bldr,t)+ 2g \left( |\wfn|^2 + \tilde{n} \right) \right] \right] \nopa 
\label{HF_ZNG}
\;.
\eea
In other words, the single-particle excitation energies are dressed by interactions in the manner discussed in Sec.\ \ref{HF_Regime}. Alternatively, one can think of the
condensate and thermal densities as providing an additional mean field potential through which the atoms propagate. This leads naturally to the definition of a {\em generalized mean field potential} given by
\be
U(\bldr,t) = V_{\rm ext}(\bldr,t) + 2g \left[ |\wfn(\bldr,t)|^2 + \tilde{n}(\bldr,t) \right]\;.
\label{U_MF}
\ee
Note that use of such a generalized mean field potential is common in theories based on Hartree-Fock eigenenergies; for the usual harmonic traps under consideration, and repulsive effective interactions, this takes the typical form of a double-well potential.

Having identified our perturbing hamiltonian, we use Eqs.\ (\ref{exp_value})-(\ref{rho_evol}) to express by formal integration the expectation value of a general operator $\hat{O}$ as
\bea
\langle \hat{O}(t) \rangle 
&=& {\rm Tr} \hat{\rho}(t_0)  \left\{ \hat{U}_{0}^{\dag}(t,t_0) \hat{O}(t_0) \hat{U}_{0}(t,t_0) 
- i \int_{t_0}^{t} dt' \hat{U}_{0}^{\dag} (t',t_0) \right. 
\nonumber \\
&& \times
\left. \left[ \hat{U}_{0}^{\dag}(t,t') \hat{O}(t_0) \hat{U}_{0}(t,t') \, , \, \hat{H}'''(t') \right] \hat{U}_{0}(t', t_0) \right\} \;.
\label{Exp_t}
\eea
In this expression, the first term depends on the initial correlations, while the second term can be identified as dynamical collisional effects arising from the perturbing hamiltonian $\hat{H}'''$.

We are interested in using Eq.\ (\ref{Exp_t}) to evaluate anomalous correlations of non-condensate operators, which we shall assume to be negligible initially, i.e.\ we can consistently ignore the first contribution.
As the evolution of $\hat{U}$ is number conserving, in evaluating such anomalous averages one need only maintain terms which explicitly conserve the total atom number (thus discarding non-atom-number-conserving contributions arising here).
We also assume as before that the collisional duration is short compared to characteristic evolution times of the parameters we are interested in. This enables us to make the Markov approximation $\hat{U}_{0}(t,t_0) \approx {\rm exp} \{ -i \hat{H}_{HF}(t) (t-t_0) \}$.

As mentioned earlier, the novel feature of our current discussion is the formulation of the problem in terms of a distribution function $f({\bf p}, \bldr, t)$, for which the description of the gas can be seen as a direct extension of the kinetic theory treatment of a classical gas \cite{Landau_Lifshitz} below the transition point. In their original presentation \cite{Kirkpatrick_1,Kirkpatrick_2}, Kirkpatrick and Dorfman formulated their discussion in terms of the full quasiparticle distribution function, although for most temperatures of experimental relevance for trapped Bose gases\footnote{The usual criterion for this to be valid is that $n a \lambda_{dB}^2 \ll 1$, which is certainly well-satisfied in current dilute atomic gas experiments.} a discussion in terms of dressed single-particle eigenstates should suffice. Thus, the formulation of Zaremba, Nikuni and Griffin is based on a single-particle Wigner distribution function.

For an atom of momentum ${\bf p}$, at location ${\bf r}$ and time $t$ the distribution function $ f({\bf p}, \bldr, t)$ is defined as the expectation value 
\be
f({\bf p},\bldr,t) = \langle \hat{f}({\bf p},\bldr,t) \rangle 
\label{Wigner_ZNG_1}
\ee
where $\hat{f}({\bf p},\bldr,t) $ is the Wigner operator
\be
\hat{f}({\bf p},{\bf r},t) = \int d\bldr' e^{i {\bf p} \cdot {\bf r'} / \hbar}
\hat{\delta}^{\dag} \left( {\bf r} + \frac{ {\bf r'}}{2}, t_0 \right)
\hat{\delta} \left( {\bf r} - \frac{ {\bf r'}}{2}, t_0 \right)\;.
\label{Wigner_ZNG_2}
\ee
The above formulation enables us to obtain the non-condensate density via 
\be
\tilde{n}(\bldr,t) = \int \frac{d {\bf p}}{(2 \pi \hbar)^3} f({\bf p}, \bldr, t)\;.
\label{nf}
\ee
Momentum variables are introduced into the system by expanding the non-condensate operators into Fourier modes (over non-zero momenta)
\cite{Kirkpatrick_2,Kirkpatrick_3}
\be
\hat{\delta}(\bldr,t_0) = \frac{1}{\sqrt{V}} \sum_{{\bf p}}' \hat{c}_{\bf p} e^{i {\bf p} \cdot {\bf r}},
\hspace{0.5cm}
\hat{\delta}^{\dag}(\bldr,t_0) = \frac{1}{\sqrt{V}} \sum_{{\bf p}}' \hat{c}_{\bf p}^{\dag} e^{-i {\bf p} \cdot {\bf r}}\;.
\ee
We now proceed explicitly to calculate the value of  $\langle \nopc \nopa \nopa \rangle$.
The temporal integrals arising in the second contribution of Eq.\ (\ref{Exp_t}) are evaluated as in Eq.\ (\ref{Integral}) via 
\bea
\int_{-\infty}^{t} dt' e^{i ( \varepsilon_c + \tilde{\varepsilon}_2-\tilde{\varepsilon}_3-\tilde{\varepsilon}_4)} &\approx &\pi \delta 
(\varepsilon_c + \tilde{\varepsilon}_2-\tilde{\varepsilon}_3-\tilde{\varepsilon}_4) \nonumber \\
&+& iP \frac{1}{(\varepsilon_c + \tilde{\varepsilon}_2-\tilde{\varepsilon}_3-\tilde{\varepsilon}_4)}
\;.
\eea
Importantly, due to the choice of Eq.\ (\ref{HF_ZNG}) as the unperturbed system hamiltonian, the energies $\tilde{\varepsilon}_i=\tilde{\varepsilon}_i(\bldr,t)=\tilde{\varepsilon}({\bf p_i}, \bldr_i,t)$ appearing in this and all subsequent expressions are evaluated `semi-classically' in the Hartree-Fock limit 
via (see Eq.\ (\ref{HF_Energy}))
\bea
\tilde{\varepsilon}_{i} (\bldr,t)
= \frac{|{\bf p_i}|^2}{2m} + V_{\rm ext}(\bldr_i,t) + 2g \left[ |\wfn(\bldr_i,t)|^2 + \tilde{n}(\bldr_i,t) \right]
\;.
\label{E_HF}
\eea
Correspondingly, the condensate energy, $\varepsilon_c$, is given by
\be
\varepsilon_c = \mu_c + \frac{1}{2} m v_c^2 \;,
\label{E_c}
\ee
where $\mu_c$ is the condensate chemical potential of Eq.\ (\ref{mu_T})  and
$v_c$ denotes the condensate speed, which should be determined self-consistently from the quantum-mechanical current
\be
{\bf v_c}(\bldr,t) = i \left( \frac{\hbar}{2m} \right) \frac{ \wfn {\bf \nabla} \wfn^* - \wfn^* {\bf \nabla} \wfn }{|\wfn|^2}
\;.
\label{v_c}
\ee 

We proceed by replacing momentum sums $(1/V) \sum_{{\bf p}}$ by integrals $\int d{\bf p}/(2 \pi \hbar)^{3}$ with the Kronecker delta symbols arising from the commutation relations replaced by Dirac delta functions, i.e.\ $\delta_{{\bf p}, {\bf 0}} \rightarrow (1/V) (2 \pi \hbar)^{3} \delta ({\bf p})$.
To obtain a closed system of coupled equations, we decorrelate higher-order correlations in the fluctuation operators $\hat{c}_{\bf p}$ by means of Wick's theorem, and
express these decorrelated averages in terms of the Wigner distribution function via
$
\langle \hat{c}_{{\bf p_{1}}}^{\dag} \hat{c}_{{\bf p_2}} \rangle \approx \delta_{{\bf p_1}\, , \, {\bf p_2}} f({\bf p_1}, {\bf r_1}, t)\;.
$

Thus, in direct analogy to Eqs.\ (\ref{triplet})-(\ref{Integral}) \cite{Proukakis_Burnett_Stoof,Proukakis_Thesis} the triplet contribution acquires to lowest order in $g$ the value \cite{ZNG}
\bea
\langle \nopc \nopa \nopa \rangle = & & -i 2 \pi g \wfn
\int \frac{d {\bf p_2}}{(2 \pi \hbar)^3}  \int \frac{d {\bf p_3}}{(2 \pi \hbar)^3}  
\int \frac{d {\bf p_4}}{(2 \pi \hbar)^3}  
\nonumber \\
&& \times (2 \pi \hbar)^3 \delta \left( m{\bf v}_c + {\bf p_2} - {\bf p_3} - {\bf p_4} \right) 
 \delta \left( \varepsilon_c + \tilde{\varepsilon}_2-\tilde{\varepsilon}_3-\tilde{\varepsilon}_4 \right) 
\nonumber \\ & &
\times \left[ f_2 (f_3 +1) (f_4 +1) - (f_2 +1) f_3 f_4 \right]
\eea
where $f_{i}=f({\bf p_{i}},{\bf r_{i}},t)$. 
Note that the same procedure can in principle be used to derive an expression for the pair anomalous average $\tilde{m}(\bldr)$, whose real contribution yields the correction to the scattering amplitude due to many-body effects discussed in Sec.\ \ref{many-body}. 

The equation for the condensate evolution (Eq.\ (\ref{genGP})) now becomes
\begin{eqnarray}
i\hbar \frac{\partial \wfn}{\partial t}(\bldr,t) 
= \left[
\hat{h}_0 + g \left( |\wfn(\bldr,t)|^{2} +2 \tilde{n}(\bldr,t) \right) - iR(\bldr,t) \right] \wfn (\bldr,t) 
\label{GPE_iR}
\end{eqnarray}
where the contribution $iR$ arising from the triplet correlation function describes the exchange of particles between the condensate and the thermal cloud (c.f.\ Eq.\ (\ref{cond_growth}))
\bea
R(\bldr,t) &=& - i g  \frac{ \langle \nopc \nopa \nopa \rangle(\bldr,t)}{ \wfn(\bldr,t)} 
\nonumber \\
&=&  2 \pi g^2
\int \frac{d {\bf p_2}}{(2 \pi \hbar)^3}  \int \frac{d {\bf p_3}}{(2 \pi \hbar)^3}  
\int \frac{d {\bf p_4}}{(2 \pi \hbar)^3}  
\nonumber \\
&& \times (2 \pi \hbar)^3 \delta \left( m{\bf v}_c + {\bf p_2} - {\bf p_3} - {\bf p_4} \right) 
 \delta \left( \varepsilon_c + \tilde{\varepsilon}_2-\tilde{\varepsilon}_3-\tilde{\varepsilon}_4 \right) 
\nonumber \\ 
& &
\times \left[ f_2 (f_3 +1) (f_4 +1) - (f_2 +1) f_3 f_4 \right]
\label{R}
\eea
The term $iR$ can be interpreted as follows: 
Out of equilibrium this term leads to transfer of atoms between the two subsystems; once local equilibrium is (re-)established
between the condensed and thermal components, $iR = 0$ identically, and the condensate is described by the simpler Hartree-Fock expression of Eq.\ (\ref{GPE_HF}).

Having identified the `growth' term in the condensate evolution equation, we must also derive the corresponding evolution of the thermal cloud in order to obtain a closed system of equations.
Following the usual formulation of the kinetic theory of gases
\cite{Kirkpatrick_1,Kirkpatrick_2,Kirkpatrick_3,Eckern}, the evolution of $f({\bf p}, {\bf r},t)$ in the presence of a slowly-varying external field $U({\bf r},t)$ (for which we can make a gradient expansion) takes the form
\be
\frac{df}{dt} = \frac{ \partial f}{\partial t} + {\bf v} \cdot {\bf \nabla_{\bldr}} f 
- ( {\bf \nabla_{\bldr}} U ) \cdot ( {\bf \nabla}_{\bf p} f)  \, ,
\ee
where ${\bf \nabla_r}$ and ${\bf \nabla_p}$ denote gradients in position and momentum variables respectively.
Although in the absence of collisions $f$ would obey Liouville's equation $df/dt=0$, 
in the presence of collisions, the value of $f$ changes at a rate $C[f]$, which is called the collision integral. 
Thus, the problem of transport in a gas reduces to the identification and calculation of the appropriate collision integrals. 
The effect of collisions between the atoms is included into the evolution of the distribution function via the contribution
\be
\frac{df}{dt} = \frac{1}{i \hbar} {\rm Tr} \tilde{\rho}(t,t_0) \left[ \hat{f}({\bf p},{\bf r},t_0) \, , \, \hat{H}'''(t) \right]\;.
\label{R_C12}
\ee
This yields the following expression for the evolution of the distribution function
\be
\frac{ \partial f}{\partial t} + {\bf v} \cdot {\bf \nabla_{\bldr}} f 
- ( {\bf \nabla_{\bldr}} U ) \cdot ( {\bf \nabla}_{\bf p} f)
= C_{12}[f] + C_{22}[f] \,\,\,.
\label{QBE}
\ee
Note that due to the presence of the mean fields of condensed and thermal atoms, all atoms experience the generalized potential $U(\bldr,t)$ defined by Eq.\ (\ref{U_MF}).
Equation (\ref{QBE}), first derived in slightly modified form in this context by Kirkpatrick and Dorfman \cite{Kirkpatrick_2}, is known as the `Quantum Boltzmann Equation' (see, e.g., \cite{QK_II,QBE_Luiten,QBE_Holland}); this equation reduces to the classical `Uehling-Uhlenbeck' equation \cite{Uhling_Uhlenbeck} when $C_{12}[f]=0$, i.e.\ above the transition temperature when there is no condensate present.
To strengthen the link to our introductory discussion (Sec.\ \ref{Prouk_QK}), we note that 
the collisional integrals  $C_{12}[f]$ and $C_{22}[f]$ appearing in Eq.\ (\ref{QBE}) are the direct analogues of Eqs.\ (\ref{thermal_exchange}) and (\ref{thermal_growth}) (and their more general expressions given in \cite{Proukakis_JPhysB,JILA_Kinetic_1} in terms of single-particle variables),  now expressed explicitly in terms of $f$. 
They describe the following processes:
\begin{itemize}
\item
A collision involving the transfer of an atom from the thermal cloud into the condensate and its inverse process (c.f.\ Eq.\ (\ref{thermal_exchange}))
\bea
C_{12}[f] & = &
\frac{4 \pi}{\hbar} g^{2} |\phi|^{2} \int \frac{ d {\bf p}_{2}}{(2 \pi \hbar)^{3}} \int \frac{ d {\bf p}_{3}}{(2 \pi \hbar)^{3}} \int \frac{ d {\bf p}_{4}}{(2 \pi \hbar)^{3}}
\nonumber \\
& \times & (2 \pi \hbar)^{3}
\delta \left( m{\bf v}_{c}+{\bf p}_{2} - {\bf p}_{3} - {\bf p}_{4} \right) 
\times \delta \left( \varepsilon_{c} + \tilde{\varepsilon}_{2} - \tilde{\varepsilon}_{3} - \tilde{\varepsilon}_{4} \right) \nonumber \\
& \times & (2 \pi \hbar)^{3}
\left[ \delta({\bf p}-{\bf p}_{2})-\delta({\bf p}-{\bf p}_{3})-\delta({\bf p}-{\bf p}_{4}) \right] \nonumber \\
& \times & \left[ (f_{2}+1)f_{3}f_{4} - f_{2} (f_{3}+1)(f_{4}+1) \right]\;.
\label{C_12}
\eea
Clearly this term must be related to the condensate source term (c.f.\ Eq.\ (\ref{cond_growth})), and indeed one finds
\be
R(\bldr,t) = \frac{\hbar}{|\wfn(\bldr,t)|^2} \int \frac{ d {\bf p}}{(2 \pi \hbar)^{3}} C_{12} [f({\bf p}, {\bf r}, t)] \;.
\ee
As a result, the final theory does actually preserve the total number of atoms in the system.

\item A collision between two thermal atoms (c.f.\ Eq.\ (\ref{thermal_growth})) given by
\bea
C_{22}[f] &=& \frac{4 \pi}{\hbar} g^{2} \int \frac{ d {\bf p}_{2}}{(2 \pi \hbar)^{3}} \int \frac{ d {\bf p}_{3}}{(2 \pi \hbar)^{3}} \int \frac{ d {\bf p}_{4}}{(2 \pi \hbar)^{3}} \nonumber \\ 
& \times & (2 \pi \hbar)^{3}
\delta \left( {\bf p}+{\bf p}_{2} - {\bf p}_{3} - {\bf p}_{4} \right) 
\delta \left( \tilde{\varepsilon} + \tilde{\varepsilon}_{2} - \tilde{\varepsilon}_{3} - \tilde{\varepsilon}_{4} \right) \nonumber \\
& \times & \left[ (f+1)(f_{2}+1)f_{3}f_{4} - f f_{2} (f_{3}+1)(f_{4}+1) \right] \;.
\label{C_22}
\eea
\end{itemize}

\begin{figure}[t]
\centering \scalebox{0.5}
 {\includegraphics{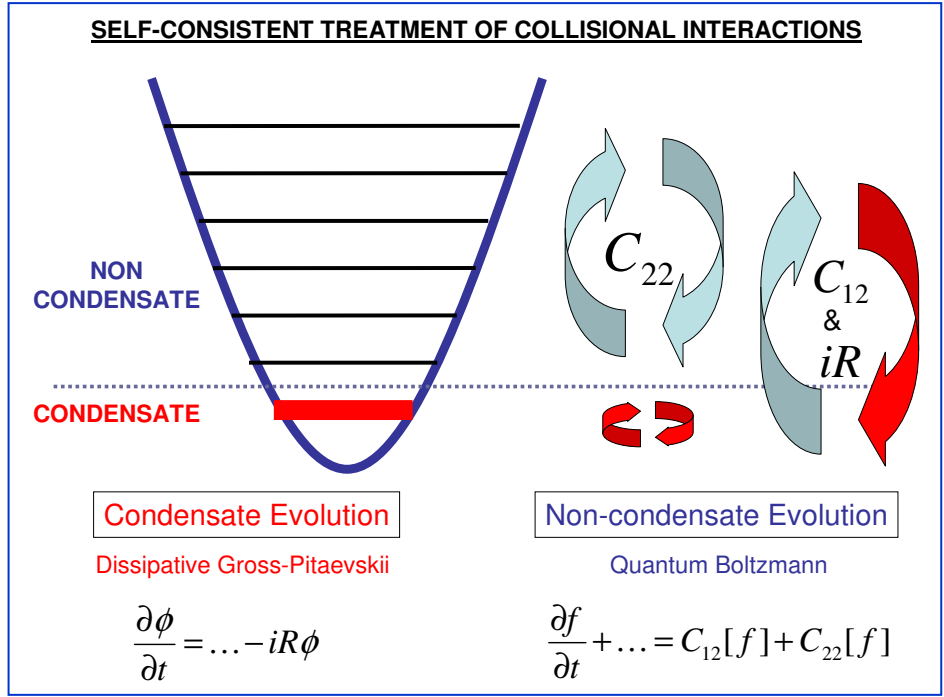}}
 \caption{
(colour online)
Schematic of characteristic collisions included in the self-consistent Gross-Pitaevskii-Boltzmann formalism of Eqs.\ (\ref{GPE_iR})-(\ref{QBE}), often referred to as the `ZNG' formalism .
}
\label{Schematic_ZNG}
\end{figure}

We have thus arrived at a self-consistent second order description of a partially Bose-condensed gas in terms of a dissipative Gross-Pitaevskii Equation (Eq.\ (\ref{GPE_iR})) for the condensate mode, coupled to a dynamical reservoir of thermal atoms obeying a `Quantum Boltzmann Equation'  (Eq.\ (\ref{QBE})), as depicted diagrammatically in Fig.\ \ref{Schematic_ZNG}.
These equations
should be solved simultaneously with Eqs.\ (\ref{nf}), (\ref{U_MF}), (\ref{R}), (\ref{C_12}), and (\ref{C_22}) with the energies of the condensate and its excitations determined self-consistently from Eqs.\ (\ref{E_c}) and (\ref{E_HF}).
An important advantage of this theory is that it incorporates thermal cloud dynamics in a self-consistent manner.
In the context of generalized mean fields and symmetry breaking, this is the most advanced model that is currently amenable to direct numerical simulations.

The first key step towards such a derivation was made possible by the pioneering work of Kirkpatrick and Dorfman \cite{Kirkpatrick_1,Kirkpatrick_2,Kirkpatrick_3}; another crucial step in the context of a mean field formulation was played by the identification of the role of the triplet correlation $\langle \nopc \nopa \nopa \rangle$ in the description of condensate kinetics (highlighted by one of us (NPP) in collaboration with Burnett and Stoof \cite{Proukakis_Burnett_Stoof,Proukakis_NIST,Proukakis_Thesis}). These elements were beautifully tied in together in important work by Zaremba, Nikuni and Griffin \cite{ZNG}, whose arguments were presented (in slightly modified form) in this section.
Over the past decade, these latter authors, along with their collaborators (including one of us - BJ) have used these coupled equations extensively to study the non-equilibrium behaviour of ultracold Bose gases (details given below). In these works, they introduced the shorthand notation `ZNG' (after their initials), and for notational convenience we shall occasionally refer to the theory presented in this section as the `ZNG' theory.
%
%
For completeness, we note that the theoretical number-conserving description formulated by Stoof \cite{Stoof_PRL,Stoof_JLTP,Stoof_Duine} and Gardiner and Zoller \cite{QK_V} presented in Secs.\ \ref{Stoof_Theory}-\ref{Gardiner_Zoller_Theory} include the above equations as a limiting case of their formalism.


In terms of the physics contained in these equations,
depending on the collisional rate, or equivalently on the collisional mean free path, one can identify two distinct regimes in the response of the system:
In the collisionless regime, which encompasses most of the current experiments with ultracold Bose gases, the effects of the mean fields dominate and collisions can be treated as perturbative corrections; here the full Wigner distribution function $f({\bf p}, {\bf r}, t)$ is needed to accurately describe the thermal cloud. In the opposite collisional regime, strong interactions lead to rapid local equilibration, and the thermal cloud can be described by a few local macroscopic variables, as in ordinary fluid mechanics; in our current context, however, these variables for the thermal cloud are also coupled to corresponding ones for the condensate parameters, thus leading to a hydrodynamic description which can be visualized as an extension of Landau's two-fluid hydrodynamics \cite{Pitaevskii_StatisticalPhysics_2}.
The formalism presented above has been used extensively both in the collisionless and in the collisional regimes (also by the authors - mainly BJ), and some examples are briefly mentioned below. 

\begin{figure}
  \begin{center}
      \resizebox{53mm}{!}{\includegraphics{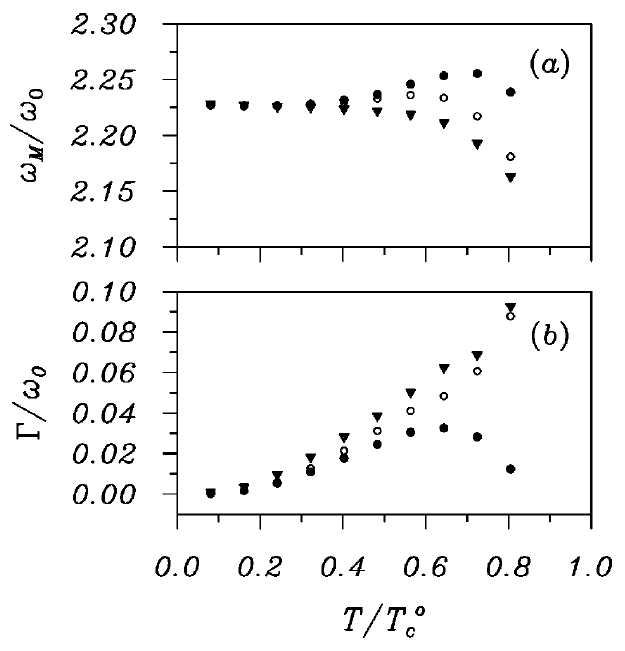}} 
      \resizebox{67mm}{!}{\includegraphics{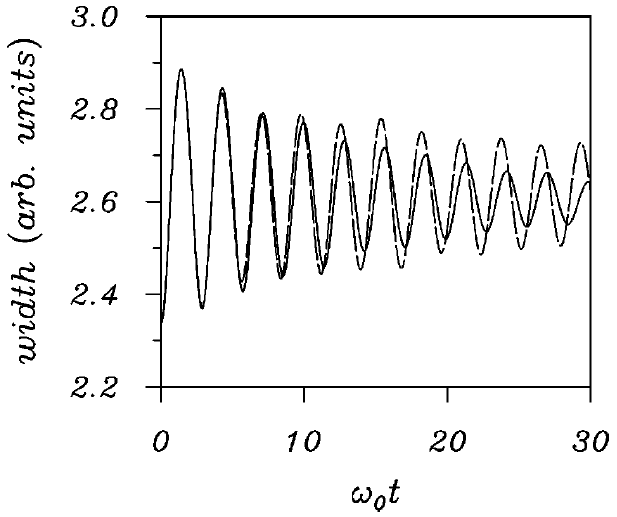}} 
    \caption{
Role of collisional proceses involving thermal atoms on the condensate dynamics following its initial excitation into a monopole, or breathing, mode.
Left: (a) Frequency ($\omega_{M}$) and (b) damping rates ($\Gamma$) vs.\ reduced temperature in the absence of any collisions involving thermal atoms ($C_{12}=C_{22}=0$; filled circles), including thermal-thermal collisions ($C_{12}=0$, $C_{22} \neq 0$; open circles) and including all collisional processes ($C_{12}$, $C_{22} \neq 0$; triangles), where $\omega_0$ is the radial trap frequency and $T_c^0$ is the critical temperature of the non-interacting gas.
Right: Corresponding damped oscillations of the condensate width at $T/T_c^0 = 0.64$ in the absence of any thermal collisions (dashed) and with all collisional processes included (solid), corresponding respectively to the filled circles and triangles in the image on the left.
Results are based on $5 \times 10^4$ $^{87}$Rb atoms in an isotropic trap of frequency $\omega_0=2 \pi \times 187$ HZ (Reprinted figures with permission from 
B.\ Jackson and E.\ Zaremba, Phys. Rev. A {\bf 66}, 033606 (2002). Copyright (2002) by the American Physical Society.)
}
    \label{ZNG_Damping}
  \end{center}
\end{figure}

An important application of this theory performed by Bijlsma, Zaremba and Stoof \cite{Bijlsma_Zaremba_Stoof} addressed the issue of condensate formation (see Sec.\ \ref{Growth}) by relying on
the ergodic approximation, which assumes that equilibration is rapid for atoms with similar energies, thus enabling the phase-space variable $ f_{i}=f({\bf r}_{i}, {\bf p}_{i},t)$  to be expressed only in terms of energy variables. 
These calculations were subsequently generalized to strongly non-equilibrium regimes (and the requirement for ergodicity lifted) by one of us (BJ) \cite{Jackson_Zaremba_1,Jackson_Zaremba_2,Jackson_Zaremba_4,Jackson_Zaremba_5,Excitations_ZNG,Jackson_LasPhys}, by representing the phase-space density by a collection of N discrete `test particles', with collisions between them handled via an appropriate Monte Carlo sampling scheme, thereby extending earlier work \cite{Monte_Carlo_Adams}. 

Importantly, 
coupling a dissipative Gross-Pitaevskii equation to a Quantum Boltzmann equation as in the `ZNG' approach
enables an assessment of the relative importance of the various collisional processes arising in ultracold gases, as first discussed by Eckern \cite{Eckern}, and analyzed more extensively by Zaremba, Nikuni and Griffin and their collaborators (including one of us - BJ).
These collisional processes
 involve both particle exchange collisions between the condensate and the thermal cloud, and `redistribution' collisions within the thermal cloud. A typical example of their effect on the excitation of a `breathing' mode, in which the condensate exhibits damped radial oscillations, is shown in Fig.\ \ref{ZNG_Damping}; the observed increase in the damping rate with temperature should be contrasted to the undamped oscillations predicted by the GPE.
Although the above formulation ignores many-body effects encompassed in the static value of the anomalous average (since $\tilde{m}_{0}=0$ here), it nonetheless yields remarkable agreement with various experiments, as evident from the study of the dynamics of various types of excitations, such as scissor's mode \cite{Jackson_Zaremba_1}, quadrupole excitations (see Sec.\ \ref{Excitations})
and transverse breathing modes of elongated condensates \cite{Jackson_Zaremba_4}. 
The `ZNG' theory also made the first quantitative predictions of vortex nucleation at finite temperatures \cite{ZNG_Vortex_Williams}.

\begin{figure}
  \begin{center}
      \resizebox{120mm}{!}{\includegraphics{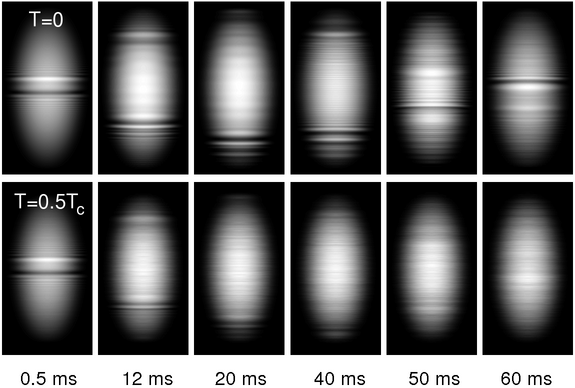}} 
    \caption{
Decay of a dark soliton due to finite temperatures, as witnessed in a sequence of expansion images,
corresponding to a different soliton propagation time in the original trap (as indicated) before that atomic gas is released from the trap. The self-consistent Gross-Pitaevskii-Boltzmann, or `ZNG' theory correctly reproduces the decay observed in the Hannover experiment \cite{Hannover_Soliton_Exp} as soon as the soliton reaches the condensate edge (top images), whereas the same soliton in a $T=0$ condensate would execute undamped oscillations in the trap (bottom images). (Reprinted figure with permission from 
B.\ Jackson, N.P.\ Proukakis and C.F.\ Barenghi, Phys. Rev. A {\bf 75} 051601(R) (2007). Copyright (2007) by the American Physical Society.)
}
    \label{ZNG_Images}
  \end{center}
\end{figure}

Moreover, this theory was recently used by the authors to study the finite temperature dynamics of dark solitons \cite{ZNG_Soliton} and vortices \cite{ZNG_Vortex}. 
In the Hannover experiment \cite{Hannover_Soliton_Exp}, performed at a temperature of $\approx 0.5T_c$ the phase imprinted soliton was found to decay when it reached the edge of the thermal cloud, a result perfectly supported by our numerical simulations (see Fig.\ \ref{ZNG_Images}). 
Application of this formalism to vortices shows that the core of an off-centred vortex spiralling out of the condensate due to dissipation is filled up by thermal atoms which closely follow the vortex trajectory, highlighting the importance of a fully-dynamical self-consistent calculation \cite{ZNG_Vortex}.

\subsubsection{Hydrodynamic Description at Finite Temperatures:}
\label{Hydro_T}

We now give the corresponding finite temperature hydrodynamic equations.
%
In particular, the continuity equation (cf. Eq.\ (\ref{mass})) within the `ZNG' formalism acquires a `source term' from the imaginary part of the triplet contribution, i.e 
\be
\frac{ \partial n_0}{\partial t} + {\bf \nabla} \cdot (n_0 {\bf v_c}) = 
\frac{2g}{\hbar} Im \left[ \wfn^* \langle \nopc \nopa \nopa \rangle \right]
= - \int \frac{ d {\bf p}}{(2 \pi \hbar)^3} C_{12} \left[ f ({\bf p}, {\bf r},t) \right],
\label{mass_T}
\ee 
whereas the corresponding imaginary contribution of the pair anomalous average vanishes.
This is coupled to a `generalized Euler equation', given here by
(cf. Eq.\ (\ref{Euler}))
\be
m \left( \frac{\partial}{\partial t} + {\bf v_c} \cdot {\bf \nabla} \right) {\bf v_c} 
= - {\bf \nabla} \mu_c \;,
\label{hydro_T}
\ee
with the condensate chemical potential $\mu_c$ already defined in Eq.\ (\ref{mu_T});
note that, within the Popov approximation which is imposed in the `ZNG' approach, the latter chemical potential ignores the real contribution from the pair anomalous average which is responsible for certain many-body corrections.

These equations have been used extensively to discuss hydrodynamic features of interacting Bose gases, 
and have most notably led to the derivation of Landau's two-fluid equations \cite{ZNG_2_Fluid}, and the Landau-Khalatnikov two-fluid equations with hydrodynamic damping, including explicit expressions for all the transport coefficients \cite{ZNG_Transport}.


{\it
\subsection{Brief Summary}

Particle-exchange collisions between the condensate and the thermal cloud, as well as binary collisions within the thermal cloud, can be included in a generalized time-dependent mean field by going beyond the mean-field approximations of Eqs.\ (\ref{mf-quad}), (\ref{mf-trip}). This can be achieved by treating the remaining part of the full system hamiltonian, namely $(\hat{H}_3-\delta \hat{H}_1)$ and $(\hat{H}_4 - (\delta H_0 + \delta \hat{H}_2))$ perturbatively, where $\delta \hat{H}_i$ are the contributions arising from the generalized mean-field approximations. 
Various such formulations have been discussed, differing essentially in the partitioning of the system hamiltonian into an unperturbed part which specifies the nature of the excitation spectrum, and a perturbing hamiltonian which governs the system dynamics. 

Kirkpatrick and Dorfman were the first to obtain a Quantum Boltzmann Equation for the quasiparticle distribution function in the presence of a condensate.
Combining this with the crucial inclusion of the anomalous triplet correlation $\langle \nopc \nopa \nopa \rangle$ (whose importance was also identified earlier by one of us - NPP - in collaboration with Burnett and Stoof), Zaremba, Nikuni and Griffin successfully formulated an important `self-consistent Gross-Pitaevskii-Boltzmann' approach which they termed the `ZNG' theory. In this theory, which has been formulated such that the effective hamiltonian fully reproduces the exact equations of motion for the condensate wavefunction and the non-condensate operator, 
excitations are specified by Hartree-Fock energies.
The resulting theory (which also arises as a limiting case of formalism yet to be presented),
fully accounts for the dynamics both of the condensate (via a dissipative Gross-Pitaevskii equation) and of the thermal cloud (via a Quantum Boltzmann equation) in a self-consistent manner; the predictions of this theory appear
 to be in excellent agreement with experiments both in the collisional and collisionless regimes. Despite ignoring many-body effects contained within the static anomalous average contribution $\tilde{m}_0^R(\bldr)$ (implicitly included in alternative formulations), this `ZNG' theory constitutes the best numerically implemented mean field theory to date.
}

\section{Phase Fluctuations and Number-Conserving Approaches}

\label{Phase_Number}

In our treatment thus far we have replaced the condensate operator $\hat{\wfn}(\bldr,t)$ by the classical field $\wfn(\bldr,t)$. 
As a result, the condensate is no longer invariant under global phase changes, and thus breaks this symmetry of the full system hamiltonian by acquiring a definite phase.
Two important implications of this are: (i) a proper inclusion of fluctuations in the condensate phase is precluded from the outset; and (ii) the total particle number in the system is no longer conserved. In this section we systematically address both these issues.

\subsection{Low Dimensional Systems: Fluctuations in the Condensate Phase}

\label{Phase_Fluct}

Under conditions of reduced effective dimensionality, Bose-Einstein condensation cannot occur in a homogeneous system \cite{Mermin,Hohenberg}; instead, quantum degeneracy leads to the appearance of a  so-called `quasi-condensate', which can be thought of as a condensate with a fluctuating phase \cite{Popov}.
Such low-dimensional geometries can now be routinely produced in numerous laboratories, under conditions of tight confinement in one or two directions, which effectively freezes the motion of the atoms in the remaining tightly-confining direction(s) 
(see, e.g.\ \cite{Low_D_Experiment_1,Low_D_Experiment_2,Low_D_Experiment_3}).
We stress that our present discussion is limited to weakly-interacting systems whose kinematic behaviour is low-dimensional, but scattering can still be described as effectively three-dimensional \cite{Olshanii_Scattering}, i.e.\ we are still far from the strongly-correlated `Tonks-Girardeau' regime arising in purely 1D Bose gases \cite{Low_D_Petrov} (whose treatment lies beyond the scope of this Tutorial)

For such systems, one can define two characteristic temperatures \cite{Low_D_Petrov}: a `degeneracy' temperature signalling the onset of macroscopic occupation of a given quantum state (i.e.\ the low-dimensional analogue of the critical temperature); and a `phase fluctuation' temperature signalling the transition to a regime where the phase of the `condensate' begins to fluctuate. 
Our preceeding discussion is based on symmetry-breaking, which 
automatically assigns a phase to the condensate; as a result, fluctuations in the phase cannot be treated properly, and so far our analysis has only accounted for the local gradient of the condensate phase (superfluid velocity); in order to account for phase fluctuations properly, as required for systems of dimensionality less than three, a modification of the previous formalism is required.
Note that phase fluctuations may also arise in three-dimensional geometries in the early stages of the formation of a Bose-Einstein condensate, although their effect is only expected to be important in a very narrow temperature region near the critical point.

Fortunately, the inclusion of phase fluctuations can be done fairly straightforwardly, essentially by
replacing Eq.\ (\ref{sep-fieldops}) by the more general expression \cite{Low_D_Stoof}
\begin{equation}
\fopart = \sqrt{n_{0}(\bldr,t)} e^{i \hat{\theta}(\bldr,t)} + \hat{\psi}'(\bldr,t)\;.
\label{Op_1D}
\end{equation}
To allow for the fact that the `condensed' part of the system (first contribution) does not have a unique phase, we have introduced here an operator $\hat{\theta}(\bldr,t)$ to account for {\em fluctuations} in the phase of the `coherent' part of the system. This quantity is useful when looking at correlations of such operators, e.g.\ evaluating correlations of the phase (contained in the one-body density matrix) at different locations, or times; thus, Eq.\ (\ref{Op_1D}) provides a mathematical tool for computing correlation functions of the gas. Density fluctuations have also been included by introducing a corresponding operator $\hat{\psi}'(\bldr,t) $.
The low temperature limit of this equation was discussed in \cite{Low_D_Petrov}, with the current extension presented in \cite{Low_D_Stoof}, with one of us (NPP) playing an instrumental part in its numerical implementation \cite{Low_D_PRA,Low_D_PhaseDiagram,Proukakis_MF_2006}. As this issue digresses slightly from the main emphasis of this Tutorial, we only give some general arguments here, and refer the reader to the above-mentioned works for more details.

To explain how Eq.\ (\ref{Op_1D}) leads to the inclusion of phase fluctuations, we discuss how it relates to the symmetry-breaking expression of Eq.\ (\ref{sep-broken}): In our preceeding analysis, we have chosen to describe the system in terms of
correlations of fluctuations about the condensate mean field up to quadratic order, i.e.\ ignoring (or only treating perturbatively) the contributions from $\hat{H}_3$ and $\hat{H}_4$. 
This is equivalent to making a lowest order expansion of the exponential in Eq.\ (\ref{Op_1D}), such that
$e^{i \hat{\theta}(\bldr,t)} \approx 1+i \hat{\theta}(\bldr,t)$,
In this limit, Eq.\ (\ref{Op_1D}) can be re-cast in the more familiar form
\begin{eqnarray}
\fopart &=& \sqrt{n_{0}(\bldr,t)} +  \left[ i \sqrt{n_{0}(\bldr,t)} \hat{\theta}(\bldr,t) + \hat{\psi}'(\bldr,t) \right] \nonumber \\
&=& \wfn(\bldr,t)+\nopa(\bldr,t)\;.
\label{Op_1D_Appr}
\end{eqnarray}
%
%
%
Why is such an approximation insufficient for describing low-dimensional systems?
%
Based on Eq.\ (\ref{Op_1D}), the total system density at position ${\bf r}$ and time $t$ becomes
\bea
& & \langle \fopcrt \fopart \rangle 
= n_0 \langle e^{-i \hat{\theta}(\bldr,t)} e^{i \hat{\theta}(\bldr,t)} \rangle 
+ \langle [ \hat{\psi}'(\bldr,t) ]^\dag \hat{\psi}'(\bldr,t)\rangle
\nonumber \\
& & \hspace{2.0cm} + \sqrt{n_0} \langle e^{-i \hat{\theta}(\bldr,t)} \hat{\psi}'(\bldr,t) \rangle
+ \sqrt{n_0} \langle [ \hat{\psi}'(\bldr,t) ]^\dag e^{i \hat{\theta}(\bldr,t)} \rangle.
\label{dens_contr}
\eea
The top line should reproduce the total system density, with the bottom line vanishing under the assumption of no correlations between the condensate and the non-condensate. Indeed locally this yields the identity
$
n(\bldr,t) = n_0(\bldr,t) + \tilde{n}(\bldr,t).
$
If however one were to make the approximation 
$e^{i \hat{\theta}(\bldr,t)} \approx 1+i \hat{\theta}(\bldr,t)$,
then the system density would incorrectly acquire an additional contribution from 
$n_0(\bldr,t) \langle \hat{\theta}(\bldr,t) \hat{\theta}(\bldr,t) \rangle$.
Although the latter contribution is small for three-dimensional homogeneous condensates, it blows up as $p \rightarrow 0$ for dimensions $d \le 2$ (except at $T=0$ and $d=2$), thus leading to well-known infrared divergences in the expression for the total density \cite{Popov,Mermin,Hohenberg}.

\begin{figure}[t]
  \begin{center}
    \begin{tabular}{cc}
	\resizebox{68mm}{!}{\includegraphics{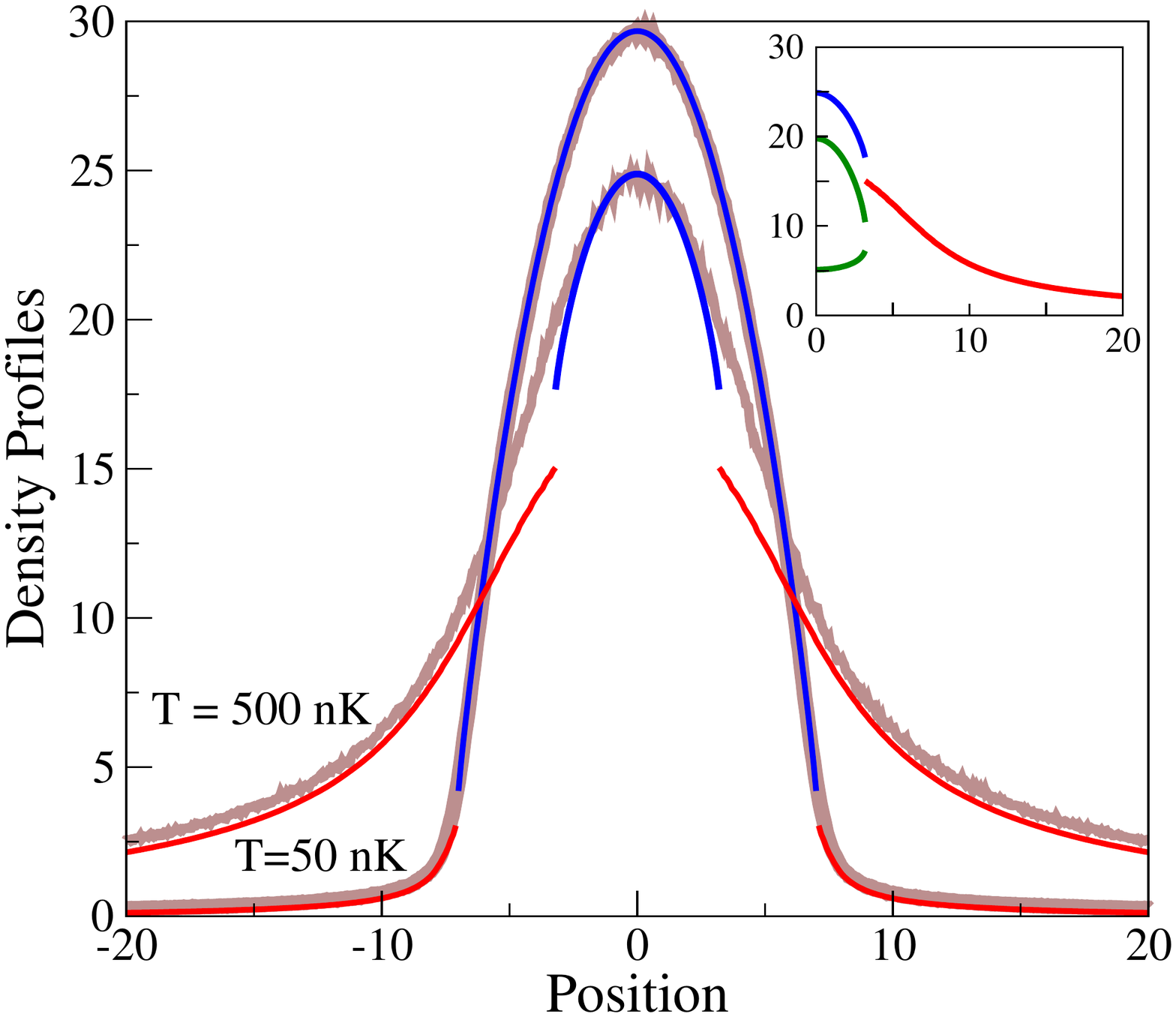}} &
	\resizebox{68mm}{!}{\includegraphics{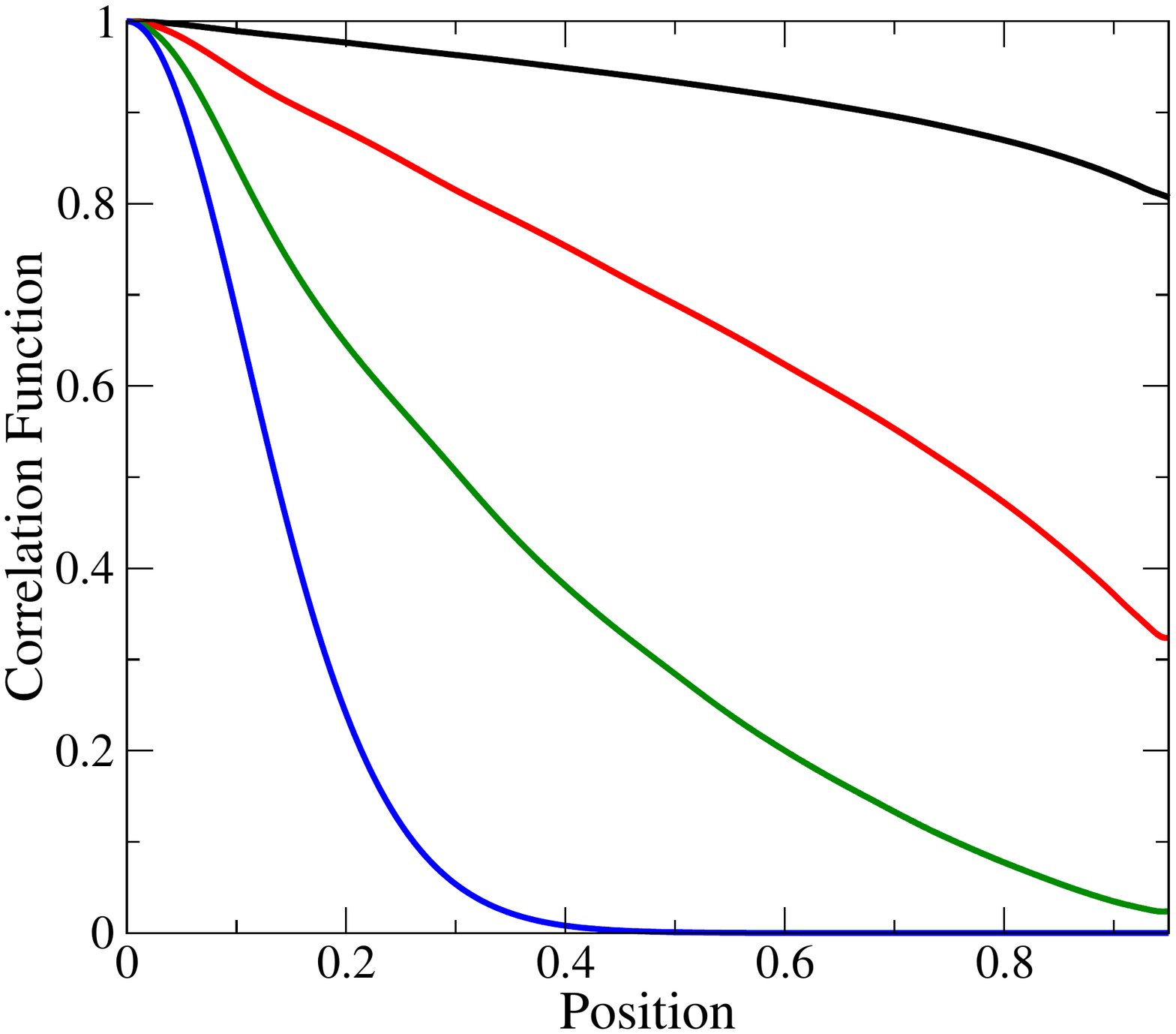}}
    \end{tabular}
 \caption{
(colour online)
Left: {\em Ab initio} determined density profiles for a trapped one-dimensional Bose gas via the modified method proposed in this section (split blue/red profiles) and the stochastic method of Sec.\ \ref{Stoof_Theory} (continuous brown curves) for two different temperatures.
Inset shows how a separation into `quasi-condensate' (upper green) and `thermal' (lower green) profiles within the inner `degenerate' region ($z < 4 l_z$ for $T=500$nk) enables the determination of the total density in that region (blue). Outside the `degenerate' regime (red curves) the density is computed by a semi-classical equation of state for a thermal gas in the many-body T-matrix approximation; the small discontinuity in the plotted profiles arises due to the use of two different equations of state. (Densities plotted as $gn(z)$ scaled to the typical harmonic oscillator energy $\hbar \omega_z$, where $g$ is the effective one-dimensional coupling constant, while positions are scaled to the harmonic oscillator length $l_z$).
Right: Normalized first order correlation functions $g^{(1)}(0,z)$ plotted against $z$ with distances scaled to the corresponding quasi-condensate size for each temperature (often referred to as the `temperature-dependent Thomas-Fermi radius' \cite{Low_D_PRA}). Displayed temperatures, from top to bottom: $T =$ $10$ (black), $50$ (red), $150$ (green) and $500$ nK (blue), all below the transition temperature. The loss of phase coherence characteristic of low-dimensional systems is evident in the decay of the correlation function to zero, even at distances smaller than the effective quasi-condensate spatial extent.
(Simulations performed for  $\approx$ 20000 $^{23}$Na atoms with $\omega_z=2 \pi \times 3.5$ Hz - see \cite{Low_D_Stoof,Low_D_PRA,Proukakis_MF_2006} for further details).
}
\label{1D_Images}
\end{center}
\end{figure}

The above discussion (which has been put on firmer ground \cite{Low_D_PRA}) suggests that one can proceed with the usual established (one-loop) mean field treatment \cite{Stoof_Variational}, and nonetheless obtain the correct equation of state by  omitting the infrared-divergent contributions; this is  based on the physical argument that such contributions have only entered the expression for the total density as a result of inappropriate handling of the condensate phase fluctuations (which leads to double-counting). 
This procedure has been shown to reproduce all known results for (weakly-interacting) Bose gases in one, two and three dimensions \cite{Low_D_PRA}, thus providing a modified mean-field theory and a general equation of state valid in all dimensions.
In fact, this simple trick enables the direct calculation of equilibrium densities in the degenerate regime \cite{Low_D_PRA}, while the identification of the expression for $\langle \hat{\theta}(\bldr,t) \hat{\theta}(\bldr,t) \rangle$ enables the direct computation of (non-local) phase fluctuations, and thus of correlation functions.
This is illustrated in Fig.\ \ref{1D_Images} which plots both density profiles and correlation functions for a one-dimensional Bose gas at temperatures where phase fluctuations are present. Note that the predictions of this theory have been further shown to
compare favourably to more exact stochastic treatments discussed in Sec.\ \ref{Stoof_Theory}.
%


\subsection{Number-Conserving Approaches}

\label{Number}


In this section we critically revisit the notion of Bose broken symmetry, and essentially repeat the analysis of Secs. \ref{finitet:static}-\ref{finitet:dynamic} without ever resorting to such an approximation.
This is important from a fundamental point of view, and leads to certain non-local corrections whose importance is still under investigation.

The concept of symmetry-breaking is familiar in many branches of physics, including condensed matter, particle physics and optics.
However, in the case of 
 `closed systems', as is relevant for finite trapped atomic gases, and because atoms, unlike photons, can neither be created nor destroyed, the notion of broken symmetry becomes somewhat `ill-defined'. This does not mean that approaches based on symmetry-breaking are incorrect, and indeed such approaches turn out to be very useful, providing in many cases excellent agreement with experiments.
However, one may not be entirely satisfied with this picture, based on the argument that at any one time the atomic state consists of a {\em definite} number of atoms which one should be able, at least in principle, to determine experimentally.
For a closed system with a fixed number, $N$, of bosonic atoms, the average $\langle N | \fopa | N \rangle = \langle N | \fopc | N \rangle = 0$ {\em identically} (since the bosonic field operators, $\fopa^{(\dag)}$, change the total number of atoms in the system from $N$ to $(N \pm 1)$, and $\langle N | N \pm 1 \rangle = 0$ in the orthogonal number state basis). It is then conceptually hard to interpret a non-vanishing average of the field operators. Usually one overcomes this problem by  
representing the state of the quantum system by a {\it coherent} superposition of states with different numbers of particles, $\cdots |N-1 \rangle, \,\, |N \rangle, \,\, |N+1 \rangle \cdots$, i.e.\ describing the system by a so-called `coherent state' $|\alpha \rangle$; such a model has proven very successful, for example, in describing the laser \cite{QO_Book_1,QO_Book_2,QO_Book_3}.
Although experimental knowledge of the atom number is usually obtained statistically (in the sense of ensemble averages from multiple realizations) it is still hard to see how shot-to-shot number coherences would build up \cite{Morgan_Gardiner}.
%
The issue of whether broken symmetry is a `necessity', a `reality', or a `convenient mathematical tool' (or none of the above) remains however somewhat controversial; for example, recent work by Yukalov \cite{Yukalov_SB} concluded that spontaneous gauge symmetry breaking is the necessary and sufficient condition for the existence of BEC in any kind of system, based on the notion of representative statistical ensembles \cite{Yukalov_Ensemble}.

In any case,
significant progress can in fact be made {\em without} 
invoking the concept of Bose broken symmetry, i.e.\ without
relaxing the requirement of number conservation.
It turns out that the physically more intuitive (yet mathematically more cumbersome) number-conserving approaches lead to essentially the same equations, but with some (non-local) corrections which are not always important.

\subsubsection{Essential Methodology:}

\label{number-cons}

The problem considered thus far can actually be reformulated
in terms of operators which have very similar properties to operators considered in symmetry-breaking treatments, but yield different physical interpretations.
For these operators,  the total number of {\em atoms}, $N$, in the system is explicitly conserved, whereas the number of {\em excitations}, $n_k$ is not. In such treatments, the number of condensate atoms is actually not a separate variable, but is determined self-consistently from $N_0 = N - \sum_{k} n_k$.
The first discussion of such number-conserving approaches was given by Girardeau and Arnowitt in 1958 \cite{Girardeau_Arnowitt}, with this approach `rediscovered' in a slightly modified form by C.W. Gardiner \cite{Gardiner_NC} and subsequently extended by Castin and Dum \cite{Castin_Dum} and Morgan and S.A. Gardiner \cite{Morgan_Gardiner}. 
While all these approaches are based on the same underlying theme, their precise implementation of number-conservation is not identical. In our subsequent discussion we briefly review the main ideas behind these works, with particular emphasis given to the recent dynamical approach of Morgan and S.A. Gardiner \cite{Morgan_Gardiner}.


\subsubsection{Approach of C.W. Gardiner:}
\label{CW_Gardiner}

This approach was based on defining an annihilation operator $\hat{A}$ for the total number of particles and a separate operator for phonons. In particular, the operator
$\hat{\alpha}_k$ which annihilates an excitation in mode $k$ was defined in terms of the usual single-particle operators $\hat{a}_k$ by
\begin{equation}
\hat{\alpha}_{k} = \frac{1}{\sqrt{\hat{N}}} \hat{a}_{0}^\dag \hat{a}_{k}\;.
\label{phonon}
\end{equation}
%
While $\hat{A}$ and thus $\hat{N}$ act on the total number of particles, the operators $\hat{\alpha}_{k}$ actually commute with $\hat{N}$. This enables approximations to be made in the treatment of the phonon operators, without the need for a change in the total particle number. 
Although such a treatment leads to explicit number conservation, the newly-introduced phonon operators only satisfy bosonic commutation relations approximately (for large $N$) via
\be
\left[ \hat{\alpha}_k , \hat{\alpha}_{k'}^{\dag} \right]
= \delta_{kk'} - \frac{1}{N} \hat{a}_{k} \hat{a}_{k'}^{\dag} \approx \delta_{kk'}\;. 
\ee

By an appropriate reformulation of the system hamiltonian, the Bose field operator $\fopart$ is expanded in terms of these new operators, with the hamiltonian split into different contributions according to the number of phonon operators present - analogously to the separation in terms of the number of fluctuation operators present in Eqs.\ (\ref{H_0})-(\ref{H_4}).
Next, Gardiner
introduced a redefined interaction parameter $g_N=gN$, which provides a measure of the interaction energy of the system. Keeping the parameter $g_N$ fixed, while increasing the atom number $N$ essentially corresponds to taking the thermodynamic limit of the system under consideration \cite{Castin_Dum,Morgan_Gardiner}.
As a result, the separation of the full system hamiltonian into these contributions can now be visualized as a systematic expansion in powers of $(1/\sqrt{N})$, via the expansion
\bea
\hat{H} = N \sum_{m=0}^{4} \left( \frac{1}{\sqrt{N}} \right)^m {\cal H}_m &\approx& 
N \hat{\cal H}_{0}+ \sqrt{N} \hat{\cal H}_{1} + \hat{ \cal H}_{2} \nonumber \\
&+& \frac{1}{\sqrt{N}} \hat{\cal H}_{3} + \frac{1}{N} \hat{\cal H}_{4}\;, 
\label{H-ncons}
\eea
where calligraphic notation has been used to highlight the fact that the hamiltonians appearing in the above expression are given in terms of the redefined phonon operators of Eq.\ (\ref{phonon}).
Each of these hamiltonians $\hat{\cal H}_m$ is now expressed in terms of $g_N$,
with the index $m$ indicating the number of phonon operators present in the system, e.g.
\be
\hat{\cal H}_0 = \int d\bldr [\wfn^N(\bldr)]^* \hat{h}_0 \wfn^N(\bldr)
+ \frac{g_N}{2} \int d \bldr \left| \wfn^N(\bldr) \right|^4\;,
\ee
where the condensate wavefunction has acquired a superscript $N$ to highlight that all calculations are performed in a number-conserving manner.

As $N \rightarrow \infty$, higher than quadratic terms become negligible, thus
yielding a number-conserving hamiltonian which can be routinely diagonalized by a Bogoliubov transformation.
Keeping only the first contribution leads to the time-independent GPE
\be
\hat{h}_0 \wfn_0^{N}(\bldr) + g_N \left| \wfn_0^{N}(\bldr) \right|^{2} \wfn_0^{N}(\bldr) = \lambda \wfn_0^{N}(\bldr)
\ee
The parameter $\lambda$ appearing here is essentially the chemical potential of the system, only in this treatment it arises naturally as a Lagrange multiplier, and should therefore be computed self-consistently. 

This generalizes straightforwardly to the time-dependent case, with a slight variant of this approach presented in more detail in Sec.\ \ref{MorganGardiner}.

\subsubsection{Number-Conserving Static Finite Temperature Bogoliubov Equations:} 

\label{Morgan}

A related analysis, based on a slightly modified version of the number-conserving formalism of Sec.\ \ref{CW_Gardiner},
 was performed by Morgan \cite{Morgan_JPhysB} in the context of a self-consistent second order perturbation theory, which focuses on the change in the system energy when a single quasiparticle is added to a particular mode.
The aim of this analysis was to determine the modified form of the static Bogoliubov equations under explicit number-conservation upon additionally including the back-action of the thermal cloud on the static properties of the condensate.

In brief, this approach is based on the following reasoning: The entire system hamiltonian, Eq.\ (\ref{H-g}) is re-expressed in terms of explicitly number-conserving operators, and broken down to contributions with different numbers of non-condensate operators in each term (c.f.\ Eq.\ (\ref{H-ncons})). In this treatment, the number-conserving operators are defined somewhat differently to Sec.\ \ref{CW_Gardiner}, by
$\hat{\alpha}_k = \hat{\beta}_0^\dag \hat{a}_k$, where $\hat{\beta}_0 = (\hat{N}_0 + 1 )^{1/2} \hat{a}_0$ and $\hat{N}_0 = \hat{a}_0^\dag \hat{a}_0$ is the operator for the condensate number. Similarly to Eq.\ (\ref{H-ncons}) this leads to an expansion of the full system hamiltonian in terms of contributions $\hat{\cal H}_m$ expressed in order of decreasing factors of condensate atom numbers. In the limit of large $N_0$, such an expansion justifies treating the quadratic part of the hamiltonian exactly, while maintaining additional cubic and quartic contributions perturbatively, analogously to the symmetry-breaking beyond-HFB treatment presented in Sec.\ \ref{Beyond_HFB}.  
In particular, one calculates the change induced in the system energy due to the effect of $(\hat{\cal H}_3 + \hat{\cal H}_4)$ in second order perturbation theory. Apart from the inclusion of normal and anomalous averages (and therefore many-body effects), this approach additionally includes 
effects arising from 
quasiparticle collisions in a self-consistent manner. The perturbative inclusion of $\hat{\cal H}_3 + \hat{\cal H}_4$ leads to a change in the static properties of the system (e.g. condensate shape, energy), which induce corrections to the basic hamiltonian $(\hat{\cal H}_0 + \hat{\cal H}_1 + \hat{\cal H}_2 )$, and such corrections are also treated perturbatively in this approach.
Combined with the enforcement of total number conservation, this leads to the inclusion of finite size effects into the formalism.
We wish to focus here on the main modifications induced by the requirement of explicit number conservation, and therefore omit the technical details expounded in \cite{Morgan_Thesis}.

%
The main conclusion of this work was the derivation of a generalized set of Bogoliubov equations of the form
\begin{eqnarray}
\left( 
\begin{array}{cc} 
\hat{\cal L}^{N}(\bldr, \epsilon_i)  & \hat{\cal M}^{N}(\bldr, \epsilon_i) \\ -\left[ \hat{\cal M}^N(\bldr, -\epsilon_i) \right]^{*} & -\left[ \hat{\cal L}^N( \bldr, -\epsilon_i) \right]^*
\end{array}
\right)
\left(
\begin{array}{l}
{\bf u}_i(\bldr) \\ {\bf v}_i(\bldr)
\end{array}
\right)
\nonumber \\
\hspace{5.9cm}  = \epsilon_i
\left(
\begin{array}{l}
{\bf u}_i(\bldr) \\ {\bf v}_i(\bldr)
\end{array}
\right)\;.
\end{eqnarray}
In comparison to the expressions encountered earlier, the operators $\hat{\cal L}^N$ and $\hat{\cal M}^N$ introduced here (denoted by $\tilde{L}$ and $\tilde{M}$ in the literature) contain additional terms.
%
%
In particular we find
\bea
\hat{\cal L}^N (\bldr, \epsilon_i) &=& \hat{h}_0 - \lambda + 2 g |\wfn_0^N|^2 + 2g \tilde{n}_0(\bldr) + \delta \hat{\cal L}(\bldr, \epsilon_i)
\nonumber \\
&\approx& \hat{\cal L}(\bldr) + \delta \hat{\cal L}(\bldr, \epsilon_i)
\eea
\bea
\hat{\cal M}^N(\bldr, \epsilon_i) &=& g  [\wfn_0^{N}(\bldr)]^2 + g \tilde{m}_0^R(\bldr) + \delta \hat{\cal M}(\bldr, \epsilon_i)
\nonumber \\
&\approx& \hat{\cal M}(\bldr) + \delta \hat{\cal M}(\bldr, \epsilon_i)
\;,
\eea
where the approximate sign has been introduced to highlight that the parameter $\lambda$ appearing in number-conserving approaches is a self-consistently determined Lagrange multiplier, rather than the chemical potential, $\mu$.
This is obtained from
\be
\left[ \hat{h}_0 + g \left| \wfn_0^N(\bldr) \right|^{2} +2g \tilde{n}_0(\bldr) \right] \wfn_0^N(\bldr) + g \tilde{m}_{0}^R \left(\wfn_0^N(\bldr)\right)^{*} = \lambda \wfn_0^N(\bldr)\;. 
\ee
%
The terms $\delta \hat{\cal L}$ and $\delta \hat{\cal M}$ are corrections due to the change in energy when a quasiparticle is created in a mode $i$. In addition to describing the interaction of two quasiparticles (Landau and Beliaev damping discussed in Sec.\ \ref{Linear_Response}), such corrections also include processes referring to the simultaneous annihilation or creation of three quasiparticles. The latter 
are expressed in terms of contributions of the form \cite{Morgan_JPhysB}
\begin{equation}
\sum_{k,m \neq 0} \left[ {\cal A}_{km}(\epsilon_i) (1+n_k+n_m) +{\cal B}_{km}(\epsilon_i)(n_m-n_k) \right]
\label{LMextra}
\end{equation}
where ${\cal A}_{km}(\epsilon_i)$ and ${\cal B}_{km}(\epsilon_i)$ depend on the energies of such excitations.

\subsection{Dynamical Finite Temperature Bogoliubov Equations}
\label{MorganGardiner}

We now wish to construct a dynamical number-conserving approach from first principles, essentially by generalizing the arguments presented in the preceeding sections.
%
%
We thus seek to construct a formalism in which the Bose-field operator $\fopart$ is expanded in a symmetry-preserving manner. Assuming this could be achieved in terms of a
slightly modified non-condensate operator $\hat{\delta}^N(\bldr,t)$, we would like this operator ideally to have the following properties:
(i) explicitly ensure orthogonality,
(ii) have zero average, i.e.\ $\langle \hat{\delta}^N \rangle = 0$,
(iii) exactly satisfy bosonic commutation relations
$[ (\hat{\delta}^N(\bldr,t)), \hat{\delta}^N(\bldr',t)^\dag ] = \delta(\bldr-\bldr')$, 
and (iv) guarantee conservation of the system total atom number. Unfortunately, not all requirements are mutually compatible, and in defining $\hat{\delta}^N(\bldr,t)$ one is typically faced with a choice of relaxing one (or more) of the above conditions. The choice is often motivated by the intended use of the resulting formalism, and leads to a range of similar number-conserving approaches \cite{Morgan_JPhysB,Girardeau_Arnowitt,Gardiner_NC,Morgan_Gardiner,Castin_Dum}.
Once the most suitable form of $\hat{\delta}^N(\bldr,t)$ has been identified, one can perform an expansion of the system hamiltonian in ascending powers of this new fluctuation operator, in direct analogy to Eqs.\ (\ref{H_0})-(\ref{H_4}) and (\ref{H-ncons}).

We now discuss how such a number-conserving formalism can be constructed.
Firstly, we wish to make a comment about orthogonality:
In our symmetry-breaking discussion of Secs.\ \ref{zeroT}-\ref{finitet:dynamic}, we have argued that the quasiparticle amplitudes $u_i(\bldr)$ and $v_i(\bldr)$ obtained from the solutions of the symmetry-breaking Bogoliubov equations (Eq.\ (\ref{BdG_Matrix_T0}) and finite temperature generalizations) are orthogonal to the condensate; this statement was actually based on the relation
$
\int d\bldr \left[ \wfn_0^*(\bldr) u_i(\bldr) + \wfn_0(\bldr) v_i(\bldr) \right] =0.
$
However, Morgan has argued \cite{Morgan_JPhysB} that this is only true in a generalized sense, whereas
 each of these two integrals is not separately zero. This implies that the quasiparticle functions defined by the symmetry-breaking Bogoliubov equations are not {\em individually} orthogonal to the condensate $\wfn_0(\bldr)$. (This is actually a direct consequence of the fact that the Bogoliubov equations are not strictly speaking necessary conditions for the hamiltonian to be diagonalized \cite{Salazar}).

We now wish to develop a formalism which explicitly {\em guarantees} the {\em orthogonality} by construction. In doing so, we shall maintain the condensate operator $\hat{a}_0$ in the condensate part of the Bose field operator defined by Eqs.\ (\ref{sep-fieldops})-(\ref{cn-fieldops}).
Orthogonality between condensate and non-condensate can be guaranteed by 
explicitly introducing projectors onto the two orthogonal subspaces.
In particular, we define the condensate annihilation operator, $\hat{a}_{0}$, by the projection $\hat{P}$ of the full Bose field operator $\fopart$ onto the `condensate state' $\wfn^{N}(\bldr,t)$, i.e.
\be
\hat{a}_0(t) = \hat{P} \fopart = \int d \bldr \wfn^{N}(\bldr,t) \fopart \;,
\label{ncons-cond}
\ee
and correspondingly we define a non-condensate operator by the {\em orthogonal projection}
\begin{equation}
\nopa(\bldr,t)= \hat{Q} \fopart = 
\int d \bldr' Q(\bldr, \bldr', t) \hat{\Psi}(\bldr', t)
\label{ncons-fluct}
\end{equation}
where $Q(\bldr, \bldr', t) = \delta(\bldr - \bldr ') - \wfn^N(\bldr, t) [\wfn^N(\bldr',t)]^*$.
%


A key role in our subsequent development is played by the Penrose-Onsager criterion for Bose-Einstein Condensation \cite{Penrose_Onsager}. This states that for a system described by a single-body density matrix
$
\rho(\bldr,\bldr',t) = \langle \fopc(\bldr',t) \fopart \rangle
$
the condensate wavefunction corresponds to the mode of the system which has the largest eigenvalue (which is macroscopically large). The condensate number, $N_0$, is then related to the above density matrix via
\be
N_0 \wfn^N(\bldr,t) = \int d \bldr' \rho(\bldr, \bldr',t) \wfn^N(\bldr',t)\;.
\ee
Next, we expand the right-hand side of this expression into an appropriately defined orthogonal set $\{\wfn_0^N(\bldr), \spwa_i^N(\bldr) \}$ via the following decomposition
\be
\fopart = \hat{a}_0 \wfn_0^N(\bldr,t) + \hat{\delta}(\bldr,t) =
\hat{a}_0 \wfn_0^N(\bldr,t) + \sum_{i \neq 0} \spwa_i(\bldr,t) \hat{a}_i(t)\;.
\label{Psi_NC}
\ee
This generates four contributions, as given by Eqs.\ (\ref{P1})-(\ref{P4}); each of these terms can be reduced to a simpler expression as indicated by the arrows, and the arguments for such reduction are presented below \cite{Gardiner_Thesis}. We thus have
\numparts
\begin{eqnarray}
N_0 \wfn^N(\bldr,t) =
\nonumber \\
\int d \bldr' [\wfn^N(\bldr',t)]^* \wfn^N(\bldr',t) \langle \hat{a}_0^\dag \hat{a}_0 \rangle \wfn^N(\bldr,t) 
&\rightarrow& \langle \hat{a}_0^\dag \hat{a}_0 \rangle \wfn^N(\bldr,t)
\label{P1} \\
+ \int d \bldr' [\wfn^N(\bldr',t)]^* \wfn^N(\bldr',t) \langle \hat{a}_0^\dag \hat{\delta}(\bldr,t) \rangle 
&\rightarrow& \langle \hat{a}_0^\dag \hat{\delta}(\bldr,t) \rangle
\label{P2} \\
+ \int d \bldr' \langle \hat{\delta}^\dag (\bldr',t) \hat{a}_0 \rangle \wfn^N(\bldr,t) \wfn^N(\bldr',t) 
&\rightarrow& 0
\label{P3} \\
+ \int d \bldr' \langle \hat{\delta}^\dag(\bldr',t) \hat{\delta}(\bldr,t) \rangle \wfn^N(\bldr',t) 
&\rightarrow& 0
\;.
\label{P4}
\end{eqnarray}
\endnumparts
Let us now explain how we have obtained the indicated simplifications:
Firstly, the expressions appearing on the right hand side of Eqs.\ (\ref{P1})-(\ref{P2}) are obtained directly by using the 
completeness relation $\int d \bldr' [\wfn^N(\bldr',t)]^* \wfn^N(\bldr',t) =1$.
Moreover, expanding the non-condensate operator $\hat{\delta}^\dag(\bldr',t)$ as 
$\sum_{i \neq 0} \spwa_i^*(\bldr',t) \hat{a}_i^\dag(t)$, the contribution of Eq.\ (\ref{P3})
becomes
$\int d \bldr' \sum_{i \neq 0} \spwa_i^*(\bldr',t) \wfn^N(\bldr',t) \wfn^N(\bldr,t) \langle \hat{a}_i^\dag \hat{a}_0 \rangle$.
This quantity is identically zero, because by construction $\spwa_i$ and $\wfn^N$ are orthogonal, i.e.\ $\int d \bldr' \spwa_i^*(\bldr',t) \wfn^N(\bldr',t) = 0$.
The same argument holds for Eq.\ (\ref{P4}), which is re-expressed as
$\int d \bldr' \sum_{i \neq 0} \spwa_i^*(\bldr',t) \wfn^N(\bldr',t) \langle \hat{a}_i^\dag \hat{\delta}(\bldr,t) \rangle$.
With these simplifications, we arrive at
\be
N_0 \wfn^N(\bldr,t) = \langle \hat{a}_0^\dag \hat{a}_0 \rangle \wfn^N(\bldr,t) + \langle \hat{a}_0^\dag \hat{\delta}(\bldr,t) \rangle 
\label{PO}
\;.
\ee
If we now multiply this by $[\wfn^N(\bldr,t)]^*$ and integrate over $\bldr$, we find that
$N_0 = \langle \hat{a}_0^\dag \hat{a}_0 \rangle$ because the other contribution identically vanishes due to orthogonality, since
\be
\int d \bldr [\wfn^N(\bldr,t)]^* \langle \hat{a}_0^\dag \hat{\delta}(\bldr,t) \rangle 
= \sum_{i \neq 0} \int d \bldr  [ \wfn^N(\bldr,t)]^* \spwa_i(\bldr,t) \langle \hat{a}_0^\dag \hat{a}_i \rangle 
= 0 \;.
\ee
The identification $N_0 = \langle \hat{a}_0^\dag \hat{a}_0 \rangle$, 
when substituted back into Eq. (\ref{PO}) yields directly the result
\be
\langle \hat{a}_0^\dag \hat{\delta}(\bldr,t) \rangle = 0\;,
\label{coherences}
\ee
i.e.\ 
there are no direct coherences between the condensate and the non-condensate within our specified number-conserving scheme.
Eq.\ (\ref{coherences}) is a rather significant result: it tells us that we could potentially choose the quantity $\hat{a}_0^\dag \hat{\delta}(\bldr,t)$ as our fluctuation operator;
this is a plausible choice due to the fact that it has a zero average
$\langle \hat{a}_0^\dag \hat{\delta}(\bldr,t) \rangle = 0 $, directly as a consequence of the implemented orthogonality between the condensate and non-condensate subspaces and the Penrose-Onsager criterion for BEC \cite{Castin_Dum,Morgan_Gardiner}.

However, for our separation into condensate and non-condensate contributions to be most useful, there is one {\em additional} property that we would like the desired non-condensate operator $\hat{\delta}^N(\bldr,t)$ to satisfy: such an operator should ideally scale (at least approximately) as
the square root of the number of non-condensate atoms - this is effectively the same scaling as in the symmetry-breaking treatments.

Recalling that, to lowest order $\hat{a}_0^\dag \approx \sqrt{N_0}$ (Bogoliubov substitution), we thus {\em define}
\be
\hat{\delta}^N(\bldr,t) = \frac{\hat{a}_0^\dag}{\sqrt{N_0}} \hat{\delta}(\bldr , t)
\label{delta_N}
\ee
as the fluctuation operator in the number-conserving formalism presented here, which clearly satisfies $\langle \hat{\delta}^N(\bldr,t) \rangle = 0$.
So, what have we actually achieved by the above considerations? For example, in the symmetry-breaking picture, we also had a fluctuation operator with a vanishing average $\langle \hat{\delta} \rangle =0$. The important additional feature in our current treatment is that the condition $\langle \hat{\delta}^N \rangle =0$ does {\em not} arise in an {\em ad hoc} manner (as was the case up to now), but rather it manifests itself as a {\em direct consequence of the orthogonality} between the two parts of the system.


We should also make a comment about our particular choice for the number-conserving fluctuation operator.
Defining the operator via Eq.\ (\ref{delta_N}) leads to only approximately-satisfied bosonic commutation relations via (c.f.\ Sec.\ \ref{CW_Gardiner}) 
\bea
\left[ \hat{\delta}^N(\bldr,t), (\hat{\delta}^N(\bldr',t))^\dag \right]
&=& \frac{ \hat{N}_0}{N_0} Q(\bldr,\bldr',t) - \frac{1}{N_0} (\hat{\delta}^N(\bldr',t))^\dag \hat{\delta}(\bldr,t) \nonumber \\
&\approx& Q(\bldr,\bldr',t)\;.
\eea
Different choices of $\hat{\delta}^N(\bldr,t)$ have been discussed in
the literature,
e.g. $( \hat{a}_0^\dag /\sqrt{\hat{N}_0}) \hat{\delta}(\bldr,t)$ \cite{Castin_Dum}, or
$(\hat{a}_0^\dag /\sqrt{\hat{N}}) \hat{\delta}(\bldr,t)$ \cite{Gardiner_NC}
which is the position representation of the phonon field operator $\hat{\alpha}_k$ of Eq.\ (\ref{phonon}); each of these approaches has distinct benefits and drawbacks, and a more detailed discussion of the main factors affecting the optimal choice are given in \cite{Morgan_Gardiner}.

Having identified the fluctuation operator $\hat{\delta}^N(\bldr,t)$ in terms of which we wish to expand the system hamiltonian, we can now rewrite our initial expression for the Bose field operator, Eq.\ (\ref{Psi_NC}), in terms of this new operator.
From Eq.\ (\ref{delta_N}), we have\footnote{Here it becomes apparent how use of an operator under the square root (i.e.\ $\sqrt{\hat{N}_0}$, or $\sqrt{\hat{N}}$) would complicate matters considerably, as one would need to carefully consider the inverse of such operators, and corresponding operator ordering in the above expression.}
\be
\hat{\delta}(\bldr,t) = \sqrt{N_0} \left[ \hat{a}_0^\dag \right]^{-1} \hat{\delta}^N(\bldr,t)\;.
\ee
The Bose field operator can now be written in the following number-conserving form
\be
\fopart = \wfn^N(\bldr,t) \hat{a}_0 + \sqrt{N_0} \left[ \hat{a}_0^\dag \right]^{-1} \hat{\delta}^N(\bldr,t)\;,
\ee
where we note that creation and annihilation operators do not commute, i.e.\ $[\hat{a}_0^\dag ]^{-1} \neq \hat{a}_0$, and $\langle \hat{\delta}^N(\bldr,t) \rangle = 0$.
We now substitute this into the system hamiltonian, Eq.\ (\ref{H-g}), and consider contributions of an ascending number of $\hat{\delta}^N(\bldr,t)$ operators.
We also define the new effective interaction parameter $g_N$ via $g_N=gN_0$
(rather than $gN$ discussed in Sec.\ \ref{number-cons}).
To illustrate the procedure, we give here explicitly the first two terms, with calligraphic notation highlighting the fact that they now depend on the new fluctuation operators $\hat{\delta}^N(\bldr,t)$. So,
\bea
\hat{\cal H}_0 &=& 
\int d \bldr [ \wfn^N(\bldr,t) ]^* \hat{h}_0 \wfn^N(\bldr,t) \hat{a}_0^\dag \hat{a}_0 +
\frac{g}{2} \int d \bldr \left| \wfn^N(\bldr,t) \right|^4 \hat{a}_0^\dag \hat{a}_0^\dag \hat{a}_0 \hat{a}_0 \nonumber \\
& = & N_0  \left( \frac{\hat{N}_0}{N_0} \right) 
\int d \bldr [\wfn^N(\bldr,t) ]^* \nonumber \\
&& \hspace{1.5cm} \times  
\left\{ \hat{h}_0  + \frac{g_N}{2} \left( \frac{\hat{N}_0-1}{N_0} \right) 
\left| \wfn^N(\bldr,t) \right|^{2}
\right\} \wfn^N(\bldr,t)   \;,
\label{H0_NS}
\eea
where we have used the fact that
$
\hat{a}_0^\dag \hat{a}_0^\dag \hat{a}_0 \hat{a}_0 = \hat{a}_0^\dag ( \hat{a}_0 \hat{a}_0^\dag -1) \hat{a}_0 = ( \hat{a}_0^\dag \hat{a}_0 )^2 - ( \hat{a}_0^\dag \hat{a}_0 )
= \hat{N}_0^2 - \hat{N}_0
$.
%
%
Similarly, we find
\bea
\hat{\cal H}_1 & = &
\int d \bldr [ \wfn^N(\bldr,t) ]^* \hat{h}_0  \hat{a}_0^\dag \left( \sqrt{N_0} [\hat{a}_0^\dag]^{-1} \hat{\delta}^N(\bldr,t) \right) \nonumber \\
& + &
g \int d \bldr \left| \wfn^N(\bldr,t) \right|^2 [\wfn^N(\bldr,t)]^* \hat{a}_0^\dag \hat{a}_0^\dag \hat{a}_0  \left( \sqrt{N_0} [\hat{a}_0^\dag]^{-1} \hat{\delta}^N(\bldr,t) \right)
\nonumber \\
& + & {\rm h.c.}
\nonumber \\
& = & \sqrt{N_0} \int d \bldr [ \wfn^N(\bldr,t) ]^* 
\left\{ \hat{h}_0  + g_N \left( \frac{\hat{N}_0-1}{N_0} \right) 
\left| \wfn^N(\bldr,t) \right|^{2} \right\} \hat{\delta}^N(\bldr,t)  
\nonumber \\
& + & {\rm h.c.} \;,
\label{H1_NS}
\eea
upon re-expressing
$
\hat{a}_0^\dag \hat{a}_0^\dag \hat{a}_0 [ \hat{a}_0^\dag ]^{-1} = \hat{a}_0^\dag ( \hat{a}_0 \hat{a}_0^\dag -1) [ \hat{a}_0^\dag ]^{-1} =  \hat{a}_0^\dag \hat{a}_0 - 1 = \hat{N_0} -1 \;.
$
In Eqs.\ (\ref{H0_NS})-(\ref{H1_NS}) we have divided all contributions involving a single condensate number operator, $\hat{N}_0$, by its mean value $\langle \hat{N}_0 \rangle = N_0$.
The above analysis shows that $\hat{\cal H}_0$ is of order $N_0$, and $\hat{\cal H}_1$ is of order $\sqrt{N_0}$.
One can express all higher-order terms ($\hat{\cal H}_2$, $\hat{\cal H}_3$ and $\hat{\cal H}_4$) in a similar manner, with each subsequent contribution
$\hat{\cal H}_{m+1}$ containing an additional factor of $(1/ \sqrt{N_0})$ and an additional fluctuation operator $[ \hat{\delta}^N (\bldr,t)]^{(\dag)}$ compared to $\hat{\cal H}_m$ ($0 \le m \le 3)$ - for detailed expressions see \cite{Morgan_Gardiner}. 
This justifies our earlier claim that we can expand the system hamiltonian systematically in terms of decreasing atom numbers, with the approach presented in this section effectively corresponding to an expansion in the ratio $(N_{NC}/N_0)$.

Following the {\em methodology} introduced in the number-conserving treatments of Secs.\ \ref{zeroT}-\ref{finitet:dynamic}, we now proceed with suitable approximations for
 $\hat{\cal H}_3$ and $\hat{\cal H}_4$, in order to obtain a more `manageable' {\em number-conserving} hamiltonian.
Although we cannot fully justify such approximations in the brief subsequent discussion, Gardiner and Morgan have argued in detail \cite{Morgan_Gardiner} that these approximations lead to the lowest non-trivial order which facilitates a consistent finite temperature description for a fixed finite number of atoms.

Firstly, we entirely ignore the $\hat{\cal H}_4$ contribution, and, for consistency
(since the omitted contributions are of similar order of magnitude \cite{Morgan_JPhysB}) we simultaneously impose quadratic mean-field approximations - analogous to those of Eq.\ (\ref{mf-trip}) - for the products of three fluctuation operators as
\bea
&&[\hat{\delta}^N(\bldr)]^\dag \hat{\delta}^N(\bldr') \hat{\delta}^N(\bldr'') \approx \langle [\hat{\delta}^N(\bldr)]^\dag \hat{\delta}^N(\bldr') \rangle \hat{\delta}^N(\bldr'') \nonumber \\
& & \hspace{1.0cm} + \langle [\hat{\delta}^N(\bldr)]^\dag \hat{\delta}^N(\bldr'') \rangle \hat{\delta}^N(\bldr') + [\hat{\delta}^N(\bldr)]^\dag \langle \hat{\delta}^N(\bldr') \hat{\delta}^N(\bldr'') \rangle 
\label{ncons-gauss}
\eea
Simultaneously, we replace $\hat{N}_0$ by $N_0$ in $\hat{\cal H}_2$ and $\hat{\cal H}_3$. The `dominant' contributions $\hat{\cal H}_0$ and $\hat{\cal H}_1$  should however be treated more accurately. In these expressions, we set
\be
\hat{N}_0 = N_0 - \int d\bldr  
\left\{ [\hat{\delta}^N(\bldr)]^\dag \hat{\delta}^N(\bldr) 
- \langle [\hat{\delta}^N(\bldr)]^\dag \hat{\delta}^N(\bldr)\rangle \right\} \;,
\ee
which physically corresponds to setting the number fluctuations between the condensate and the non-condensate to be equal and opposite \cite{Morgan_Gardiner}. 

The above approximations lead to a quadratic number-conserving hamiltonian. This generates the following set of {\em non-local} equations for the condensate and the non-condensate, which can be thought of as the number-conserving analogues of the HFB equations.
The condensate energy evolves according to
%
\bea
& i \hbar \frac{\partial}{\partial t} \wfn^N(\bldr,t) & = 
\left[ \hat{h}_0(\bldr,t) - \lambda(t) \right] \wfn^N(\bldr,t) \nonumber \\
&& + g \left( N_0(t) + \Delta N_0 \right) \left| \wfn^N(\bldr,t) \right|^{2} \wfn^N(\bldr,t)
\nonumber \\
&&  +2g \tilde{n}(\bldr,t) \wfn^N(\bldr,t) + g \tilde{m}^{R}(\bldr,t) \left[ \wfn^N(\bldr,t) \right]^* - f(\bldr,t)\;.
\label{GPE_Ncons}
\eea
The quantity $f(\bldr,t)$ appearing above is related to the dynamical interaction between the condensate and the non-condensate \cite{Gardiner_Private}; more specifically, it refers to the back-action of the (changes in the) normal and anomalous averages on the temporal evolution of the condensate mode, and is defined by
\bea
f(\bldr,t) &=& 
\int d \bldr' g \left| \wfn^N(\bldr',t) \right|^2
\left\{ \tilde{n}(\bldr, \bldr', t) \wfn^N(\bldr',t) \right.
\nonumber \\
&& \hspace{3.0cm} + \left. \tilde{m}^{R}(\bldr, \bldr', t) \left[ \wfn^N(\bldr',t)\right]^* \right\}\;.
\eea
The physical justification for such a term is to ensure orthogonality is maintained throughout the system evolution.
Moreover, $\Delta N_0 = (\langle \hat{N}_0^2 \rangle - \langle \hat{N}_0 \rangle^2)/N_0 - 1$ is a typically small contribution from statistical fluctuations included here for completeness.
The number-conserving versions of the normal and anomalous averages are respectively given by
$\tilde{n}(\bldr,\bldr',t) = \langle ( \hat{\delta}^N (\bldr,t))^{\dag} \hat{\delta}^N(\bldr',t) \rangle$ 
and
$\tilde{m}(\bldr,\bldr',t) = \langle \hat{\delta}^N(\bldr,t) \hat{\delta}^N(\bldr',t) \rangle$. 
The fluctuation operators evolve according to
\begin{eqnarray}
i \hbar \frac{\partial}{\partial t}
\left(
\begin{array}{c}
\hat{\delta}^N(\bldr,t) \\ (\hat{\delta}^N(\bldr,t))^{\dag}
\end{array}
\right)
\nonumber \\
\hspace{0.5cm} =
\left( 
\begin{array}{cc} 
\hat{\mathcal{L}}^{N}(\bldr,t) & \hat{\mathcal{M}}^{N}(\bldr,t) \\ -\left[ \hat{\mathcal{M}}^{N} (\bldr,t) \right]^{*}  & -\left[ \hat{\mathcal{L}}^{N}(\bldr,t) \right]^{*}
\end{array}
\right)
\left(
\begin{array}{c}
\hat{\delta}^N(\bldr,t) \\ (\hat{\delta}^N (\bldr,t))^{\dag}
\end{array}
\right)\;.
\label{BdG_Ncons}
\end{eqnarray}
where we have defined the corresponding generalized number-conserving expressions
\bea
\hat{\mathcal{L}}^N(\bldr,t) &=& \hat{h}_0(\bldr,t) - \lambda(t) +gN_0 \left| \wfn^{N}(\bldr,t) \right|^{2}
\nonumber \\
&& + gN_0 \hat{Q}(t) \left| \wfn^{N}(\bldr,t) \right|^{2} \hat{Q}(t),
\\
\hat{\mathcal{M}}^{N}(\bldr,t) &=& g N_0 \hat{Q}(t) \left[ \wfn^{N}(\bldr,t) \right]^{2} \hat{Q}^{*}(t) \;.
\eea
One can make two immediate observations which are a direct consequence of orthogonality and number-conservation: (i) firstly, these expressions explicitly include the time-dependent projection operators $\hat{Q}(t)$ of Eq.\ (\ref{ncons-fluct}), which guarantee that the orthogonality of the considered subspaces is maintained during the entire evolution. 
(ii) secondly, the derived equations of motion are highly non-local, with double projectors implying the following relation \cite{Morgan_PRA_2004}
\bea
\hat{Q}(t) \left| \wfn^{N}(\bldr,t) \right|^{2} \hat{Q}(t) \hat{\delta}^N(\bldr,t) =
\nonumber \\
\int \int d\bldr' d\bldr'' Q(\bldr, \bldr', t) \left| \wfn^{N}(\bldr',t) \right|^{2}
Q(\bldr', \bldr'', t) \hat{\delta}^N(\bldr'',t) \;.
\eea


Having obtained the full evolution equations for condensate and non-condensate operators, Eqs.\ (\ref{GPE_Ncons}) and (\ref{BdG_Ncons}), we 
can also investigate the frequencies of collective excitations in the presence of a dynamical thermal cloud, a regime in which this theory has been applied to with remarkable success (see Sec.\ \ref{Excitations}). This is achieved by simplifying the full dynamical evolution in the context of 
a perturbative linear response analysis, analogous to that presented in Sec.\ \ref{Linear_Response}.
In particular, we consider perturbing the system by an external force which induces the mean fields to oscillate around their static values.
i.e.
$\wfn^{N}(\bldr,t) = \wfn^N_0(\bldr) + \delta \wfn^N(\bldr,t)$, $\tilde{n}(\bldr,t)= \tilde{n}_0(\bldr) + \delta \tilde{n}(\bldr,t)$, and $\tilde{m}^R(\bldr,t)= \tilde{m}_0^R(\bldr) + \delta \tilde{m}(\bldr,t)$.
As before, we obtain the linearized equation of motion for the condensate fluctuations $\delta \wfn^N(\bldr,t)$. In order to study the response of the system to an external perturbation, we Fourier transform
the {\em temporal} variables of $\delta \wfn^N(\bldr,t)$ and $( \delta \wfn^N(\bldr,t))^*$, and expand the resulting expressions in terms of quasiparticle amplitudes via \cite{Morgan_PRA_2004,Excitations_Burnett_2003,Morgan_PRA_2005,Morgan_RoyalSoc}
\begin{eqnarray}
\left( 
\begin{array}{c} 
\delta \wfn^N(\bldr, \omega)  \\ ( \delta \wfn^N (\bldr, \omega))^*  
\end{array}
\right)
= \sum_{i} b_{i} (\omega)
\left(
\begin{array}{l}
u_i(\bldr) \\ v_i(\bldr)
\end{array}
\right)\;.
\end{eqnarray}
The response amplitudes $b_{i}(\omega)$ appearing above contain information on the condensate density fluctuations.
These amplitudes are proportional to the response function which, in general, is made up of two contributions: a static one, arising from mean fields, and a dynamic one, associated either with direct excitation from the perturbation or indirect excitation from fluctuations induced in the other mean fields as a result of the perturbation.
At low temperatures, the former contribution generally dominates, with the condensate excited directly from the applied perturbation. However, under certain conditions at finite temperatures, the perturbation may predominantly excite the non-condensate part, whose response to the perturbation leads to a secondary mechanism for the excitation of the condensate (see Sec.\ \ref{Excitations}).

\subsection{Brief Summary}

{\em
While mean field theories have been shown to produce generally very good results, their formulation is still based on the assumption that the condensate can be described as a macroscopic classical entity, thus crudely discarding its operator nature, which is generally `allowed' for systems with a large number of atoms.
Such treatment however automatically precludes the description of low-dimensional gases, where the simple picture of a coherent condensate is replaced by a so-called `quasi-condensate', which can be thought of as a condensate which exhibits large phase fluctuations; an ab initio modified mean field theory has been formulated to deal with such cases at equilibrium.

\begin{sidewaysfigure}
 \scalebox{0.7}
 {\includegraphics{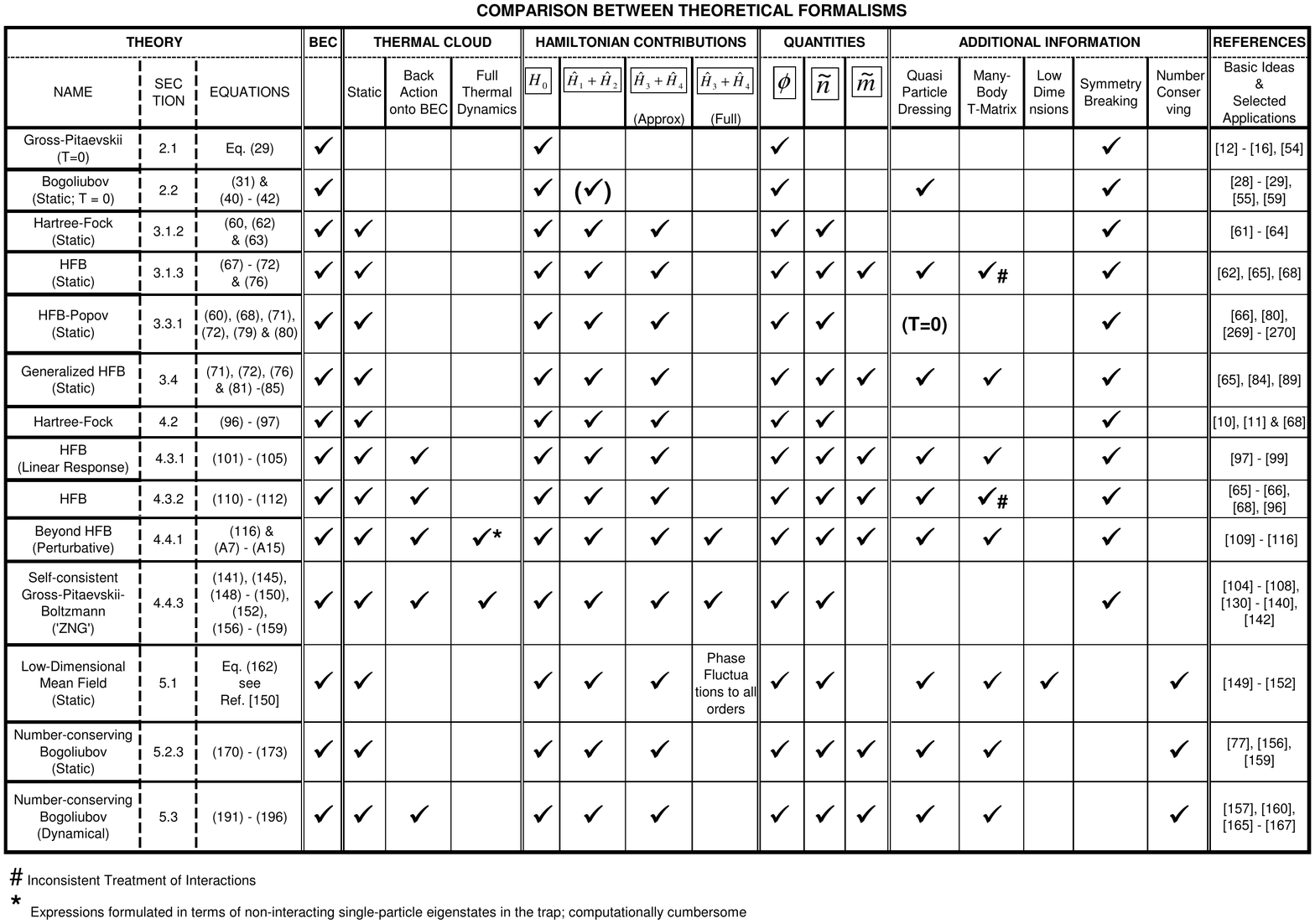}}
 \caption{
Summary of comparison of main theories discussed so far for ultracold Bose gases at finite temperatures.
}
\label{Table_1}
\end{sidewaysfigure}

Moreover, the formal development of the preceeding sections where the system hamiltonian is separated into contributions according to the number of non-condensate operators appearing in them, can actually be generalized rather straightforwardly to the number-conserving case. To do this, one must first ensure the orthogonality between condensate and excitations;
this can be achieved by expanding the Bose field operator as
$\hat{\Psi}(\bldr,t)=\wfn^N(\bldr,t) \hat{a}_0 + \hat{\delta}(\bldr,t)$,
where  $\hat{a}_0$ denotes an operator annihilating a condensate atom which is defined as the projection of the full Bose field operator $\hat{\Psi}(\bldr,t)$ onto the condensate state $\wfn^N(\bldr,t)$ (where $N$ stands for Number-conserving); the non-condensate operator $\hat{\delta}(\bldr,t)$  is specified by the orthogonal projection. Identification of the condensate wavefunction as the state with the largest eigenvalue via the Penrose-Onsager criterion, combined with the desire for the fluctuation operator to scale as the number of non-condensate atoms, leads to the definition of an appropriate non-condensate operator $\hat{\delta}^N(\bldr,t)=(\hat{a}_0^\dag/\sqrt{N}) \hat{\delta}(\bldr,t)$,
which is guaranteed to satisfy $\langle \hat{\delta}^N(\bldr,t) \rangle = 0$, and
 in terms of which the system hamiltonian can be expanded in a number-conserving fashion. 
Following pretty much the same procedure as in the mean-field treatments, the full system hamiltonian is now expanded in contributions of ascending orders of $\hat{\delta}^N(\bldr,t)$; as each such term additionally contains a prefactor of $1/\sqrt{N_0}$, 
such an expansion can in fact be visualized as an expansion in powers of the ratio of non-condensate to condensate atoms; this justifies the perturbative treatment of hamiltonian contributions $( \hat{\cal H}_3 + \hat{\cal H}_4)$ containing three or four number-conserving non-condensate operators $\hat{\delta}^N(\bldr,t)$. This approach leads to a finite temperature Gross-Pitaevskii and corresponding Bogoliubov equations which have similar structure as their mean field counterparts, but also contain additional non-local terms arising from the orthogonality requirement. While this theory has been successfully implemented to the temperature dependence of the shift of elementary excitation frequencies in the linear response limit, a full self-consistent treatment of the thermal cloud is lacking at present.\\
}

This concludes our perturbative treatment of generalized mean-field approaches and the related number-conserving formalism, on which this Tutorial has been largely based.
The treatments presented thus far are summarized in Fig.\ \ref{Table_1}, which makes a systematic classification of all previous approaches in terms of how the thermal cloud is treated, which hamiltonian contributions and (where appropriate) which generalized mean fields are included in each theory, what is the excitation spectrum predicted by each model, and what additional (e.g.\ many-body, low-dimensional) effects can be described by each such approach.

The most advanced current approaches presented from each `class' of models, namely the symmetry-breaking treatment of Sec.\ \ref{Beyond_HFB} and the number-conserving approach of Sec.\ \ref{CW_Gardiner} lead both to a successful numerical implementation and to a very good agreement with ultracold gas experiments at finite temperatures for a wide range of experimental conditions. However, it would be misleading to conclude this Tutorial without describing alternative rather distinct approaches that have been employed to study trapped Bose gases at finite temperatures. Such approaches, presented in Sec.\ \ref{scho-class}, {\em are} actually required to accurately tackle various experimental issues, such as the onset of condensate growth from a purely thermal cloud, 
and fluctuations in low-dimensional systems. Most of the approaches presented below, which are based on very different theoretical formulations, are simply appropriate generalizations of the ideas of earlier sections, and can thus be shown to reduce to theories already discussed. For example, the successful approach of Zaremba, Nikuni and Griffin described in Sec.\ \ref{ZNG} arises as a special case of the approaches of Stoof (Sec.\ \ref{Stoof_Theory}) and Gardiner-Zoller and co-workers (Sec.\ \ref{Gardiner_Zoller_Theory}) in the appropriate limits. It is important to stress that while such approaches are presented last here, as appropriate from a pedagogical point of view, they have been actually derived over a large number of years, with some of the most important results {\em preceeding} those of alternative treatments of Secs.\ \ref{Beyond_HFB} and \ref{CW_Gardiner}.

Each of the topics mentioned below would merit a Tutorial in its own right, and, for example, a Review on Classical Field Theories has been published recently \cite{Polish_Review}. For the purposes of this Tutorial, the subsequent treatment is limited to a brief overview, highlighting the main physical ideas, final equations and applicability of such approaches. Numerous references are given to aid the reader who may wish to pursue further studies on these topics. An excellent set of presentation transparencies (with additional references) on these topics can be found in the websites of two recent meetings on finite temperature models for Bose-Einstein condensation held in Sandbjerg (Denmark) \cite{Sandbjerg_Website} and Heidelberg (Germany) \cite{RupertoCarola_Website} in the summer of 2007.

\section{Alternative Beyond Mean Field Approaches}
\label{scho-class}

The approaches presented below are explicitly number-conserving, and are already widespread in the study of ultracold atomic gases (see also \cite{Davis_Review} for a related review of such approaches).

\subsection{Classical Field Approaches \& The Projected Gross-Pitaevskii Equation}
\label{Classical_Fields}

In our preceeding discussion we have argued against the validity of the Gross-Pitaevskii equation at finite temperatures, on the basis that it ignores both mean field effects and dynamical coupling to the thermal cloud.
However, the Gross-Pitaevskii equation is actually an equation for a {\em classical} field; as such it should be able to describe all classical aspects of a finite temperature system of ultracold gases. For example, it is well-known from the classical theory of electromagnetic radiation that the Rayleigh-Jeans model provides a good approximation to the full quantum system for modes with energies less than $k_B T$, provided all such modes are `highly occupied', such that the Bose-Einstein distribution function of Eq.\ (\ref{BE-dist}) can be well approximated by the `classical' expression $f(\epsilon_i) \approx (k_B T) / \epsilon_i$. Formalisms based on this approximation are termed `classical field' approaches and have recently been reviewed in \cite{Polish_Review}. The discussion of this approximation in the context of ultracold Bose gases has been driven by work of Svistunov, Kagan and Shlyapnikov on the formation and dynamics of Bose-Einstein condensation \cite{Svistunov_1,Svistunov_2,Svistunov_3,Svistunov_4}.
Such an approximation was also used early on in other contexts, such as the electroweak phase transition \cite{Moore_Turok} and the equilibration of a Bose gas to a superfluid state \cite{Damle_Classical}, with a qualitative two-dimensional study of evaporative cooling performed in \cite{Marshall_Classical}.

To study an ultracold Bose gas, one seeds the initial state of the system with arbitrary initial conditions. 
Typically one expands the initial wavefunction in appropriate eigenstates $\spwa_k(\bldr)$ via $\wfn(\bldr,t=0)=\sum_k c_k \spwa_k(\bldr)$; in doing so, the amplitudes and phases are appropriately chosen under the constraint of fixed total atom number and energy: for example, for the homogeneous gas, one typically chooses $\spwa_k(\bldr) = e^{i {\bf k} \cdot \bldr}$, with the populations $|c_k|^2$ chosen such that the distribution is as flat as possible, and the phases of $c_k$ are selected randomly \cite{Svistunov_4,PGPE_Homogeneous}. 
Such an initial state is then propagated by the GPE of Eq.\ (\ref{GP-T0}). Due to its intrinsic nonlinearity, this equation mixes different modes and relaxes rapidly to a classical thermal distribution. In fact, the precise initial conditions are largely irrelevant, as they are lost after a short temporal evolution, and it is therefore not even important to set the initial conditions to be equilibrium ones for the particular system.
In these simulations, the temperature of the system is not set a priori, but it is actually subsequently determined by extracting the temperature upon fitting the number occupation of the relaxed system with a classical thermal distribution. The parameter $\wfn(\bldr,t)$ in such simulations does not simply model the condensate, but actually it describes the entire multi-mode `classical' atomic gas, thus enabling the study of various parameters, such as correlation functions. Clearly such simulations require an upper energy (and momentum) cut-off to ensure that the system remains in the `classical regime' throughout the simulations \cite{Polish_Review}. 

\begin{figure}
  \begin{center}
      \resizebox{120mm}{!}{\includegraphics{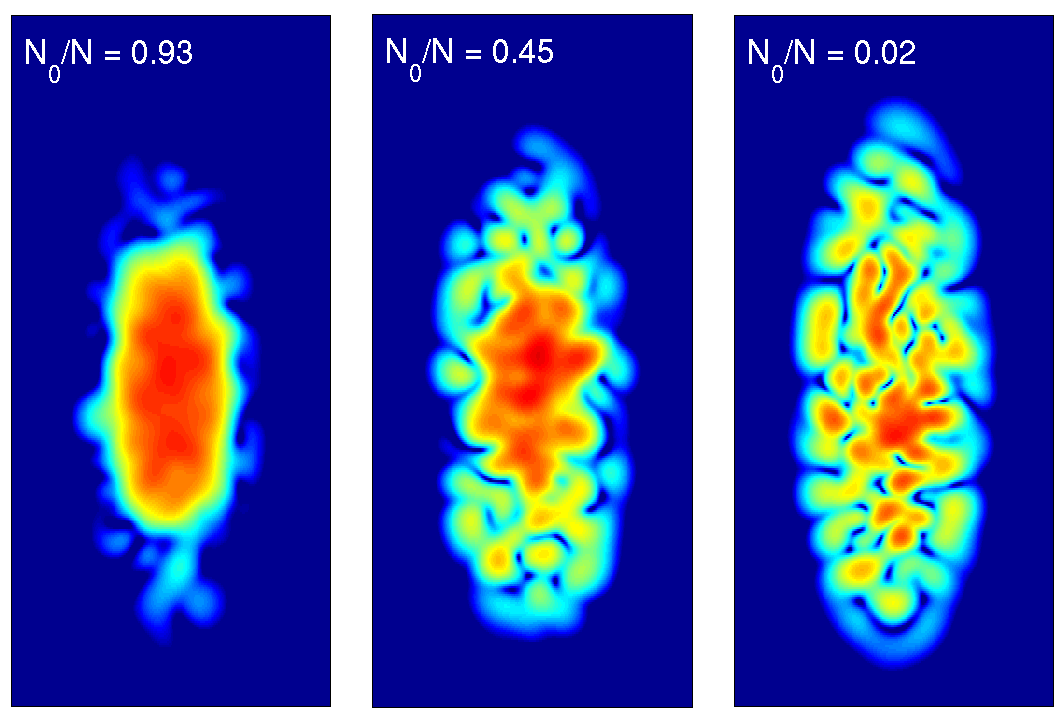}} 
    \caption{
(colour online)
Typical thermalized (equilibrium) images of a classical field consisting of a fixed number of $2,000$ atoms at three different energies (i.e.\ temperatures).
Plotted are the density profiles $|\wfn(x,y,0)|^2$ of an anisotropic ($\omega_\perp/\omega_z = \sqrt{8}$) trapped 3D Bose gas arising from a single run of the PGPE, with colour representing the atomic density (plotted on logarithmic scale - black/blue: zero density, red/dark brown: maximum density).
The enhancement of fluctuations at higher temperatures corresponding to a lower condensate fraction $N_0/N$ (indicated on figure) is evident.
(Images provided by Matt Davis - see also \cite{PGPE_Trap}.)
}
    \label{PGPE_Images}
  \end{center}
\end{figure}

Although the use of a numerical grid in performing simulations imposes itself a cut-off, the most accurate implementation of this approach explicitly introduces a projector into the GPE, to ensure that only momentum-conserving processes consistent with the large occupation approximation are considered, and no numerical `aliasing' arises \cite{Norrie_Thesis,PGPE}.
The projector used in this theory is defined by
$\hat{P} \{ \phi({\bf r,t}) \} = \sum_{k \epsilon C} \spwa_{k}({\bf r}) \int d{\bf r'} \spwa_{k}^{*} ({\bf r'}) \phi({\bf r' },t)$,
where $k$ is restricted within the coherent (classical) region $C$ \cite{PGPE}.
This projector is only required in the nonlinear term, and leads to the replacement of the nonlinear term $|\wfn|^2 \wfn$ appearing in the original GPE by $\hat{P} \{ |\wfn|^2 \wfn \}$. The resulting equation, introduced by Davis, Morgan and Burnett is known as the {\em Projected Gross-Pitaevskii Equation} (PGPE) \cite{PGPE_Homogeneous,PGPE,PGPE_Trap}. While the implementation of the projector is very straightforward in the homogeneous gas \cite{PGPE_Homogeneous}, its extension to the trapped case has only been discussed recently \cite{PGPE_Trap}, by expanding the classical fields in harmonic oscillator eigenstates.
Thermal fluctuations are included into the treatment, as evident from characteristic single-shot images of the density profiles of a trapped gas at three different temperatures, which is plotted in Fig.\ \ref{PGPE_Images}. However, other results such as condensate fraction and correlation functions which require suitable averaging over the fluctuations, are evaluated by replacing ensemble averages by time-averages, based on an ergodic hypotehsis (see e.g., \cite{Polish_Review,PGPE_Trap}).
%
Although such theories appear to work well in diverse contexts, an
%
extension of the PGPE has also been formulated, in which the coherent region is explicitly coupled to a heat bath of non-condensate atoms \cite{PGPE_T}; in this treatment, additional terms are introduced into the PGPE to describe the coupling of the modes in the coherent region to a heat bath of non-condensate atoms.
This idea of splitting the description of the system into low- (coherent) and high-lying modes (above a carefully selected cut-off) essentially constitutes a (multi-mode) generalization of the ZNG formalism, and will be further discussed in Secs.\ \ref{Stoof_Theory}-\ref{Gardiner_Zoller_Theory}.

There have been numerous applications of the classical field method 
\cite{Svistunov_4,Berloff_Classical,Polish_Classical_1,Polish_Classical_2,Polish_Classical_3,Polish_Classical_4,Polish_Classical_5,Polish_Classical_6,Adams_Classical,Lobo_Classical} and of the PGPE \cite{PGPE_Homogeneous,PGPE,PGPE_Trap,PGPE_Shift}. 
The PGPE is a non-perturbative method and has also been used to study the process of non-equilibrium condensation \cite{Berloff_Classical,Berloff_Youd}, to determine the shift in the critical temperature $T_{c}$ \cite{PGPE_Shift}, and to study spontaneous vortex-antivortex pair production in quasi-2D gases \cite{PGPE_BKT_PRL}. 
Despite their success, classical field approaches discard quantum fluctuations {\em by construction}, and therefore more accurate treatments may be needed in certain contexts - as developed below.

\subsection{Stochastic Methods}
\label{Stochastic}

We thus discuss next various schemes which introduce quantum fluctuations into the theoretical description in different levels of approximations.

\subsubsection{The Truncated Wigner Approximation:}
\label{TWA}

Closely related to the above approach is the so-called {\em Truncated Wigner} approximation \cite{Steel_TWA}. 
In this approach, quantum fluctuations are {\em approximately} included in the formalism as follows: in particular, the system evolution is monitored as before by the usual (Projected) Gross-Pitaevskii equation, only the initial conditions have been modified to additionally contain `the right amount' of quantum noise, i.e.\ quantum fluctuations of half a particle per mode (on average),
such that they appropriately sample the full Wigner function.
The effect of quantum fluctuations is thus introduced here by averaging over numerous distinct numerical realizations.

This theory can be put on firm ground as follows (an excellent detailed discussion of this can be found in \cite{Norrie_Thesis}):
Consider a system described by a density operator $\hat{\rho}(t)$. To investigate dynamical effects, instead of studying the evolution of the full density operator, one can equivalently monitor the evolution of a suitably-constructed quantum {\em quasi-probability distribution function} \cite{QO_Book_1,QO_Book_2,QO_Book_3}, known as the Wigner function, $W$. This is constructed in terms of a suitable integral depending on $\hat{\rho}(t)$  and on the eigenvalue, $\alpha$, of the single-particle annihilation operator in a coherent state $|\alpha \rangle$ (in the general case of a multi-mode field applicable here, this would be replaced by multiple integrals). One must thus consider the equation for the evolution of the Wigner function, which acquires a relatively straightforward form of a nonlinear partial differential equation in time. Since the phase space occupied by the Wigner function is quite large, one would instead prefer to track the trajectories of single realizations of the system through phase space, which would be described by some sort of stochastic differential equations. Knowledge of a large number of such trajectories, would then essentially enable the reconstruction of the Wigner function, and thus provide details of the system evolution. 

Consider for the sake of our current discussion a single-mode system with a corresponding Wigner function $W(\alpha,\alpha^*,t)$ (generalizes trivially to multi-mode systems): the full evolution of this Wigner function is known to contain, among other terms, third-order derivative terms of the form 
$\partial W /\partial t \propto [\partial^3/\partial^2 \alpha \partial \alpha^*](\alpha W)
- [\partial^3/\partial \alpha \partial^2 \alpha^*](\alpha^* W)$, 
which do not facilitate an immediate correspondence of this equation to a stochastic differential equation
\cite{QO_Book_1,QO_Book_2}.
Fortunately, it turns out that these terms are quite small provided essentially that all modes of the system are highly occupied ($f_i \gg 1$) (for details and precise conditions see \cite{Castin_TWA}). In this limit, one can `truncate' the exact equation of motion by discarding such terms. The evolution of the Wigner function 
is thus mapped onto the evolution of a field, whose dynamics is governed by an equation
formally identical to the Gross-Pitaevskii equation of Eq.\ (\ref{GP-T0}) (and the same arguments regarding the projector apply as in last section) \cite{Norrie_Thesis}. Having made this identification, one can now routinely solve the (P)GPE as usual, and simply seed the initial states with noise, such that spontaneous processes also become feasible \cite{Norrie_Thesis}. To further justify the above, we note that Polkovnikov recently performed a systematic perturbation theory in quantum fluctuations around the classical system evolution, obtaining the GPE to lowest order, and the truncated Wigner approximation to next order \cite{Polkovnikov}.

Although the truncated Wigner method has been used extensively in quantum optics \cite{TWA_QO}, the
first numerical implementation of the Truncated Wigner approximation in utlracold gases was performed in 1998 in \cite{Steel_TWA}; a sampling technique based on the number-conserving Bogoliubov theory \cite{Gardiner_NC,Castin_Dum} was then subsequently analysed in  
\cite{Castin_TWA,Castin_JModOpt} (see also \cite{TWA_Castin_PRL}). 
The Truncated Wigner has since been used to study various phenomena, including colliding condensates \cite{TWA_Norrie_PRA}, condensate reflection from a steep barrier \cite{Scott_Hutchinson_Gardiner}, three-body recombination processes \cite{TWA_3_Body}, collapsing condensates \cite{TWA_Collapsing_BEC}, atom interferometers \cite{Scott_Interferometer} and optical lattices 
\cite{TWA_Ruostekoski_1,TWA_Ruostekoski_2,TWA_Ruostekoski_3,TWA_Ferris,TWA_Beata}.

 
Both treatments presented above describe an ultracold atomic gas by a single equation for the highly-occupied modes of the system. 
At low temperatures, high mode occupation is restricted to the
lower part of the energy spectrum, corresponding to the condensate and the states dressed by their proximity to the condensate. 
Thermalization is actually achieved by the nonlinear mixing of different modes. 
Due to their similarities, we should now briefly comment on the physical interpretation and validity of the above approaches. In the classical field approaches, one assumes that all relevant modes of the system are highly-occupied, such that thermal fluctuations entirely dominate over quantum fluctuations. This automatically restricts its validity above a minimum temperature (which is nonetheless relatively small), with the same method clearly also applicable above the critical region, where the gas {\em is} classical anyway. On the other hand, the Truncated Wigner method only differs from the PGPE in that it additionally includes numerical noise in the initial conditions of the simulations, which is aimed at mimicking quantum fluctuations.
It would thus be natural to assume that the validity regime of the Truncated Wigner is the same as that of the PGPE, with an additional extension to the very low temperature regime - this is however not the case.

As mentioned earlier, the Truncated Wigner method is based on an approximate equation for the evolution of the full Wigner quasi-probability distribution (which discards certain contributions), an approximation which should generally be valid as long as all relevant modes of the system are highly occupied. Importantly, in the Truncated Wigner method, noise is only included in the initial conditions in a manner which on average corresponds to the addition of half a particle per mode. However, evolution of this initial quantum distribution by the classical (P)GPE results in the
initial quantum distribution thermalizing into a classical field at a slightly higher temperature than the initial distribution, thus giving rise to unphysical heating \cite{Castin_TWA}. The net result of this is to limit the applicability of the theory either to relatively short times (before any such classical thermalization takes place), or to the low-temperature regime defined by the condition that the maximum energy of the Bogoliubov modes should not exceed a few times the thermal energy $k_B T$ \cite{Castin_TWA}. As a result, the Truncated Wigner method is more useful for studying situations where the quantum processes largely dominate over thermal effects.
On the other hand, one could interpret the (P)GPE as the thermodynamic limit of the equation for the dynamics of the Wigner function, for which spontaneously initiated (as opposed to stimulated) processes are negligible. 
In other words, the Truncated Wigner method yields a more accurate description of relatively low temperatures, where quantum effects provide a noticeable contribution, with higher temperature generally described quite well by the classical field method, or the PGPE.


How can such an unphysical thermalization arising within the Truncated Wigner method be avoided? For this purpose, it is important to note that the low-lying modes we have been discussing so far are actually coupled to a range of higher-lying `thermal' modes which are occupied more sparsely, and hence above the chosen cut-off. In fact, coupling the previously-considered part of the system to a thermal gas provides the necessary irreversibility for the system to relax, while simultaneously  guaranteeing it relaxes to the correct equilibrium. This additional coupling gives rise both to dissipative terms and to {\em dynamical} noise terms which modify the Gross-Pitaevskii equation to an appropriate {\em Langevin}, or {\em Stochastic Gross-Pitaevskii} equation. 
%

Next, we describe two seemingly very distinct approaches, formulated respectively by Stoof \cite{Stoof_PRL,Stoof_JLTP,Stoof_GreenBook}, and Gardiner, Zoller and co-workers \cite{QK_V,QK_I,QK_III,SGPE_I,SGPE_II}, which are however essentially both based on a physical description of the system in terms of an appropriate probability distribution function. 
Such treatments combine the ideas of the formalism of Zaremba, Nikuni and Griffin (Sec.\ \ref{ZNG}) whereby the gas is split into two dynamically coupled subsystems, with the inclusion of quantum fluctuations discussed previously.
This is achieved by a generalization of the earlier picture of splitting the system into condensate and non-condensate contributions, to a separation between low- and high-lying modes.
The low-lying modes forming the coherent region refer to the condensate and modes affected by it due to thermal and quantum fluctuations and are treated exactly.
Such modes are dynamically coupled to the incoherent part of the system (high-lying modes), as in the spirit of the approach discussed by Zaremba, Nikuni and Griffin, only their evolution is now {\em stochastic}, i.e.\ it includes {\em explicit noise terms} to account for fluctuations \cite{Stoof_JLTP,SGPE_I,SGPE_II,Stoof_Langevin}.
This description arises by mapping the effective equation for the probability distribution of such low-lying modes onto a stochastic equation for their evolution.
Although the non-condensate modes are usually treated as a thermal particle reservoir, the full description models the high-lying modes by an appropriate quantum Boltzmann equation.
This set of equations, which in the authors' opinion constitutes the `next generation' kinetic theory of ultracold bosonic gases, is mathematically closely related to the treatment of Sec.\ \ref{ZNG}, although the interpretation of $\wfn(\bldr,t)$ is now distinct, as it constitutes the `order parameter' of the (multi-mode) system, rather than the condensate wavefunction.
These theories can be shown to reduce in the appropriate limits to either of the self-consistent Gross-Pitaevskii-Boltzmann, or `ZNG' theory (and thus to all more elementary mean field theories), the classical field method and the truncated Wigner approximation.


\subsubsection{Stoof's Non-equilibrium Theory:}
\label{Stoof_Theory}

Stoof derived an equation for the evolution of a Wigner probability distribution $P [ \phi^{*}, \phi; t ]$ based on functional integration techniques \cite{Stoof_PRL,Stoof_JLTP,Stoof_Duine,Stoof_GreenBook}.
This distribution expresses the probability of the system to be in a coherent state $| \phi ({\bf r}) ;t \rangle $, 
derived from the vacuum state $|0 \rangle$ by 
$| \phi ({\bf r}) ;t \rangle = {\rm exp} \{ \int d {\bf r} \phi({\bf r}) \hat{\Psi}^{\dag}({\bf r},t) \} |0 \rangle$. 
By expanding the density matrix in coherent states, the probability distribution is cast in a form which requires the calculation of a functional integral containing the quantity 
$\left| \langle \phi ; t | \phi_{0} ; t_{0} \rangle \right|^{2} = \langle \phi ; t | \phi_{0} ; t_{0} \rangle \langle \phi_0 ; t_0 | \phi ; t \rangle$. 
Using standard techniques, each of these matrix elements $\langle \phi ; t | \phi_{0} ; t_{0} \rangle$ can be written as a `{\em path integral}' \cite{Negele_Book} over all complex field evolutions $\psi(\bldr,t)$ 
(with $\psi(\bldr,t_0)=\phi_0(\bldr)$ and $\psi^*(\bldr,t)=\phi^*(\bldr)$),
via 
$\int d[\psi^{*}] d [\psi] {\rm exp} \left\{i S_{\rm EFF} [ \psi^{*},\psi ] / \hbar \right\}$, where $S_{\rm EFF}$ is the effective action of the system (essentially analogous to the choice of an effective hamiltonian in our preceeding formalism). The determination of 
$\left| \langle \phi ; t | \phi_{0} ; t_{0} \rangle \right|^{2} $ requires one to
study all possible paths that the evolution of the field may follow from some time $t$ back to the initial time $t_0$ (evolution backward in time), and then back again to $t$ (forward evolution), leading directly to the introduction of the Keldysh contour \cite{Keldysh}, over which the paths are to be calculated.
Such an approach is explicitly number-conserving, and typically discards any details of the initial quantum state (Markov approximation).
After performing a standard transformation in the field variables $\psi$ to explicitly separate the semi-classical dynamics from the effect of fluctuations, this treatment 
leads to a quadratic effective action in the fluctuations, in which the interactions have been renormalized to include many-body effects. 
The evolution of the probability distribution takes the form of a Fokker-Planck equation (see Eq.\ (\ref{FP})), which is the generalized analogue of the `quantum' evolution described by the truncated Wigner approximation. 
Fluctuations around $\phi$ can be systematically calculated, with correlation functions of any order obtained from suitable moments of $P[\phi^*,\phi;t]$.
An excellent non-technical discussion of this approach can be found in \cite{Stoof_NATO}, whereas a more mathematical paedagogical presentation is given in \cite{Stoof_LesHouches}.

The system may be separated into two `components' by expressing the probability distribution of the system as
$P \left[ \phi^{*}, \phi; t \right] = P_{0} \left[ \Phi^{*}, \Phi; t \right] P_{1} \left[ \phi'^{*}, \phi'; t \right]$, which amounts to a Hartree-Fock-type approximation.
This procedure leads to two coupled equations: firstly, substituting this ansatz into the full Fokker-Planck equation and integrating over the non-condensate degrees of freedom $\phi'$ leads to a Fokker-Planck equation (given by Eq.\ (\ref{FP})) for the low-lying modes of the system, which constitute the `coherent region'. Performing the related integration over the condensate contribution $\Phi$ leads to the corresponding Fokker-Planck equation for the non-condensate, whose `semi-classical' treatment yields a Quantum Boltzmann equation (Eq.\ \ref{QBE_2})) describing the incoherent system dynamics.

The low-lying modes of the system are described by the Fokker-Planck equation \cite{Stoof_JLTP,Stoof_Duine}
\begin{eqnarray}
&& i  \hbar \frac{ \partial}{\partial t} P_{0} \left[ \Phi^{*}, \Phi ; t \right]  = 
\nonumber \\
&& -  \int d{\bf r} \frac{ \delta}{\delta \Phi({\bf r})} \left[ \hat{h}_{0} - \mu(t) - i R({\bf r},t)+g \left| \Phi({\bf r}) \right|^{2} \right] \Phi({\bf r}) P_{0} \left[ \Phi^{*}, \Phi ; t \right] \nonumber \\
&& +  \int d{\bf r} \frac{ \delta}{\delta \Phi^{*}({\bf r})} \left[ \hat{h}_{0} - \mu(t) + i R({\bf r},t)+g\left| \Phi({\bf r}) \right|^{2} \right] \Phi^{*}({\bf r}) P_{0} \left[ \Phi^{*}, \Phi ; t \right] \nonumber \\
&& - \frac{1}{2} \int d{\bf r} \frac{ \delta^{2}}{\delta \Phi({\bf r})\delta      \Phi^{*}({\bf r})} \Phi^{*}({\bf r}) \hbar \Sigma^{K}({\bf r},t) P_{0} \left[ \Phi^{*}, \Phi ; t \right]\;, 
\label{FP} 
\end{eqnarray}
where $\delta / \delta \Phi^{(*)}(\bldr) $ denote functional derivatives \cite{Blaizot_Ripka,Negele_Book}.
Let us now briefly interpret this equation:
Both $iR(\bldr,t)$ and $\Sigma^K(\bldr,t)$ arise from the coupling to the higher-lying modes (the `reservoir').
Let us firstly ignore the last contribution of Eq.\ (\ref{FP}). If we were to additionally set $iR=0$, then the above equation would be equivalent to the GPE for $\Phi(\bldr,t)$ (first line), and the corresponding evolution for $\Phi^*(\bldr,t)$ (second line), which would thus be describing the classical evolution of the low-lying modes of the system. The $iR$ term is precisely the contribution that was included in our perturbative beyond-HFB approach and describes the transfer of particles between the coherent part of the system and the higher-lying (thermal) modes of the system.
It is given by Eq.\ (\ref{R}), upon a slight redefinition whereby we drop the ${\bf v_c}$ term from the momentum-conserving delta functions and absorb the kinetic energy into $\mu_c$, such that $\varepsilon_c$ of Eq.\ (\ref{E_c}) is replaced here by $\varepsilon_c=\mu_c$; moreover the non-condensate energies $\tilde{\epsilon}_i$ of Eq.\ (\ref{E_HF}) are replaced by
$\tilde{\varepsilon}_i(\bldr,{\bf p}) = ( |{\bf p}_{i}|^{2}/2m + V_{\rm ext}({\bf r}_i) + 2 g \langle |\Phi(\bldr_i)|^2 \rangle  )$.
The explicit inclusion of $iR$ modifies the `reactive' GPE to the dissipative GPE of Eq.\ (\ref{GPE_iR}).
The latter contribution in Eq.\ (\ref{FP}) contains the so-called {\em Keldysh Self-energy}, $\Sigma^K(\bldr,t)$, given by \cite{Stoof_JLTP,Stoof_Duine}
\begin{eqnarray}
\Sigma^{K}({\bf r},t) & = &
- i \left( \frac{4 \pi}{\hbar} \right) g^{2} \int \frac{ d {\bf p}_{2}}{(2 \pi \hbar)^{3}} \int \frac{ d {\bf p}_{3}}{(2 \pi \hbar)^{3}} \int \frac{ d {\bf p}_{4}}{(2 \pi \hbar)^{3}} \nonumber \\
& \times & (2 \pi \hbar)^{3} \delta \left( {\bf p}_{2} - {\bf p}_{3} - {\bf p}_{4} \right) 
\delta \left( \varepsilon_{c}+\tilde{\varepsilon}_{2}-\tilde{\varepsilon}_{3}-\tilde{\varepsilon}_{4} \right)
\nonumber \\
& \times & \left[ (N_{2}+1)N_{3}N_{4} + N_{2} (N_{3}+1)(N_{4}+1) \right] \;,
\label{Sigma_K_Full}
\end{eqnarray}
where $N_{i} = N(\tilde{\varepsilon}_{i})$ denote the thermal populations.
%
$\Sigma^K$ corresponds to fluctuations associated with the {\em same collisional processes} described by $iR$ (i.e.\ incoherent collisions between condensate and non-condensate atoms), so the last contribution of Eq.\ (\ref{FP}) should also be included whenever collisions are included into the treatment via $iR \neq 0$.
Its inclusion maps Eq.\ (\ref{FP}) onto a stochastic partial differential differential equation for $\Phi$; this equation is essentially equivalent to the dissipative Gross-Pitaevskii equation discussed in our second order mean field approaches of Sec.\ \ref{Beyond_HFB} (see, e.g.\ `ZNG' Eq.\ (\ref{GPE_iR})) but with the further addition of an appropriate noise term.

However, both the dissipation $iR(\bldr,t)$ and the self-energy $\Sigma^K(\bldr,t)$ depend implicitly on $\Phi(\bldr,t)$ through their dependence on the condensate energy $\varepsilon_c=\hat{h}_0+g|\Phi(\bldr,t)|^2$ and through the induced change in the non-condensate factors $N_i$.
Thus, the corresponding stochastic equation is a Langevin equation with multiplicative noise and a prefactor with a complicated dependence on $\Phi(\bldr,t)$ \cite{Stoof_Duine}, which is hard to simulate numerically \cite{Stoof_Duine}. To reduce this equation to a more manageable form, we
restrict our discussion to the regime near equilibrium, for which one can show that
the Fokker-Planck equation reduces to the following stochastic equation \cite{Stoof_Duine,Stoof_Langevin}
\begin{eqnarray}
i \hbar \frac{\partial \Phi({\bf r},t)}{\partial t} &=& 
\left[ 1 + \frac{\beta}{4} \hbar \Sigma^K(\bldr,t) \right] 
\left( \hat{h}_{0} + g |\Phi({\bf r},t)|^{2} - \mu \right) \Phi({\bf r},t)
\nonumber \\
&+& \eta({\bf r},t) 
\label{SGPE_Eq}
\end{eqnarray}
with the noise term having gaussian correlations of the form
\be
\langle \eta^{*}({\bf r},t) \eta({\bf r'},t') \rangle = i (\hbar^{2}/2) \Sigma^{K}({\bf r},t) \delta (t-t') \delta ({\bf r}-{\bf r'})\;.
\label{Noise}
\ee

\begin{figure}[t]
\centering \scalebox{0.4}
{\includegraphics{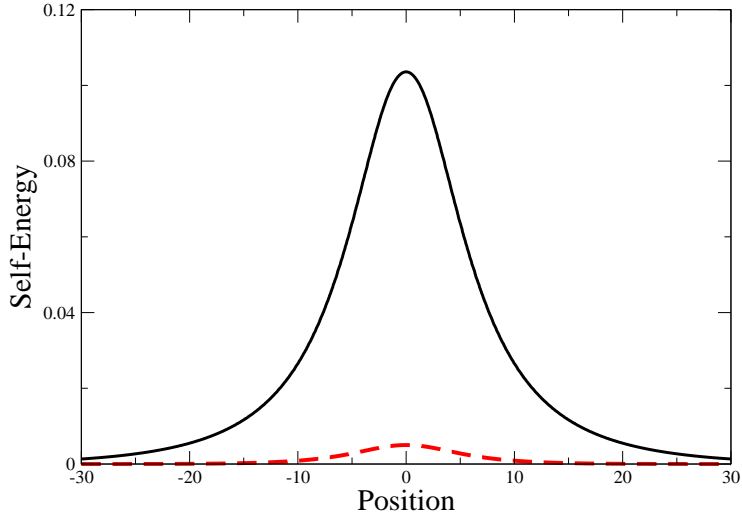}}
 \caption{
(colour online)
Typical position and temperature dependence of the self-energy for a trapped one-dimensional Bose gas ($\omega_z=2 \pi \times 10$Hz, $^{87}$Rb). Plotted is the scaled self-energy $\beta |\hbar \Sigma_K(z)| /4$ appearing in Eq.\ (\ref{SGPE_Eq}) (where $\beta=1/k_B T$) against position (measured in harmonic oscillator units) for two different temperatures $T=400$nK (solid black, top) and $T=100$nk (dashed red, bottom).
}
\label{Self_Energy}
\end{figure}

How does such an equation come about?
From the formulation of the theory, close to equilibrium where the thermal cloud can be described by the Bose distribution function, one can derive a fluctuation-dissipation theorem linking the strength of the fluctuations, $\hbar \Sigma^K$, with the dissipation, $iR$.
Both of there depend on the energy $\varepsilon_c$ for promoting a condensate atom (located at position $\bldr$) out of the low-lying part of the system, with their mathematical relation given by
$iR(\bldr,\varepsilon_c) = - (\hbar/2) \Sigma^K(\bldr,\varepsilon_c) [ 1+ 2N(\varepsilon_c)]^{-1}$.
Using now the classical field approximation
$N(\varepsilon_c) \approx (k_B T)/(\varepsilon_c-\mu)$
valid for the low-lying modes of the system which are highly populated ($N(\varepsilon_c) \gg1$), one can re-express 
$[1+2N(\varepsilon_c)]^{-1} \approx (\beta/2) (\varepsilon_c-\mu)$.
Finally we note that, within the present formalism, $\varepsilon_c$ is actually an {\em operator} in the configuration space of the order parameter in contrast to the semi-classical treatment of Sec.\ \ref{ZNG}; this leads directly to the approximate expression for the dissipative term
%
$iR(\bldr,t) \approx - (\beta/4) \hbar \Sigma^K(\bldr,t)
[ \hat{h}_0 +g | \Phi(\bldr,t)|^2 -\mu ]$,
which has already been implicitly assumed in Eq.\ (\ref{SGPE_Eq}).
A plot of the position and temperature dependence of the self-energy , $\Sigma^K(z)$, for a harmonically trapped one-dimensional Bose gas is shown in Fig.\ \ref{Self_Energy}.

Having already identified the origin of the different contributions, we can now re-interpret the quantity $\Phi(\bldr,t)$ appearing in Eq.\ (\ref{SGPE_Eq}) as the {\em order parameter} of the system, since it describes not only the condensate, but also incorporates thermal and (in its full version) quantum fluctuations, i.e.\ accurately describes the low-lying modes of the system \cite{Stoof_JLTP,Stoof_Langevin}.
%
We can now identify the important new physical feature introduced by this equation. All mean field approaches discussed earlier had the fundamental restriction of no spontaneous initiation, i.e if the condensate mean field were initially zero, it would remain so at all subsequent times. Contrary to that picture, the presence of the noise term in Eq.\ (\ref{SGPE_Eq}) provides a `quantum mechanical seed' to {\em initiate} condensate growth.
(In fact, the critical region has been studied in more detail by Stoof, who showed explicitly that the system acquires the necessary irreversibility required for the onset of condensation \cite{Stoof_PRL,Stoof_JLTP}.)

The derivation sketched above relies on the quadratic nature of the effective action, and some comments are necessary here: Although the cubic and quartic contributions of fluctuations around $\phi$ do not {\em explicitly} appear in the treatment, their effect {\em has} actually {\em been taken into consideration} in generating both the effective many-body T-matrix interaction and the coupling to the high-lying modes. This argument is essentially equivalent to the elimination of the anomalous average to yield many-body effects (Sec.\ \ref{many-body}), and the interpretation of the triplet correlations $\langle \nopc \nopa \nopa \rangle$ as providing the required coupling to the non-condensate (Sec.\ \ref{Beyond_HFB}).

\begin{figure}
  \begin{center}
    \begin{tabular}{ccc}
      \resizebox{41mm}{!}{\includegraphics{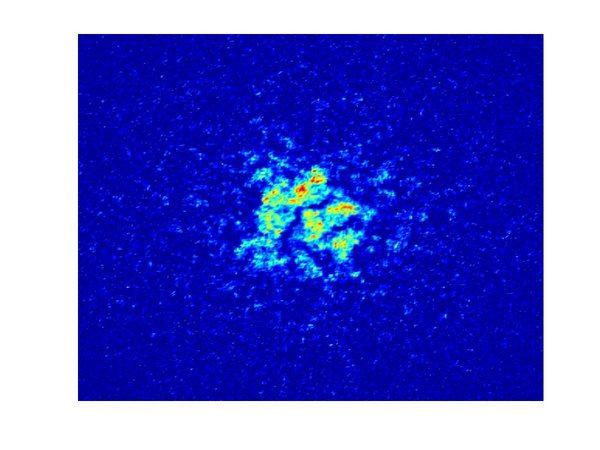}} &
      \resizebox{41mm}{!}{\includegraphics{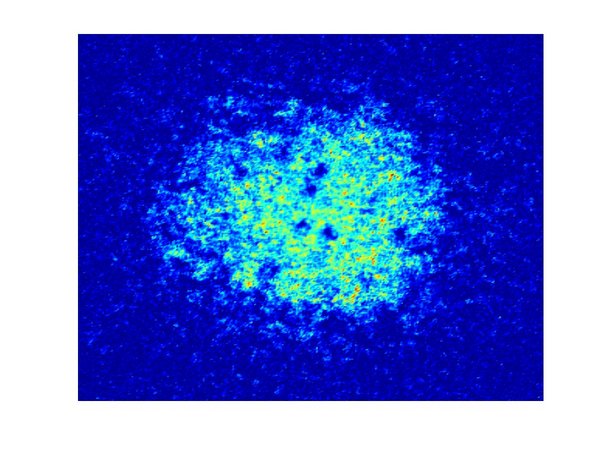}} &
      \resizebox{41mm}{!}{\includegraphics{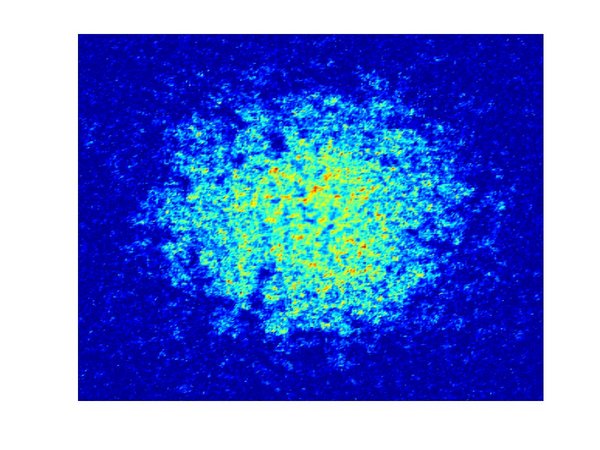}}  \\
      \resizebox{41mm}{!}{\includegraphics{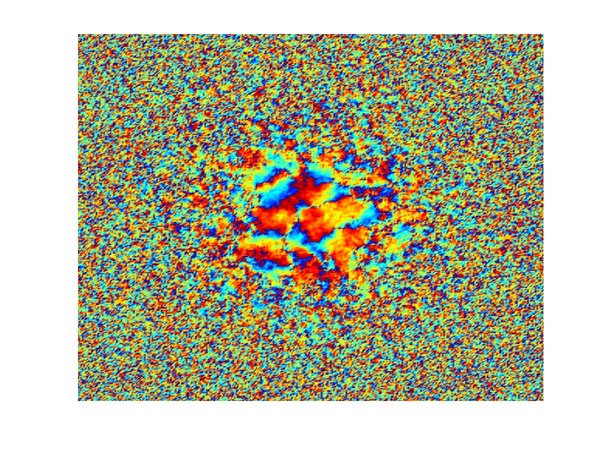}} &
      \resizebox{41mm}{!}{\includegraphics{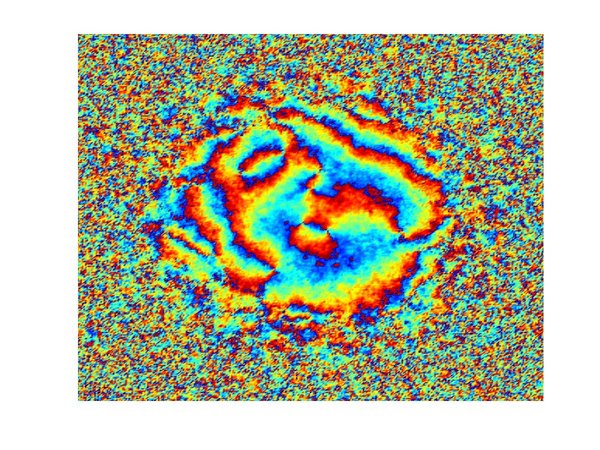}} &
      \resizebox{41mm}{!}{\includegraphics{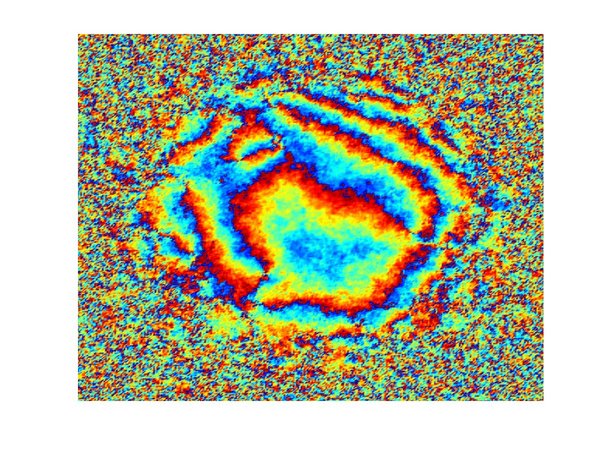}}  \\
      \resizebox{41mm}{!}{\includegraphics{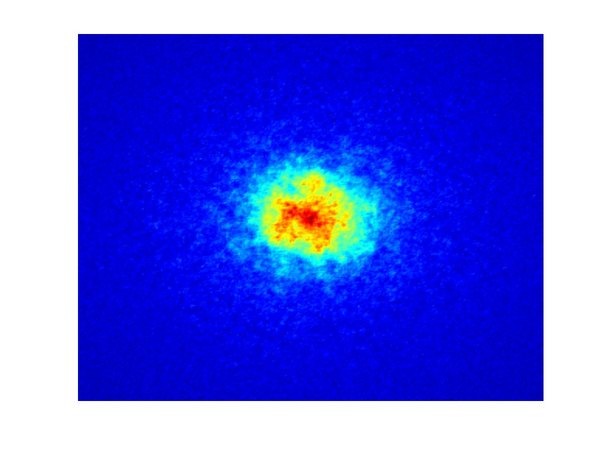}} &
      \resizebox{41mm}{!}{\includegraphics{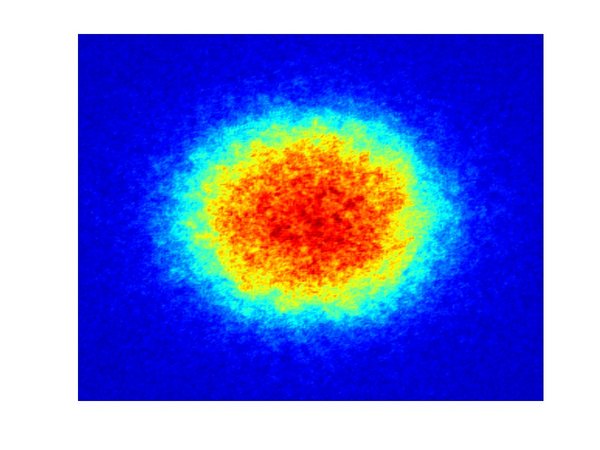}} &
      \resizebox{41mm}{!}{\includegraphics{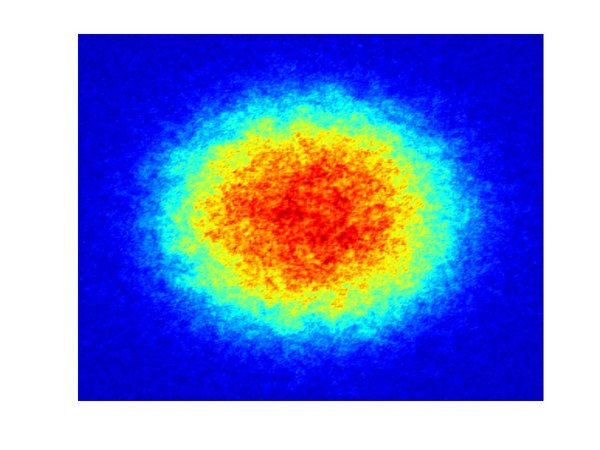}}  \\
    \end{tabular}
    \caption{
(colour online)
Typical evolution predicted by the solution of the stochastic Gross-Pitaevskii equation in two dimensions.
Top/Middle: Single-run results showing the density/phase, clearly highlighting the spontaneous vortex formation in the ab initio growth of a quasi-condensate coupled to a static heat bath. The quasi-condensate is formed in  a harmonic trap with $\omega_x=\omega_y=2 \pi \times 200$Hz, and contains approximately $65000$ $^{23}$Na atoms at equilibrium for a heat bath at $T=500$nK.
Bottom: Corresponding profiles obtained after averaging over a small number of runs ($\approx 25$), as might be done experimentally; in the latter profiles spontaneous singularities are largely washed out.
(Images provided by Stuart Cockburn - see also \cite{Davis_SGPE_New}; note that in the above images the colourbar is redefined from column to column for optimal visualization.)
}
    \label{Stochastic_Images}
  \end{center}
\end{figure}

To complete our treatment, we must also discuss
the evolution of the occupation numbers $N$.
This can be done by introducing the Wigner distribution via \cite{Stoof_JLTP}
\be
\int d\bldr' e^{-i {\bf p} \cdot {\bf r'} / \hbar}
\langle \phi' \left( {\bf r} + \frac{ {\bf r'}}{2} \right)
\phi'^* \left( {\bf r} - \frac{ {\bf r'}}{2} \right) \rangle
= N({\bf p},{\bf r},t) + \frac{1}{2}\;,
\ee
and using the corresponding Fokker-Planck equation for the non-condensate $\phi'$ (obtained by integrating out the condensate degrees of freedom) to extract
the evolution of $\langle \phi' \phi'^* \rangle$. 
Comparison to the non-condensate populations $f({\bf p},{\bf r},t)$ introduced in  Eqs.\ (\ref{Wigner_ZNG_1})-(\ref{Wigner_ZNG_2}) 
reveals that the field $\phi'(\bldr,t)$ now contains both thermal {\em and} quantum fluctuations, with quantum-mechanical fluctuations of half a particle per mode added to the thermal occupation $N({\bf p},\bldr,t)$ of each mode. 
Although such a noise contribution was previously introduced in the {\em initial} conditions in the Truncated Wigner approach, here the noise is {\em dynamical}, i.e.\ it is added onto $\Phi(\bldr,t)$ at {\em each time-step} in the simulations \cite{Stoof_Langevin}.
The non-condensate populations evolve according to a quantum Boltzmann equation. Making the usual gradient expansion (assuming the non-condensate varies on a much larger lengthscale than the external trapping potential), one obtains \cite{Stoof_JLTP}
\be
\frac{ \partial N}{\partial t} 
+ ( {\bf \nabla_p} \tilde{\varepsilon} ) \cdot ({\bf \nabla_r} N) 
- ( {\bf \nabla_r} \tilde{\varepsilon} ) \cdot ( {\bf \nabla}_{\bf p} N)
= {\cal C}_{12}[N] + {\cal C}_{22}[N] \;.
\label{QBE_2}
\ee
The two collisional integrals ${\cal C}_{12}[N]$ and ${\cal C}_{22}[N]$ appearing above are practically identical to 
those discussed in Sec.\ \ref{ZNG} in the context of the `self-consistent Gross-Pitaevskii-Boltzmann', or `ZNG' theory.
(Eqs.\ (\ref{C_12})-(\ref{C_22})).
One important difference here is that they are expressed in terms of the
quantity $\Phi(\bldr,t)$, instead of the condensate density $\wfn(\bldr,t)$:
\bea
{\cal C}_{12}[N] &=& \frac{4 \pi}{\hbar} g^{2} |\Phi|^2 \int \frac{ d {\bf p}_{2}}{(2 \pi \hbar)^{3}} \int \frac{ d {\bf p}_{3}}{(2 \pi \hbar)^{3}} \int \frac{ d {\bf p}_{4}}{(2 \pi \hbar)^{3}} \nonumber \\ 
& \times & (2 \pi \hbar)^{3}
\delta \left( {\bf p}_{2} - {\bf p}_{3} - {\bf p}_{4} \right) 
\delta \left( \varepsilon_c + \tilde{\varepsilon}_{2} - \tilde{\varepsilon}_{3} - \tilde{\varepsilon}_{4} \right) \nonumber \\
&\times& (2 \pi \hbar)^3
\left[ \delta({\bf p}-{\bf p_2}) - \delta({\bf p}-{\bf p_3}) - \delta({\bf p}-{\bf p_4}) \right] \nonumber \\
& \times & \left[ (N_{2}+1)N_{3}N_{4} - N_{2} (N_{3}+1)(N_{4}+1) \right] \;,
\label{C12}
\eea
\bea
{\cal C}_{22}[N] &=& \frac{4 \pi}{\hbar} g^{2} \int \frac{ d {\bf p}_{2}}{(2 \pi \hbar)^{3}} \int \frac{ d {\bf p}_{3}}{(2 \pi \hbar)^{3}} \int \frac{ d {\bf p}_{4}}{(2 \pi \hbar)^{3}} \nonumber \\ 
& \times & (2 \pi \hbar)^{3}
\delta \left( {\bf p}+{\bf p}_{2} - {\bf p}_{3} - {\bf p}_{4} \right) 
\delta \left( \tilde{\varepsilon} + \tilde{\varepsilon}_{2} - \tilde{\varepsilon}_{3} - \tilde{\varepsilon}_{4} \right) \nonumber \\
& \times & \left[ (N+1)(N_{2}+1)N_{3}N_{4} - N N_{2} (N_{3}+1)(N_{4}+1) \right] \;.
\label{C22}
\eea
%
%

The stochastic approach presented in this section amounts to solving self-consistently the stochastic Gross-Pitaevskii equation, Eq.
(\ref{SGPE_Eq}), with dissipation and noise respectively defined by Eqs.\ (\ref{R}), (\ref{Sigma_K_Full}) and (\ref{Noise}), coupled to the dynamical Quantum Boltzmann Equation of Eq.\ (\ref{QBE_2}) with the collisional integrals of Eq.\ (\ref{C12})-(\ref{C22}).
The non-condensate density is obtained via $n_{NC} = \int d {\bf p}/(2 \pi \hbar)^{3} N({\bf p}, \bldr, t)$ (as in Sec.\ \ref{ZNG}).
Applications of this approach (in suitable limits) can be found in 
\cite{Stoof_Duine,Low_D_PRA,Stoof_Langevin,Stoof_Vortex,Proukakis_AtomChip,Proukakis_Equilibrium,SGPE_AtomLaser}
(see also Secs.\ \ref{Excitations}-\ref{Growth}).

How are the solutions of the Stochastic GPE to be interpreted?
As the formalism explicitly includes dynamical noise, the main idea behind it is to extract information from an averaging over a large number of numerical `runs' (ranging from a few tens to a few hundred - depending on required accuracy and regime of simulation). Due to the random nature of fluctuations included, averaging over many runs is equivalent to averaging over different experimental realizations; such an approach yields, for example, smooth density profiles and correlation functions. 
It should however be remarked that as the Stochastic GPE describes the whole matter-wave field, it makes no distinction between condensate and thermal cloud, a common `artificial' feature of all treatments presented in Secs.\ \ref{zeroT}-\ref{Phase_Number}: To highlight this point the images of the `modified low-dimensional theory' presented in  Sec.\ \ref{Phase_Fluct} were compared to results of the stochastic code for the same parameters (see earlier Fig.\ (\ref{1D_Images})); while the overall agreement was very good, the stochastic images automatically produce the total atomic density, rather than needing to sum over the two independently-determined (but coupled)  sub-contributions. Thus simulations of the stochastic GPE can be seen as corresponding precisely to `numerical experiments', and therefore an (artificial) partitioning of the atomic gas to a `condensate' and a `non-condensate' contribution should be performed in the same way as when analyzing experimental data - e.g.\ via bimodal density fits \cite{Ketterle_Varenna} (or via a manipulation of system correlations \cite{Proukakis_Equilibrium}).

At the same time, like with the Projected GPE discussed earlier, one can also extract useful information from a single run of the Stochastic GPE, which would resemble results from a single experimental realization - such analysis is relevant when looking at processes which are sensitive to individual runs, such as the collapse of a BEC \cite{BEC_Exp_3,Stoof_Duine}, or the visualization of images of phase and density fluctuations. For example, a single run of the Stochastic GPE in two dimensions (Fig.\ \ref{Stochastic_Images}, top images) leads to the spontaneous appearance of vortices as the (quasi)condensate grows from the thermal cloud, with the location (and the number) of the vortices differing from realization to realization (see also \cite{Davis_SGPE_New}) - the spontaneous appearance of vortices in the context of a second order phase transition induced by a rapid quench was originally proposed in the cosmological constant \cite{Spontaneous_Vortex_Kibble,Spontaneous_Vortex_Zurek}, and subsequently discussed in the context of an atomic BEC in \cite{Spontaneous_Vortex_BEC}. However, as soon as one averages over a large number of independent realizations (runs) spontaneous excitations are rapidly  `washed out', leading to much smoother density profiles (Fig.\ \ref{Stochastic_Images}, bottom images), as also relevant for calculations of correlation functions \cite{Proukakis_Equilibrium}.

A simplified form of the Stochastic GPE of Eq.\ (\ref{SGPE_Eq}) with time-independent occupation numbers in the non-condensate, and thus a time-independent self-energy $\Sigma^{K}({\bf r})$ (see subsequent Eq.\ (\ref{Sigma_K_Eqm})) was used to study reversible condensate formation when cycling through the phase transition \cite{MIT_Dimple}, thus providing the very first application \cite{Stoof_Langevin} of such a stochastic equation to the study of ultracold gases. 
One of us (NPP) has been heavily involved in subsequent simulations, investigating
fluctuations of one-dimensional Bose gases \cite{Low_D_PRA,Proukakis_Equilibrium}, the growth of coherence of an atom laser \cite{SGPE_AtomLaser}, and quasi-condensate growth on an atom chip \cite{Proukakis_AtomChip}.
It is also important to remark that instead of solving the full SGPE numerically, one can resort to variational calculations.
Such a technique was used to discuss collisional frequencies and damping rates of collective excitations, growth-collapse cycles in attractive condensates \cite{Stoof_Duine}, and finite temperature dynamics of a single vortex \cite{Stoof_Vortex}.

{\em Stochastic Hydrodynamics: }
Consideration of the stochastic effects leads to a modification in the corresponding finite temperature hydrodynamic equations discussed in Sec.\ \ref{Hydro_T}.
To study these, we use the Madelung transformation $\Phi(\bldr,t)=\sqrt{n(\bldr,t)}e^{i \theta(\bldr,t)}$ directly {\em within} the effective action, which is thus re-expressed as $S_{\rm EFF}[n,\theta]$ in terms of the density $n(\bldr,t)$ and the phase $\theta(\bldr,t)$ of the system. This procedure leads to the following two coupled stochastic equations of motion \cite{Stoof_Duine}:
\bea
&& \frac{\partial n}{\partial t} +{\bf \nabla} \cdot ( n {\bf v_c}) 
= -i \frac{\beta}{2} \Sigma^K \left( \tilde{\mu}_c-\mu \right) n
+ 2 \sqrt{n} \xi_n \\
&& \frac{ \partial \theta}{\partial t} - i \frac{\beta}{4}  \Sigma^K 
\left( \frac{\hbar^2}{2mn} \right) {\bf \nabla} \cdot \left( n {\bf \nabla} \theta \right)
= - \frac{(\tilde{\mu}_c - \mu)}{\hbar} + \frac{1}{\sqrt{n}} \xi_\theta
\label{Hydro_Stoch}
\eea
where the above noise sources ($\xi_n$, $\xi_\theta$) have gaussian correlations of the form
\be
\langle \xi_n(\bldr,t) \xi_n(\bldr',t') \rangle =
\langle \xi_\theta(\bldr,t) \xi_\theta(\bldr',t') \rangle =
i \frac{1}{4} \Sigma^K(\bldr,t) \delta(\bldr-\bldr') \delta(t-t').
\ee
Here $\tilde{\mu}_c = \mu_0 + gn + (1/2)m v_s^2 
= - (\hbar^2 \nabla^2 \sqrt{n})/(2m \sqrt{n}) + V_{\rm TRAP} +gn + (1/2)m v_s^2 $
is the chemical potential of the condensate (the condensate kinetic energy contribution has been absorbed into the chemical potential here, contrary to the corresponding definition of Eq.\ (\ref{mu_T}), and $\mu$ is the chemical potential of the thermal cloud.
Comparing these equations to the hydrodynamic description given earlier, we find the following modifications \cite{Stoof_Duine}: (i) The `continuity equation' acquires an additional {\em drift} term due to $\xi_n$; (ii) The equation for the condensate phase (from which the Euler-like equation for the superfluid velocity can be obtained via ${\bf v}(\bldr,t) = (\hbar/m) {\bf \nabla} \theta(\bldr,t)$)  acquires both a {\em spatial diffusion-like} contribution proportional to $i \Sigma^K(\bldr,t)$ due to collisions between condensate and thermal atoms, and a noise contribution inversely proportional to the square root of the density.

Next, we briefly review the results of Gardiner, Zoller and co-workers who used a different methodology, which is nonetheless similar in spirit to the above treatment, and also yields a similar Stochastic Gross-Pitaevskii equation. Before doing so, we should however also highlight here the related work of Ramos and co-workers \cite{Ramos_1,Ramos_2,Ramos_3} on the non-equilibrium field-theoretic formulation and numerical simulation of the equilibration of a homogeneous Bose gas following a rapid temperature quench.

\subsubsection{The Gardiner-Zoller Kinetic Theory:}
\label{Gardiner_Zoller_Theory}

The theoretical description of Gardiner, Zoller and coworkers 
\cite{QK_V,QK_I,QK_III}
is also explicitly number-conserving and is based on techniques established in the quantum optics community, and described in various textbooks \cite{QO_Book_1,QO_Book_2,QO_Book_3}.  In their treatment, the system is also split into two parts: the first one, termed the `condensate band' ($R_C$) corresponds to the low-lying modes of the system, which include the condensate and those modes which are affected by its presence; states lying above a particular energy, belong to the `non-condensate band' ($R_{NC}$). Note that, as before, the non-condensate band includes all modes which may be occupied during a collision, and should not be confused with the yet higher-lying modes which have been implicitly eliminated in order to introduce the effective interaction in the binary collisions.
These two bands, apart from having their own internal dynamics, are also allowed to interact and exchange both particles and energy, pretty much as in the schematic of Fig.\ \ref{Schematic_ZNG}.
The system hamiltonian is thus split into three parts, describing the contributions within each band ($\hat{H}_{C}$ and $\hat{H}_{NC}$), and their respective interaction, 
 $\hat{H}_{INT}$. One can write down an equation of motion for the density operator of the system, $\hat{\rho}$, via \cite{QK_V,QK_I,QK_III}
\be
i \hbar \frac{d \hat{\rho}}{dt} = \left[ \left( \hat{H}_C + \hat{H}_{NC} + \hat{H}_{INT} \right) \, , \, \rho \right] \;.
\ee

By making the assumption that there are many more atoms in $R_{NC}$ compared to those in $R_C$, one may initially assume that the non-condensate remains practically unaffected by the interaction process, and can thus be treated as a `reservoir'. In order to obtain the evolution of populations in the condensate band, the standard procedure then is to eliminate the reservoir by tracing over its degrees of freedom, so that one ends up with a density operator for the condensate, $\hat{\rho}_C = {\rm Tr_{NC}}(\hat{\rho})$. 
This is similar in spirit to the elimination of the non-condensate modes $\phi'$ performed by Stoof in order to obtain a Fokker-Planck equation for the low-lying modes of the system.
One then uses the Laplace transform method (and also ignores interactions in the kernel and makes the Markov approximation \cite{QK_V}) to obtain the so-called `master equation' for the evolution of $\hat{\rho}$. This is a fully quantum-mechanical equation describing the evolution of the low-lying modes in contact with the non-condensate which is treated as a heat bath.
The derivation of the resulting master equation is quite lengthy and will not be given here, with the final form of this equation presented in \ref{GZ_ME}. 
Suitable limiting cases of this master equation have already been used in numerical simulations 
\cite{QK_II,QK_IV,QK_VI,QK_VII,QK_PRL_I,QK_PRL_II,QK_PRL_III,QK_JPhysB,QK_PRL_IV,SGPE_IV}, and are briefly presented below.
We should also mention here the related work of Anglin \cite{Anglin_PRL} who formulated a master equation for a single trapped mode of an ultracold weakly-interacting Bose gas.

{\em The `Quantum-optical Master Equation':} 
In its simplest application, the master equation can be used to generate rate equations for the average populations of each mode - as done in textbook discussions of laser theory \cite{QO_Book_1,QO_Book_2,QO_Book_3}.
The simplest approximation that can be used to study condensate dynamics, is to `shrink down' the entire condensate band to simply one mode describing the condensate (as in the treatment of Sec.\ \ref{ZNG}), and study its interaction with the reservoir of thermal particles.

Close to equilibrium, the difference in the rates of scattering into and out of the condensate are related by the factor $[1-e^{\beta(\mu_c-\mu)}]$, where $(\mu_c -\mu)$ denotes the difference between the condensate and the thermal chemical potentials, with $\mu_c$ determined from a time-independent GPE.
This gives rise to the following rate equation for the {\em mean} number of atoms, $N_0$ in the condensate \cite{QK_III,QK_PRL_I}
\be
\frac{dN_0}{dt} = 2 W^+ \left[ \left( 1 - e^{(\mu_c - \mu)/k_BT} \right) N_0 + 1 \right] \;.
\label{Rate_BEC}
\ee
The quantity $W^+$ denotes the scattering rate {\em into} the condensate, 
which, upon ignoring all spatial dependence, can be approximated by \cite{QK_III,QK_V,QK_VI,QK_PRL_I,QK_PRL_II,QK_PRL_IV}
\bea
W^+ &=& \frac{g^2}{(2 \pi)^5 \hbar^2} \int d{\bf k_2} \int d{\bf k_3} \int d{\bf k_4}
\delta({\bf k_2}+{\bf k_3}-{\bf k_4}) \nonumber \\
&\times& \delta(\varepsilon_2+\varepsilon_3-\varepsilon_4 -\mu_c) f_1 f_2 (f_3+1)
\approx {\rm few} \times \frac{4m (a k_B T)^2}{\pi \hbar^3}\;,
\label{W+}
\eea
as briefly explained below.

Although such a rate can be calculated more precisely, initial studies \cite{QK_I} were restricted to somewhat crude approximations: in first instance, apart from ignoring all spatial dependence, the non-condensate band distribution $f_i$ was approximated by a classical Maxwell-Boltzmann distribution $e^{-\beta (\varepsilon_i-\mu)}$, and the integrals were calculated over the entire energy range (instead of only within the non-condensate band). Making the additional approximations that $f_i \ll 1$, such that $f_i+1 \approx 1$ led to a constant rate fixed only by the temperature and the interaction strength, and explicitly given by $4m(a k_BT)^2/\pi \hbar^3$, in which the (weak) dependence on the number of atoms has been suppressed \cite{QK_III,QK_PRL_I,Lee_Masters}. Despite the drastic approximations made, this study nonetheless revealed good qualitative agreement with experiments.
%
%
This model was subsequently improved to also account for low-lying modes within the condensate band, with the full Bose-Einstein distribution used and the range of integration restricted to outside the condensate band \cite{QK_PRL_II}. An analytical expression obtained in the limit when all spatial dependence was ignored yielded a similar qualitative description, but with an additional prefactor of ${\rm few} \approx 3$ in Eq.\ (\ref{W+}).

In the ergodic approximation, a number $g_k$ of distinct levels is grouped in groups of mean energy $e_k$ with $n_k = n(e_k) = g_k f(e_k)$, where $e_k$ denote the energies of the dressed levels in the condensate band.
In this limit, the occupation numbers evolve according to \cite{QK_VII,QK_PRL_II}
\be
\frac{dn_m}{dt} = 2 W^+ \left[ \left( 1 - e^{(e_m - \mu)/k_BT} \right) n_m + g_m \right] \;,
\label{Rate_QP}
\ee
where the rate $W^+$ is now re-expressed as an {\em energy} integral over the energies, $e_i$. The distribution functions appearing in the integral now become time-dependent, and their evolution can be calculated by the ergodic quantum Boltzmann equation \cite{QBE_Holland}
\bea
g_n \frac{ \partial f_n}{\partial t} 
= \frac{8m a^2 \bar{\omega}^2}{\pi \hbar} 
&&\sum_{ijk} \left[ (f_i +1) (f_n +1)f_j f_k - f_i f_n (f_j +1) (f_k +1) \right]
\nonumber \\
&&\times \delta(e_i + e_n - e_j - e_k)\;,
\label{Rate_QBE}
\eea
where $\bar{\omega}$ is the mean harmonic oscillator frequency of the trap.
%

The above equations only deal explicitly with the  `{\em growth}' terms, describing collisions which lead to a transfer of atoms from one band to the other. Actually, there is another process that should be considered here, corresponding to `{\em scattering}' terms whereby an atom in the condensate band interacts with another atom in the non-condensate band, with population redistribution taking place within {\em each} band, but without any population transfer between bands \cite{QK_VI,QK_PRL_II}. These treatments were not explicitly dealt with in previous kinetic formulations. While we do not give their respective expressions here (see \ref{Appendix_Gardiner}), we note that their presence may lead to an acceleration of the initial (spontaneous) growth phase i.e.\ before growth due to bosonic stimulation dominates the dynamics \cite{QK_II}.

This approach led to the first quantitative predictions of condensate growth \cite{QK_PRL_I}, and to good agreement with various experiments \cite{QK_PRL_II,QK_JPhysB,MIT_Formation,QK_PRL_III,Growth_Orsay} (see Sec.\ \ref{Growth}).

{\em The `High-temperature Master Equation:'}
The exact master equation of Gardiner-Zoller for the condensate band (\ref{GZ_ME}) may be exploited further, by mapping it onto a probabilistic Fokker-Planck equation \cite{SGPE_I,SGPE_II,SGPE_III,SGPE_IV} which is similar, but not identical, to that of Stoof. This equation is valid in the regime when the eigenfrequencies of the condensate band operator are small compared to the temperature, i.e.\ $\hbar \omega \ll k_B T$, which roughly coincides with the criterion for large occupation per mode.
Mapping onto a stochastic differential equation yields an equation similar to that of Eq.\ (\ref{SGPE_Eq}), but with {\em additional} contributions due to the scattering terms mentioned earlier 
\cite{SGPE_I,SGPE_II,SGPE_III}.
By modelling these non-local scattering terms approximately \cite{SGPE_II} (although more accurate expressions can also be given \cite{SGPE_II,SGPE_III}), one arrives at a local stochastic GPE for the condensate band which takes the form
\bea
d \alpha(\bldr,t) = &-& \frac{i}{\hbar} \bar{L}_c \alpha(\bldr,t) dt
+ {\cal P}_c \left[ K_G(\bldr,t)dt + dW_G(\bldr,t) \right] \nonumber \\
&+& {\cal P}_c \left[ K_M(\bldr,t)dt + i \alpha(\bldr) dW_M(\bldr,t) \right]
\;,
\label{SGPE_Gardiner}
\eea
where ${\cal P}_c$ is a suitable projector into the condensate band, and $\bar{L}_c$ is essentially the unperturbed energy in the condensate band (modified by the mean field of the non-condensate).
The noise sources appearing here are independent, with $ dW_G(\bldr,t)$ being complex and $dW_G(\bldr,t)$ real.
The precise forms of the noise correlations and expressions for the redistribution ($K_G(\bldr,t)$) and scattering ($K_M(\bldr,t)$) contributions are given in \ref{GZ_SGPE}.
The formulation of this stochastic equation essentially merges the ideas of quantum kinetic theory with those of the Projected GPE formalism, whose finite temperature generalisations \cite{PGPE_T} also include (qualitatively) the coupling to a heat bath \cite{SGPE_II}.
%
The Stochastic GPE has been applied to the dynamics of hydrogen condensates \cite{SGPE_I}, to the spontaneous formation of vortices and vortex arrays \cite{SGPE_I,SGPE_III,Davis_SGPE_New} and to equilibrium properties of finite temperature Bose gases \cite{SGPE_IV}.

At this stage, we wish to point out that, in the limit where the scattering contributions are ignored, this equation is identical to that of Stoof (Eq.\ \ref{SGPE_Eq}), as shown explicitly in \ref{GZ_SGPE}.
Note however that additional subtle differences do exist between the approaches of Stoof and Gardiner {\em et al.} (e.g. determination of energy `cut-off' between the two bands, use of projectors), and the reader is referred to the discussion given by Gardiner and Davis \cite{SGPE_II}.
The equivalence between the corresponding stochastic GPEs enables us to briefly present a simple damped GPE which arises as a limiting case of both above formalisms \cite{SGPE_I,Zoutekow}, and which has been used by numerous authors (including one of us - NPP) to study finite temperature properties of Bose gases 
\cite{QK_PRL_IV,Pitaevskii_Phenomenology,Choi_Phenomenology,Vortex_Lattice_Ueda_1,Vortex_Lattice_Ueda_2,Proukakis_Parametric}.

We note here in passing that a related Langevin equation for a single-mode condensate was formulated by Graham \cite{Graham_Langevin} and used to study fluctuations around the equilibrium state of the condensate after it had been formed.

\subsubsection{The Damped Gross-Pitaevskii Equation:}
\label{Damp_GP}

If one heuristically ignores the `noise term' $\eta$ in Eq.\ (\ref{SGPE_Eq}),  the condensate evolution can be cast (in the classical approximation) in the form
\begin{equation}
i \hbar \frac{\partial}{\partial t} \Phi({\bf r},t) = \left[ 1 - i \gamma(\bldr,t,T) \right]  \left( \hat{h}_{0} +g |\Phi(\bldr,t)|^2 - \mu \right) \Phi({\bf r},t), \label{Damped_GPE}
\end{equation}
where $\gamma(\bldr,t,T)$ denotes a dynamical temperature- and position-dependent damping rate, given by $\gamma = i (\beta / 4) \hbar \Sigma^{K}({\bf r,t})$
\cite{QK_PRL_IV,Zoutekow}.  
The addition of a phenomenological damping coefficient $\gamma$ onto the GPE  was originally proposed using general arguments by Pitaevskii \cite{Pitaevskii_Phenomenology}, and first implemented to trapped Bose gases by Choi {\em et al.} \cite{Choi_Phenomenology} who used a constant, position-independent rate $\gamma$ to discuss damping of excitations. A similar, but not identical, phenomenologically-damped equation with the factor $(1-i \gamma)$ appearing on the {\em left} hand side of the equation has been used in diverse studies, including vortex lattice growth 
\cite{Vortex_Lattice_Ueda_1,Vortex_Lattice_Ueda_2}, 
and dark soliton decay \cite{Proukakis_Parametric}.
Although the constant value of $\gamma$ chosen in such approaches was such that it agreed qualitatively, and to order of magnitude, with experiments, the values used had no microscopic justification. The arguments of the preceeding section show how such 
a `phenomenologically-damped GPE' can be justified from a {\em microscopic} perspective, while simultaneously providing an explicit expression for this damping coefficient.

Close to equilibrium, the full self-energy expression of Eq.\ (\ref{Sigma_K_Full}) can be simplified to an approximate version which is simpler to deal with. To achieve this, we use the identity
$N(\tilde{\varepsilon}) +1 = e^{\beta(\tilde{\varepsilon}-\mu)}/(e^{\beta(\tilde{\varepsilon}-\mu)}-1) = - N(-\tilde{\varepsilon}) ,
$
where $N(\tilde{\varepsilon})=[e^{\beta(\tilde{\varepsilon}-\mu)} -1 ]^{-1}$ is the Bose-Einstein distribution function.
By noting that expressions of the form $[ N_2 (N_3 +1)(N_4+1) \pm (N_2+1)N_3N_4 ]$ appear always within integrals containing energy and momentum conservation factors (see, e.g.\ Eqs.\ (\ref{Sigma_K_Full}), (\ref{C12})), we can immediately relate these factors via
$
N_2 (N_3 +1)(N_4+1) \pm (N_2+1)N_3N_4 = 
\left[ 1 \pm e^{-\beta(\tilde{\varepsilon}-\mu)} \right] (N_2+1)N_3N_4 \;,
$
a result which was used independently by Zaremba, Nikuni and Griffin \cite{ZNG}, Stoof \cite{Stoof_Langevin} and Gardiner, Zoller and coworkers \cite{QK_V,QK_III}.
For the purposes of our present discussion, we note that, upon considering the expression of Eq.\ (\ref{Sigma_K_Full}) for the self-energy sufficiently close to equilibrium, we can use the above relation for the sum of the `in' and `out' rates, and simultaneously approximate the exponential $e^{-\beta(\tilde{\varepsilon}-\mu)} \approx 1 -\beta(\tilde{\varepsilon}-\mu) \approx 1 $, to obtain
\begin{eqnarray}
& & \Sigma^{K}({\bf r},t) \approx 
- i \frac{8 \pi}{\hbar} g^2 \int \frac{ d {\bf p}_{2}}{(2 \pi \hbar)^{3}} \int \frac{ d {\bf p}_{3}}{(2 \pi \hbar)^{3}} \int \frac{ d {\bf p}_{4}}{(2 \pi \hbar)^{3}} (2 \pi \hbar)^{3}  \nonumber \\
&& \hspace{1.0cm} \times  \delta \left( {\bf p}_{2} - {\bf p}_{3} - {\bf p}_{4} \right) 
\delta \left( \varepsilon_{c}+\tilde{\varepsilon}_{2}-\tilde{\varepsilon}_{3}-\tilde{\varepsilon}_{4} \right)
 (N_{2}+1)N_{3}N_{4}\;. 
\label{Sigma_K_Eqm}
\end{eqnarray}
For the purposes of the damped GPE it suffices to good approximation to use the 
damping rate $\gamma = i (\beta / 4) \hbar \Sigma^{K}({\bf r,t}) \approx {\rm few} \times 4m (a k_B T)^2 / \pi \hbar^3$ deduced by
Gardiner, Davis and co-workers, where the prefactor has a typical value of around 3 \cite{QK_VI,QK_PRL_II}.

Before concluding this section on alternative beyond mean field approaches, we briefly mention two additional methodologies which are currently receiving increasing attention in the ultracold atom physics community, particularly due to their ability to handle both weakly- and strongly-interacting regimes.

\subsection{Positive P-Representation:}

Sec.\ \ref{Stochastic} discussed the formulation of a stochastic differential equation in the Wigner representation. Alternative choices are also possible, as well-known in the quantum optics context \cite{QO_Book_1,QO_Book_2,QO_Book_3}, and one could for example choose to map the master equation into a related stochastic differential equation using the Glauber P function, instead of the Wigner function \cite{SGPE_I}. Although such a function cannot always be interpreted in a probabilistic sense, a variant of this technique, in which the number of independent variables is doubled, removes this restriction. Such a positive P-representation \cite{Positive_P1,Positive_P2,Positive_P3} has the appealing feature that it generates {\em exact} differential equations, with the absence of explicit vacuum noise contributions in the final equations implying that one does not require a mean occupation per mode much larger than one, which is a limitation of Wigner-function-based approaches. This technique has the advantage that it is exact, and can also handle the strongly-correlated regime. 
The first discussion of this method in the context of BEC was done in \cite{Steel_TWA}.
Although the validity of its predictions is limited to relatively short times (with the sampling error in the simulations growing exponentially after that), there has been recent success in studying numerous topics of experimental relevance \cite{PP_Evaporative_Cooling,PP_Molecular_BEC}, including condensate collisions \cite{PP_Collisions}, with recent developments \cite{PP_Gauge} promising a wider applicability in the future.

\subsection{2-Path Irreducible Effective Action:}

Far from equilibrium dynamics in which initial non-markovian effects become crucial can also be described in the context of the 2-Path-Irreducible (2PI) closed time path effective action, a non-perturbative technique which has led to significant progress in the description of strongly interacting relativistic systems (see, e.g. \cite{2PI_Relativistic}). 
This approach is also based on a formalism in terms of a Schwinger-Keldysh effective action and preserves (at any truncation) important conservation laws such as total particle number and energy. Specifically, Rey {\em et al.} \cite{2PI_Rey_I,2PI_Rey_II} and Gasenzer {\em et al.} \cite{2PI_Gasenzer_I,2PI_Gasenzer_II} applied a systematic expansion of the effective action in powers of the inverse numbers of field components, which amounts to an expansion of the theory about a strong quasi-classical field. This approach, which is also valid for strongly-correlated systems, has already been shown to reduce \cite{2PI_Rey_I} , in the appropriate limits, to the mean field regime, the Bogoliubov (one-loop) approximation, and the time-dependent HFB formalism.

\subsection{Brief Summary}

{\em
Further approaches have been formulated for the description of ultracold Bose gases at finite temperatures which behave either predominantly classically, or are additionally affected by quantum effects:

For the low-lying modes of the system which are typically highly-occupied, one can assume that thermal fluctuations largely overwhelm quantum fluctuations, and so the system can be described by the (classical) Gross-Pitaevskii Equation. As the evolution of this equation is largely independent of the initial conditions (except for very short times), one typically starts from easy-to-implement initial non-equilibrium conditions. Fluctuations are evident in single numerical runs, whereas temporal averaging (assuming ergodicity) produces smooth density profiles and correlation functions. The most detailed implementation also includes a projector to ensure that only modes within the `classical' region under study contribute to the system evolution.

In the opposite limit where quantum fluctuations are important, one can still propagate the system by the same classical equation, while approximately accounting for quantum effects by the inclusion of random fluctuations of, on average, half a particle per mode in the initial conditions, in the so-called Truncated Wigner approximation. At higher temperatures, this approach leads to spurious damping, and its validity is therefore restricted to relatively short times.

In general, one can envisage separating the modes of the trapped gas into low-lying modes that should be treated accurately, and higher-lying (thermal) modes that can be treated semi-classically. The low-lying modes are found to obey a so-called Stochastic Gross-Pitaevskii equation; this is a dissipative Gross-Pitaevskii equation where dissipation arises from the coupling of these modes to the reservoir of high-lying modes;
unlike the semiclassical treatments discussed earlier, this equation additionally
includes a dynamical noise contribution with gaussian correlations to account for fluctuations. Simultaneously, the high-lying modes are described by a Quantum Boltzmann Equation.
On the formulation side, this latter approach constitutes the appropriate generalization of the self-consistent Gross-Pitaevskii-Boltzmann, or `ZNG' formalism - note however that the stochastic Gross-Pitaevskii equation has so far only been numerically implemented when coupled to a static heat bath, with the implementation of the required coupling to the Quantum Boltzmann Equation currently being pursued. In the context of the stochastic Gross-Pitaevskii equation,
single-run results indicate typical fluctuations present in experiments, while suitable averaging over a number of runs is required to generate smooth density profiles and correlation functions.
}
\\

This concludes our formal presentation of both mean field and more advanced formalisms used to describe ultracold Bose gases at finite temperatures, and the next section puts these theories into context
by briefly comparing their respective predictions for experimentally relevant cases.

\section{Applications and Comparison of Above Approaches}
\label{applications}

The theoretical study of ultracold atoms has been enhanced by the possibility of immediate comparison to experiments, and the theories presented above have been applied to diverse experimental conditions, as mentioned throughout this Tutorial.
Initial applications of these theories focused on static properties, such as condensate fractions and density profiles, which are described {\em fairly} well already at the Hartree-Fock level \cite{Stringari_Review}, with relatively small (but measurable) differences provided one is far from the critical region in 3D, or the corresponding regimes of large phase fluctuations in 1D and 2D. 
As there are unfortunately no `benchmark' tests available to date in the ultracold gas community (despite recent efforts by the authors), we restrict our comparison below to two key experimental issues which both stimulated and assisted in the development of the above approaches; we do not give any new results here, but merely reproduce previously published results. 
The subsequent discussion is therefore {\em not} intended as a detailed {\em overview} of those research topics, but rather as a brief introduction to highlight some key issues mentioned in our preceeding discussion; accordingly, the list of references given below is far from complete.

\subsection{Finite Temperature Excitation Spectrum}
\label{Excitations}

Studies of the response of a system to external perturbations
has been the subject of detailed investigation in `traditional' condensed matter systems over the past decades \cite{Griffin_Book}. 
The excitations can be split into `collisional' (or hydrodynamic) and `collisionsless' regimes, depending on the density of the system, or equivalently on the collisional mean free path:
Most of the (early) experiments with ultracold gases probed the latter regime, in which 
collisions between thermal atoms must be explicitly considered, and the dominant effects arise from self-consistent mean fields.
These can be further classified depending on the relative size of the excitation wavelength $\lambda_{\rm ex}$ compared to the healing length $\xi \propto (\mu m)^{-1}$ of the system: for $\lambda_{\rm ex} \ll \xi$ (i.e.\ large momenta) one obtains `single-particle' excitations, whereas in the opposite regime of small momenta ($\lambda_{\rm ex} \gg \xi$) one obtains phonon-like, or collective excitations (see Sec.\ \ref{Bog-dG} and Eq.\ (\ref{Bog_Spectrum})). In confined systems, the size of the trap sets an additional lengthscale to the problem \cite{Ketterle_Varenna}. When $\lambda_{\rm ex}$ approaches the condensate size, the excitation spectrum becomes discretized, i.e.\ the low energy collective modes of the system are standing sound waves at specific frequencies, whereas excitations with energies larger than the typical trap frequency behave semi-classically \cite{Ketterle_Sound}.

\begin{figure}
 \begin{center}
      \resizebox{120mm}{!}{\includegraphics{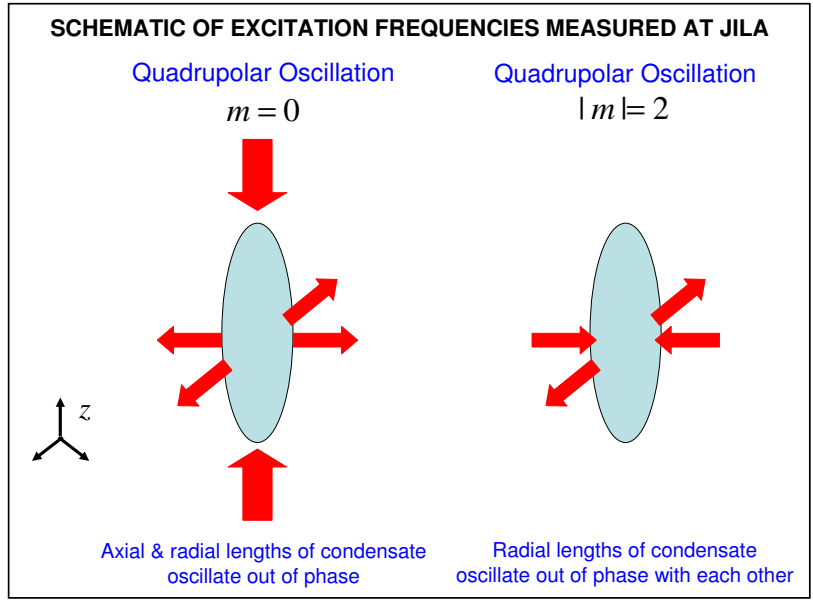}} 
    \caption{
(colour online)
Schematic of two different quadrupolar modes of excitations observed in early BEC experiments \cite{JILA_Exc_T_0,MIT_Exc_T_0,JILA_Exc_T,MIT_Exc_T}, with emphasis here on the modes observed in the slightly elongated condensates at JILA. 
Each excitation is identified by the angular momentum of the excitation about the $z$-axis, denoted by
$m$. The $m=0$ mode corresponds to a cylindrically-symmetric quadrupolar excitation, whereas the cylindrical symmetry is broken for $|m|=2$, with axial and radial motion taking place out of phase with each other.
While the $m=0$ mode was well understood, experimental attempts to excite the $m=2$ mode at finite temperatures generated some `anomalous' features discussed in the text. (Image modified from similar picture presented in \cite{Ketterle_Varenna}).
}
    \label{Excitations_Schematic}
  \end{center}
\end{figure}

The study of discrete collective excitations provides a very stringent precision test for the validity of theoretical approaches. In the early years of atomic BEC experiments, most of the measurements (density profiles, expansion dynamics), could be reasonably explained by simple mean field theory \cite{Stringari_Review,Burnett_GPE_3}, with the collective oscillation frequencies at low temperatures measured at JILA \cite{JILA_Exc_T_0} and MIT \cite{MIT_Exc_T_0} well-explained by the zero-temperature Bogoliubov equations (Eq.\ (\ref{BdG_Matrix_T0})) 
\cite{Burnett_GPE_3,Stringari_Modes}. 
However, subsequent measurements performed in the presence of a thermal cloud 
\cite{JILA_Exc_T,MIT_Exc_T} posed significant challenges to theorists.
In particular, an experiment at JILA \cite{JILA_Exc_T} which measured the temperature dependence of the oscillation frequencies of a slightly-elongated condensate 
was not in agreement with existing theoretical models, and the theoretical effort to understand the physics of this experiment was an important drive
for the establishment of dynamical finite temperature theories.

\begin{figure}[t]
 \begin{center}
      \resizebox{120mm}{!}{\includegraphics{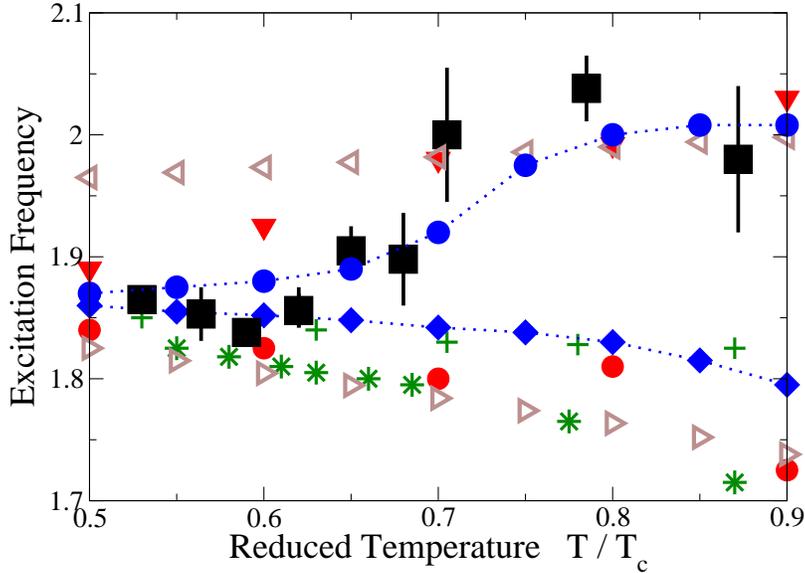}} 
    \caption{
(colour online)
Comparison of the predictions for the temperature dependence of the excitation frequency of the $m=0$ mode measured at JILA \cite{JILA_Exc_T} (shown by black squares) based on different theoretical models.
Plotted figure is restricted to the regime $T \ge 0.5T_c$, with the different theories at lower temperatures generally converging to the experimentally observed (and theoretically predicted) $T=0$ value of $\approx 1.86$.
Blue (filled): Number-conserving formalism with (circles) or without (diamonds) direct excitation of the thermal cloud from the probe \cite{Excitations_Burnett_2003}.
Green: Static thermal cloud theories with (`$*$', generalized HFB with $g_{t}=g(\bldr)$ \cite{Hutchinson_GHFB}) or without (`$+$', HFB-Popov \cite{HFB_Popov_Burnett}) inclusion of the anomalous average.
Red (filled): Predictions of ZNG approach for different excitation probe frequencies aimed at exciting primarily the condensate (circles, $\omega=1.75 \omega_z$) or the thermal cloud (inverted triangles, $\omega=2 \omega_z$) \cite{Excitations_ZNG}.
Brown: In-phase (left triangles, top) and out of phase (right triangles, bottom) mode of excitation between condensate and thermal cloud \cite{Excitations_Bijlsma_Stoof,Excitations_Al_Khawaja_Stoof}.
(Note that all data have been extracted {\em manually} from corresponding published works; each data point has an independent small error which should be on the order of the size of the symbol).
}
    \label{Excitation_Frequencies}
  \end{center}
\end{figure}

The JILA experiment, in which a time-dependent sinusoidal perturbation distorted the trap transversally, measured the quadrupole oscillation modes of the condensate shown schematic in Fig.\ \ref{Excitations_Schematic}; these are characterized by the projection $m$ of the angular momentum onto the weakly-confining trap axis as follows: (i) in the $m=0$ cylindrically-symmetric mode, axial and radial directions oscillate out of phase, and (ii) in the $m=2$ mode the cylindrical symmetry is broken, with the condensate radial oscillations out of phase with {\em each other}.
The oscillation frequencies of both modes were observed to decrease with increasing temperature, with the $m=0$ mode however additionally displaying an `anomalous' upward shift at temperatures $T > 0.7T_c$, with the observed frequencies differing by about $10-20\%$ from the corresponding zero-temperature results
\cite{JILA_Exc_T}.
The experimental data of the $m=0$ mode, along with predictions based on various theories applied to this particular problem are shown in Fig.\ \ref{Excitation_Frequencies}, and are explained in detail below.

The first finite temperature attempts to model this experiment were based on a static thermal cloud. Initially, the static HFB theory was applied in the so-called `Popov' limit of no anomalous correlations (Sec.\ \ref{HFB_Popov}); this was found to provide an excellent description for temperatures up to $0.6T_c$ (Fig.\ \ref{Excitation_Frequencies}, `$+$' symbols); remarkably, the HFB-Popov predictions were found to be only slightly different from corresponding $T=0$ results when one additionally accounted for the different number of condensate atoms at the corresponding temperatures \cite{HFB_Popov_Hutchinson,HFB_Popov_Burnett}.
The need to understand the system behaviour at higher temperatures stimulated the development of improved theoretical models:
motivated by analytical work performed also by one of us (NPP) \cite{Proukakis_Morgan}, Hutchinson, Dodd and Burnett carried out simulations which explicitly included the anomalous average and approximate many-body effects \cite{Hutchinson_GHFB} via the so-called generalized HFB models (Sec.\ \ref{many-body}). 
From the two models proposed, the one based on replacing the two-body effective interaction strength $g$ by
$g(\bldr)$ throughout Eqs.\ (\ref{BdG_GHFB})-(\ref{GPE_GHFB}) (i.e.\ both for condensate-condensate and condensate-thermal collisions) led to a correct prediction of the downward shift of the $m=2$ mode; nonetheless, such a theory incorrectly predicted a similar shift for the $m=0$ mode
(Fig.\ \ref{Excitation_Frequencies}, `$*$' symbols).
Similar results were obtained by Minguzzi and Tosi based on a time-dependent linearized Hartree-Fock
model \cite{Excitations_Minguzzi}, by Shi and Zheng in the context of a finite temperature variational method \cite{Excitations_Shi_Zheng} and by Reidl {\em et al.} \cite{Excitations_Dielectric} by means of the dielectric formalism. As already mentioned,
these models had the common feature of a static thermal cloud, which was quickly realized to be too restrictive for the particular experiments.
In order to overcome this limitation, Giorgini performed a linear response treatment of the coupled dynamics (Sec.\ \ref{Linear_Response}) but nonetheless found a similar downward shift for both modes \cite{Giorgini_2}.

A qualitative explanation of the observed `anomalous' temperature dependence of the $m=0$ mode was given by
Bijlsma, Al Khawaja and Stoof in the context of a finite temperature GPE in the Hartree-Fock approximation coupled to a collisionless Quantum Boltzmann equation
\cite{Excitations_Bijlsma_Stoof}, a study subsequently generalized to the collisional regime \cite{Excitations_Al_Khawaja_Stoof};
although these works were motivated by the full theory of Stoof (Sec.\ \ref{Stoof_Theory}), the analysis was performed here in the simplified semi-classical mean field limit which essentially amounts to the `ZNG' theory presented in Sec.\ \ref{ZNG}. However, instead of solving these equations self-consistently, their analysis was based on a variational approach. Their conclusion was that the reported $m=0$ excitation frequency actually corresponded to simultaneous measurements of two distinct excitations
(brown left/right triangles in Fig.\ \ref{Excitation_Frequencies})
which become coupled above a certain temperature; 
a similar interpretation was also given by Olshanii \cite{Excitations_Olshanii}.
Bijlsma, Al Khawaja and Stoof 
suggested that these modes corresponded to in-phase and out-of-phase oscillations of the condensate and the thermal cloud. This picture was later confirmed by one of us (BJ) by detailed numerical simulations based on the `ZNG' theory (Sec.\ \ref{ZNG}) which accounts for the full dynamical coupling between condensate and thermal cloud; this work identified the two distinct modes in which the condensate is excited as corresponding to a damped condensate oscillation in contact with a quasi-static heat bath, and an oscillation coupled to the motion of the thermal cloud \cite{Excitations_ZNG}. 
This work highlighted the extreme sensitivity of the observed system response to the precise details of the imposed perturbation (filled red inverted triangles/circles in Fig.\ \ref{Excitation_Frequencies}), such as the driving frequency which was not very accurately determined in the experiments.
The temperature dependence of the damping rates based this theory was also found to closely match the experimental observations.
The magnitude of the observed damping rates was also interpreted by means of a perturbative approach \cite{Excitations_Fedichev_1,Excitations_Fedichev_2} and field theoretic techniques \cite{Excitations_Liu}.

The full proof of the simultaneous excitation of the two modes in the experiment, was later presented by Morgan, Rusch, Hutchinson and Burnett 
\cite{Morgan_PRA_2004,Excitations_Burnett_2003,Morgan_PRA_2005}
by an extension of their earlier work \cite{Morgan_JPhysB,Excitations_Burnett_2000} which accounts for all second order processes within a number-conserving formalism (Sec.\ \ref{MorganGardiner}).
They explicitly showed (blue circles/diamonds joined with dotted lines in Fig.\ \ref{Excitation_Frequencies})) that the `anomalous' behaviour of the $m=0$ mode is a result of a temperature-dependent interplay in the {\em method of excitation} of the condensate: at low temperatures, this is dominated by direct excitation of the condensate from the probe, whereas at higher temperatures the condensate is excited predominantly by its coupling to the thermal cloud, which is itself excited from the probe. This detailed study led further to the identification of novel dynamical resonances which could be experimentally observed \cite{Morgan_RoyalSoc}.

In addition to the above comparison to experiments, another crucial test of the consistency of a kinetic theory, is provided by the so-called Kohn mode. Kohn's theorem for electrons states that the cyclotron frequency is not affected by interactions in a static magnetic field \cite{Kohn,Dobson}. By analogy, the dipole mode of the condensate, which corresponds to the centre of mass oscillation, should occur at the harmonic oscillator frequency, without being affected by 2-body interactions, since in a harmonic trap the centre of mass is explicitly decoupled from the internal degrees of freedom. It turns out that the Kohn mode is only accurately reproduced by theories which treat the thermal cloud dynamics on an equal footing as that of the condensate. These include the second order excitation theory of Morgan and Burnett \cite{Morgan_PRA_2004,Excitations_Burnett_2003,Morgan_PRA_2005}, the `self-consistent Gross-Pitaevskii-Boltzmann' or `ZNG' approach \cite{ZNG} and the fully dynamical stochastic approaches of Stoof \cite{Stoof_JLTP} and Gardiner-Zoller-Davis-Ballagh and co-workers \cite{QK_V,SGPE_II}.

Our preceeding discussion was restricted to a specific mode of excitation, as a large number of diverse theories was applied to its study, thus enabling a direct comparison between them. Numerous other excitation frequencies were studied in detail using diverse theories, and we briefly remark here that, where applied, the self-consistent Gross-Pitaevskii-Boltzmann, or semi-classical `ZNG' theory of Sec.\ \ref{ZNG} has yielded excellent agreement with such experiments \cite{Jackson_Zaremba_1,Jackson_Zaremba_4}.

\subsection{Condensate Growth}
\label{Growth}

The study of quantum phase transitions has been an active and challenging research topic in diverse areas of physics. In the context of ultracold Bose gases one is interested in understanding the process of condensate formation and the associated evolution of coherence, leading to the establishment of off-diagonal long-range order. Apart from being an interesting topic from a fundamental point of view, such studies pose the most stringent validity test of the various theoretical models of ultracold Bose gases.
It is thus appropriate to discuss here to what extent the presented models facilitate an understanding of the strongly non-equilibrium features observed experimentally. 
Early theoretical work on this issue (initiated even before the pioneering experiments in 1995) focused mainly on (i) a qualitative identification and understanding of the distinct stages of the process of condensate formation and equilibration starting from a thermal gas
\cite{Stoof_PRL,Svistunov_1,Svistunov_2,Svistunov_3,Svistunov_4,Stoof_GreenBook,Levich_Yakhot,Snoke_Wolfe,Stoof_1991,Stoof_1992,Kagan_GreenBook,Semikoz_Tkachev,Arias_Smerzi},
and, more specifically (ii) on the evaporative cooling process 
\cite{QBE_Luiten,QBE_Holland,Ketterle_vanDruten,Berg,Drummond_Corney,Yamashita}
which plays a dominant role in the experimental route towards condensation.
Clearly the topic of condesate formation is challenging from a theoretical perspective, and only the most advanced approaches presented in this Tutorial may be realistic candidates for succeeding in such a task. 
%

Pioneering work on this issue was carried out in 1998 at MIT, where in situ studies of condensate formation were performed \cite{MIT_Formation}. This experiment introduced the technique of `shock cooling', whereby a trapped thermal cloud is suddenly quenched below the transition point, and the evolution of the resulting `supersaturated cloud' into a condensate is studied. Theoretically, this initial condition can be modelled by truncating the Bose distribution at some temperature $T$, corresponding to an energy $k_BT$ above which all atoms are assumed to be efficiently removed by evaporative cooling.
The condensate was observed to grow slowly initially (spontaneous growth), before the bosonic enhancement set in, leading to exponential growth. This process was shown to be distinct from a simple relaxation mechanism, with the condensate eventually equilibrating when its chemical potential approached that of the thermal cloud \cite{MIT_Formation}. 

Two theoretical approaches have been applied to model this experiment, as discussed below,
although at this point we should also highlight the related work of Barci, Fraga, Gleiser and Ramos on the growth and equilibration of a homegeneous Bose condensate \cite{Ramos_1,Ramos_2,Ramos_3}.
 The first numerical studies of condensate growth came from Gardiner, Zoller, Ballagh and Davis \cite{QK_PRL_I}, before any specific condensate formation data were available. Their analysis was performed in the context of the `simple' quantum-optical rate equation (Eq.\ (\ref{Rate_BEC})), and appeared to be in qualitative agreement with existing experiments
\cite{BEC_Exp_1,BEC_Exp_2}.  Following the pioneering MIT controlled growth experiment \cite{MIT_Formation}, their analysis was improved to additionally include the dynamics of the occupations of low-lying trap levels via Eqs.\ (\ref{Rate_QP})-(\ref{Rate_QBE}); this provided the {\em first quantitative} results which yielded good overall agreement with the experiment \cite{QK_PRL_II}; furthermore, treatment of the scattering terms was found to speed up the {\em initial} period of growth. Their model was further improved by the inclusion of temporal depletion of the non-condensate \cite{QK_VII}, which effectively modified the simple rate equation by the replacement of the static thermal cloud chemical potential by a dynamical {\em effective} chemical potential determined self-consistently, leading to improved agreement \cite{QK_VII}.

%

\begin{figure}[t]
 \begin{center}
      \resizebox{120mm}{!}{\includegraphics{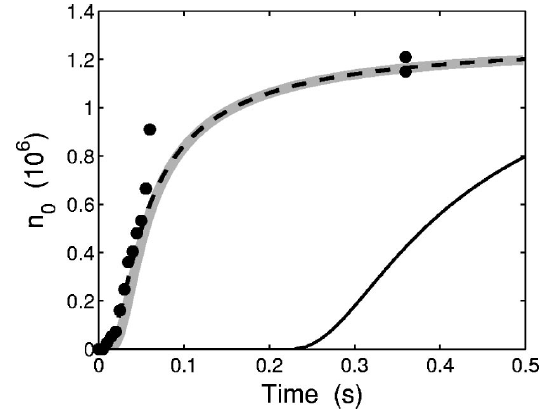}} 
 \caption{
Comparison of a set of experimental data from the MIT group (filled circles) \cite{MIT_Formation} and theory. Shown are apparent fits to the data generated by Bijlsma, Zaremba and Stoof in the context of the self-consistent Gross-Pitaevskii-Boltzmann or `ZNG' theory (dashed line), and by Davis, Gardiner and Ballagh based on the condensate master equation (solid grey); both these simulations - which are practically identical except for small differences in the initial growth characteristics due to the different initial condensate numbers assumed in each implementation - are performed for an initial thermal cloud of $40 \times 10^6$ $^{23}$Na atoms at an initial temperature of $T_i = 765$nK, and a relatively severe truncation of the energy distribution.
Equilibrium predictions resulting from these initial parameters appear to be in disagreement with the observed static (equilibrated) MIT data. Choosing instead the parameters so as to match the static experimental data and performing simulations with the same condensate master equation actually leads to very different growth dynamics, as shown by the solid black line.
(Reprinted figure with permission from M.J.\ Davis, C.W.\ Gardiner and R.J.\ Ballagh, Phys. Rev. A {\bf 62}, 063608 (2000). Copyright (2000) by the American Physical Society.)
}
\label{Growth_Figure}
  \end{center}
\end{figure}

Parallel work was undertaken by Bijlsma, Zaremba and Stoof in the context of the `self-consistent Gross-Pitaevskii-Boltzmann' or `ZNG' theory (which constitutes a special case of the theory put forward by Stoof in Sec.\ \ref{Stoof_Theory}); these authors arrived at similar conclusions \cite{Bijlsma_Zaremba_Stoof}, with their predictions found to be compatible with the results of Davis, Gardiner and Ballagh for approximately the same initial parameters \cite{QK_VII}.
Before comparing the predictions of these two theories to the experiments, it is worth remarking that 
although the semi-classical mean-field model cannot describe the {\em onset} of condensation, it can nonetheless accurately describe the subsequent dynamics; thus the semi-classical theory can still provide good agreement with experiments provided the condensate is initially artificially seeded with a small number of atoms (typically determined by the number of atoms in the lowest harmonic oscillator state at the temperature of the truncated Bose distribution \cite{Bijlsma_Zaremba_Stoof}).

The predictions of these theories are compared against each other and a particular set of experimental data from the MIT group (corresponding to Fig.\ 4 in \cite{MIT_Formation}) in Fig.\ \ref{Growth_Figure}. Although the agreement between these two numerical implementations
(denoted respectively by solid grey and black dashed lines)
was found to be very good \cite{QK_VII}, and even {\em appears} to reproduce the experimental data (dots) rather accurately, we note that the displayed simulations are not actually for parameters consistent with the experimental data quoted by the MIT group (or inferred from their data). 
Here, it should be noted that, in theoretically analyzing the experiment, it was not always clear what input parameters the model should use in order to correspond to the experimental conditions, as not all such parameters had been measured in (or extracted correctly from \cite{QK_VI}) the experiment. This creates certain ambiguity in performing detailed comparisons. While most of the experimentally obtained curves seemed to be generated by ab initio application of the theory (using `admissible' experimental parameters), there was at least one `puzzling' case (shown in the figure) where the theoretical model could not reproduce the observed behaviour for parameters consistent with the experimental values \cite{QK_VII}. In particular,
a simulation performed by Davis, Gardiner and Ballagh for parameters extracted from the MIT static experimental data (after equilibration) actually generated the solid black curve (instead of the solid grey one) - in clear disagreement to the measured data.
Despite the good qualitative and even in some cases quantitative agreement, the fact that not all observed growth curves could be reproduced {\em ab initio} from the experimental parameters, required further study, such as removing the condition of ergodicity (as undertaken by one of us - BJ - in collaboration with Zaremba) \cite{Jackson_LasPhys}, incorporating the quasiparticle nature of the spectrum of low-lying modes, or allowing for a non-adiabatic condensate growth exhibiting shape oscillations.


The next systematic experimental study of condensate growth was performed in 2002 by K\"{o}hl {\em et al.} for a $^{87}$Rb condensate \cite{QK_PRL_III}. This study was different from the MIT one, in that it was performed under {\em continuous} evaporative cooling into the quantum degenerate regime, thus enabling (for suitable parameters) to observe the formation process in `slow motion'. In this work, {\em all} relevant parameters required for an ab initio modelling of the experiment were measured. In the case of strong cooling (as in the MIT experiment), this led to remarkable agreement with the theory of Davis, Gardiner and Ballagh with {\em no free parameters} (once the modification of the trap potential due to gravity was taken into account \cite{QK_JPhysB}). However, K\"{o}hl {\em et al.} also studied the opposite regime of {\em slow cooling}, in which the system spends a longer time around the transition region, and found a `two-stage' growth \cite{QK_PRL_III}: the evolution featured a slow initial growth which could {\em not} be explained by the above model, with this regime followed at later times by the usual exponential growth to equilibrium. As these features occured very close to $T_c$, it was suggested that the slowing down of the initial growth rate could arise as a result of the appearance of a quasi-condensate, which is thought to preempt condensate formation even in three-dimensional systems \cite{Svistunov_1,Svistunov_2,Svistunov_4,Kagan_GreenBook}. Subsequent related experiments performed at Amsterdam \cite{Growth_Amsterdam}, Orsay \cite{Growth_Orsay} and 
Zurich \cite{Growth_Zurich} provided contradictory indications regarding the appearance and role of the quasi-condensate in such systems, which is believed to be associated with the appearance of condensate shape oscillations.

Such shape oscillations were indeed observed, and strong phase fluctuations measured in an experiment studying the formation of condensation into nonequilibrium states at Amsterdam, which highlighted the role of {\em local} thermalization \cite{Growth_Amsterdam}. The Orsay experiment measured the evolution of both population and coherence in a highly elongated trap. While they found evidence of shape oscillations, there was no indication of a two-stage growth despite being deep within the quasi-condensation regime \cite{Growth_Orsay}.
In this limit, theoretical models which do {\em not} fully account for phase fluctuations cannot a priori be assumed to be valid. Nonetheless, (perhaps somewhat unexpectedly) the model of Eqs.\ (\ref{Rate_BEC})-(\ref{Rate_QBE}) led to good agreement with experiments, up to an arbitrary delay time in the onset of condensate growth \cite{Growth_Orsay}; while such a delay time was introduced in an ad hoc manner in order to match the observed growth curves, it is still unclear whether this delay is related to phase fluctuations. Moreover, recent work in the real-time observation of the formation of long-range order performed in Esslinger's group found no evidence of a quasi-condensate stage \cite{Growth_Zurich}.
Thus, there appears to be some controversy regarding the appearance and effect of shape oscillations and phase fluctuations in the evolution of a quenched strongly-non-equilibrium Bose gas into an equilibrium condensate. This and related issues require more advanced studies and may be resolved by simulations based on the {\em stochastic} techniques of Secs.\ \ref{Stoof_Theory}-\ref{Gardiner_Zoller_Theory}. 

Simulations of the stochastic GPE of Stoof (Eq.\ (\ref{SGPE_Eq})) have in fact been performed in collaboration with one of us (NPP) in the context of condensate growth in a dimple microtrap \cite{Proukakis_AtomChip} (and in the related study of atom laser dynamics \cite{SGPE_AtomLaser}); in such experiments, the onset of condensation is induced by an entirely different mechanism, based on adiabatic local phase-space compression \cite{Compression}.
The first such experiment was performed at MIT \cite{MIT_Dimple}: The harmonic confinement of a thermal cloud at equilibrium was perturbed by the addition of a narrow gaussian dimple trap at the centre of the cloud, yielding a maximum increase in phase space density by a factor of 50. Condensate formation in the limit of slow trap addition was studied by means of non-destructive imaging, revealing condensate fractions of up to 20\% for different dimple depths. Importantly, application of a sinusoidal modulation of the trap depth led to a controlled and reversible crossing through the phase transition, with the condensate formation found to lag the trap modulation by about $70$ ms \cite{MIT_Dimple}.

The very first application of the stochastic GPE of  Eq.\ (\ref{SGPE_Eq}) to ultracold gases was in fact undertaken by Bijlsma and Stoof \cite{Stoof_Langevin} in a purely one-dimensional context in an attempt to analyze this experiment; this yielded good qualitative agreement with the experimental results with respect to the magnitude of the previously mentioned lagging time. However, a detailed quantitative comparison could not be performed, as the experiment was undertaken in a fully-three-dimensional regime, whereas the simulations were limited to the one-dimensional regime. 
It is important to remark here that attempts to model the dynamics by a semi-classical theory consisting of a finite temperature GPE dynamically coupled to a reservoir of thermal atoms, i.e.\ the stochastic GPE of Eq.\ (\ref{SGPE_Eq}) without any noise, failed to describe the experiment accurately; in particular, such simulations predicted cycles of successively decreasing numbers of condensate atoms, with the rate of decrease of the condensate atom number between such cycles depending on the initial conditions. The above scenario highlights one characteristic example where stochastic theories are {\em required} in order to accurately describe the observed behaviour.

In the same experimental publication, Stamper-Kurn {\em et al.}\ \cite{MIT_Dimple} suggested studying non-equilibrium (quasi)condensate growth induced by the sudden addition of a dimple trap. Preliminary experimental work along this line was undertaken in the group of      J\"{o}rg Schmiedmayer by the addition of a deep dimple microtrap on the weakly-confining axis of an atomic gas trapped close to an atom chip \cite{Schmiedmayer_Personal}. Theoretical modelling of this scenario was performed by one of us (NPP) in the context of the stochastic GPE (Eq.\ (\ref{SGPE_Eq})), focusing on the case of a weakly-interacting quasi-one-dimensional Bose gas \cite{Proukakis_AtomChip}. This led to the identification of a range of competing dynamical phenomena (shock wave propagation, direct quasi-condensate growth) which could be detected in future experiments.

Other characteristic dynamical examples where beyond mean field theories are required include (but are not restricted to) (i) the spontaneous process of vortex formation (see Fig.\ \ref{Stochastic_Images}) studied recently in more detail in \cite{Davis_SGPE_New}, and the related issue of the dynamics of the Berezinskii-Kosterlitz-Thouless (BKT) transition \cite{Critical_Book} in two dimensions to a state which does not possess long-range order but is instead associated with the unbinding of vortex pairs \cite{PGPE_BKT_PRL}.


While it is believed that the stochastic approaches, along with the appropriate time-dependent treatment of the thermal cloud should successfully address these issues, this 
presumption will need to stand the test of time, particularly as experimentalists are continuously exploring novel interesting phenomena which generally provide more stringent tests to the existing theories.


\section{Conclusions and Outlook}
\label{conclusions}

This Tutorial has presented a description of the most common theoretical methods available for describing weakly-interacting ultracold Bose gases at finite temperatures, and some concluding remarks are needed here.

Before doing so however, we feel compelled to note that, although there is consensus on a limited number of points (e.g.\ the evident importance of the thermal cloud, or the need to account for phase fluctuations in low dimensions), {\em there is no universally accepted `optimal' theory for the description of such systems}, with researchers coming from different communities typically taking different `stands' on this topic.
This makes such a concluding discussion hard and even possibly somewhat `controversial'.
In our preceeding presentation we have gone to great length to ensure objectivity of discussion and avoid introducing any personal bias into this Tutorial, and we hope our conclusions reflect the same impartiality.
To aid the less experienced reader, we have attempted here a rough classification in terms of the Table shown in Fig.\ \ref{Table_OptimalChoice}. Clearly such a presentation, while indicative, should not be taken too literally, as it is impossible to summarize an entire theoretical development in a single sentence, let alone classify each theory by `ticking relevant boxes'.
With these points in mind, we proceed with some concluding remarks, all of which relate to dilute weakly-interacting gases with $na^3 \ll1$ and $a \ll \lambda_{dB}$, such that only elastic binary collisions are relevant for describing the system properties.

\subsection{Classification of Theoretical Approaches: Role of Symmetry-breaking}

The first point to make is that, loosely speaking, one can classify existing theoretical formalisms into three different `classes' of approaches, based on certain common conceptual notions shared between them. 

The main such notion is the idea of symmetry-breaking: In both high-energy physics and condensed matter physics it is convenient to assume that the operator describing a relevant quantum field reduces, below the critical temperature for some phase transition, to an appropriate macroscopic complex (wave)function. 
This implies that the macroscopic variable no longer exhibits the symmetry (or all symmetries) of the original system hamiltonian, which (for the systems we are interested in) is invariant under a change of phase of the operator; in this case one says that the symmetry has been (spontaneously) broken. While views are generally divided as to whether this is a `physical reality' or simply a convenient `mathematical concept', in some systems this makes no difference as there is no definite (fixed) number of particles - that is certainly the case in high energy physics, and the same is essentially true for conventional (macroscopic) condensed matter systems. In such cases, the concept of symmetry breaking has proven extremely useful in explaining the system's behaviour. The notion of broken symmetry however becomes somewhat more `problematic' in the case of a small (finite) system of atoms, as is relevant for experiments with ultracold trapped gases, which typically consist of no more than a few (tens of) millions of atoms; the reason is that if one were to accept the view that at any time the system consists of a {\em definite} number of atoms, then the mean value of the Bose field operator $\hat{\Psi}(\bldr,t)$ should be {\em identically} zero.

The simplest - what one might call generic - approaches for ultracold gases rely on symmetry-breaking, i.e.\ on the assumption that the Bose field operator $\hat{\Psi}(\bldr,t)$ can be split into a mean-field condensate contribution $\wfn(\bldr,t)$ (denoting the condensate wavefunction) and an operator describing the (quantum and thermal) fluctuations about this mean field.
Approaches relying on this separation are known as `mean-field' approaches, whereas treatments which explicitly maintain the operator nature of the condensate contribution (i.e.\ $\wfn(\bldr,t) \rightarrow \hat{\wfn}(\bldr,t)$) automatically conserve the total atom number, and are therefore known as `number-conserving'. In a large coherent three-dimensional system far from the regime of critical fluctuations such a distinction becomes an `academic issue', i.e.\ one is essentially `free to choose' the `class' which one is most comfortable with, with predictions between the two classes of approaches believed to lead to negligible differences (which may not even be observable experimentally). The same is however {\em not} true in {\em all} experimentally-relevant regimes.

\subsection{Mean Field Theories}

Mean field theories are typically based on identifying a set of slowly-varying mean-field quantities and formulating a closed system of equations to be solved self-consistently either in a static or a dynamical manner. However, such a closed system is not exact, and to arrive at such a formulation one has to impose certain decorrelation approximations, i.e.\ assume that averages of products of operators denoting the fluctuations about this mean field can be split into products of `lower-order' averages, each containing a smaller number of operators. Depending on the `crudeness' of the imposed approximations and the related choice of the `mean field variables' of the system, one can generate all different mean field theories, as discussed in Secs.\ \ref{zeroT}-\ref{finitet:dynamic}, with such treatments essentially formulated in first or second order perturbation theory. 
First order perturbation theory corresponds to interactions of static self-consistent mean fields, whereas formulations to second order in the effective interaction strength also allow for dynamical (particle) exchange between such generalized mean fields.
Note that second order perturbation theory works well here when combined with a pseudopotential approximation to the upgraded effective interaction (T-matrix), as it can essentially be interpreted as an expansion in terms of the difference of the actual effective interaction experienced by two atoms in the medium of condensed and thermal atoms from the corresponding one (two-body or many-body T-matrix) used within the particular theory under consideration.

Mean field theories can be typically grouped into different categories, depending on the following three `classification issues', with each of these corresponding to different physical content:


(i) {\em Nature of Excitation Spectrum:} Are the system excitations treated as single-particle ones (modified by mean fields to the level of approximation), or does one need to consider `particle mixing' into quasiparticles, which exhibit a different (phononic) excitation spectrum at low momenta? Inclusion of quasiparticle physics is a straightforward, yet lengthy process, and is lacking in most current implementations; nonetheless, this does not necessarily place huge restrictions on the validity of these theories for dilute weakly-interacting gases, unless they are applied to systems at extremely low temperatures 
In the finite temperature theories of interest, this choice respectively makes a distinction into `Hartree-Fock'- and `Hartree-Fock-Bogoliubov'-type of approaches.

(ii) {\em Anomalous Average and Many-Body Effects:} At finite temperatures, one should at least describe the system by a mean field for the condensate and a corresponding one for the non-condensate, or thermal cloud. If one additionally chooses to include the so-called (pair) anomalous average as a `fundamental' system variable, the theory will then incorporate additional physical content, in the sense that many-body effects, which lead to a modification of the effective scattering length due to the presence of the medium, are largely accounted for. Note however that extreme care is needed when doing so, as not all resulting theories treat atom-atom interactions consistently, and may even lead to worse predictions than `more basic' theories for certain quantities of interest.

(iii) {\em Particle-exchanging Collisions:} Most importantly, does the theory only include mean-field coupling between the condensate and the thermal cloud (formulation to lowest order in the effective interaction), or does it additionally allow for the dynamical exchange of particles between these two sub-systems (second order formulation)? In the former case of purely mean field coupling, the theory is limited essentially to equilibrium properties, e.g.\ density profiles and condensate fraction: there is a diverse range of theories to choose from here, depending on the issues raised earlier, with possible choices (starting from the most basic) being Hartree-Fock, Hartree-Fock-Bogoliubov(Popov), or generalized Hartree-Fock-Bogoliubov. If one is only interested in bulk quantities or qualitative features, then Hartree-Fock is an excellent starting point - although a detailed comparison to experiments will typically require (at least) use of the other more generalized mean field theories.
In the other extreme, i.e.\ when one is interested in quantities which depend critically on the dynamics of either (or both) of the sub-systems, one should necessarily include atom transfer collisions between the condensate and the thermal atoms, and, in general, also redistribution collisions within the thermal cloud. In this case, one should best resort to a description in terms of a dissipative Gross-Pitaevskii equation which is self-consistently coupled to a `reservoir' of high-lying modes described by the Quantum Boltzmann equation; the latter, is an extension of the usual Boltzmann equation for the distribution function of a gas when it is dynamically coupled to a condensate, and is somtimes referred to as the `ZNG' theory.
In addition to constituting a suitable choice as long as one is not too close to the regime of critical fluctuations, such a theoretical description also enables one to `switch off' at will either, or both, of these collisional contributions to investigate their relative role. Such investigations have been undertaken in various studies (also by the authors - mainly BJ), and in general the mechanism of transfer of atoms between the condensate and the thermal cloud is crucial in order to accurately predict experiments looking at dynamical effects, such as damping of elementary or macroscopic excitations. 

In order to derive all mean field theories mentioned above, one essentially separates the full system hamiltonian accounting for elastic binary interactions into contributions depending on the number of non-condensate operators (from zero to four) appearing within each contribution. Terms up to quadratic order only accurately take account of the condensate mean field and small excitations on top of this, and are generally limited to $T\approx0$; treating the remaining higher order contributions purely within `suitable' mean field approximations leads to an approximate inclusion of thermal effects, but only within the context of static variables, i.e.\ self-consistent mean field coupling. Inclusion of particle-exchange collisions additionally requires the perturbative treatment of the difference between the exact higher-order contributions to the hamiltonian from their corresponding approximate expressions based on the selected mean field approximations.

However, it is important to note that no mean field theory is able to predict the behaviour of the system at the critical point, or to simulate condensate growth from a purely thermal initial gas (although a self-consistent Gross-Pitaevskii-Boltzmann or `ZNG' model nonetheless works rather accurately as soon as there is a small condensate `seed' present in the initial conditions). In this case, one requires consideration of beyond-mean-field, or number-conserving approaches, in which the condensate contribution is described by an appropriate operator, rather than by a corresponding mean field. There are essentially two `types' of such approaches, as highlighted below:

\subsection{Number-conserving Perturbative Treatments}

The first such approach is very similar in {\em procedural} development to the treatment of individual contributions to the main system hamiltonian mentioned above, although there are important conceptual differences. The main difference here is that the excitations are by construction orthogonal to the condensate (not strictly true in mean field approaches), and the non-condensate operator is defined in a slightly different manner which explicitly conserves the total number of atoms. Apart from not requiring the use of symmetry-breaking (i.e.\ the condensate contribution is still an {\em operator}), at the expense of reasonable algebraic complexity, such treatments are well-suited for small trapped gases where such effects and issues associated with precise number-conservation may become important. The power of such approaches has been demonstrated in distinguishing between direct and indirect excitation mechanisms for condensate oscillations in a particular experiment; however, such a formalism has not yet been widely applied and 
(in light of the significant extra effort involved in formulating/simulating the theory) 
its benefit over other approaches remains unclear; this issue will presumably be resolved when such approaches are generalized to the full dynamical inclusion of the thermal cloud within the context of a fully self-consistent second order perturbation theory which is, however, presently lacking.




\subsection{Alternative Number-conserving Approaches: Classical Field and Stochastic}

This brings us to the third class of approaches, which incorporates classical-field and stochastic methods. 
A key difference of these to all earlier approaches is that such treatments include fluctuations (in the form of added numerical noise for the stochastic approaches), and the results for various parameters should be extracted via suitable averaging - either temporally, or over a number of different numerical realizations.
In addition, these theories describe the low-lying modes of the system in a `unified' manner, so the identification of the `condensate', or `quasi-condensate' in these theories is slightly different to the perturbative mean field, or number-conserving approaches discussed previously - an important point to bear in mind when attempting direct comparisons between different theoretical formulations.

There are three main approaches\footnote{Note that our discussion omits two other powerful theories (Positive-P Representation and 2-Path-Irreducible Effective Action) which are currently gaining ground against some of the above approaches, and whose main added benefit is that they can handle strongly-correlated regimes that this Tutorial was not concerned with.} we wish to highlight here:

(i) {\em The Classical Field Method:}  
In this approach, as suggested by its name,
the entire evolution of the system is described classically. Such a treatment should be valid as long as thermal effects become so large, that they essentially `wipe out' all quantum corrections. The classical field describing the system is propagated via the Gross-Pitaevskii Equation (GPE), with the possible addition of a projector to ensure that the bounds of the classical region are correctly specified throughout the simulations - in which case the theory is referred to as the Projected Gross-Pitaevskii Equation (or PGPE). Starting from appropriately selected (but largely irrelevant) initial non-equilibrium conditions, one studies the system evolution to equilibrium; this approach is non-perturbative, and has even been used by some researchers at/near the critical region.

(ii) {\em The Truncated Wigner Approximation:} This approach additionally includes quantum fluctuations, but these are only introduced in the initial conditions of the simulation. 
The subsequent evolution is again based on the (P)GPE, which here arises however only as an approximate equation for the full quantum evolution of the Wigner quasiprobability distribution function.
As a result, such evolution actually leads to spurious damping, unless the study is limited to relatively short times or low temperatures. Contrary to the classical field methods, this approach is good for studying circumstances where quantum effects dominate the behaviour of the system, e.g.\ in optical lattices.

(iii) {\em The Stochastic Gross-Pitaevskii Equation (SGPE):} 
This approach combines all the `essential' elements of the successful previous theories (self-consistent Gross-Pitaevskii-Boltzmann or `ZNG', classical field and Truncated Wigner) into a `single-package', whose full potential has yet to be numerically unleashed.
The SGPE resembles the ordinary GPE, but with the addition of a dissipative term (as appropriate for perturbative mean field theories) and a dynamical noise term. 
In brief, the SGPE describes the dynamical coupling of the low-lying modes of a trapped gas to the `reservoir' of higher-lying (thermal) modes, including both coherent and incoherent mechanisms.
In principle, the higher-lying modes also evolve according to the quantum Boltzmann equation which further accounts for particle redistribution within the thermal cloud, as well as particle-exchanging collisions.

\subsubsection{Comparison between above Approaches:}

The stochastic formulation
has the formal advantage over the various mean field theories that it is an explicitly number-conserving theory, which additionally includes a stochastic term.
This theory essentially reduces to a description in terms of a dissipative Gross-Pitaevskii equation, in which the condensate energy is treated as a differential operator, coupled to a Quantum Boltzmann equation.
Removing the operator nature of this energy, and imposing a `symmetry-breaking' approximation leads to a semi-classical theory which Zaremba, Nikuni and Griffin have termed the `ZNG' theory.
Note that the latter theory introduces the additional simplification that interaction energies and chemical potentials are not treated fully self-consistently, but only to first order in the effective interaction strength.
However, unlike that latter theory, instead of dealing with a coherent condensate, the SGPE provides a different physical interpretation, with $\Phi(\bldr,t)$ referring to the full matter-wave field; this additionally includes both thermal and quantum fluctuations, and thus encompasses both the classical field theory and the Truncated Wigner method. In fact, the early formulation of the PGPE was as the limiting case of an equation for the low-lying modes of the system when the coupling to a reservoir of higher-lying modes normally present is artificially removed. Inclusion of this coupling would essentially lead to the SGPE.

\begin{sidewaysfigure}
\centering \scalebox{0.7}
 {\includegraphics{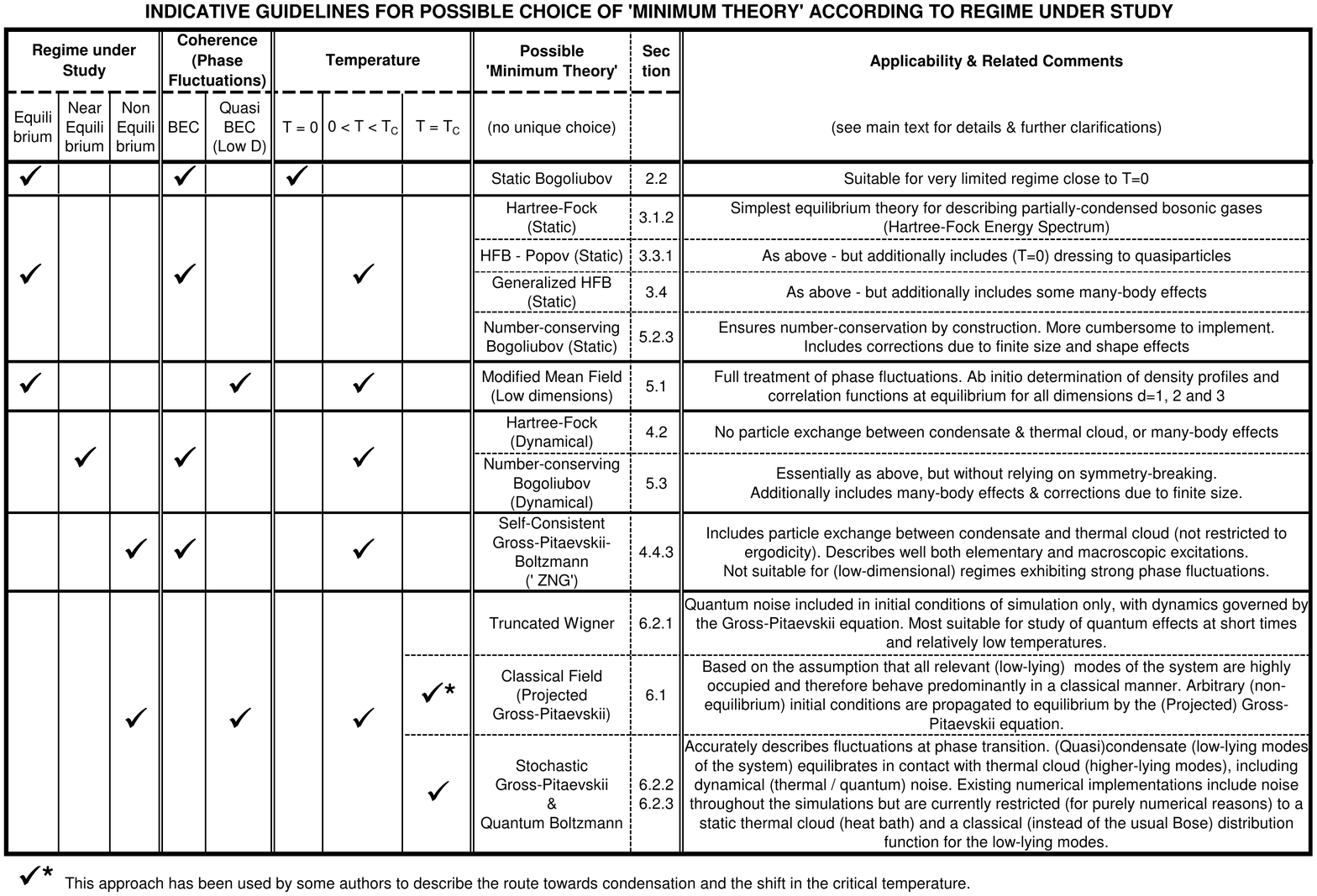}}
 \caption{
Indicative discussion of possible choices for a 
`minimum' theory to be used in the description of relevant experimental studies. Such a classification is suggestive and should not be taken too literally, as other factors may affect the choice (e.g. computational effort, particulars of each case, etc.). Also, there are more than one possible `minimal choices' for certain scenarios, and the choice for what constitutes a simpler theory is objective and may depend on the particular tools available to each researcher. This table is therefore only intended to accompany the detailed and summative discussions given in this Tutorial.
Naturally, improved theories can always be used to describe simpler phenomena (although that is not necessarily required unless improved accuracy is needed) - e.g.\ the Projected Gross-Pitaevskii, or the Stochastic Gross-Pitaevskii equation could be used to discuss finite temperature equilibrium properties of 3D Bose gases.
}
\label{Table_OptimalChoice}
\end{sidewaysfigure}

The role of the coupling of the low-lying modes of the system described by the SGPE is to ensure that
the system relaxes to the correct thermal equilibrium.
%
Due to its coupling to the reservoir of higher-lying modes, the SGPE
does not suffer from the low-temperature limitations of Truncated Wigner which is prone to spurious damping (heating) as a result of classical thermalization during the simulations.
Although by its mathematical construction the SGPE formalism, coupled to a Quantum Boltzmann Equation, is in our opinion the most advanced formalism to date, its full potential has not yet been explored.
In particular, in terms of its numerical implementation, most simulations to date have been performed 
(i) in the classical approximation, i.e.\ the thermal modes are not treated by the full Bose-Einstein distribution function, and quantum effects are discarded, and
(ii) in coupling to a {\em static} heat bath, rather than to a dynamical thermal cloud, 
%
It is anticipated that both of these difficulties will be gradually overcome in the future.


\subsection{The Quest for an `Optimal' Theory}

So, where does that leave us regarding a choice for an `optimal' theory?
We hope that the preceeding discussion has convinced the reader that the answer to this question
actually depends on the details of the particular problem under consideration. While the most advanced theories can also be used to study much simpler problems, clearly one normally seeks the simplest (and computationally fastest) approach that will describe each particular case; we have thus attempted an approximate - but neither unique, nor free from controversy - classification along those lines in Fig.\ \ref{Table_OptimalChoice}. In brief:
\begin{itemize}
\item If dynamics are not an issue, then the self-consistent Hartree-Fock theory is a very good starting point, with Hartree-Fock-Bogoliubov-Popov or generalized Hartree-Fock-Bogoliubov perhaps constituting a `safer bet'. However, if one is investigating low-dimensional systems at equilibrium, one should fully account for phase fluctuations by means of the appropriately modified mean field theory.
\item If one is interested in coupled condensate-thermal cloud dynamics (beyond the linear response limit), then one should in first instance resort to a `two-component' formulation in terms of a dissipative Gross-Pitaevskii equation, coupled to a Quantum Boltzmann equation - sometimes referred to as the `ZNG' theory.
\item If one is interested in studying regimes with large phase fluctuations, e.g.\ effectively low-dimensional geometries, spontaneous processes and related phenomena, or even the physics of the phase transition itself, then one should resort either to classical field or to stochastic methods.
An added feature of these approaches, which are computationally more demanding, is that they do not make an `unphysical' separation of the atomic gas into condensate and thermal cloud (which is nonetheless desired by many researchers for ease of visualization), but rather they produce images reminiscent of experimental density profiles. The task of extracting information from such theories is thus somewhat more involved than for other types of theories, as one typically needs to perform a similar data analysis as would be done for analyzing an experiment (e.g. bimodal fits, averages over different realizations).
\end{itemize}
In light of our preceeding analysis, our personal bias is to recommend a description based on the Stochastic Gross-Pitaevskii Equation for the predominantly coherent modes of the system, coupled to a Quantum Boltzmann Equation for the higher-lying modes of the system, a view generally shared by numerous - but certainly not all - researchers. 
Such a `complete theory' has not yet been numerically implemented, although all essential features have been individually tested. Here we note that 
(i) the Stochastic Gross-Pitaevskii equation has been numerically simulated in coupling to a static heat bath by Stoof and collaborators (including one of us - NPP) and by Davis-Gardiner and collaborators, with all such treatments currently limited to one or two dimensions;
(ii) moreover, a dissipative Gross-Pitaevskii equation, which is closely related to the stochastic Gross-Pitaevskii equation in the absence of the stochastic noise term, has been numerically solved (also by the present authors) in full coupling to a dynamical thermal cloud in three dimensions beyond the ergodic approximation. We believe that the implementation of such a theoretical framework constitutes the next significant development in the modelling of finite temperature trapped Bose gases.\\


The issue of non-equilibrium dynamics of weakly-interacting Bose gases remains a fascinating topic of research, with  a diverse range of theories covering different elements of relevance to current experiments. 
More challenging experiments are bound to be undertaken in the near future, particularly at/near the transition point, and the comparison of existing / improved theoretical models to such experiments  over the course of time will determine which (if any) of these theories will live up to their expectation and whether one day one of these theories may be `universally' recognized as the `standard' theory for such systems.

\ack
%
We are indebted to Charles Adams, Carlo Barenghi, Keith Burnett, Henk Stoof, Sandro Stringari and Eugene Zaremba for introducing us to this fascinating topic and for insightful discussions and collaborations during the past decade. 
We are also particularly grateful to Stuart Cockburn, Rembert Duine, Simon Gardiner and Allan Griffin for detailed comments on this manuscript. We also thank Stuart Cockburn and Matt Davis for providing us with some data/images which have been included in our presentation.

We also acknowledge discussions and/or collaboration with:
Usama Al Khawaja, Jens Andersen, Jan Arlt, Aidan Arnold, Rob Ballagh, Michiel Bijlsma, Blair Blakie, Joachim Brand, Stephen Choi, Charles Clark, Matt Davis, Dennis Dickerscheid, Jacob Dunningham, Mark Edwards, Sandy Fetter, Chris Foot, Dimitris Frantzeskakis, Mark Fromhold, Crispin Gardiner, Thomas Gasenzer, Stefano Giorgini, Masud Haque, Carsten Henkel, Ed Hinds, Ifan Hughes, David Hutchinson, Panos Kevrekidis, Karen Kheruntsyan, Wolf von Klitzing, Thorsten K\"{o}hler, Peter Lambropoulos, Mark Lee, Tony Leggett, Jim McCann, Anna Minguzzi, Klaus M{\o}lmer, Sam Morgan, Antonio Negretti, Tetsuro Nikuni, Patrik \"{O}hberg, Maxim Olshanii, Nick Parker, Lev Pitaevskii, Jocelyn Retter, Ana-Maria Rey, Janne Ruostekoski, Peter Ruprecht, Martin Rusch, Luis Santos, J\"{o}rg Schmiedmayer, Robin Scott, Gora Shlyapnikov, Marzena Szymanska, Masahito Ueda, Jochen Wachter, Reinhold Walser, Jamie Williams and Slava Yukalov.

Over the past decade, both authors have been actively involved in the development and implementation of numerous theories described in this Tutorial. We would therefore like to take this opportunity to acknowledge continuing finanical support from the following sources during this exciting period:  Alexander S. Onassis Public Benefit Foundation (Greece), the Max-Planck-Gessellschaft (Germany), the Ultraviolet Laser Facility (Greece), the Natural Sciences and Engineering Research Council of Canada (Canada), the Engineering and Physical Sciences Research Council (U.K.), the Nederlandse Organisatie voor Wetenschappelijk Onderzoek (Netherlands) and the Ministero dell'   Istruzione, dell' Universit{\'a} e della Ricerca (Italy), as well as the hospitality of the Lorentz Centre (Netherlands), Benasque Centre for Sciences (Spain), European Centre for Theoretical Physics (Italy) and Institut Henri Poincare (France).

\appendix

\section{The JILA Kinetic Theory}
\label{JILA_Theory}

The extended beyond-HFB second order perturbation theory discussed in Sec.\ \ref{Prouk_QK} was originally developed by Walser, Williams, Cooper and Holland in 1999 \cite{JILA_Kinetic_1,JILA_Kinetic_2}. Their treatment is based on the construction of a generalised kinetic theory for a `coarse-grained' many-particle density operator which depends only on a few carefully selected `master' variables; these were identified as the condensate mean field and gaussian fluctuations around it, i.e.\ normal and anomalous averages, which are precisely the HFB variables. Like the treatment presented in Sec.\ \ref{Prouk_QK}, their derivation relies on a separation of timescale argument, under the assumption that the duration of a typical collisional event is much smaller than the inverse collision rate; this enables the development of a systematic perturbative expansion. 
Below we briefly review this formalism, highlighting the fact that
all expressions are formulated in terms of a single-particle energy basis.
The notation presented here explicitly maintains spatial dependence and preserves geometrical transformation properties.

The Bose field operator $\hat{\Psi}(\bldr,t)$ is expanded in a complete single-particle basis $|1\rangle$ via 
$\fopart = \sum_{1} \langle r | 1 \rangle \hat{a}_1$.
The binary interaction hamiltonian can be expressed in symbolic compact notation as
\be
\hat{H} = \langle 1 | \hat{h}_0 - \mu | 2 \rangle \hat{\Psi}_1^\dag \hat{\Psi}_2
+ V_{1234}  \hat{\Psi}_1^\dag  \hat{\Psi}_2^\dag \hat{\Psi}_3 \hat{\Psi}_4\;,
\ee
where an implicit summation over repeated indices is assumed. Here
$\langle 1 | \hat{h}_0 - \mu | 2 \rangle  = \int d \bldr \langle 1 | \bldr \rangle [ \hat{h}_0(\bldr) - \mu ] \langle \bldr | 2 \rangle$, and
$V_{1234} = (g/2) \int d \bldr \langle 1 | \bldr \rangle  \langle 2 | \bldr \rangle  \langle \bldr | 3 \rangle  \langle \bldr | 4 \rangle  $. The latter expression has made use of the pseudopotential approximation, with resulting expressions being symmetrised, i.e.\ $V_{1234}=V_{1243}=V_{2134}=V_{2143}$.
We introduce the slowly-varying `master' variables describing the system as follows
\cite{JILA_Kinetic_1,JILA_Kinetic_2}:

\noindent The {\em condensate} is defined by 
\be
| \phi \rangle = \phi_1 | 1 \rangle = \langle \hat{\Psi} \rangle
\hspace{1.0cm}
\langle \hat{\Psi}^\dag \rangle = \langle \phi | = \phi_1^* \langle 1 |
\ee

\noindent The {\em normal average} is defined by
\be
f = \langle \hat{\Psi}^\dag \hat{\Psi} \rangle = f^c + \tilde{f} = | \phi \rangle \langle \phi |
+\tilde{f}_{12} |1 \rangle \langle 2 |
\ee
where $f^c$ the coherent and $\tilde{f}$ the incoherent contribution (the latter quantity is the analogue of $\rho$ defined by Eq.\ (\ref{rhokappa}) and used in Sec.\ \ref{Prouk_QK}). This notation can be understood by expressing the normal average in position representation via
\bea
f(\bldr_1,\bldr_2) &=& \langle \bldr_1 | f | \bldr_2 \rangle 
= \langle \hat{\Psi}^\dag(\bldr_2) \hat{\Psi}(\bldr_1) \rangle
= \phi^*(\bldr_2) \phi(\bldr_1) + \langle \hat{\delta}^\dag(\bldr_2) \hat{\delta}(\bldr_1) \rangle
\nonumber \\
&=& \langle \bldr_1 | f^c | \bldr_2 \rangle + \langle \bldr_1 | \tilde{f} | \bldr_2 \rangle
= f^c(\bldr_1,\bldr_2) + \tilde{f}(\bldr_1, \bldr_2) 
\;.
\eea
Similarly, the {\em anomalous average} corresponding to pair correlations is defined by
\be
m = \langle \hat{\Psi} \hat{\Psi} \rangle = m^c + \tilde{m} = | \phi \rangle | \phi \rangle
+\tilde{m}_{12} |1 \rangle | 2 \rangle
\ee
where $m^c$ the coherent and $\tilde{m}$ the incoherent contribution (analogue of $\kappa$ of Eq.\ (\ref{rhokappa})). In position representation, we find
\bea
m(\bldr_1,\bldr_2) &=& \langle \bldr_1 | \langle \bldr_2 | m |  
= \langle \hat{\Psi}(\bldr_2) \hat{\Psi}(\bldr_1) \rangle 
= \phi(\bldr_2) \phi(\bldr_1) + \langle \hat{\delta}(\bldr_2) \hat{\delta}(\bldr_1) \rangle
\nonumber \\
&=& \langle \bldr_1 | \langle \bldr_2 | m^c + \langle \bldr_1 | \langle \bldr_2 | \tilde{m}  
= m^c(\bldr_1,\bldr_2) + \tilde{m}(\bldr_1, \bldr_2) 
\;.
\eea
As in Sec.\ \ref{tdep_HFB}, we express the coherent ($| \phi \rangle$) and fluctuation parts of the system ($\tilde{f}$, $\tilde{m}$) in generalised matrix notation via (c.f.\ expressions of $R_{C}$ and $R_{NC}$ in Eq.\ (\ref{RC_RNC}))
\begin{eqnarray}
\chi = \left( 
\begin{array}{c} 
\phi \\ \phi^*
\end{array}
\right)\;,
\hspace{1.5cm}
G^{<} = \left( 
\begin{array}{cc} 
\tilde{f} & \tilde{m} \\ \tilde{m}^{*} & \left( \tilde{f}^* +  1 \right)
\end{array}
\right)\;.
\end{eqnarray}
The corresponding generalized hamiltonians for the condensate $\Pi$ (c.f.\ $H_c$) and the non-condensate $\Sigma$ (c.f.\ $H_{NC}$ of Eq.\ (\ref{HC_HNC})) are defined by
\begin{eqnarray}
\Pi = \left( 
\begin{array}{cc} 
\Pi_N & \Pi_A \\ -\Pi_A^* & -\Pi_N ^{*}
\end{array}
\right),
%
\hspace{0.3cm}
%
\Sigma = \left( 
\begin{array}{cc} 
\Sigma_N & \Sigma_A \\ -\Sigma_A^{*} & -\Sigma_N^{*}
\end{array}
\right)\;,
\end{eqnarray}
where we have introduced
\bea
\Pi_N = \hat{h}_0 - \mu + U_{f^c} + 2 U_{\tilde{f}} 
\hspace{1.0cm}
\Pi_A = U_{\tilde{m}}
\nonumber \\
\Sigma_N = \hat{h}_0 - \mu + 2 U_{f^c} + 2 U_{\tilde{f}} 
\hspace{1.0cm}
\Sigma_A = U_{m}
\eea
with $N$ and $A$ labelling normal and anomalous averages.
The above equations make use of
the shorthand notation (c.f.\ expressions for $h$ and $\Delta$ of Eqs.\ (\ref{h}) and (\ref{delta}))
\bea
U_f = 2 V_{12'3'4'} f_{3'2'} |1 \rangle \langle 4'| = 2 V_{12'3'4'} \left( f_{3'2'}^c+\tilde{f}_{3'2'} \right) |1 \rangle \langle 4'|
\nonumber \\
U_m = 2 V_{12'3'4'} m_{3'4'} |1 \rangle |2' \rangle = 2 V_{12'3'4'} \left( m_{3'4'}^c+\tilde{m}_{3'4'} \right) |1 \rangle |2' \rangle|
\;.
\eea

The full second order dynamical theory which includes anomalous averages can thus be cast in the form (setting $\hbar=1$)
\bea
\frac{d}{dt} \chi = -i \Pi \chi + \left( Y^{<} - Y^{>} \right) \chi \;, \nonumber \\
\frac{d}{dt} G^{<} = -i \Sigma G^{<} + \left( \Gamma^{<} G^{>} - \Gamma^{>} G^{<} \right) +{\rm h.c.} 
\;.
\label{JILAHFB}
\eea
(Note the slight differences in notation between \cite{JILA_Kinetic_2} and \cite{JILA_Kinetic_3,Wachter_Masters}.)
In their simplest {\em first order} limit, where only the first term (and its conjugate) is maintained, these give rise to the time-dependent HFB equations of Eq.\ (\ref{HFB_Temporal}).
The remaining contributions appearing in Eq.\ (\ref{JILAHFB}) include {\em all} second order collisional integrals (expressed in terms of single-particle operators). In particular,
we have defined
\begin{eqnarray}
Y^{<} = \left( 
\begin{array}{cc} 
Y_N^< & Y_A^< \\ -Y_A^{>*} & -Y_N^{>*}
\end{array}
\right),
%
\hspace{0.3cm}
%
\Gamma^{<} = \left( 
\begin{array}{cc} 
\Gamma_N^{<} & \Gamma_A^{<} \\ -\Gamma_A^{>*} & -\Gamma_N^{>*}
\end{array}
\right)
\end{eqnarray}
with the time-reversed matrices generated from these via
$Y^{>} = - \sigma_1 Y^{<*} \sigma_1$ and
$\Gamma^{>} = - \sigma_1 \Gamma^{<*} \sigma_1$, where $\sigma_1$ the usual Pauli matrix exchanging the two components of the vectors, defined by
$\sigma_1 = \left( \begin{array}{cc} 
0 & 1 \\ 1 & 0 \end{array} \right)$.

Considering the {\em condensate} evolution first, the forward and backward transition rates ($Y_N^<$,
$Y_A^<$, $Y_N^>$ and $Y_A^>$) describe the scattering of non-condensate particles into and out of the condensate; they are given by the following scattering processes $\Gamma$:
\be
Y_N^< = \Gamma_{\tilde{f} \tilde{f} (\tilde{f}+1)}+2 \Gamma_{\tilde{f} \tilde{m} \tilde{m}^*}
\hspace{1.0cm}
Y_A^< = \Gamma_{\tilde{m} \tilde{m} \tilde{m}^*}+2 \Gamma_{\tilde{f} \tilde{m}(\tilde{f}+1)}
\ee
with the `out' rates denoted by the superscript `$>$' defined by the corresponding expressions, obtained directly from the above by replacing $(\tilde{f}+1)$ by $\tilde{f}$ (and vice versa) in all expressions. This highlights the fact that only {\em normal} fluctuations $\tilde{f}$ become bosonically enhanced, with the condensate and anomalous fields not exhibiting such an effect.

The rates entering the corresponding expressions for the non-condensate are
\bea
\Gamma_N^< &=& 
\Gamma_{f \tilde{f} (\tilde{f}+1)}+\Gamma_{\tilde{f} f^c (\tilde{f}+1)}
+\Gamma_{\tilde{f} \tilde{f} f^c} \nonumber \\
&& + 2 \left[ \Gamma_{ f \tilde{m} \tilde{m}^*} + \Gamma_{ \tilde{f} m^c \tilde{m}^*}  
+ \Gamma_{ \tilde{f} \tilde{m} (m^{c})^*} \right]
\nonumber \\
\Gamma_A^< &=& 
\Gamma_{m \tilde{m} \tilde{m}^*}+\Gamma_{\tilde{m} m^c \tilde{m}^*}
+\Gamma_{\tilde{m} \tilde{m} (m^{c})^*} \nonumber \\
&& + 2 \left[ \Gamma_{f \tilde{m} (\tilde{f}+1)} + \Gamma_{ \tilde{f} m^c (\tilde{f}+1)}  
+ \Gamma_{ \tilde{f} \tilde{m} f^c} \right]
\;.
\eea
Note that the first contribution, $\Gamma_{f \tilde{f} (\tilde{f}+1)} $, to $\Gamma_N^<$  explicitly contains the quantity $f=f^c + \tilde{f}$ (unlike the remaining contributions which are expressed in terms of $f^c$ or $\tilde{f}$).
The corresponding `$>$' rates are obtained by replacing $\tilde{f}$ by $(\tilde{f}+1)$ {\em and} vice versa (and correspondingly $f \leftrightarrow (f+1)$).
The collisional processes introduced above are defined by
\bea
\Gamma_{fff} &=& 8 V_{12'3'4'} \tilde{V}_{1''2''3''4''} f_{3'1''} f_{4'2''} f_{4''2'} |1 \rangle \langle 3''|
\nonumber \\
\Gamma_{fmf} &=& 8 V_{12'3'4'} \tilde{V}_{1''2''3''4''} f_{3'1''} m_{4'3''} f_{4''2'} |1 \rangle |2'' \rangle
\nonumber \\
\Gamma_{fmm^*} &=& 8 V_{12'3'4'} \tilde{V}_{1''2''3''4''} f_{3'1''} m_{4'3''} m^*_{2''2'} |1 \rangle \langle 4''|
\nonumber \\
\Gamma_{mmm^*} &=& 8 V_{12'3'4'} \tilde{V}_{1''2''3''4''} m_{3'4''} m_{4'3''} m^*_{2''2'} |1 \rangle | 1'' \rangle
\;,
\eea
where we have introduced the notation
\[
\tilde{V}_{1''2''3''4''} = V_{1''2''3''4''} \left[ \pi \delta(\Delta \varepsilon'') + i {\cal P} \left( \frac{1}{\Delta \varepsilon''}\right)
\right]
\;,
\]
where
$\Delta \varepsilon'' = \varepsilon_1'' + \varepsilon_2'' - \varepsilon_3'' - \varepsilon_4''$ denotes the difference in the single-particle energies between incoming and outgoing states.

To gain some insight into the above lengthy expressions, we now highlight those contributions to the collisional integrals which involve only `normal averages' $\tilde{f}$ or $f^c$ (i.e.\ we ignore anomalous averages). We thus find\footnote{For simplicity, we ignore here the second and third contributions to $\Gamma_N^<$.}
\bea
\frac{d \phi}{dt} = \cdots + \left( \Gamma_{\tilde{f} \tilde{f} ( \tilde{f}+1)} - \Gamma_{(\tilde{f}+1)(\tilde{f}+1) \tilde{f}} \right) \phi 
\eea
\bea
\frac{d \tilde{f}}{dt} = \cdots + \Gamma_{f \tilde{f} ( \tilde{f}+1)} (\tilde{f}+1)- \Gamma_{(f+1)(\tilde{f}+1) \tilde{f}} \tilde{f} 
\;.
\label{df_dt}
\eea
These equations contain the multi-mode generalisations of the condensate and non-condensate evolution given within our `toy model' of Sec.\ \ref{Prouk_QK}. 
In particular: 

\noindent (i) the expression for the condensate mean field $\phi$ corresponds to Eq.\ (\ref{cond_growth}).

\noindent (ii) Upon noting that $f=f^c + \tilde{f}$, Eq.\ (\ref{df_dt}) can be re-expressed as
\bea
\frac{d \tilde{f}}{dt} = \cdots &+& \Gamma_{\tilde{f} \tilde{f} ( \tilde{f}+1)} (\tilde{f}+1)- \Gamma_{(\tilde{f}+1)(\tilde{f}+1) \tilde{f}} \tilde{f} \nonumber \\
&+& \Gamma_{f^c \tilde{f} ( \tilde{f}+1)} (\tilde{f}+1)- \Gamma_{f^c(\tilde{f}+1) \tilde{f}} \tilde{f}
\;.
\eea
These two contributions correspond respectively to
Eqs.\ (\ref{thermal_exchange}) and (\ref{thermal_growth}), i.e.\ the collisional integrals of the
Quantum Boltzmann Equation describing both population transfer collisions between condensate and thermal atoms, and collisions between non-condensate atoms which lead to thermal (incoherent) population redistribution.

\section{Method of Non-commutative Cumulants}
\label{Cumulants}

The full system evolution can be derived
from the fully {\em non-local} Hamiltonian of Eq.\ (\ref{H}) via the study of suitably truncated equations of motion for averages of the field operator $\fopart$, and its products:
\[
i \hbar \frac{\partial}{\partial t} \langle \fopa(\bldr) \rangle (t)= 
\langle \left[ \hat{\Psi}(\bldr), \hat{H} \right] \rangle (t)
\]
\[
\cdots \cdots \cdots \cdots \cdots \cdots \cdots \cdots \cdots \cdots \cdots \cdots \cdots
\]
\[
i \hbar \frac{\partial}{\partial t} \langle \hat{\Psi}^{\dag}(\bldr) \cdots \hat{\Psi}(\bldr) \rangle (t) = 
\langle \left[  \hat{\Psi}^{\dag}(\bldr) \cdots  \hat{\Psi}(\bldr), \hat{H} \right] \rangle (t)
\]
A possible consistent truncation scheme was discussed by one of us (NPP) in \cite{Proukakis_NIST}, whereby all averages of products of up to {\em three} fluctuation operators were maintained, with averages of products of more than three operators decorrelated into products of averages of up to cubic terms. Such equations are to be solved self-consistently, with the requirement that the {\em full} interatomic potential is maintained.
(It would be incorrect to use a pseudopotential here since it is precisely the role of some of these correlations to upgrade the exact interatomic potential to an effective one).
This Tutorial has focused on the Markovian limit of such equations, where
collisional processes are assumed to occur much faster than the typical evolution timescales of mean fields, thus rendering the state of the system independent of any initial correlations that may have arisen at some (distant) time in the past, i.e.\ there are no `memory' effects. While this may be true for a large range of experiments with ultracold atoms, this condition is clearly violated in the experimentally-relevant case of attractive condensates \cite{BEC_Attractive_Exp_1,BEC_Attractive_Exp_2} or in the vicinity of Feshbach resonances \cite{BEC_Feshbach}, where a virtual bound state is supported close to the dissociation threshold of a two-body potential, both of which have been realized in recent experiments.
(Note however that a discussion on non-Markovian effects was given in the context of the formalism of Sec.\ \ref{Beyond_HFB} in \cite{JILA_NonMarkovian}).
The method of non-commuting cumulants \cite{Fricke,Thorsten_Method} provides a mathematically well-defined procedure for incorporating such effects within a finite temperature scheme based on a suitable truncation of the hierarchy of coupled equations of motion.

Mathematically, the cumulants can be defined as functional derivatives of a generating functional via \cite{Weinberg}
\bea
&& \langle \fopc(\bldr_n) \cdots \fopa(\bldr_1) \rangle_c \nonumber \\
&& = \frac{ \delta}{\delta J(\bldr_n)} \cdots 
\frac{ \delta}{\delta J^* (\bldr_1)}
{\rm ln} \langle e^{\int d \bldr J^*(\bldr) \fopart + J(\bldr) \fopcrt} \rangle \large|_{J=J^*=0}\;,
\eea
in the limit when the externally-added symmetry breaking field $J(\bldr)$ goes to zero.
The cumulants may be obtained recursively (for general operators $\hat{O}_1$, $\hat{O}_2$, $\hat{O}_3$, $\cdots$) via 
\bea
\langle \hat{O}_1 \rangle &=& \langle O_1 \rangle_c
\nonumber \\
\langle \hat{O}_1 \hat{O}_2 \rangle &=& \langle O_1 O_2 \rangle_c + 
\langle O_1 \rangle_c \langle O_2 \rangle_c
\nonumber \\
\langle \hat{O}_1 \hat{O}_2 \hat{O}_3 \rangle &=& \langle O_1 O_2 O_3 \rangle_c  
+ \langle O_1 \rangle_c \langle O_2 \rangle_c \langle O_3 \rangle_c
\nonumber \\
&+&
\langle O_1 \rangle_c \langle O_2 O_3 \rangle_c 
+ \langle O_2 \rangle_c \langle O_1 O_3 \rangle_c  
+ \langle O_3 \rangle_c \langle O_1 O_2 \rangle_c 
\eea
Cumulants of order higher than two provide a measure of how far the system is from the ideal Bose gas in thermal equilibrium (where such higher order correlations vanish).
An approximate method to truncate the hierarchy of coupled equations of motion for the non-cummutative cumulants was proposed by Fricke \cite{Fricke}:
Expanding the field operators in terms of single-mode operators one can decompose products of multiple operators into sums of pairwise contractions by means of Wick's theorem. Writing down the time-dependent equations for the correlation functions up to a given order of cumulants yields a closed system of coupled equations. 
The simplest possible scenario would be to include
only first order cumulants ($n=1$), thus obtaining the following evolution for the condensate mean field defined via $\Psi_c(\bldr,t)=\langle \fopa(\bldr) \rangle_c(t)$:
\bea
i \hbar \frac{ \partial}{\partial t}  \Psi_c(\bldr,t)  
= \left[ \hat{h}_0 + \int d \bldr' V(\bldr-\bldr') |\Psi_c(\bldr',t)|^2 \Psi_c(\bldr,t) \right] 
\;.
\eea
This equation should {\em not} be misinterpreted as the GPE of Eq.\ (\ref{GP-T0}), as it is explicitly expressed here in terms of the {\em actual} interatomic potential $V(\bldr-\bldr')$.
The required upgrade to an effective interaction (by the inclusion of multiple scattering) can be performed by the following self-consistent scheme: For a cumulant of order $n$, one must include the `free dynamics' of normal-ordered cumulants of orders $(n+1)$ and $(n+2)$, i.e.\ their evolution for which cumulants of order $(n+3)$ and $(n+4)$ are {\em ignored} within their respective equations of motion.
We follow K\"{o}hler and Burnett in explaining this in more detail \cite{Thorsten_Method}.
Assuming we are mainly interested in the condensate dynamics (i.e.\ $n=1$), consistent implementation of this scheme requires a systematic consideration of the cumulants of up to order $(n+2)=3$.
The {\em exact} dynamics of the condensate becomes \cite{Thorsten_Method}
\bea
&& i \hbar \frac{\partial}{\partial t} \Psi_c(\bldr,t)= \hat{h}_0(\bldr) \Psi_c(\bldr,t) \nonumber \\
&& + \int d\bldr' V(\bldr-\bldr') \Psi_c^*(\bldr',t) \left[ \Psi_c(\bldr,t) \Psi_c(\bldr',t) + \Phi_c(\bldr,\bldr',t) \right] \nonumber \\
&& +  \int d\bldr' V(\bldr-\bldr') 
\left[ \Psi_c(\bldr',t) \Gamma_c(\bldr,\bldr',t) + \Psi_c(\bldr,t) \Gamma_c(\bldr',\bldr',t) \right]
\nonumber \\
&& + \int d\bldr' V(\bldr-\bldr')  \langle \fopc(\bldr') \fopa(\bldr') \fopa(\bldr) \rangle_c(t) 
\;,
\eea 
where we have defined the second order cumulants
$ \Phi_c(\bldr_1,\bldr_2,t) = \langle \fopa(\bldr_2) \fopa(\bldr_1) \rangle_c(t)$ (pair function), and
$\Gamma_c(\bldr_1,\bldr_2,t) = \langle \fopc(\bldr_2) \fopa(\bldr_1) \rangle_c(t)$
(non-condensate one-body density matrix) (at the level $n+1=2$). 
Note that the third order cumulant 
$\langle \fopc(\bldr') \fopa(\bldr') \fopa(\bldr) \rangle_c(t)$ 
also appears in this expression, consistent with Eq.\ (\ref{zn_triplet}).

To proceed further, one should formulate equations of motion for these cumulants to the desired level of truncation. We find \cite{Thorsten_Method}
\bea
&& i \hbar \frac{\partial }{\partial t} \Phi_c(\bldr,\bldr',t) = 
\left[ \hat{h}_0(\bldr) + \hat{h}_0(\bldr') +V(\bldr-\bldr') \right] \Phi_c(\bldr,\bldr',t)
\nonumber \\
&& \hspace{2.5cm} + V(\bldr-\bldr') \Psi_c(\bldr,t) \Psi_c(\bldr',t) \;, \nonumber \\
&& i \hbar \frac{\partial }{\partial t} \Gamma_c(\bldr,\bldr',t)
= \left[ \hat{h}_0(\bldr) - \hat{h}_0(\bldr') \right] \Gamma_c(\bldr,\bldr',t) \;, \nonumber \\
&& i \hbar \frac{\partial}{\partial t} \langle \fopc(\bldr'') \fopa(\bldr') \fopa(\bldr) \rangle_c(t)
\nonumber \\
&&= \left[ \hat{h}_0(\bldr) + \hat{h}_0(\bldr') +V(\bldr-\bldr') -\hat{h}_0(\bldr'') \right]
\langle \fopc(\bldr'') \fopa(\bldr') \fopa(\bldr) \rangle_c(t)
\nonumber \\
&& + V(\bldr-\bldr') \left[ \Psi_c(\bldr',t) \Gamma(\bldr,\bldr'',t) + \Psi_c(\bldr,t) \Gamma(\bldr',\bldr'',t) \right]
\;.
\eea
(In obtaining the corresponding equation of motion for 
$\langle \fopc(\bldr'') \fopa(\bldr') \fopa(\bldr) \rangle_c(t)$ we have ignored products of normal-ordered cumulants containing $n+4=5$ Bose field operators.) 
Solving these equations formally by means of two-body Green's functions (in the basis of trap eigenstates) generates an exact closed {\em non-Markovian} nonlinear Schr\"{o}dinger equation for the condensate wavefunction, in which the interaction term has been {\em formally renormalised} to the two-body T-matrix - see Ref.\ \cite{Thorsten_Method} for detailed expressions and further details. Note that the usual GPE of Eq.\ (\ref{GP-T0}) arises as a limiting case of this equation upon imposing the Markov approximation and using the contact potential approximation for the two-body T-matrix.
The above procedure generalises straightforwardly for larger $n$: for example,
for $n=2$ one recovers a generalised non-local non-Markovian form of the HFB equations. 

The method of non-commutative cumulants has been successfully applied to diverse problems including the scattering of two condensates \cite{Thorsten_Method} and the atomic-molecular BEC problem in the vicinity of Feshbach resonances \cite{BEC_Feshbach} - for a recent review see \cite{Thorsten_Review}.
An important advantage of this approach is that it can incorporate three-body interactions by means of non-local pairwise potentials, thus yielding generalised three-body T-matrices satisfying Fadeev equations \cite{Thorsten_3_Body}.

\section{The Gardiner-Zoller Kinetic Theory}
\label{Appendix_Gardiner}

\subsection{The Condensate Band Master Equation}
\label{GZ_ME}

The full master equation for the evolution of the {\em condensate band} density operator developed by Gardiner and Zoller is given by \cite{QK_III,QK_V,SGPE_I,SGPE_II,SGPE_III}
\be
\dot{\rho}_c = \dot{\rho}_c^{\rm Ham}+\dot{\rho}_c^{\rm Growth}+\dot{\rho}_c^{\rm Scatt}\;.
\ee
The first contribution arises entirely from internal condensate band dynamics, via $\dot{\rho}_c^{\rm Ham} = -(i/\hbar) [ H_c, \rho_c ]$, where $\hat{H}_c$ is the condensate band hamiltonian which also includes the effect of the average non-condensate density on the condensate. The remaining two terms arise from the interaction between the condensate and the non-condensate band, and are obtained by consideration of the interaction hamiltonian $\hat{H}_{\rm int}$. In presenting the explicit form of these contributions, we use for simplicity the notation $\hat{\Psi} = \hat{\phi}+\hat{\delta}$, although it should be understood that these operators are actually the corresponding number-conserving operators discussed in Sec.\ \ref{CW_Gardiner}, which have been already implicitly projected onto orthogonal subspaces. 
To write down the full master equation, we additionally define the centre of mass and relative coordinates $\bldr_0 = (\bldr+\bldr')/2$ and $\bar{\bldr}=(\bldr - \bldr')$, where $\bldr$ and $\bldr'$ the coordinates of the colliding atoms.

Like in our beyond-HFB mean field discussion (Sec.\ \ref{Beyond_HFB}), the {\em growth terms} corresponding to particle transfer {\em between} the two bands arise from the contributions to the system hamiltonian which involve {\em three} non-condensate operators $\hat{\delta}^\dag \hat{\delta} \hat{\delta}$ (i.e.\ from the corresponding number-conserving generalisation of $\hat{H}_3$), which yield a contribution of the form 
\bea
\dot{\rho}_c^{\rm Growth} = \int d\bldr_0 \int d\bar{\bldr} \left\{ \left( {\cal A}_1-{\cal A}_2 \right) + \left( {\cal B}_1-{\cal B}_2 \right) \right\} \;.
\eea
Each of the four contributions appearing in the integrand are given in terms of a suitable commutator of a product involving the condensate band density matrix $\rho_c$, a creation (annihilation) operator $\hat{\delta}^{(\dag)}$, and a matrix 
$ G^{(+)}$ or $ G^{(-)}$ corresponding to scattering rates {\em into} and {\em out of} the condensate band.
In particular
\bea
{\cal A}_1 = \left[ G^{(-)}(L_c) \, \hat{\phi} \left (\bldr_0-\frac{\bar{\bldr}}{2} \right) \rho_c \, \, , \, \, \hat{\phi}^{\dag} \left (\bldr_0+\frac{\bar{\bldr}}{2} \right)  \right] \nonumber \\
{\cal A}_2 = \left[ \rho_c \, G^{(-)}(-L_c) \, \hat{\phi}^\dag \left (\bldr_0-\frac{\bar{\bldr}}{2} \right) \, \, , \, \, \hat{\phi} \left (\bldr_0+\frac{\bar{\bldr}}{2} \right)  \right] \nonumber \\
{\cal B}_1 = \left[ G^{(+)}(-L_c) \, \hat{\phi}^\dag \left (\bldr_0-\frac{\bar{\bldr}}{2} \right) \, \rho_c \, \, , \, \, \hat{\phi} \left (\bldr_0+\frac{\bar{\bldr}}{2} \right)  \right] \nonumber \\
{\cal B}_2 = \left[ \rho_c \, G^{(+)}(L_c) \, \hat{\phi} \left (\bldr_0-\frac{\bar{\bldr}}{2} \right) \, \, , \, \, \hat{\phi}^\dag \left (\bldr_0+\frac{\bar{\bldr}}{2} \right)  \right] 
\;.
\eea
Here $ G^{(+)}$ and $ G^{(-)}$ depend both on $\bldr_0$ and $\bar{\bldr}$ (suppressed in  above expressions)  and on energy, with $L_c$ defined by $L_c \hat{\phi}(\bldr)=[\hat{\phi}(\bldr) \, , \, H_c]$. 
They are given by integrals of the form
\bea
G^{(\pm)}(\bldr_0, \bar{\bldr}, \omega) &=& \frac{g^2}{(2 \pi)^8 \hbar^2} 
\int d {\bf k_2} \int d{\bf k_3} \int d {\bf k_4} \delta (\omega_2 + \omega_3 - \omega_4 - \omega ) \nonumber \\
&\times& e^{-i ({\bf k_2}+{\bf k_3}+{\bf k_4}) \cdot \bar{\bldr} } {\cal F}^{\pm}(\bldr_0, {\bf k_i})
\;.
\label{G_pm}
\eea
Here
$ {\cal F}^{+}(\bldr_0, {\bf k_i}) = F_1 F_2 (F_3 +1)$ and 
$ {\cal F}^{-}(\bldr_0, {\bf k_i}) = (F_1+1)( F_2+1) F_3 $, and $F_i$ denotes the one-particle Wigner function introduced by
\be
\langle \hat{\delta}^{\dag} \left (\bldr_0+\frac{\bar{\bldr}}{2} \right)
\hat{\delta} \left (\bldr_0-\frac{\bar{\bldr}}{2},t \right) \rangle
= \frac{1}{(2 \pi)^3} \int d{\bf k} F(\bldr_0,{\bf k})
e^{-i ({\bf k} \cdot \bldr_0+ \omega t) }
,
\ee 
where $\omega = \hbar k^2 / 2m + V_{\rm ext}(\bldr_0 )$ and the range of integration has been restricted to the non-condensate band.
If the non-condensate band is in thermal equilibrium, the `in' and `out' factors are related (as discussed in Sec.\ \ref{Damp_GP}) by 
$G^{(+)}(\omega)=e^{- \beta (\hbar \omega - \mu)} G^{(-)}(\omega)$.
Note the strong similarity of the two terms of Eq.\ (\ref{G_pm}) to the contributions (corresponding to `in' and `out' rates) contained in $iR(\bldr,t)$ of Eq.\ (\ref{R}) and $\Sigma^K(\bldr)$ of Eq.\ (\ref{Sigma_K_Full}).

The {\em scattering terms} describe collisions of atoms in different bands, which lead to population redistribution within each {\em band}, but cause no net transfer between bands.
They arise from contributions to the system hamiltonian which contain one non-condensate creation ($\hat{\delta}^{\dag}$) and one non-condensate annihilation ($\hat{\delta}$) operator (and also a product $\hat{\phi}^\dag \hat{\phi}$ of condensate creation and annihilation operators) - i.e.\ effectively from the `resonant' part of the non-condensate form of the number-conserving $\hat{H}_2$ contribution. Their evolution takes the form
\bea
\dot{\rho}_c^{\rm Scatt} = \int d\bldr_0 \int d\bar{\bldr} \left\{ {\cal C}_1 + {\cal C}_2 \right\} \;,
\eea
where
\bea
{\cal C}_1 = \left[ \hat{\phi}^\dag \left (\bldr_0+\frac{\bar{\bldr}}{2} \right) 
\hat{\phi} \left (\bldr_0+\frac{\bar{\bldr}}{2} \right) \, \, , \, \, \rho_c \, M(L_c) \, \hat{\phi}^{\dag} \left (\bldr_0-\frac{\bar{\bldr}}{2} \right)  
\hat{\phi} \left( \bldr_0-\frac{\bar{\bldr}}{2} \right)  \right] 
\nonumber \\
{\cal C}_2 = \left[ M(-L_c) \, \hat{\phi}^\dag \left (\bldr_0-\frac{\bar{\bldr}}{2} \right) 
\hat{\phi} \left (\bldr_0-\frac{\bar{\bldr}}{2} \right) \rho_c \, \, , \, \, \hat{\phi}^{\dag} \left (\bldr_0+\frac{\bar{\bldr}}{2} \right)  
\hat{\phi} \left (\bldr_0+\frac{\bar{\bldr}}{2} \right)  
\right] \nonumber,
\eea
with
\bea
M(\bldr_0, \bar{\bldr}, \pm \omega) = \frac{2g^2}{(2 \pi)^5 \hbar^2} 
&& \int d {\bf k_2} \int d{\bf k_3} 
e^{i ({\bf k_2}-{\bf k_3}) \cdot \bar{\bldr} } \nonumber \\
&& \times \delta (\omega_2 - \omega_3 \mp \omega ) F_1 (F_2 +1)
\;.
\label{M_pm}
\eea
Again, at thermal equilibrium 
$M(\bldr_0, \bar{\bldr}, \omega) = e^{-\beta \hbar \omega} M(\bldr_0, \bar{\bldr}, -\omega)$.

\subsection{The Gardiner-Davis Stochastic Gross-Pitaevskii Equation}
\label{GZ_SGPE}

Here we give explicit expression for the terms appearing in the local form of the Gardiner-Davis stochastic GPE of Eq.\ (\ref{SGPE_Gardiner}) \cite{SGPE_II}. The {\em growth term} is given by
\cite{SGPE_I,SGPE_II,SGPE_III,SGPE_IV}
\be
K_G (\bldr,t) = \beta \bar{G}(\bldr) \left( \mu - \hbar \bar{L}_c \right) \alpha(\bldr,t)
\ee
where
$\bar{G}(\bldr) = \int d \bar{\bldr} G^{(+)} (\bldr, \bar{\bldr}, \omega=0)$
with $G^{(+)}$ defined in Eq.\ (\ref{G_pm}) , and $\bar{L}_c$ is the condensate eigenvalue obtained from the projected finite temperature GPE via
$\bar{L}_c \alpha(\bldr) = {\cal P}_c \left\{ \hat{h}_0 + 2 g \bar{n}_{NC}(\bldr) + g | \alpha(\bldr)|^2 \right\} \alpha(\bldr)$.
Here  ${\cal P}_c$ is a projector onto the subspace orthogonal to the trap basis, defined via ${\cal P}_c(\bldr,\bldr')=1-\sum_n Y_n^*(\bldr) Y_n(\bldr')$, where $Y_n(\bldr)$ are a complete set of trap eigenfunctions and the summation is restricted to modes of the non-condensate band. In the above expression, we have also
defined the average non-condensate band density via
$ \bar{n}_{NC}(\bldr) = {\rm Tr}_{NC} \left\{ \hat{\delta}^\dag(\bldr) \hat{\delta}(\bldr) \rho_{NC} \right\}$.
The corresponding {\em scattering term}, $K_M(\bldr,t)$ is defined by \cite{SGPE_II}
\be
K_M(\bldr,t) = - \frac{\beta}{2} \bar{M}(\bldr) \alpha(\bldr)
\left\{ \alpha^*(\bldr,t) \bar{L}_c \alpha(\bldr,t)
- \alpha(\bldr,t) \left[ \bar{L}_c \alpha(\bldr,t) \right]^* \right\},
\ee
where the parameter $\bar{M}$ is an {\em approximate} local form of the quantity $M(\bldr_0, \bar{\bldr}, \pm \omega)$ of Eq.\ (\ref{M_pm}).
%
The correlations of the noise appearing in Eq.\ (\ref{SGPE_Gardiner}) are
\bea
dW_G^*(\bldr,t) dW_G(\bldr',t) = 2 \bar{G}(\bldr) \delta(\bldr-\bldr') dt \nonumber \\
dW_G(\bldr,t) dW_G(\bldr',t) = dW_G^*(\bldr,t) dW_G^*(\bldr',t) = 0 \nonumber \\
dW_M(\bldr,t) dW_M(\bldr',t) = 2 \bar{M}(\bldr) \delta(\bldr-\bldr') dt \;.
\label{Noise_Gardiner}
\eea

We now demonstrate the equivalence between the stochastic equations of Gardiner, Anglin, Fudge and Davis (Eq.\ (\ref{SGPE_Gardiner})) and that of Stoof (Eq.\ (\ref{SGPE_Eq})). 
For this purpose, we firstly ignore the scattering contributions (which are cumbersome to deal with) and recast Eq.\ (\ref{SGPE_Gardiner}) in the simpler form
\bea
d \alpha(\bldr,t) = &-& \frac{i}{\hbar} {\cal P}_c L_{GP} \alpha(\bldr,t) dt \nonumber \\
&-& \frac{i}{\hbar}  \left[-i \beta \hbar G(\bldr) \right] {\cal P}_c \left(L_{GP}-\mu \right)
\alpha(\bldr,t) dt + dW_G(\bldr,t).
\eea
Firstly, we eliminate the `trivial' time-dependence by re-expressing the equation in  terms of $\tilde{\alpha}(\bldr,t)=e^{i \mu t /\hbar} \alpha(\bldr,t)$ to obtain
\bea
d \tilde{\alpha}(\bldr,t) = &-& \frac{i}{\hbar} \left\{ \left[ 1 - i \beta \hbar G(\bldr) \right]
{\cal P}_c \left( L_{GP} - \mu \right) \right\} \tilde{\alpha}(\bldr,t) dt \nonumber \\
& + & \tilde{dW_G}(\bldr,t) 
\;,
\eea
where $L_{GP}$ is the eigenvalue of the ordinary Gross-Pitaevskii equation, defined by
$L_{GP} \alpha(\bldr) =  ( \hat{h}_0  + g | \alpha(\bldr)|^2 ) \alpha(\bldr)$
and
$\tilde{dW_G}(\bldr,t) = e^{i \mu t / \hbar} dW_G(\bldr,t)$.
Upon noting that $\Sigma_K(\bldr) = -4i G(\bldr)$ we see that this equation is identical to Eq.\ (\ref{SGPE_Eq}). It only remains to prove that the noise sources are the same; the noise contribution corresponding to the differential equation $i \hbar \partial \tilde{\alpha}(\bldr,t)/\partial t$ becomes $i \hbar \partial \tilde{W_G}(\bldr,t)/\partial t$; using the expressions of Eq.\ (\ref{Noise_Gardiner}), its correlations thus become 
\bea
\langle \left( - i \hbar \tilde{dW_G}^*(\bldr,t) \right) && \left( i \hbar \tilde{dW_G}(\bldr',t) \right) \rangle 
= \hbar^2 \langle dW_G^*(\bldr,t) dW_G(\bldr',t) \rangle \nonumber \\
&&= 2 \hbar^2 G(\bldr) \delta_C(\bldr,\bldr') 
\rightarrow \frac{i}{2} \hbar^2 \Sigma^K(\bldr) \delta(\bldr-\bldr')
\;,
\eea
demonstrating the direct equivalence between the stochastic equations of Stoof and Gardiner and co-workers in the limit where the scattering processes can be ignored.

\section*{References}

\end{document}